\documentclass[reqno, eucal]{amsart}

\usepackage{epic,eepic}
\usepackage[mathscr]{eucal} 
\usepackage{psfrag}     

\numberwithin{equation}{section}

\usepackage{amsmath}
\usepackage{amsthm}

\usepackage[dvips]{graphicx,color}  

\usepackage{amssymb}
\usepackage{epic,eepic}
\usepackage{amscd}
\usepackage{longtable}
\usepackage{array}

\newtheorem{theorem}{Theorem}[section]
\newtheorem{lemma}[theorem]{Lemma}
\newtheorem{conjecture}[theorem]{Conjecture}
\newtheorem{proposition}[theorem]{Proposition}

\theoremstyle{definition}

\newtheorem{example}[theorem]{Example}
\newtheorem{remark}[theorem]{Remark}
\newtheorem{assumption}[theorem]{Assumption} 

\newcommand{\Z}{{\mathbb Z}}
\newcommand{\R}{{\mathbb R}}
\newcommand{\C}{{\mathbb C}}

\newcommand{\hf}{\textstyle\frac{1}{2}}
\newcommand{\tai}{\textstyle\frac{1}{t_a}}
\newcommand{\re}{\Re{\rm e}}
\newcommand{\im}{\Im{\rm m}}

\begin{document}
\begin{center}
\begin{LARGE}

{\bf T-systems and Y-systems in integrable systems}

\end{LARGE}

\vspace{1cm}
\begin{large}
Atsuo Kuniba$^a$, Tomoki Nakanishi$^b$ and Junji Suzuki$^c$
\end{large}

\vspace{0.6cm}

\begin{center}
{\it Dedicated to the memory of Professor Morikazu Toda}
\end{center}

\vspace{0.6cm}

${}^a$ Institute of Physics,
University of Tokyo, Komaba, 
Tokyo, 153-8902, Japan
\vspace{0.2cm}

${}^b$ Graduate School of Mathematics, Nagoya University,
Nagoya, 464-8604, Japan
\vspace{0.2cm}

${}^c$ Department of Physics, Faculty of Science,
Shizuoka University,
Ohya,~836,~Japan

\end{center}
\vspace{3cm}
\begin{center}{\bf Abstract}
\end{center}
\vspace{0.4cm}
T and Y-systems are ubiquitous structures in classical and quantum 
integrable systems. 
They are difference equations having a variety of aspects 
related to commuting transfer matrices in solvable lattice models, 
$q$-characters of Kirillov-Reshetikhin modules of quantum affine algebras, 
cluster algebras with coefficients,
periodicity conjectures of Zamolodchikov and others, 
dilogarithm identities in conformal field theory,
difference analog of $L$-operators in KP hierarchy, 
Stokes phenomena in 1D Schr\"odinger problem,
AdS/CFT correspondence,
Toda field equations on discrete spacetime, 
Laplace sequence in discrete geometry, 
Fermionic character formulas and 
combinatorial completeness of Bethe ansatz,
Q-system and ideal gas with exclusion statistics,
analytic and thermodynamic Bethe ans\"atze, 
quantum transfer matrix method and so forth.
This review article is a collection of short reviews 
on these topics which can be read more or less independently.


\newpage 
\tableofcontents

\section{Introduction}\label{s:intro}

\subsection{T and Y-systems}
The T-system is a difference equation among 
commuting variables $T^{(a)}_m(u)$, most typically appearing as
($m \in \Z_{\ge 0}$)
\begin{equation*}
T^{(a)}_m(u-1)T^{(a)}_m(u+1)
=T^{(a)}_{m-1}(u)T^{(a)}_{m+1}(u)
+T^{(a-1)}_{m}(u)T^{(a+1)}_{m}(u).
\end{equation*}
Originally it was found as
a functional relation in 2D solvable lattice models 
in statistical mechanics \cite{KNS2}.
In this context, $T^{(a)}_m(u)$ is a 
commuting row transfer matrix in the sense of Baxter \cite{Ba3} 
labeled with $(a,m)$ and 
having the spectral parameter $u$\footnote{By T we meant 
Transfer matrices,  
but it can either be thought as Toda or Tau. }.

The Y-system is another difference equation,
typically like ($m \in \Z_{\ge 1}$)
\begin{equation*}
Y^{(a)}_m(u-1)Y^{(a)}_m(u+1)
=\frac{(1+Y^{(a-1)}_{m}(u))(1+Y^{(a+1)}_{m}(u))}
{(1+Y^{(a)}_{m-1}(u)^{-1})(1+Y^{(a)}_{m+1}(u)^{-1})}.
\end{equation*}
It was extracted as a universal functional relation in 
thermodynamic Bethe ansatz (TBA) for
solvable lattice models as well as 
$(1\!+\!1)$D integrable quantum field theory models \cite{Z1,KN1,RTV}.
In this context, $Y^{(a)}_m(u)$ stands for the Boltzmann factor 
of an excitation mode in the sense of Yang-Yang \cite{YY} 
labeled with $(a,m)$ and having the rapidity $u$.

As such, the both systems originate in 
Yang-Baxter quantum integrable systems
but are apparently concerned with the objects 
that are not related too directly.
The first curiosity is nevertheless that the formal substitution 
\begin{equation*}
Y^{(a)}_m(u) 
= \frac{T^{(a-1)}_m(u)T^{(a+1)}_m(u)}
{T^{(a)}_{m-1}(u)T^{(a)}_{m+1}(u)}
\end{equation*}
provides a solution to the Y-system in terms of the T-system.
Moreover, such a canonical pair of companion systems
can be formulated uniformly 
for all the classical simple Lie algebras 
$\mathfrak{g}$ \cite{KNS2}\footnote{Actually to be understood as 
Yangian $Y(\mathfrak{g})$ or 
untwisted quantum affine algebra $U_q(\hat{\mathfrak{g}})$.
Twisted case is also known.
See Remark \ref{re:u}. }.
Now we can give a deferred explanation of the superscript $a$;
it runs over the vertices of the Dynkin diagram of $\mathfrak{g}$. 
The above formulas are just the examples from type $A$\footnote{The 
T-system for type $A$ formally coincides with what is known as 
the Hirota-Miwa equation in soliton theory,
which was an unexpected link also to classical integrable systems.}, 
where the case $\mathfrak{g}=A_1$ goes back to \cite{KP2}.

In the relevant developments across the centuries,
the T and Y-systems have turned out to be ubiquitous structures
with a wealth of applications.
For instance, they emerge in
$q$-characters for Kirillov-Reshetikhin modules of quantum affine algebras,
exchange relations in cluster algebras with coefficients,
periodicity conjectures of Zamolodchikov and others,
dilogarithm identities in conformal field theory (CFT) 
and their functional generalizations,
dressed vacuum forms in analytic Bethe ansatz,
Stokes phenomena in ordinary differential equations,
anomalous scaling dimensions of 
${\mathcal N}=4$ super Yang-Mills operators,  
area of minimal surface in AdS, 
Laplace sequence of quadrilateral lattice in discrete geometry,
tau functions in lattice Toda field equations,
Fermionic formulas for branching coefficients and weight multiplicities
for Lie algebra characters, 
combinatorial completeness of string hypothesis in Bethe ansatz,
Q-system and grand partition function of ideal gas with exclusion statistics, 
quantum transfer matrix approach to finite temperature problems and so on.

This review is a collection of brief expositions of these topics 
where the T and Y-systems have played key roles. 
It consists of sections of moderate length 
which are not too mutually dependent.
A more detailed account of the contents can be found 
in Section \ref{ss:cbg}.

As an overview,  
T-systems are fundamental structures reflecting symmetries 
and algebraic aspects of the problems rather directly.
They can also accommodate 
various gauge/normalization freedom of concrete models.
On the other hand, Y-systems are more universal being 
more or less free from such degrees of freedom.
They are suitable for practical applications with appropriate analyticity input.  
In fact, the connection between the T and Y-systems mentioned previously
has opened a route to establish TBA type integral equations 
directly from transfer matrices without recourse to the TBA itself.
In this sense,  Y-systems are the format in which 
the symmetries encoded in the T-systems are most efficiently 
utilized as a practical implement.

In the light of ever growing perspectives, 
what sort of equations or structures are to be recognized 
as T or Y-systems is actually a matter of time-dependent option.
For instance from an algebraic point of view (leaving analytic aspects), 
T-systems have been generalized broadly 
to the quantum affinization of quantum Kac-Moody algebras 
by Hernandez \cite{Her3} (Section \ref{ss:qaq}).
Cluster algebra with coefficients 
by Fomin and Zelevinsky \cite{FZ4} 
offers a comprehensive scheme 
to generalize and control the T and Y-systems simultaneously 
by quivers (Section \ref{s:ca}).
Nonetheless, this paper is mostly devoted to 
the description of basic results 
concerning the aforementioned ``classic" T and Y-systems  
associated with $\mathfrak{g}$.
We therefore look forward to the next review to come,
hopefully someday by some author, 
bringing a delightful renewal.

\subsection{Contents and brief guide}\label{ss:cbg}

Here are abstracts of the subsequent sections.
They will be followed by another brief guide to the paper.

Section \ref{s:presen}.
The T and Y-systems for untwisted and twisted quantum affine algebras
are presented.
They have unrestricted and level $\ell$ restricted versions. 
Those for Yangian are formally the same with the unrestricted ones for 
the untwisted quantum affine algebra $U_q(\hat{\mathfrak{g}})$,
where $\mathfrak{g}$ denotes a finite dimensional simple Lie 
algebra throughout the paper.
We also include the $U_q(sl(r|s))$ case.
This section is meant to be the reference of these systems 
throughout the paper.
The first property, T-system provides a solution to Y-system, is stated.
Subsequent sections will mainly be concerned 
with the untwisted case $U_q(\hat{\mathfrak{g}})$\footnote{
Thus in most situations we will say simply T and Y-systems for $\mathfrak{g}$
instead of $U_q(\hat{\mathfrak{g}})$.}.

Section \ref{s:ctm}.
The T-system was originally discovered 
as functional relations among commuting transfer matrices for 
solvable lattice models in statistical mechanics.
We give an elementary exposition of such contexts 
for the both vertex and restricted solid-on-solid (RSOS) models 
along with their fusion procedure.
The two types of models are related to 
the unrestricted and restricted T-systems, respectively.

Section \ref{s:qg}.
We describe the background of the T-system 
in the representation theory of quantum affine algebras
such as classification of irreducible finite dimensional representations, 
Kirillov-Reshetikhin modules and $q$-characters.
The fundamental results are that $q$-characters of the Kirillov-Reshetikhin
modules satisfy the T-system (Theorem \ref{th:nh}) and 
the description of the Grothendieck ring 
$\mathrm{Rep}\, U_q(\hat{\mathfrak{g}})$ by the T-system
(Theorem \ref{th:28}). 
A broad extension of the T-system to the
quantum affinization of quantum Kac-Moody algebras is also mentioned.
The results of this section are not necessary elsewhere 
except the basics of $q$-characters 
which will be mentioned in tableau sum formulas (Section \ref{s:tab}), 
analytic Bethe ansatz (Section \ref{s:aba})
and Q-system (Section \ref{s:q}).

Section \ref{s:ca}.
The cluster algebra with coefficients is built upon 
cluster variables and coefficient tuples obeying certain 
exchange relations controlled by a quiver.
We demonstrate how such a setup encodes
the T and Y-systems simultaneously in an essential way.
It opens a fruitful link with the cluster category theory, which led to
a final proof of the dilogarithm identities in conformal field theory 
and the periodicity conjecture on the both systems 
for arbitrary level and $\mathfrak{g}$.

Section \ref{s:jt}.
Jacobi-Trudi type determinant formulas are listed  
for T-systems for non exceptional $\mathfrak{g}$.
The type $C_r$ and $D_r$ cases involve Pfaffians as well.

Section \ref{s:tab}.
Tableau sum formulas are presented for 
T-systems for non exceptional $\mathfrak{g}$ along the 
context of $q$-characters.

Section \ref{s:aba}.
We argue the relation between $q$-characters 
and eigenvalue formulas (dressed vacuum forms) of transfer matrices in 
solvable lattice models by analytic Bethe ansatz.
Combined with the results in Section \ref{s:tab}, 
it leads to solutions of T-systems in terms of the Baxter Q-functions. 
We mainly concern vertex models 
and include a brief argument on RSOS models.

Section \ref{s:tq}.
We introduce a difference analog of 
$L$-operators in soliton theory to construct 
solutions to the T-systems for $\mathfrak{g}=A_r$ and $C_r$ 
by Casoratians (difference analog of Wronskians).
The Baxter Q-functions are identified with a special class of Casoratians  
and generalized to a wider family of functions that admit
B\"acklund transformations.
Analogous difference $L$-operators are presented also for 
$B_r, D_r$ and $sl(r|s)$.

Section \ref{s:ODE}.
A restricted T-system for $A_1$ emerges 
in Stokes phenomena of 
1D Schr\"odinger equation with a specific potential.
Similar facts hold also for the T-system for $A_r$ and a class of 
$(r+1)$th order ordinary differential equation (ODE).
Wronskians for these equations evaluated at the 
origin play an analogous role
to the Casoratians in Section \ref{s:tq}
(Wronskian-Casoratian duality).
We describe these features that stay within 
an elementary algebraic part in the so-called 
ODE/IM (integrable models) correspondence.

Section \ref{s:ags}.
This section is most hep-th oriented.
We briefly digest applications of some specific T and Y-systems 
in the two topics from the AdS/CFT correspondence.
The first is from the gauge theory about 
the anomalous scaling dimensions (planar AdS/CFT spectrum)
of ${\mathcal N}=4$ super Yang-Mills operators. 
The second is the area of the minimal surface in AdS 
from the string theory, 
which is relevant to gluon planar scattering amplitudes.
The analysis in the latter topic involves the 
Stokes phenomena related to a
generalized sinh-Gordon equation, 
which may be viewed as a generalization of the
ODE/IM correspondence mentioned in Section \ref{s:ODE}.

Section \ref{s:cis}.
Continuous limits of the 
T-system for $\mathfrak{g}$ yield the difference-differential
or 2D differential equations known as the (lattice) Toda field equation.
Their Hamiltonian structure is presented for general $\mathfrak{g}$.
We also discuss an aspect from classical discrete geometry, where
the Y-system for $A_\infty$ arises as the Laplace sequence
of quadrilateral lattice, the discrete geometry analog of the conjugate net. 

Section \ref{s:q}.
T-system without spectral parameter is called Q-system\footnote{
This Q is unrelated with Baxter's Q-functions.
See Section \ref{ss:xm} for the origin of the name.}.
We systematically construct certain power series solutions to 
the (generalized) Q-system by multi-variable Lagrange inversion. 
As a corollary of this and results from Section \ref{s:qg}, 
the so-called Fermionic character formula for 
the Kirillov-Reshetikhin modules is fully established
for all $\mathfrak{g}$.
Physically, this problem is also connected to 
the grand partition function of ideal gas with exclusion statistics.
These results are reviewed in conjunction with 
the intimately related subject known as
combinatorial completeness of Bethe ansatz for 
$U_q(\hat{\mathfrak g})$ both at $q=1$ and $q=0$,
where the case $q=1$ goes back to Bethe \cite{Be},
the godfather of the subject, himself.

Section \ref{s:y}.
We explain how the Y-system for ${\mathfrak g}$ emerges from 
the TBA equation 
associated to $U_q(\hat{\mathfrak g})$ with $q$ being a root of unity
derived in Section \ref{s:tbarsos}.
Various relations among the TBA kernels are summarized.
The constant Y-system is introduced and related to the Q-system.
They are essential ingredients in the dilogarithm identity (Section \ref{t:ss:di}) 
and the TBA analysis of RSOS models (Section \ref{s:tbarsos}).
As a related issue, we briefly discuss the Q-system at root of unity 
including Conjecture \ref{conj:rq}.

Section \ref{s:tbarsos}.
The $U_q(\hat{\mathfrak g})$ Bethe equation with $q$ a root of unity 
is relevant to the critical RSOS models sketched in Section \ref{ss:rsos}.
We outline the TBA analysis to evaluate the
high temperature entropy by the level restricted Q-system 
(Section \ref{ss:rwq}--\ref{ss:qru})
and central charges 
by the dilogarithm identity (Section \ref{t:ss:di}).
The TBA equation obtained here uniformly for general ${\mathfrak g}$
is the origin of our Y-system as shown in 
Section \ref{ss:ydc} and \ref{ss:ytba}.

Section \ref{s:app}.
The finite size or finite temperature problems
in solvable lattice models are analyzed efficiently 
by the use of T and Y-systems without relying on TBA approach
and string hypothesis.
We illustrate various such methods 
along the simplest vertex and RSOS models based on 
$\mathfrak g=A_1$.
We also include a simple application of the periodicity 
of the level $0$ restricted T-system to the calculation 
of correlation lengths of vertex models in Section \ref{correlationlength}.

\smallskip
Let us close the introduction with yet another brief guide of the contents.  
As we already mentioned, 
Section \ref{s:presen} is the collection of the basic data; 
concrete forms of the T and Y-systems that will be considered in the review
and definitions/notations concerning the root system of $\mathfrak{g}$. 
With regard to the subsequent sections,  
it is too demanding to assume the familiarity of the 
contents in earlier sections.
So we have avoided such a style 
and tried to make each section into a more or less independently readable 
review on a specific topic around ten pages.
Most of them contain bibliographical notes at the end, 
which hopefully help the readers gain more perspectives
into the subjects and activities around.

There are nevertheless several sections that are intimately related 
or partly dependent of course.
Roughly, they may be grouped (non exclusively) under the following theme.

\smallskip\noindent
$\bullet$ Solvable lattice models and their analysis: 
Sections \ref{s:ctm}, \ref{s:aba}, \ref{s:tbarsos}, \ref{s:app}.

\noindent
$\bullet$ Kirillov-Reshetikhin modules and their $q$-characters:
Sections \ref{s:qg}, \ref{s:tab}, \ref{s:aba}, \ref{s:q}.

\noindent
$\bullet$ Variety of solutions to the T-system:
Sections \ref{s:jt}, \ref{s:tab}, \ref{s:aba}, \ref{s:tq}.

\noindent
$\bullet$ Stokes phenomena:
Sections \ref{s:ODE}, \ref{s:ags}.

\noindent
$\bullet$ Q-system and constant Y-system:
Sections \ref{s:q}, \ref{s:y}.

\noindent
$\bullet$ Y-system and TBA:
Sections \ref{s:ags}, \ref{s:y}, \ref{s:tbarsos}.

\section{T and  Y-systems for  quantum affine algebras and 
Yangians}\label{s:presen}

We present the T-system and Y-system associated with 
untwisted and twisted quantum affine algebras.
They have unrestricted and level restricted versions.
Those for Yangian are formally the same with the unrestricted ones for 
the untwisted quantum affine algebras.
We also include the case $U_q(sl(r|s))$.
This section is devoted to the presentation of these systems
with the basic data on root systems.
Thus we will only state their first property,
T-system provides a solution to Y-system, in Theorem \ref{th:ty1}.
leaving the exposition of 
variety of aspects in subsequent sections. 

\subsection{Untwisted case}\label{ss:utw}

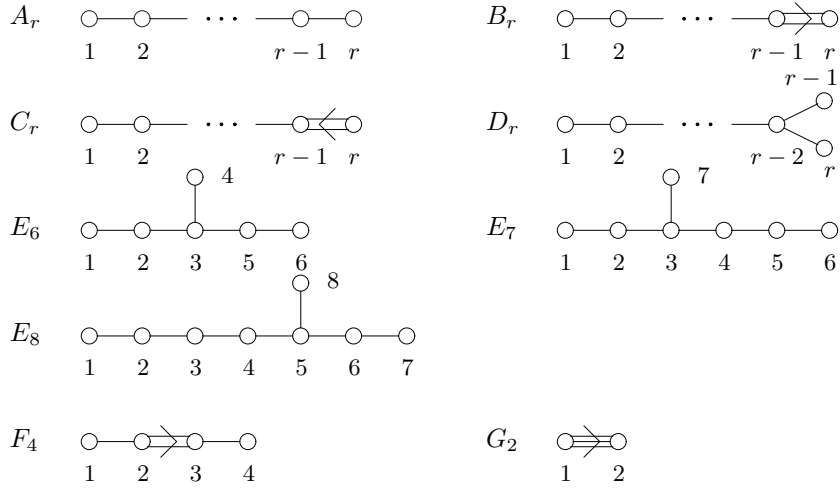
\begin{figure}[h]
\begin{picture}(283,185)(-10,-175)
%
\put(0,0){\circle{6}}
\put(20,0){\circle{6}}
\put(80,0){\circle{6}}
\put(100,0){\circle{6}}
\put(45,0){\circle*{1}}
\put(50,0){\circle*{1}}
\put(55,0){\circle*{1}}
\drawline(3,0)(17,0)
\drawline(23,0)(37,0)
\drawline(63,0)(77,0)
\drawline(83,0)(97,0)
\put(-30,-2){$A_r$}
\put(-2,-15){\small $1$}
\put(18,-15){\small $2$}
\put(70,-15){\small $ r-1$}
\put(98,-15){\small $r$}
%
\put(180,0){
\put(0,0){\circle{6}}
\put(20,0){\circle{6}}
\put(80,0){\circle{6}}
\put(100,0){\circle{6}}
\put(45,0){\circle*{1}}
\put(50,0){\circle*{1}}
\put(55,0){\circle*{1}}
\drawline(3,0)(17,0)
\drawline(23,0)(37,0)
\drawline(63,0)(77,0)
\drawline(82,-2)(98,-2)
\drawline(82,2)(98,2)
\drawline(87,6)(93,0)
\drawline(87,-6)(93,0)
\put(-30,-2){$B_r$}
\put(-2,-15){\small $1$}
\put(18,-15){\small $2$}
\put(70,-15){\small $ r-1$}
\put(98,-15){\small $r$}
}
%

\put(0,-40){
\put(0,0){\circle{6}}
\put(20,0){\circle{6}}
\put(80,0){\circle{6}}
\put(100,0){\circle{6}}
\put(45,0){\circle*{1}}
\put(50,0){\circle*{1}}
\put(55,0){\circle*{1}}
\drawline(3,0)(17,0)
\drawline(23,0)(37,0)
\drawline(63,0)(77,0)
\drawline(82,-2)(98,-2)
\drawline(82,2)(98,2)
\drawline(87,0)(93,-6)
\drawline(87,0)(93,6)
\put(-30,-2){$C_r$}
\put(-2,-15){\small $1$}
\put(18,-15){\small $2$}
\put(70,-15){\small $ r-1$}
\put(98,-15){\small $r$}
}

%
\put(180,-40){
\put(0,0){\circle{6}}
\put(20,0){\circle{6}}
\put(80,0){\circle{6}}
\put(98,9){\circle{6}}
\put(98,-9){\circle{6}}
\put(45,0){\circle*{1}}
\put(50,0){\circle*{1}}
\put(55,0){\circle*{1}}
\drawline(3,0)(17,0)
\drawline(23,0)(37,0)
\drawline(63,0)(77,0)
\drawline(82.8,1.5)(95.5,7.8)
\drawline(82.8,-1.5)(95.5,-7.8)
\put(-30,-2){$D_r$}
\put(-2,-15){\small $1$}
\put(18,-15){\small $2$}
\put(83,15){\small $ r-1$}
\put(70,-15){\small $ r-2$}
\put(98,-20){\small $r$}
}
%
\put(0,-80){
\put(0,0){\circle{6}}
\put(20,0){\circle{6}}
\put(40,0){\circle{6}}
\put(60,0){\circle{6}}
\put(80,0){\circle{6}}
\put(40,20){\circle{6}}
\drawline(3,0)(17,0)
\drawline(23,0)(37,0)
\drawline(43,0)(57,0)
\drawline(40,3)(40,17)
\drawline(63,0)(77,0)
\put(-30,-2){$E_6$}
\put(-2,-15){\small $1$}
\put(18,-15){\small $2$}
\put(38,-15){\small $3$}
\put(58,-15){\small $5$}
\put(78,-15){\small $6$}
\put(50,18){\small $4$}
}
%
\put(180,-80){
\put(0,0){\circle{6}}
\put(20,0){\circle{6}}
\put(40,0){\circle{6}}
\put(60,0){\circle{6}}
\put(80,0){\circle{6}}
\put(100,0){\circle{6}}
\put(40,20){\circle{6}}
\drawline(3,0)(17,0)
\drawline(23,0)(37,0)
\drawline(43,0)(57,0)
\drawline(40,3)(40,17)
\drawline(63,0)(77,0)
\drawline(83,0)(97,0)
\put(-30,-2){$E_7$}
\put(-2,-15){\small $1$}
\put(18,-15){\small $2$}
\put(38,-15){\small $3$}
\put(58,-15){\small $4$}
\put(78,-15){\small $5$}
\put(98,-15){\small $6$}
\put(50,18){\small $7$}
}
%
\put(0,-120){
\put(0,0){\circle{6}}
\put(20,0){\circle{6}}
\put(40,0){\circle{6}}
\put(60,0){\circle{6}}
\put(80,0){\circle{6}}
\put(100,0){\circle{6}}
\put(120,0){\circle{6}}
\put(80,20){\circle{6}}
\drawline(3,0)(17,0)
\drawline(23,0)(37,0)
\drawline(43,0)(57,0)
\drawline(80,3)(80,17)
\drawline(63,0)(77,0)
\drawline(83,0)(97,0)
\drawline(103,0)(117,0)
\put(-30,-2){$E_8$}
\put(-2,-15){\small $1$}
\put(18,-15){\small $2$}
\put(38,-15){\small $3$}
\put(58,-15){\small $4$}
\put(78,-15){\small $5$}
\put(98,-15){\small $6$}
\put(118,-15){\small $7$}
\put(90,18){\small $8$}
}
%
\put(0,-160){
\put(0,0){\circle{6}}
\put(20,0){\circle{6}}
\put(40,0){\circle{6}}
\put(60,0){\circle{6}}
\drawline(3,0)(17,0)
\drawline(43,0)(57,0)
\drawline(22,-2)(38,-2)
\drawline(22,2)(38,2)
\drawline(27,6)(33,0)
\drawline(27,-6)(33,0)
\put(-30,-2){$F_4$}
\put(-2,-15){\small $1$}
\put(18,-15){\small $2$}
\put(38,-15){\small $3$}
\put(58,-15){\small $4$}
}
%
\put(180,-160){
\put(0,0){\circle{6}}
\put(20,0){\circle{6}}
\drawline(3,0)(17,0)
\drawline(2,-2)(18,-2)
\drawline(2,2)(18,2)
\drawline(7,6)(13,0)
\drawline(7,-6)(13,0)
\put(-30,-2){$G_2$}
\put(-2,-15){\small $1$}
\put(18,-15){\small $2$}
}
\end{picture}
\caption{The Dynkin diagrams for ${\mathfrak g}$ and their enumerations.}
\label{fig:Dynkin}
\end{figure}

Let ${\mathfrak g}$ be a simple Lie algebra
associated with a Dynkin diagram of finite type.
We set $I=\{1,\dots, r\}$ with $r = \mathrm{rank}\,\mathfrak{g}$ 
and enumerate the vertices of the Dynkin diagrams as 
Figure \ref{fig:Dynkin}.
We follow \cite{Ka} {\em except for $E_6$\/},
for which we choose the one
naturally corresponding to 
the enumeration of the twisted affine diagram
$E^{(2)}_6$ in Section \ref{ss:twi}.
With a slight abuse of notation, we will 
write for example $\mathfrak{g} = A_r$ to mean that 
$\mathfrak{g}$ is the one associated with the Dynkin diagram of type 
$A_r$.
The cases $A_r, D_r, E_6, E_7$ and $E_8$ are referred to as simply laced.

We set  numbers $t$ and $t_a$ ($a\in I$) by
\begin{equation}
\label{eq:t1}
 t=
\begin{cases}
1 & {\mathfrak g}: \text{simply laced},\\
2 & {\mathfrak g}=B_r, C_r, F_4,\\
3 & {\mathfrak g}=G_2,
\end{cases}
\quad t_a=
\begin{cases}
1 & {\mathfrak g}: \text{simply laced},\\
1 & {\mathfrak g}: \text{nonsimply laced, $\alpha_a$: long root},\\
t & {\mathfrak g}: \text{nonsimply laced, $\alpha_a$: short root}.
\end{cases}
\end{equation}
Let $\alpha_a, \omega_a\,(a \in I)$ be 
the simple roots and the fundamental weights of ${\mathfrak g}$.
We fix a bilinear form $(\; | \; )$ on the dual space of the 
Cartan subalgebra normalized as
\begin{equation}\label{aaw}
(\alpha_a|\alpha_a) = \frac{2}{t_a},\quad
(\alpha_a|\omega_b) = \frac{\delta_{a b}}{t_a}.
\end{equation}
Let $C=(C_{ab})$, 
$C_{ab}=2(\alpha_a|\alpha_b)/(\alpha_a|\alpha_a)$,
be the Cartan matrix of ${\mathfrak g}$.
We have
$C_{a b}= t_a(\alpha_a|\alpha_b)$,
$\alpha_a = \sum_{b=1}^r C_{b a}\omega_b$ 
and $(C^{-1})_{ab}= t_a(\omega_a|\omega_b)$.
We denote by $h$ and $h^\vee$ the Coxeter number 
and the dual Coxeter number of $\mathfrak{g}$, respectively.
They are listed as follows with the dimension of ${\mathfrak g}$.
\begin{equation}\label{hhd}
\begin{tabular}{c|c|c|c|c|c|c|c|c|c}
$\mathfrak{g}$ &  $A_r$  &  $B_r$  &  $C_r$  &  $D_r$  
&  $E_6$  &  $E_7$  &  $E_8$  &  $F_4$  &  $G_2$  \\
\hline
$\dim \mathfrak{g}$ & $r(r+2)$  &  $r(2r+1)$  &  $r(2r+1)$  &  $r(2r-1)$  
&  $78$  &  $133$  &  $248$  &  $52$  &  $14$  \\
$h$ &  $r+1$  &  $2r$  &  $2r$  &  $2r-2$  
&  $12$  &  $18$  &  $30$  &  $12$  &  $6$  \\
$h^\vee$ &  $r+1$  &  $2r-1$  &  $r+1$  &  $2r-2$  
&  $12$  &  $18$  &  $30$  &  $9$  &  $4$ 
\end{tabular}
\end{equation}
The relation $\dim\mathfrak{g}=(1+h)\mathrm{rank}\,\mathfrak{g}$ holds
as is well known.

The unrestricted T-system for ${\mathfrak g}$ 
is the following relations among the commuting 
variables $\{T^{(a)}_m(u)\mid a \in I, 
m \in \Z_{\ge 1}, u \in U\}$,
where $T^{(0)}_m (u)=T^{(a)}_0 (u)= 1$ if they occur
in the RHS.

For simply laced ${\mathfrak g}$,
\begin{align}\label{tade}
T^{(a)}_m(u-1)T^{(a)}_m(u+1)
=
T^{(a)}_{m-1}(u)T^{(a)}_{m+1}(u)
+
\prod_{b\in I: C_{ab}=-1}
T^{(b)}_{m}(u).
\end{align}
For example in type $A_r$, it has the form 
\begin{equation}\label{ta}
T^{(a)}_m(u-1)T^{(a)}_m(u+1)
=T^{(a)}_{m-1}(u)T^{(a)}_{m+1}(u)
+T^{(a-1)}_{m}(u)T^{(a+1)}_{m}(u),
\end{equation}
for $1\le a \le r$ with $T^{(r+1)}_m(u)=1$.
In particular, for $A_1$ it reads 
\begin{equation}\label{ta1}
T_m(u-1)T_m(u+1) = T_{m-1}(u)T_{m+1}(u)+1
\end{equation}
with the simplified notation $T_m(u) = T^{(1)}_m(u)$.

For ${\mathfrak g}=B_r$,
\begin{align}\label{tb}
T^{(a)}_m(u-1)T^{(a)}_m(u+1)
&=
T^{(a)}_{m-1}(u)T^{(a)}_{m+1}(u)\\
&\qquad
+T^{(a-1)}_{m}(u)T^{(a+1)}_{m}(u)
\quad
 (1\leq a\leq r-2),\notag\\
T^{(r-1)}_m(u-1)T^{(r-1)}_m(u+1)
&=
T^{(r-1)}_{m-1}(u)T^{(r-1)}_{m+1}(u)
+
T^{(r-2)}_{m}(u)T^{(r)}_{2m}(u),\notag\\
T^{(r)}_{2m}\left(u-\textstyle\frac{1}{2}\right)
T^{(r)}_{2m}\left(u+\textstyle\frac{1}{2}\right)
&=
T^{(r)}_{2m-1}(u)T^{(r)}_{2m+1}(u)\notag\\
&\qquad
+
T^{(r-1)}_{m}\left(u-\textstyle\frac{1}{2}\right)
T^{(r-1)}_{m}\left(u+\textstyle\frac{1}{2}\right),
\notag\\
T^{(r)}_{2m+1}\left(u-\textstyle\frac{1}{2}\right)
T^{(r)}_{2m+1}\left(u+\textstyle\frac{1}{2}\right)
&=
T^{(r)}_{2m}(u)T^{(r)}_{2m+2}(u)
+
T^{(r-1)}_{m}(u)T^{(r-1)}_{m+1}(u).
\notag
\end{align}

{\allowdisplaybreaks
For ${\mathfrak g}=C_r$,
\begin{align}\label{tc}
T^{(a)}_m\left(u-\textstyle\frac{1}{2}\right)
T^{(a)}_m\left(u+\textstyle\frac{1}{2}\right)
&=
T^{(a)}_{m-1}(u)T^{(a)}_{m+1}(u)\\
&\qquad
+T^{(a-1)}_{m}(u)T^{(a+1)}_{m}(u)
\quad
 (1\leq a\leq r-2),\notag\\
T^{(r-1)}_{2m}\left(u-\textstyle\frac{1}{2}\right)
T^{(r-1)}_{2m}\left(u+\textstyle\frac{1}{2}\right)
&=
T^{(r-1)}_{2m-1}(u)T^{(r-1)}_{2m+1}(u)\notag\\
&\qquad
+
T^{(r-2)}_{2m}(u)
T^{(r)}_{m}\left(u-\textstyle\frac{1}{2}\right)
T^{(r)}_{m}\left(u+\textstyle\frac{1}{2}\right)
,\notag\\
T^{(r-1)}_{2m+1}\left(u-\textstyle\frac{1}{2}\right)
T^{(r-1)}_{2m+1}\left(u+\textstyle\frac{1}{2}\right)
&=
T^{(r-1)}_{2m}(u)T^{(r-1)}_{2m+2}(u)\notag\\
&\qquad
+
T^{(r-2)}_{2m+1}(u)
T^{(r)}_{m}(u)T^{(r)}_{m+1}(u),
\notag\\
T^{(r)}_{m}(u-1)
T^{(r)}_{m}(u+1)
&=
T^{(r)}_{m-1}(u)T^{(r)}_{m+1}(u)
+
T^{(r-1)}_{2m}(u).
\notag
\end{align}
}

For ${\mathfrak g}=F_4$,

\begin{align}\label{tf}
T^{(1)}_m(u-1)T^{(1)}_m(u+1)
&=
T^{(1)}_{m-1}(u)T^{(1)}_{m+1}(u)
+T^{(2)}_{m}(u),\\
T^{(2)}_m(u-1)T^{(2)}_m(u+1)
&=
T^{(2)}_{m-1}(u)T^{(2)}_{m+1}(u)
+
T^{(1)}_{m}(u)T^{(3)}_{2m}(u),\notag\\
T^{(3)}_{2m}\left(u-\textstyle\frac{1}{2}\right)
T^{(3)}_{2m}\left(u+\textstyle\frac{1}{2}\right)
&=
T^{(3)}_{2m-1}(u)T^{(3)}_{2m+1}(u)\notag\\
&\qquad
+
T^{(2)}_{m}\left(u-\textstyle\frac{1}{2}\right)
T^{(2)}_{m}\left(u+\textstyle\frac{1}{2}\right)
T^{(4)}_{2m}(u),\notag\\
T^{(3)}_{2m+1}\left(u-\textstyle\frac{1}{2}\right)
T^{(3)}_{2m+1}\left(u+\textstyle\frac{1}{2}\right)
&=
T^{(3)}_{2m}(u)T^{(3)}_{2m+2}(u)
+
T^{(2)}_{m}(u)T^{(2)}_{m+1}(u)
T^{(4)}_{2m+1}(u),\notag\\
\notag
T^{(4)}_{m}\left(u-\textstyle\frac{1}{2}\right)
T^{(4)}_{m}\left(u+\textstyle\frac{1}{2}\right)
&=
T^{(4)}_{m-1}(u)T^{(4)}_{m+1}(u)
+T^{(3)}_m(u).
\notag
\end{align}

For ${\mathfrak g}=G_2$,
\begin{align}\label{tg}
T^{(1)}_m(u-1)T^{(1)}_m(u+1)
&=
T^{(1)}_{m-1}(u)T^{(1)}_{m+1}(u)
+
T^{(2)}_{3m}(u),\\
T^{(2)}_{3m}\left(u-\textstyle\frac{1}{3}\right)
T^{(2)}_{3m}\left(u+\textstyle\frac{1}{3}\right)
&=
T^{(2)}_{3m-1}(u)T^{(2)}_{3m+1}(u)\notag\\
&\qquad
+
T^{(1)}_{m}\left(u-\textstyle\frac{2}{3}\right)
T^{(1)}_m(u)
T^{(1)}_{m}\left(u+\textstyle\frac{2}{3}\right),\notag\\
T^{(2)}_{3m+1}\left(u-\textstyle\frac{1}{3}\right)
T^{(2)}_{3m+1}\left(u+\textstyle\frac{1}{3}\right)
&=
T^{(2)}_{3m}(u)T^{(2)}_{3m+2}(u)\notag\\
&\qquad+
T^{(1)}_{m}\left(u-\textstyle\frac{1}{3}\right)
T^{(1)}_{m}\left(u+\textstyle\frac{1}{3}\right)
T^{(1)}_{m+1}(u),\notag\\
T^{(2)}_{3m+2}\left(u-\textstyle\frac{1}{3}\right)
T^{(2)}_{3m+2}\left(u+\textstyle\frac{1}{3}\right)
&=
T^{(2)}_{3m+1}(u)T^{(2)}_{3m+3}(u)\notag\\
&\qquad+
T^{(1)}_{m}(u)
T^{(1)}_{m+1}\left(u-\textstyle\frac{1}{3}\right)
T^{(1)}_{m+1}\left(u+\textstyle\frac{1}{3}\right).\notag
\end{align}

We note that these relations are {\em not} bilinear in general
under the boundary condition stated before (\ref{tade}).
The second terms on the RHS can be of order 0,1,2 and 3 
in $T^{(a)}_m(u)$.

The variable $u\in U$ is called the {\em spectral parameter}.
The set $U$ can be either the complex plane $\mathbb{C}$, or
the cylinder $\mathbb{C}_{\xi}:= \mathbb{C}/
(2\pi \sqrt{-1}/\xi) \mathbb{Z}$  such that 
$2\pi \sqrt{-1}/\xi \not\in \mathbb{Q}$.
The choice will not matter seriously, but reflects the underlying algebra.

\begin{remark}\label{re:u}
In Section \ref{s:qg} we will see that 
the T-system for ${\mathfrak g}$ is actually 
associated with the untwisted quantum affine algebra
$U_q(\hat{\mathfrak g})$ with $q=e^\hbar$ when $U=\C_{t\hbar}$.
The choice $U=\mathbb{C}$ corresponds to 
the Yangian $Y({\mathfrak g})$ in a similar sense.
In this review we will mostly be concerned with the 
$U_q(\hat{\mathfrak g})$ case. 
Thus we have simply chosen to say T-system for $\mathfrak{g}$ 
rather than T-system for $U_q(\hat{\mathfrak g})$.
The latter terminology is more balanced when  
the twisted case is considered in Section \ref{ss:twi}. 
Note that the choice 
$U=\mathbb{C}_{\xi}$ effectively imposes an additional periodicity
$T^{(a)}_m(u)=
T^{(a)}_m(\textstyle u+ \frac{2\pi \sqrt{-1}}{\xi})$.
By the assumption $2\pi \sqrt{-1}/\xi\notin
\mathbb{Q}$, this does not interfere with the T-system.
Similar remarks apply to the Y-system in what follows. 
\end{remark}

The unrestricted Y-system for ${\mathfrak g}$ is
the following relations among commuting variables
$\{Y^{(a)}_m(u)\mid a \in I, m \in \Z_{\ge 1}, u\in  U \}$,
where $Y^{(0)}_m (u)=Y^{(a)}_0 (u)^{-1}=0$ if they occur
in the RHS.

For simply laced ${\mathfrak g}$,
\begin{equation}\label{yade}
Y^{(a)}_m(u-1)Y^{(a)}_m(u+1)
=
\frac{
\prod_{b\in I: C_{ab}=-1}
(1+Y^{(b)}_{m}(u))
}
{
(1+Y^{(a)}_{m-1}(u)^{-1})(1+Y^{(a)}_{m+1}(u)^{-1})}.
\end{equation}

For ${\mathfrak g}=B_r$,
\begin{align}
\label{yb}
Y^{(a)}_m(u-1)Y^{(a)}_m(u+1)
&=
\frac{
(1+Y^{(a-1)}_{m}(u))(1+Y^{(a+1)}_{m}(u))
}
{
(1+Y^{(a)}_{m-1}(u)^{-1})(1+Y^{(a)}_{m+1}(u)^{-1})
} \\
&\hskip130pt
 (1\leq a\leq r-2),\notag\\
Y^{(r-1)}_m(u-1)Y^{(r-1)}_m(u+1)
&=
\frac{
\begin{array}{l}
\textstyle
(1+Y^{(r-2)}_{m}(u))
(1+Y^{(r)}_{2m-1}(u))
(1+Y^{(r)}_{2m+1}(u))\\
\textstyle
\quad\times(1+Y^{(r)}_{2m}\left(u-\frac{1}{2}\right))
(1+Y^{(r)}_{2m}\left(u+\frac{1}{2}\right))
\end{array}
}
{
(1+Y^{(r-1)}_{m-1}(u)^{-1})(1+Y^{(r-1)}_{m+1}(u)^{-1})
},
\notag\\
Y^{(r)}_{2m}\left(u-\textstyle\frac{1}{2}\right)
Y^{(r)}_{2m}\left(u+\textstyle\frac{1}{2}\right)
&=
\frac{1+Y^{(r-1)}_{m}(u)}
{
(1+Y^{(r)}_{2m-1}(u)^{-1})(1+Y^{(r)}_{2m+1}(u)^{-1})
},\notag\\
Y^{(r)}_{2m+1}\left(u-\textstyle\frac{1}{2}\right)
Y^{(r)}_{2m+1}\left(u+\textstyle\frac{1}{2}\right)
&=
\frac{1}{(1+Y^{(r)}_{2m}(u)^{-1})(1+Y^{(r)}_{2m+2}(u)^{-1})}.
\notag
\end{align}

{\allowdisplaybreaks
For ${\mathfrak g}=C_r$,
\begin{align}
\label{yc}
Y^{(a)}_m\left(u-\textstyle\frac{1}{2}\right)
Y^{(a)}_m\left(u+\textstyle\frac{1}{2}\right)
&=
\frac{
(1+Y^{(a-1)}_{m}(u))(1+Y^{(a+1)}_{m}(u))
}{
(1+Y^{(a)}_{m-1}(u)^{-1})(1+Y^{(a)}_{m+1}(u)^{-1})
}\\
&\hskip130pt
 (1\leq a\leq r-2),\notag\\
Y^{(r-1)}_{2m}\left(u-\textstyle\frac{1}{2}\right)
Y^{(r-1)}_{2m}\left(u+\textstyle\frac{1}{2}\right)
&=
\frac{
(1+Y^{(r-2)}_{2m}(u))(1+Y^{(r)}_{m}(u))
}{
(1+Y^{(r-1)}_{2m-1}(u)^{-1})(1+Y^{(r-1)}_{2m+1}(u)^{-1})
},\notag\\
Y^{(r-1)}_{2m+1}\left(u-\textstyle\frac{1}{2}\right)
Y^{(r-1)}_{2m+1}\left(u+\textstyle\frac{1}{2}\right)
&=
\frac{
1+Y^{(r-2)}_{2m+1}(u)
}{
(1+Y^{(r-1)}_{2m}(u)^{-1})(1+Y^{(r-1)}_{2m+2}(u)^{-1})
},\notag\\
Y^{(r)}_{m}(u-1)
Y^{(r)}_{m}(u+1)
&=
\frac{
\begin{array}{l}
\textstyle
(1+Y^{(r-1)}_{2m+1}(u))
(1+Y^{(r-1)}_{2m-1}(u))\\
\textstyle
\quad\times(1+Y^{(r-1)}_{2m}\left(u-\frac{1}{2}\right))
(1+Y^{(r-1)}_{2m}\left(u+\frac{1}{2}\right))
\end{array}
}
{
(1+Y^{(r)}_{m-1}(u)^{-1})(1+Y^{(r)}_{m+1}(u)^{-1})
}.
\notag
\end{align}
}

For ${\mathfrak g}=F_4$,
\begin{align}\label{yf}
Y^{(1)}_m(u-1)Y^{(1)}_m(u+1)
&=
\frac{
1+Y^{(2)}_{m}(u)
}{
(1+Y^{(1)}_{m-1}(u)^{-1})(1+Y^{(1)}_{m+1}(u)^{-1})
},\\
Y^{(2)}_m(u-1)Y^{(2)}_m(u+1)
&=
\frac{
{
\begin{array}{l}
\textstyle
(1+Y^{(1)}_{m}(u))
(1+Y^{(3)}_{2m-1}(u))
(1+Y^{(3)}_{2m+1}(u))\\
\textstyle
\quad\times(1+Y^{(3)}_{2m}\left(u-\frac{1}{2}\right))
(1+Y^{(3)}_{2m}\left(u+\frac{1}{2}\right))
\end{array}
}
}
{
(1+Y^{(2)}_{m-1}(u)^{-1})(1+Y^{(2)}_{m+1}(u)^{-1})
},
\notag\\
Y^{(3)}_{2m}\left(u-\textstyle\frac{1}{2}\right)
Y^{(3)}_{2m}\left(u+\textstyle\frac{1}{2}\right)
&=
\frac{
(1+Y^{(2)}_{m}(u))(1+Y^{(4)}_{2m}(u))
}{
(1+Y^{(3)}_{2m-1}(u)^{-1})(1+Y^{(3)}_{2m+1}(u)^{-1})
},\notag\\
Y^{(3)}_{2m+1}\left(u-\textstyle\frac{1}{2}\right)
Y^{(3)}_{2m+1}\left(u+\textstyle\frac{1}{2}\right)
&=
\frac{
1+Y^{(4)}_{2m+1}(u)
}{
(1+Y^{(3)}_{2m}(u)^{-1})(1+Y^{(3)}_{2m+2}(u)^{-1})
},\notag\\
Y^{(4)}_{m}\left(u-\textstyle\frac{1}{2}\right)
Y^{(4)}_{m}\left(u+\textstyle\frac{1}{2}\right)
&=
\frac{
1+Y^{(3)}_{m}(u)
}{
(1+Y^{(4)}_{m-1}(u)^{-1})(1+Y^{(4)}_{m+1}(u)^{-1})
}.\notag
\end{align}

For ${\mathfrak g}=G_2$,
\begin{align}\label{yg}
Y^{(1)}_m(u-1)Y^{(1)}_m(u+1)
&=
\frac{
{
\begin{array}{l}
\textstyle
(1+Y^{(2)}_{3m-2}(u))
(1+Y^{(2)}_{3m+2}(u))\\
\textstyle
\times(1+Y^{(2)}_{3m-1}\left(u-\frac{1}{3}\right))
(1+Y^{(2)}_{3m-1}\left(u+\frac{1}{3}\right))\\
\textstyle
\times(1+Y^{(2)}_{3m+1}\left(u-\frac{1}{3}\right))
(1+Y^{(2)}_{3m+1}\left(u+\frac{1}{3}\right))\\
\textstyle
\times(1+Y^{(2)}_{3m}\left(u-\frac{2}{3}\right))
(1+Y^{(2)}_{3m}\left(u+\frac{2}{3}\right))\\
\times (1+Y^{(2)}_{3m}\left(u\right))
\end{array}
}
}
{
(1+Y^{(1)}_{m-1}(u)^{-1})(1+Y^{(1)}_{m+1}(u)^{-1})
},
\\
Y^{(2)}_{3m}\left(u-\textstyle\frac{1}{3}\right)
Y^{(2)}_{3m}\left(u+\textstyle\frac{1}{3}\right)
&=
\frac{1+Y^{(1)}_m(u)}
{
(1+Y^{(2)}_{3m-1}(u)^{-1})(1+Y^{(2)}_{3m+1}(u)^{-1})
},
\notag\\
Y^{(2)}_{3m+1}\left(u-\textstyle\frac{1}{3}\right)
Y^{(2)}_{3m+1}\left(u+\textstyle\frac{1}{3}\right)
&=
\frac{1}
{
(1+Y^{(2)}_{3m}(u)^{-1})(1+Y^{(2)}_{3m+2}(u)^{-1})
},
\notag\\
Y^{(2)}_{3m+2}\left(u-\textstyle\frac{1}{3}\right)
Y^{(2)}_{3m+2}\left(u+\textstyle\frac{1}{3}\right)
&=
\frac{1}
{
(1+Y^{(2)}_{3m+1}(u)^{-1})(1+Y^{(2)}_{3m+3}(u)^{-1})
}.
\notag
\end{align}

We stress that the T and Y-systems 
for nonsimply laced ${\mathfrak g}$ are {\em not} just a folding of 
simply laced cases.

We also remark that 
T and Y-systems for $B_2$ and $C_2$  
are equivalent and transformed to each other 
by $T^{(1)}_m(u) \leftrightarrow T^{(2)}_m(u)$ and
$Y^{(1)}_m(u) \leftrightarrow Y^{(2)}_m(u)$ 
reflecting the fact $B_2 \simeq C_2$.

\subsection{Restriction}\label{ss:res1}

We fix an integer $\ell \ge 2$ called {\em level}.
Let $t_a$ be the number in (\ref{eq:t1}).
The level $\ell$ restricted T-system for ${\mathfrak g}$ 
(with the unit boundary condition) is relations 
(\ref{tade})--(\ref{tg})
naturally restricted to 
$\{T^{(a)}_m(u)\mid a \in I, 1 \le m \le t_a\ell-1, u \in U\}$
by imposing $T^{(a)}_{t_a\ell}(u)=1$.

The level $\ell$ restricted Y-system for ${\mathfrak g}$ 
is relations (\ref{yade})--(\ref{yg}) naturally restricted to 
$\{Y^{(a)}_m(u)\mid a \in I, 1 \le m \le t_a\ell-1, u \in U\}$
by imposing $Y^{(a)}_{t_a\ell}(u)^{-1}=0$.

Note that for ${\mathfrak g}$ nonsimply laced, 
the above restriction makes sense also at $\ell=1$. 
The resulting T and Y-systems become equivalent to the  
level $t$ restricted T and Y-systems  for $A_n$ with  
$n=\sharp\{a \in I\mid t_a=t\}$ under 
the rescaling of the spectral parameter $u\rightarrow u/t$.
One can also consider the level 0 case formally.
See around (\ref{j:bulkT}).

\begin{example}\label{ex:a2}
We write down the level $2$ restricted T and Y-systems for $A_2$:
\begin{align*}
T^{(1)}_1(u-1)T^{(1)}_1(u+1)&=1+T^{(2)}_1(u),\quad
T^{(2)}_1(u-1)T^{(2)}_1(u+1)=1+T^{(1)}_1(u),\\
Y^{(1)}_1(u-1)Y^{(1)}_1(u+1)&=1+Y^{(2)}_1(u),\quad
Y^{(2)}_1(u-1)Y^{(2)}_1(u+1)=1+Y^{(1)}_1(u).
\end{align*}
Thus they are identical.
\end{example}

\begin{example}\label{ex:c2}
We write down the level $2$ restricted T-system for $C_2$:
\begin{align*}
T^{(1)}_{1}(u-\textstyle\frac{1}{2})
T^{(1)}_{1}(u+\textstyle\frac{1}{2})
 &= T^{(1)}_{2}(u)  +
T^{(2)}_{1}(u),\\     
T^{(1)}_{2}(u-\textstyle\frac{1}{2})
T^{(1)}_{2}(u+\textstyle\frac{1}{2})
 &= T^{(1)}_{1}(u)T^{(1)}_{3}(u)+
T^{(2)}_{1}(u-\textstyle\frac{1}{2})
T^{(2)}_{1}(u+\textstyle\frac{1}{2}),\\     
T^{(1)}_{3}(u-\textstyle\frac{1}{2})
T^{(1)}_{3}(u+\textstyle\frac{1}{2})
 &= T^{(1)}_{2}(u)  + T^{(2)}_{1}(u),\\
T^{(2)}_1(u-1) T^{(2)}_1(u+1) &= 1 + T^{(1)}_{2}(u). 
\end{align*}
\end{example}

\begin{example}\label{ex:dua}
Level $\ell$ restricted T-system for $A_{r-1}$ has the form
\begin{equation*}
T^{(a)}_m(u-1)T^{(a)}_m(u+1)
=T^{(a)}_{m-1}(u)T^{(a)}_{m+1}(u)
+T^{(a-1)}_{m}(u)T^{(a+1)}_{m}(u)
\end{equation*}
for $1\le a \le r-1$ and $1 \le m \le \ell-1$.
It is invariant under the simultaneous transformation 
$T^{(a)}_m(u) \mapsto T^{(m)}_a(\pm u+{\rm const})$ and 
$r \leftrightarrow \ell$.
The similar property holds also for the 
level $\ell$ restricted Y-system for $A_{r-1}$.
This symmetry is called the {\em level-rank duality}.
\end{example}

\subsection{Relation between T and Y-systems}\label{ss:rty}

The unrestricted T-system for ${\mathfrak g}$ has the form
\begin{equation}\label{ttm0}
T^{(a)}_m(u-\tai)
T^{(a)}_m(u+\tai)
=T^{(a)}_{m-1}(u)T^{(a)}_{m+1}(u)+
\displaystyle\prod_{(b,k,v)}T^{(b)}_k(v)^{N(a,m,u| b,k,v)},
\end{equation}
where the last term is a finite product.
Then, it is easy to see that the unrestricted 
Y-system for the same ${\mathfrak g}$ takes the form
\begin{equation}\label{ttm1}
Y^{(a)}_m(u-\tai)
Y^{(a)}_m(u+\tai)=\displaystyle
\frac{\prod_{(b,k,v)}(1+Y^{(b)}_k(v))^{N(b,k,v|a,m,u)}}
{(1+Y^{(a)}_{m-1}(u)^{-1})(1+Y^{(a)}_{m+1}(u)^{-1})}.
\end{equation}
The same relation holds also between the level $\ell$ 
restricted T and Y-systems.

Let us write (\ref{ttm0}) simply as 
\begin{equation}\label{ttm}
\textstyle
T^{(a)}_m(u-\frac{1}{t_a})T^{(a)}_m(u+\frac{1}{t_a})
=T^{(a)}_{m-1}(u)T^{(a)}_{m+1}(u)+M^{(a)}_m(u).
\end{equation}

\begin{theorem}[\cite{KNS2}]\label{th:ty1}
Suppose $T^{(a)}_m(u)$ satisfies the unrestricted T-system
for ${\mathfrak g}$.
Then
\begin{equation}\label{ty1}
Y^{(a)}_m(u)=\frac{M^{(a)}_m(u)}
{T^{(a)}_{m-1}(u)T^{(a)}_{m+1}(u)}
\end{equation}
is a solution of the unrestricted Y-system for ${\mathfrak g}$.
The same claim holds 
between the level $\ell$ restricted T and Y-systems.
\end{theorem}

\noindent{\it Sketch of proof}. 
This can be directly verified by substituting the 
resulting relations
\begin{align}
1+Y^{(a)}_m(u)&=
\frac{T^{(a)}_{m}\bigl(u-\textstyle\frac{1}{t_a}\bigr)
T^{(a)}_{m}\bigl(u+\textstyle\frac{1}{t_a}\bigr)}
{T^{(a)}_{m-1}(u)T^{(a)}_{m+1}(u)},\\
1+Y^{(a)}_m(u)^{-1}&=
\frac{T^{(a)}_{m}\bigl(u-\textstyle\frac{1}{t_a}\bigr)
T^{(a)}_{m}\bigl(u+\textstyle\frac{1}{t_a}\bigr)}
{M^{(a)}_m(u)}
\end{align}
into the Y-system.
Here we demonstrate the calculation for simply laced ${\mathfrak g}$.
\begin{align*}
&\ Y^{(a)}_m(u-1)Y^{(a)}_m(u+1)\\
=&\
\frac{
\prod_{b:C_{ab}=-1}
T^{(b)}_{m}(u-1)T^{(b)}_{m}(u+1)
}
{
T^{(a)}_{m-1}(u-1)T^{(a)}_{m+1}(u-1)
T^{(a)}_{m-1}(u+1)T^{(a)}_{m+1}(u+1)
}\\
=&\,
\frac{
\prod_{b:C_{ab}=-1}
(T^{(b)}_{m-1}(u)T^{(b)}_{m+1}(u)
+
\prod_{c:C_{bc}=-1}T^{(c)}_{m}(u))
}
{
T^{(a)}_{m-2}(u)T^{(a)}_{m}(u)+\prod_{b:C_{ab}=-1}
T^{(b)}_{m-1}(u)
}
\\
&\quad \times
\frac{1}{T^{(a)}_{m}(u)T^{(a)}_{m+2}(u)+\prod_{b:C_{ab}=-1}
T^{(b)}_{m+1}(u)
}
\\
=&\ \frac{
\prod_{b:C_{ab}=-1}(1+Y^{(b)}_{m}(u))
}
{
(1+Y^{(a)}_{m-1}(u)^{-1})(1+Y^{(a)}_{m+1}(u)^{-1})
}.
\end{align*}
This calculation is valid also at $m=1$  by 
formally setting $T^{(a)}_{-1}(u)=0$.
For level $\ell$ restricted case, 
it is valid similarly by 
formally setting $T^{(a)}_{\ell+1}(u)=0$.

Theorem \ref{th:ty1} has a natural account from the viewpoint of  
cluster algebra with coefficients. See Remark \ref{re:ty}.

\begin{example}
We write down the relation (\ref{ty1}) for
the level 2 restricted T-system for $C_2$. 
{}From Example \ref{ex:c2}, they read
\begin{alignat*}{2}
Y^{(1)}_{1}(u)
&= \frac{T^{(2)}_{1}(u)}{T^{(1)}_{2}(u)},&\qquad   
Y^{(1)}_{2}(u)
&= \frac{T^{(2)}_{1}(u-\textstyle\frac{1}{2})
T^{(2)}_{1}(u+\textstyle\frac{1}{2})}
{T^{(1)}_{1}(u)T^{(1)}_{3}(u)},\\     
Y^{(1)}_{3}(u)
&= \frac{T^{(2)}_{1}(u)}{T^{(1)}_{2}(u)}, &
Y^{(2)}_1(u)&= T^{(1)}_{2}(u). 
\end{alignat*}
Thus the specific construction (\ref{ty1}) automatically 
imposes the condition $Y^{(1)}_1(u) = Y^{(1)}_3(u)$.
However, the level restricted Y-system alone does {\em not}
restrict itself to such a situation in general.
\end{example}

\begin{remark}\label{re:g}
Consider a slight modification of the general 
T-system relation (\ref{ttm}) into
\begin{equation}\label{tga}
\textstyle
T^{(a)}_m(u-\frac{1}{t_a})T^{(a)}_m(u+\frac{1}{t_a})
=T^{(a)}_{m-1}(u)T^{(a)}_{m+1}(u)+g^{(a)}_m(u)M^{(a)}_m(u),
\end{equation}
where $g^{(a)}_m(u)$ is any function satisfying 
\begin{equation}\label{gae}
\textstyle
g^{(a)}_m(u-\frac{1}{t_a})g^{(a)}_m(u+\frac{1}{t_a})
=g^{(a)}_{m-1}(u)g^{(a)}_{m+1}(u).
\end{equation}
Then it is easily checked that the substitution 
\begin{equation}\label{ty2}
Y^{(a)}_m(u)=\frac{g^{(a)}_m(u)M^{(a)}_m(u)}
{T^{(a)}_{m-1}(u)T^{(a)}_{m+1}(u)}
\end{equation}
is still a solution of the same Y-system.
\end{remark}

\subsection{Twisted case}\label{ss:twi}

Let us proceed to the
T and Y-systems associated with the twisted 
quantum affine algebras following \cite{KS2, Her4}.
In this subsection and the next,
$X_N$ exclusively denotes a
Dynkin diagram of type $A_N$ ($N\geq 2$),
 $D_N$ ($N\geq 4$) or $ E_6$.
We keep the enumeration of the nodes of 
$X_N$  by the set $I=\{1,\ldots, N\}$
as in Figure \ref{fig:Dynkin}.
For a pair $(X_N,\kappa)=(A_N,2)$, $(D_N,2)$, $(E_6,2)$ or $(D_4,3)$,
we define
the diagram automorphism $\sigma: I \rightarrow I$
of $X_N$ of order $\kappa$
as follows:
$\sigma(a)=a$ {\em except for the following  cases
in our enumeration\/}:
\begin{alignat}{2}
\label{eq:sigma1}
& \sigma(a)=N+1-a\quad (a\in I)&&(X_N,\kappa)=(A_N,2),\\
& \sigma(N-1)=N,\ \sigma(N)=N-1&&(X_N,\kappa)=(D_{N},2), 
 \notag\\
& \sigma(1)=6,\  \sigma(2)=5,\  \sigma(5)=2,\ \sigma(6)=1
&\quad& (X_N,\kappa)=(E_6,2), \notag\\
& \sigma(1)=3,\  \sigma(3)=4,\  \sigma(4)=1
&&(X_N,\kappa)=(D_4,3).
\notag
\end{alignat}
Let $I/\sigma$ be the set of the $\sigma$-orbits
 of nodes of $X_N$.
We choose, at our discretion,
 a complete set of representatives $I_{\sigma}\subset I$
of $I/\sigma$ as
\begin{equation}
\label{eq:sigma2}
I_{\sigma}= 
\begin{cases}
\{ 1,2,\dots, r\} & 
(X_N,\kappa)=(A_{2r-1},2), (A_{2r},2), (D_{r+1},2),\\
\{ 1,2,3,4\} &
(X_N,\kappa)=(E_6,2),\\
 \{ 1,2\} &
(X_N,\kappa)=(D_4,3).
\end{cases}
\end{equation}

\begin{figure}[h]
\begin{picture}(283,115)(-20,-90)
%
\put(0,0){\circle{6}}
\put(20,0){\circle{6}}
\put(20,20){\circle{6}}
\put(80,0){\circle{6}}
\put(100,0){\circle*{6}}
\put(45,0){\circle*{1}}
\put(50,0){\circle*{1}}
\put(55,0){\circle*{1}}
\drawline(3,0)(17,0)
\drawline(20,3)(20,17)
\drawline(23,0)(37,0)
\drawline(63,0)(77,0)
\drawline(82,-2)(98,-2)
\drawline(82,2)(98,2)
\drawline(87,0)(93,-6)
\drawline(87,0)(93,6)
\put(-30,-2){$A^{(2)}_{2r-1}$}
\put(-2,-15){\small $1$}
\put(28,17){\small $0$}
\put(18,-15){\small $2$}
\put(70,-15){\small $ r-1$}
\put(98,-15){\small $r$}
%
\put(180,0){
\put(0,0){\circle{6}}
\put(20,0){\circle{6}}
\drawline(3,1)(17,1)
\drawline(3,-1)(17,-1)
\drawline(2,-3)(18,-3)
\drawline(2,3)(18,3)
\drawline(7,6)(13,0)
\drawline(7,-6)(13,0)
\put(-30,-2){$A^{(2)}_{2}$}
\put(-2,-15){\small $0$}
\put(18,-15){\small $1$}
}
%
\put(0,-40){
\put(0,0){\circle{6}}
\put(20,0){\circle{6}}
\put(80,0){\circle{6}}
\put(100,0){\circle{6}}
\put(45,0){\circle*{1}}
\put(50,0){\circle*{1}}
\put(55,0){\circle*{1}}
\drawline(2,-2)(18,-2)
\drawline(2,2)(18,2)
\drawline(13,0)(7,-6)
\drawline(13,0)(7,6)
\drawline(23,0)(37,0)
\drawline(63,0)(77,0)
\drawline(82,-2)(98,-2)
\drawline(82,2)(98,2)
\drawline(93,0)(87,-6)
\drawline(93,0)(87,6)
\put(-30,-2){$A^{(2)}_{2r}$}
\put(-2,-15){\small $0$}
\put(18,-15){\small $1$}
\put(70,-15){\small $ r-1$}
\put(98,-15){\small $r$}
}
%
\put(180,-40){
\put(0,0){\circle{6}}
\put(20,0){\circle*{6}}
\put(80,0){\circle*{6}}
\put(100,0){\circle{6}}
\put(45,0){\circle*{1}}
\put(50,0){\circle*{1}}
\put(55,0){\circle*{1}}
\drawline(2,-2)(18,-2)
\drawline(2,2)(18,2)
\drawline(7,0)(13,-6)
\drawline(7,0)(13,6)
\drawline(23,0)(37,0)
\drawline(63,0)(77,0)
\drawline(82,-2)(98,-2)
\drawline(82,2)(98,2)
\drawline(93,0)(87,-6)
\drawline(93,0)(87,6)
\put(-30,-2){$D^{(2)}_{r+1}$}
\put(-2,-15){\small $0$}
\put(18,-15){\small $1$}
\put(70,-15){\small $ r-1$}
\put(98,-15){\small $r$}
}
%
\put(0,-80){
\put(0,0){\circle{6}}
\put(20,0){\circle{6}}
\put(40,0){\circle{6}}
\put(60,0){\circle*{6}}
\put(80,0){\circle*{6}}
\drawline(3,0)(17,0)
\drawline(23,0)(37,0)
\drawline(63,0)(77,0)
\drawline(42,-2)(58,-2)
\drawline(42,2)(58,2)
\drawline(53,6)(47,0)
\drawline(53,-6)(47,0)
\put(-30,-2){$E^{(2)}_6$}
\put(-2,-15){\small $0$}
\put(18,-15){\small $1$}
\put(38,-15){\small $2$}
\put(58,-15){\small $3$}
\put(78,-15){\small $4$}
}
%
\put(180,-80){
\put(0,0){\circle{6}}
\put(20,0){\circle{6}}
\put(40,0){\circle*{6}}
\drawline(3,0)(17,0)
\drawline(23,0)(37,0)
\drawline(22,-2)(38,-2)
\drawline(22,2)(38,2)
\drawline(33,6)(27,0)
\drawline(33,-6)(27,0)
\put(-30,-2){$D^{(3)}_4$}
\put(-2,-15){\small $0$}
\put(18,-15){\small $1$}
\put(38,-15){\small $2$}
}
\end{picture}
\caption{The Dynkin diagrams 
$X^{(\kappa)}_N$
of twisted affine type and their enumerations
by $I_{\sigma}\cup \{0\}$.
For a filled node $a$, $\sigma(a)=a$
(i.e., $\kappa_a = \kappa$) holds.
}
\label{fig:tDynkin}
\end{figure}
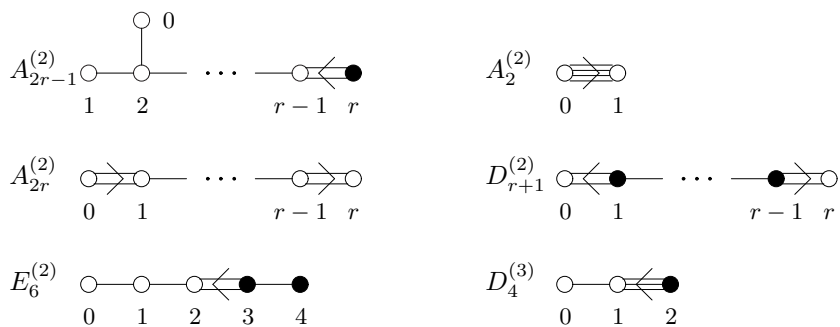

Let $X^{(\kappa)}_N = A^{(2)}_{2r-1}\;(r \ge 2), A^{(2)}_{2r}\; (r \ge 1), 
D^{(2)}_{r+1}\; (r \ge 3), E^{(2)}_6$ or $D^{(3)}_4$
be a Dynkin diagram of  twisted affine type \cite{Ka}.
We enumerate the nodes of $X^{(\kappa)}_N$
with $I_\sigma \cup \{ 0 \}$ as in
Figure \ref{fig:tDynkin},
where $I_{\sigma}$ is the one for $(X_N,\kappa)$. 
By this, 
we  have established the identification of the non-0th  nodes
of the diagram $X^{(\kappa)}_N$
with the nodes of the diagram $X_N$ belonging to
the set $I_{\sigma}$.
For example, for $E^{(2)}_6$, the correspondence is as follows:
\begin{equation*}
\raisebox{-13pt}
{
\begin{picture}(283,35)(-30,-15)
\put(0,0){
\put(0,0){\circle{6}}
\put(20,0){\circle{6}}
\put(40,0){\circle{6}}
\put(60,0){\circle*{6}}
\put(80,0){\circle*{6}}
\drawline(3,0)(17,0)
\drawline(23,0)(37,0)
\drawline(63,0)(77,0)
\drawline(42,-2)(58,-2)
\drawline(42,2)(58,2)
\drawline(53,6)(47,0)
\drawline(53,-6)(47,0)
\put(-30,-2){$E^{(2)}_6$}
\put(-2,-15){\small $0$}
\put(18,-15){\small $1$}
\put(38,-15){\small $2$}
\put(58,-15){\small $3$}
\put(78,-15){\small $4$}
\dottedline{3}(13,-7)(87,-7)
\dottedline{3}(13,7)(87,7)
\dottedline{3}(13,7)(13,-7)
\dottedline{3}(87,7)(87,-7)
}
%
%
%
\put(160,0){
\put(0,0){\circle{6}}
\put(20,0){\circle{6}}
\put(40,0){\circle{6}}
\put(60,0){\circle{6}}
\put(80,0){\circle{6}}
\put(40,20){\circle{6}}
\drawline(3,0)(17,0)
\drawline(23,0)(37,0)
\drawline(43,0)(57,0)
\drawline(40,3)(40,17)
\drawline(63,0)(77,0)
\dottedline{3}(-7,-7)(47,-7)
\dottedline{3}(-7,-7)(-7,7)
\dottedline{3}(47,-7)(47,27)
\dottedline{3}(-7,7)(33,7)
\dottedline{3}(33,7)(33,27)
\dottedline{3}(33,27)(47,27)
\put(-30,-2){$E_6$}
\put(-2,-15){\small $1$}
\put(18,-15){\small $2$}
\put(38,-15){\small $3$}
\put(58,-15){\small $5$}
\put(78,-15){\small $6$}
\put(50,18){\small $4$}
}
\end{picture}
}
\end{equation*}
The filled nodes 3,4 in $E^{(2)}_6$ correspond to the
fixed nodes by $\sigma$ in $E_6$.
We use this identification throughout.
(The 0th node of $X^{(\kappa)}_N$ is irrelevant in our setting
here.)

We define $\kappa_a$ $(a \in I_{\sigma})$ as
\begin{equation}\label{kapa}
\kappa_a = \begin{cases}
1 & \sigma(a) \neq a,\\
\kappa & \sigma(a) = a.
\end{cases}
\end{equation}
Note that $X^{(2)}_N=A^{(2)}_{2r}$ is the unique case in which 
$\kappa_a=1$ for any $a \in I_{\sigma}$.
By $U_q(X^{(\kappa)}_N)$ we mean the 
quantized universal enveloping algebra \cite{J1}
of the twisted affine Lie algebra of type $X^{(\kappa)}_N$ \cite{Ka}.

Let us proceed to the unrestricted T-systems.
Choose  $\hbar\in
\mathbb{C}\setminus 2\pi\sqrt{-1}\mathbb{Q}$
arbitrarily.
The unrestricted T-system for $U_q(X^{(\kappa)}_N)$
is the following relations for commuting variables 
$\{T^{(a)}_m(u)\mid a \in I_\sigma, m \in \Z_{\ge 1},
u \in {\mathbb C}_{\kappa_a\hbar}\}$,
where $\Omega = 2\pi \sqrt{-1}/\kappa \hbar$,
and  $T^{(0)}_m(u)=T^{(a)}_0(u)=1$ if they 
occur on the RHS in the relations: 

For $X^{(\kappa)}_N=A^{(2)}_{2r-1}$,
\begin{align}
T^{(a)}_m(u-1)T^{(a)}_m(u+1)&=T^{(a)}_{m-1}(u)T^{(a)}_{m+1}(u)
\label{ttao}\\
&\quad + T^{(a-1)}_m(u)T^{(a+1)}_m(u)
\quad (1 \le a \le r-1),\notag\\
T^{(r)}_m(u-1)T^{(r)}_m(u+1)&=T^{(r)}_{m-1}(u)T^{(r)}_{m+1}(u)
+ T^{(r-1)}_m(u)
T^{(r-1)}_m(u+\Omega).\notag
\end{align}

For $X^{(\kappa)}_N=A^{(2)}_{2r}$,
\begin{align}
T^{(a)}_m(u-1)T^{(a)}_m(u+1)&=T^{(a)}_{m-1}(u)T^{(a)}_{m+1}(u)
\label{ttae}\\
&\quad + T^{(a-1)}_m(u)T^{(a+1)}_m(u)
\quad (1 \le a \le r-1),\notag\\
T^{(r)}_m(u-1)T^{(r)}_m(u+1)&=T^{(r)}_{m-1}(u)T^{(r)}_{m+1}(u)
+ T^{(r-1)}_m(u)
T^{(r)}_m(u+\Omega).\notag
\end{align}

For $X^{(\kappa)}_N=D^{(2)}_{r+1}$,
\begin{align}
T^{(a)}_m(u-1)T^{(a)}_m(u+1)&=T^{(a)}_{m-1}(u)T^{(a)}_{m+1}(u)
\label{ttdr}\\
&\quad + T^{(a-1)}_m(u)T^{(a+1)}_m(u)
\quad (1 \le a \le r-2),\notag\\
T^{(r-1)}_m(u-1)T^{(r-1)}_m(u+1)&=T^{(r-1)}_{m-1}(u)T^{(r-1)}_{m+1}(u)
\notag\\
&\quad + T^{(r-2)}_m(u)T^{(r)}_m(u)
T^{(r)}_m(u+\Omega),\notag\\
T^{(r)}_m(u-1)T^{(r)}_m(u+1)&=T^{(r)}_{m-1}(u)T^{(r)}_{m+1}(u)
+ T^{(r-1)}_m(u).\notag
\end{align}

For $X^{(\kappa)}_N=E^{(2)}_6$,
\begin{align}
T^{(1)}_m(u-1)T^{(1)}_m(u+1)&=T^{(1)}_{m-1}(u)T^{(1)}_{m+1}(u)
+ T^{(2)}_m(u),\label{tte6}\\
T^{(2)}_m(u-1)T^{(2)}_m(u+1)&=T^{(2)}_{m-1}(u)T^{(2)}_{m+1}(u)
+ T^{(1)}_m(u)T^{(3)}_m(u),\notag\\
T^{(3)}_m(u-1)T^{(3)}_m(u+1)&=T^{(3)}_{m-1}(u)T^{(3)}_{m+1}(u)
+ T^{(2)}_m(u)T^{(2)}_m(u+\Omega)T^{(4)}_m(u),
\notag\\
T^{(4)}_m(u-1)T^{(4)}_m(u+1)&=T^{(4)}_{m-1}(u)T^{(4)}_{m+1}(u)
+ T^{(3)}_m(u).\notag
\end{align}

For $X^{(\kappa)}_N=D^{(3)}_4$,
\begin{align}
T^{(1)}_m(u-1)T^{(1)}_m(u+1)&=T^{(1)}_{m-1}(u)T^{(1)}_{m+1}(u)
+ T^{(2)}_m(u),\label{ttd34}\\
T^{(2)}_m(u-1)T^{(2)}_m(u+1)&=T^{(2)}_{m-1}(u)T^{(2)}_{m+1}(u)
\notag\\
&\qquad
+ T^{(1)}_m(u)
T^{(1)}_m(u-\Omega)
T^{(1)}_m(u+\Omega).\notag
\end{align}

The domain $\mathbb{C}_{\kappa_a\hbar}$ of the parameter
$u$ effectively imposes the following periodicity:
\begin{equation}\label{tome}
T^{(a)}_m(u)=
\begin{cases}
T^{(a)}_m(u+ \kappa\Omega)& \sigma(a)\neq a,\\
T^{(a)}_m(u+ \Omega)& \sigma(a)= a.
\end{cases}
\end{equation}

\begin{remark}\label{re:fold}
The T-system for $U_q(X^{(\kappa)}_N)$ is obtainable from 
the T-system for $\mathfrak{g}=X_N$ by a folding in the following sense.
Denoting the variable in the latter by ${\tilde T}^{(a)}_m(u)$
with $a\in I$, one imposes the condition
${\tilde T}^{(\sigma^k(a))}_m(u) = {\tilde T}^{(a)}_m(u+k\Omega)$
and identifies ${\tilde T}^{(a)}_m(u)$ with $a \in I_\sigma \subset I$
as the variable $T^{(a)}_m(u)$ in the former.
The same remark applies also to the Y-system given in what follows.
\end{remark}

The unrestricted Y-system for  
$U_q(X^{(\kappa)}_N)$
is the following relations for
the commuting variables 
$\{Y^{(a)}_m(u)\mid a \in I_\sigma, m \in \Z_{\ge 1},
u \in {\mathbb C}_{\kappa_a\hbar}\}$,
where $\Omega = 2 \pi \sqrt{-1}/\kappa\hbar$,
and $Y^{(0)}_m(u)=Y^{(a)}_0(u)^{-1}=0$ if they 
occur on the RHS in the relations: 

For $X^{(\kappa)}_N=A^{(2)}_{2r-1}$,
\begin{align}
Y^{(a)}_m(u-1)Y^{(a)}_m(u+1) & = 
\frac{(1+Y^{(a-1)}_m(u))(1+Y^{(a+1)}_m(u))}
{(1+Y^{(a)}_{m-1}(u)^{-1})(1+Y^{(a)}_{m+1}(u)^{-1})}
\label{tyao}\\
&\hskip130pt(1 \le a \le r-1),\notag\\
Y^{(r)}_m(u-1)Y^{(r)}_m(u+1) & = 
\frac{(1+Y^{(r-1)}_m(u))
(1+Y^{(r-1)}_m(u+\Omega))}
{(1+Y^{(r)}_{m-1}(u)^{-1})(1+Y^{(r)}_{m+1}(u)^{-1})}.\notag
\end{align}
For $X^{(\kappa)}_N=A^{(2)}_{2r}$,
\begin{align}
Y^{(a)}_m(u-1)Y^{(a)}_m(u+1) & = 
\frac{(1+Y^{(a-1)}_m(u))(1+Y^{(a+1)}_m(u))}
{(1+Y^{(a)}_{m-1}(u)^{-1})(1+Y^{(a)}_{m+1}(u)^{-1})}
\label{tyae}\\
&\hskip130pt(1 \le a \le r-1),\notag\\
Y^{(r)}_m(u-1)Y^{(r)}_m(u+1) & = 
\frac{(1+Y^{(r-1)}_m(u))
(1+Y^{(r)}_m(u+\Omega))}
{(1+Y^{(r)}_{m-1}(u)^{-1})(1+Y^{(r)}_{m+1}(u)^{-1})}.\notag
\end{align}
For $X^{(\kappa)}_N=D^{(2)}_{r+1}$,
\begin{align}
Y^{(a)}_m(u-1)Y^{(a)}_m(u+1) & = 
\frac{(1+Y^{(a-1)}_m(u))(1+Y^{(a+1)}_m(u))}
{(1+Y^{(a)}_{m-1}(u)^{-1})(1+Y^{(a)}_{m+1}(u)^{-1})}
\label{tydr}\\
&\hskip130pt(1 \le a \le r-2),\notag\\
Y^{(r-1)}_m(u-1)Y^{(r-1)}_m(u+1) & = 
\frac{(1+Y^{(r-2)}_m(u))(1+Y^{(r)}_m(u))
(1+Y^{(r)}_m(u+\Omega))}
{(1+Y^{(r-1)}_{m-1}(u)^{-1})(1+Y^{(r-1)}_{m+1}(u)^{-1})},
\notag\\
Y^{(r)}_m(u-1)Y^{(r)}_m(u+1) & = 
\frac{1+Y^{(r-1)}_m(u)}
{(1+Y^{(r)}_{m-1}(u)^{-1})(1+Y^{(r)}_{m+1}(u)^{-1})}.\notag
\end{align}
For $X^{(\kappa)}_N=E^{(2)}_6$,
\begin{align}
Y^{(1)}_m(u-1)Y^{(1)}_m(u+1) & = 
\frac{1+Y^{(2)}_m(u)}
{(1+Y^{(1)}_{m-1}(u)^{-1})(1+Y^{(1)}_{m+1}(u)^{-1})},
\label{tye6}\\
Y^{(2)}_m(u-1)Y^{(2)}_m(u+1) & = 
\frac{(1+Y^{(1)}_m(u))(1+Y^{(3)}_m(u))}
{(1+Y^{(2)}_{m-1}(u)^{-1})(1+Y^{(2)}_{m+1}(u)^{-1})},
\notag\\
Y^{(3)}_m(u-1)Y^{(3)}_m(u+1) & = 
\frac{(1+Y^{(2)}_m(u))
(1+Y^{(2)}_m(u+\Omega))
(1+Y^{(4)}_m(u))}
{(1+Y^{(3)}_{m-1}(u)^{-1})(1+Y^{(3)}_{m+1}(u)^{-1})},
\notag\\
Y^{(4)}_m(u-1)Y^{(4)}_m(u+1) & = 
\frac{1+Y^{(3)}_m(u)}
{(1+Y^{(4)}_{m-1}(u)^{-1})(1+Y^{(4)}_{m+1}(u)^{-1})}.\notag
\end{align}
For $X^{(\kappa)}_N=D^{(3)}_4$,
\begin{align}
Y^{(1)}_m(u-1)Y^{(1)}_m(u+1) & = 
\frac{1+Y^{(2)}_m(u)}
{(1+Y^{(1)}_{m-1}(u)^{-1})(1+Y^{(1)}_{m+1}(u)^{-1})},
\label{tyd34}\\
Y^{(2)}_m(u-1)Y^{(2)}_m(u+1) & = 
\frac{(1+Y^{(1)}_m(u))
(1+Y^{(1)}_m(u-\Omega))
(1+Y^{(1)}_m(u+\Omega))}
{(1+Y^{(2)}_{m-1}(u)^{-1})(1+Y^{(2)}_{m+1}(u)^{-1})}.
\notag
\end{align}

\subsection{Restriction and relations 
between T and Y-systems}\label{ss:res2}

Fix an integer $\ell \ge 2$ called {\em level}.
The level $\ell$ restricted T-system for $U_q(X^{(\kappa)}_N)$
(with the unit boundary condition) is the relations 
(\ref{ttao})--(\ref{ttd34}) naturally restricted to 
$\{T^{(a)}_m(u)\mid 
a \in I_\sigma, 1 \le m \le \ell-1, 
u \in {\mathbb C}_{\kappa_a\hbar}\}$
by imposing $T^{(a)}_{\ell}(u)=1$ (the unit boundary condition).

The level $\ell$ restricted Y-system for $U_q(X^{(\kappa)}_N)$ 
is the relations (\ref{tyao})--(\ref{tyd34}) naturally restricted to 
$\{Y^{(a)}_m(u)\mid 
a \in I_\sigma, 1 \le m \le \ell-1, 
u \in {\mathbb C}_{\kappa_a\hbar}\}$
by imposing $Y^{(a)}_{\ell}(u)^{-1}=1$.

The properties stated in Theorem \ref{th:ty1} and 
Remark \ref{re:g} also hold 
between the T and Y-systems of 
for $U_q(X^{(\kappa)}_N)$.
On the other hand, the correspondence like 
(\ref{ttm0}) and (\ref{ttm1}) in the untwisted case 
is not valid.

\subsection{\mathversion{bold}$U_q(sl(r|s))$ case}\label{ss:sup}

Among a variety of Lie super algebras, 
we present the T-system and the Y-system related to 
$U_q(sl(r|s))$ as a typical example. 
For brevity we employ the following notation within this subsection.
\begin{equation}\label{j:fh}
H_{r,s} = (\Z_{>0}\times \Z_{>0})\setminus
(\Z_{\ge r}\times \Z_{\ge s}),\quad
\overline{H}_{r,s} = 
(\Z_{\ge 0}\times \Z_{\ge 0})\setminus
(\Z_{> r}\times \Z_{> s}).
\end{equation}
These sets are often called fat hook.
The T-system for $U_q(sl(r|s))$ is the following relations among
the commuting variables 
$\{T^{(a)}_m(u)\mid (a,m) \in \overline{H}_{r,s}, u \in U\}$.
\begin{align}
&T^{(a)}_m(u-1)T^{(a)}_m(u+1)=
T^{(a-1)}_m(u)T^{(a+1)}_m(u)+ T^{(a)}_{m-1}(u)T^{(a)}_{m+1}(u),
\label{j:t0}\\
&T^{(r)}_{s+1}(u) = T^{(r+1)}_s(u).   \label{j:ex}
\end{align}
Relation (\ref{j:t0}) is imposed for all 
$(a,m) \in \overline{H}_{r,s}\setminus \{(0,0)\}$, where if any
$T^{(b)}_k(u)$ with $(b,k) \not\in  \overline{H}_{r,s}$ is 
contained in the RHS, it should be understood as $0$.
\begin{equation}\label{j:soto}
T^{(b)}_k(u) = 0 \quad \text{if}\;\;(b,k) \not\in  
\overline{H}_{r,s}.
\end{equation}
This leads to the simple recursion relations for the sequences 
corresponding to the boundary $\overline{H}_{r,s}\setminus H_{r,s}$.
\begin{equation}\label{j:br}
\begin{split}
T^{(a)}_m(u-1) T^{(a)}_m(u+1) 
&= T^{(a)}_{m+1} (u) T^{(a)}_{m-1} (u)\qquad
(a,m) \in (r,\Z_{>s}) \cup (0,\Z_{>0}),  \\
T^{(a)}_{m}(u-1)  T^{(a)}_{m}(u+1) 
&= T^{(a-1)}_{m} (u) T^{(a+1)}_{m} (u)    \quad
(a,m) \in (\Z_{>r},s) \cup (\Z_{>0},0).   
\end{split}
\end{equation}
The extra relation (\ref{j:ex}) leads by induction to 
\begin{equation}\label{j:15}
T^{(r)}_{s+a}(u) = T^{(r+a)}_s(u)\quad a \ge 0.
\end{equation}
In the applications, 
the variables appearing in (\ref{j:br}) and (\ref{j:15}) are
chosen appropriately reflecting the normalization of the system.
The relation (\ref{j:t0}) is the same as type $A$ case.
The essential difference from it lies in (\ref{j:soto})
and (\ref{j:15}).

Let us proceed to the Y-system.
We assume $r\ge s \ge 2$ first.
The Y-system for $U_q(sl(r|s))$ is the following relations among
the commuting variables 
$\{\Upsilon^{(a)}_1(u), \Upsilon^{(a)}_2(u)
\mid a\in \Z_{\ge 1}, u \in U\}
\cup 
 \{Y^{(a)}_m(u)\mid (a,m) \in H_{r,s}, u \in U\}$.
\begin{align}
Y^{(a)}_m(u-1)Y^{(a)}_m(u+1) 
&=\frac{(1+Y^{(a)}_{m+1}(u))(1+Y^{(a)}_{m-1}(u))}
{(1+Y^{(a-1)}_{m}(u)^{-1})(1+Y^{(a+1)}_{m}(u)^{-1})}
\quad (a,m) \in H_{r,s},\label{j:y1}\\
{\Upsilon}_1^{(1)} (u-1){\Upsilon}_1^{(1)} (u+1)  
&={\Upsilon}_2^{(2)} (u) (1+Y^{(1)}_{s-1}(u)),  \label{j:y2}\\
{\Upsilon}_1^{(a)} (u-1){\Upsilon}_1^{(a)} (u+1) 
&={\Upsilon}_1^{(a+1)} (u){\Upsilon}_1^{(a-1)} (u) 
\frac{1+Y^{(a)}_{s-1}(u)} {1+Y^{(a-1)}_{s}(u)} 
\quad a \ge 2, \label{j:y3}\\
{\Upsilon}_2^{(a)}(u-1){\Upsilon}_2^{(a)}(u+1) 
&={\Upsilon}_2^{(a+1)}(u){\Upsilon}_2^{(a-1)}(u)(1+Y^{(a)}_{s-1}(u))
\quad a \ge 2,  \label{j:y4}\\
{\Upsilon}_1^{(1)}(u)  &= {\Upsilon}_2^{(1)}(u),\quad
{\Upsilon}_1^{(r)}(u) =Y^{(r)}_s(u).\label{j:con}
\end{align}
On the RHS of these relations,  any factor 
$(1+Y^{(b)}_k(u)^{\pm 1})$ with $(b,k) \not\in H_{r,s}$ 
is to be understood as $1$. 
When $r>s=1$,  the equations (\ref{j:y2}) and (\ref{j:y4}) are absent. 
The Y-system for $s\ge r \ge 2$ 
is given by (\ref{j:y1})--(\ref{j:con}) by interchanging $r$ and $s$.

There is a simple relation between the T-system and Y-system 
analogous to Theorem \ref{th:ty1}. 
Suppose that $T^{(a)}_m(u)$ is a solution to the T-system.  
Then the combinations
\begin{align}
Y^{(a)}_m(u) &=\frac{T^{(a)}_{m+1}(u)T^{(a)}_{m-1}(u)}
{T^{(a+1)}_{m}(u)T^{(a-1)}_{m}(u)}
\qquad (a,m) \in H_{r,s},\\
{\Upsilon}_1^{(a)}(u)&=\frac{T^{(a)}_{s-1}(u)}{T^{(a-1)}_s(u)}, 
\qquad
{\Upsilon}_2^{(a)}(u) =\frac{T^{(a)}_{s-1}(u)}{T^{(0)}_{s+a-1}(u)}
\end{align}
satisfy the Y-system.
In particular, (\ref{j:con}) holds due to (\ref{j:ex}).
When $s\ge r\ge 2$,  the parallel fact holds by 
interchanging $r$ and $s$ and the role of indices $a$ and $m$ 
in $T^{(a)}_m(u)$ and $Y^{(a)}_m(u)$ everywhere. 
In view of the symmetry of the sets (\ref{j:fh}),
we do not introduce the level restriction. 

\begin{remark}
The above set of relations seems different 
from those given in \cite{JKS3}  for $gl(2|2)$, where 
a special relation $T^{(r)}_{s-2} \propto T^{(r-2)}_s$ 
valid only for this case is utilized.
Thanks to this,  ${\Upsilon}_1^{(a)} \, (a\ne r)$ and 
 ${\Upsilon}_2^{(a)}$ are not necessarily needed.  
The two sets of Y-systems nevertheless 
lead to an identical set of  
thermodynamic Bethe ansatz equations\footnote{There are 
typos in \cite{JKS3} for $gl(2|2)$,
around (5.4) and (5.5).}. 
The Y-system (\ref{j:y1})--(\ref{j:con}) is consistent with 
the thermodynamic Bethe ansatz equations in \cite{Sa} 
under the identification $N,K \leftrightarrow r,s$ and 
\begin{align*}
&Y^{(a)}_{s-m}={\rm e}^{-\zeta^{(a)}_{m}/T}\,(1\le a, 1\le m \le s-1), 
\qquad
Y^{(a)}_s= {\rm e}^{-\epsilon_{a}/T}\,(1\le a \le r),   \\
&Y^{(a)}_{s+j}={\rm e}^{\kappa^{(j)}_a/T}   \,(1\le j, 1\le a \le r-1).
\end{align*}
\end{remark}

\subsection{Bibliographical notes}\label{ss:bty}

The Hirota relation (\ref{ta}) for transfer matrices 
in the $A_r$ case first appeared in \cite{KNS2}, 
where the T-system for ${\mathfrak g}$ was introduced
as functional relations
among the commuting transfer matrices 
$\{T^{(a)}_m(u)\}$.
The models relevant to the unrestricted and restricted 
versions are the vertex 
and the restricted solid-on-solid (RSOS) type models, respectively. 
In such a setting, T-system acquires some 
scalar coefficients depending 
on the normalization of $T^{(a)}_m(u)$ as in Remark \ref{re:g}.
The unit boundary condition is also modified accordingly.
Actually in \cite{KNS2}, 
the restricted T-system was introduced by imposing 
a slightly weaker condition $T^{(a)}_{t_a\ell+1}(u)=0$.
The T-system for the twisted case was 
introduced in \cite{KS2} in a similar context.
Our presentation here follows \cite{IIKNS, Her4}.
The T-system unifies the many functional relations 
studied earlier individually. 
See Sections \ref{s:ctm}-\ref{s:qg} for more details.

The level $\ell$ 
restricted Y-system for ${\mathfrak g}$ was introduced in
\cite{Z1} for simply laced ${\mathfrak g}$ with $\ell=2$
as a universal property of the thermodynamic
Bethe ansatz (TBA) equation in the context of 
integrable perturbations of conformal field theories.
Then, it was extended to the general case in \cite{KN1}
based on the TBA equation 
related to RSOS models for $U_q(\hat{\mathfrak g})$ \cite{Ku}.
This procedure is detailed in Section \ref{s:y}.
The Y-system for simply laced ${\mathfrak g}$ was also given in \cite{RTV}
independently.
For more literatures in the similar context, see Section \ref{ss:byq}.
The transformation (\ref{ty1}) between the 
T and Y-systems first appeared in \cite{KP2}
for the simplest case ${\mathfrak g}=A_1$,
and extended in \cite{KNS2} to general ${\mathfrak g}$.
T-systems related to Lie super algebras and super symmetric models
have been studied in various contexts.
See for example \cite{JKS3,T2,T3,T4,KSZ,Heg,GKT} and references therein.
  
\section{T-system among commuting transfer matrices}\label{s:ctm}

The aim of this section is to introduce the basic examples of 
solvable lattice models, both vertex and 
restricted solid-on-solid (RSOS) type,
and demonstrate how the T-system is obtained for 
their transfer matrices in connection to the fusion procedure.
Although these issues are nowadays well recognized to be  
intimately related to the representation theory of quantum groups,
we defer such a description to Section \ref{s:qg} 
avoiding too many definitions from the beginning. 
Our presentation here is based on explicit 
calculations in trigonometric parameterization
along the simplest example from ${\mathfrak g}= A_1$
The exception is the last subsection \ref{ss:vrg},
where we will formally argue the general features of those models 
associated with general ${\mathfrak g}$
quoting known facts on 
Kirillov-Reshetikhin modules and Q-system
from Sections \ref{s:qg}, \ref{ss:qs} and \ref{ss:qru}.  

\subsection{Vertex models and fusion}\label{ss:vm}

We recall the 6 vertex model and its fusion without 
much recourse to the representation theory\footnote{
Some terminology will be refined after (\ref{esv}).}. 
Consider the two dimensional square lattice, 
where each edge is assigned with a local variable 
belonging to $\{1,2\}$.
Around each vertex, we allow the following 6 configurations 
with the respective Boltzmann weights.
\begin{equation}\label{6v}
\begin{picture}(300,62)(0,-3)
\multiput(14.5,38)(50,0){6}{
\put(-12,0){\line(1,0){24}}
\put(0,-10){\line(0,1){20}}
}

\put(12,51){1}
\put(-5,34){1}
\put(29,34){1}
\put(12,16){1}

\put(62,51){2}
\put(45,34){2}
\put(79,34){2}
\put(62,16){2}

\put(112,51){1}
\put(95,34){2}
\put(129,34){2}
\put(112,16){1}

\put(162,51){2}
\put(145,34){1}
\put(179,34){1}
\put(162,16){2}

\put(212,51){2}
\put(195,34){1}
\put(229,34){2}
\put(212,16){1}

\put(262,51){1}
\put(245,34){2}
\put(279,34){1}
\put(262,16){2}

\put(0,0){$1-q^2z$}
\put(50,0){$1-q^2z$}
\put(100,0){$q(1-z)$}
\put(150,0){$q(1-z)$}

\put(200,0){$z(1-q^2)$}

\put(255,0){$1-q^2$.}
\end{picture}
\end{equation}
The other 10 configurations are assigned with 0 Boltzmann weight.
Let $V=\C v_1 \oplus \C v_2$. 
Then (\ref{6v}) is arranged in 
the quantum $R$ matrix $R(z) \in {\rm End}(V\otimes V)$ as
\begin{equation}\label{eqa:r}
\begin{split}
&R(z) = a(z)\sum_iE_{ii}\otimes E_{ii} +
b(z)\sum_{i\neq j}E_{ii}\otimes E_{jj} 
+ c(z)\Bigl(z\sum_{i<j}+\sum_{i>j}\Bigr)
E_{ji}\otimes E_{ij},\\
&a(z) = 1-q^2z,\quad b(z) = q(1-z),\quad c(z) = 1-q^2.
\end{split}
\end{equation}
Here the indices run over $\{1,2\}$ and 
$E_{ij}$ is the matrix unit acting as 
$E_{ij}v_k = \delta_{jk}v_i$.
The $R$ matrix $R(z)$ is associated with 
the quantum affine algebra $U_q=U_q(A^{(1)}_1)$ \cite{J1}.
In fact, ${\check R}(z):=PR(z)$ commutes with $\Delta(U_q)$,
where $P$ denotes the transposition of the 
components\footnote{The asymmetry between the last two in 
(\ref{6v}) is due to our choice of the coproduct (\ref{cpr}).
It fits the crystal base theory making 
the limit $q\rightarrow 0$ of (\ref{wjk}) well defined,
although this fact will not be used in this review.}.
A more detailed account will be given in Section \ref{ss:rf}.
Schematically (\ref{eqa:r}) is expressed as
\begin{equation}\label{re}
R(z) = \sum_{ijkl}\Bigl(
\begin{picture}(35,22)(-17,-3)
\put(-2,13){$l$}\put(-17,-3){$j$}\put(3,-7){$z$}
\put(12,-3){$i$}\put(-2,-19){$k$}\drawline(0,-3)(3,0)
\put(-10,0){\line(1,0){20}}
\put(0,-10){\line(0,1){20}}
\end{picture}
\Bigr)E_{ij}\otimes E_{kl},
\qquad
{\check R}(z) = \sum_{ijkl}\Bigl(
\begin{picture}(35,22)(-17,-3)
\put(-2,13){$l$}\put(-17,-3){$j$}\put(3,-7){$z$}
\put(12,-3){$k$}\put(-2,-19){$i$}\drawline(0,-3)(3,0)
\put(-10,0){\line(1,0){20}}
\put(0,-10){\line(0,1){20}}
\end{picture}
\Bigr)E_{ij}\otimes E_{kl},
\end{equation}
where the $z$ dependence is exhibited.
The Yang-Baxter equation 
\begin{equation*}
R_{23}(z')R_{13}(z)R_{12}(z/z')
= R_{12}(z/z')R_{13}(z)R_{23}(z')
\end{equation*}
holds \cite{Ba3}, where the indices signify the 
components in the tensor product as
$\overset{1}{V}\otimes \overset{2}{V} \otimes 
\overset{3}{V}$ 
on which the both sides act.
It is depicted as  

\begin{equation}\label{ybe}
\begin{picture}(200, 60)(0,-30)

\put(0,10){\line(2,-1){60}}
\put(0,-10){\line(2,1){60}}

\put(46,-30){\line(0,1){60}}

\drawline(46,9)(50,15)

\drawline(24,2)(24,-2)

\drawline(46,-18)(50,-15)

\put(29,-1){$\scriptstyle{z\!/\!z'}$}

\put(50,7){$\scriptstyle{z}$}

\put(48,-24){$\scriptstyle{z'}$}

\put(80,0){\makebox(0,0){$=$}}

\multiput(157,0)(0,0){1}{

\put(-46,-30){\line(0,1){60}}

\put(0,10){\line(-2,-1){60}}

\put(0,-10){\line(-2,1){60}}

\drawline(-46,8)(-42,11)\put(-45,3){$\scriptstyle{z'}$}

\drawline(-16,2)(-16,-2)\put(-11,-1){$\scriptstyle{z\!/\!z'}$}

\drawline(-46,-17)(-42,-11.5)\put(-43,-18){$\scriptstyle{z}$}

}

\end{picture}
\end{equation}

Starting from the 6 vertex model \cite{Lie,Su1},
one can construct higher spin solvable vertex models by 
the fusion procedure \cite{KRS}.
Let $V_m$ be the irreducible $U_q$ module 
spanned by the $m$ fold $q-$symmetric tensors.
Concretely, $V_1=V$ and $V_m$ with 
$m \ge 2$ is realized as the quotient
$V^{\otimes m}/A$,
where 
$A= \sum_j V^{\otimes j} \otimes {\rm Im}\,{\check R}(q^{-2}) 
\otimes V^{\otimes m-2-j}$.
It is easy to see ${\rm Im}\,{\check R}(q^{-2}) 
= {\rm Ker}\,{\check R}(q^{2}) 
= \C(v_1\otimes v_2 - q v_2\otimes v_1)$.
We take the base vector of $V_m$ as  
$v_2^{\otimes x_2}\otimes v_1^{\otimes x_1} \mod A$, where 
$x_i \in \Z_{\ge 0}$ and $x_1 + x_2 = m$.
The base will also be denoted by $x=(x_1, x_2)$ for brevity. 
Obviously $\dim V_m = m+1$.

The Yang-Baxter equation (\ref{ybe}) with $z'=zq^{2}$ shows that 
${\rm Im}\,{\check R}(q^{-2})
\subset \overset{1}{V}\otimes \overset{2}{V}$ is preserved under the
action of  $R_{13}(zq^2)R_{23}(z)$.
Therefore its action on 
$(\overset{1}{V} \otimes \overset{2}{V})\otimes 
\overset{3}{V}$
can be restricted to 
$V_2\otimes V_1 
= \left((V\otimes V)/{\rm Im}\,{\check R}(q^{-2})\right)
\otimes V$.
Similarly, by using (\ref{ybe}) repeatedly, it is shown that 
the composition 
\begin{equation}\label{rcomp}
\frac{R_{1,m+1}(zq^{m-1})
R_{2,m+1}(zq^{m-3})\cdots R_{m,m+1}(zq^{-m+1})}
{a(zq^{m-3})a(zq^{m-5})\cdots a(zq^{-m+1})}
\end{equation}
can be restricted to $V_m \otimes V_1$. 
The resulting operator, the fusion $R$ matrix
$R^{(m,1)}(z) \in {\rm End}(V_m \otimes V_1)$, is given by
\begin{align}
R^{(m,1)}(z)(x\otimes v_j) 
&= \sum_{k=1,2}\Bigl(
\begin{picture}(35,22)(-17,-3)
\put(-2,14){$j$}\put(-17,-3){$x$}\put(3,-7){$z$}
\put(12,-3){$y$}\put(-2,-18){$k$}\drawline(3,0)(0,-3)
\put(-10,0.2){\line(1,0){20}}
\put(-10,0){\line(1,0){20}}
\put(-10,-0.2){\line(1,0){20}}
\put(0,-10){\line(0,1){20}}
\end{picture}
\Bigr)\;y \otimes v_k,\label{r1m}\\
\begin{picture}(35,22)(-17,-3)
\put(-2,14){$j$}\put(-17,-3){$x$}\put(3,-7){$z$}
\put(12,-3){$y$}\put(-2,-18){$k$}\drawline(3,0)(0,-3)
\put(-10,0.2){\line(1,0){20}}
\put(-10,0){\line(1,0){20}}
\put(-10,-0.2){\line(1,0){20}}
\put(0,-10){\line(0,1){20}}
\end{picture}
&=
\begin{cases}
q^{m-x_k}-q^{x_k+1}z & j=k,\\
(1-q^{2x_1})z & (j,k)=(2,1),\\
1-q^{2x_2} & (j,k)=(1,2),
\end{cases}\label{wjk}
\end{align}
where $y=(y_1,y_2)$ is specified by the weight conservation 
(so called ``ice rule") as 
$y_i = x_i + \delta_{i j}- \delta_{i k}$.
By the definition $R^{(1,1)}(z)=R(z)$ and 
(\ref{wjk}) reduces to (\ref{6v}) for $m=1$.
In the case $(j,k)=(1,2)$ for example, 
the matrix element 
$1-q^{2x_2} $ is obtained from the following calculation
($D=$denominator in (\ref{rcomp})):
\begin{equation}
\begin{picture}(280,100)(45,-48)

\multiput(0,0)(0,0){1}{
\put(22,-3){$\frac{1}{D}
\sum_{i=1}^{x_2}q^{i-1}$}
}

\put(110,-10){\color{red}\line(0,1){60}}
\put(110,-50){\color{blue}\line(0,1){40}}

\put(100,40){\color{red}\line(1,0){20}}
\drawline(110,37)(113,40)
\put(112.5,33){$\scriptstyle{zq^{-m+1}}$}

\put(104,28){\color{red}$\vdots$}
\put(100,25){\color{red}\line(1,0){20}}
\put(100,15){\color{blue}\line(1,0){20}}
\put(100,0){\color{blue}\line(1,0){20}}

\put(100,-10){\color{blue}\line(1,0){10}}
\put(110,-10){\color{red}\line(1,0){10}}

\put(100,-20){\color{blue}\line(1,0){20}}
\put(104,-32.5){\color{blue}$\vdots$}
\put(100,-35){\color{blue}\line(1,0){20}}

\drawline(110,-38)(113,-35)
\put(112,-42){$\scriptstyle{zq^{m-1}}$}

\drawline(95,41)(93,39)
\drawline(93,39)(93,26)\put(82,30){$\scriptstyle{x_1}$}
\drawline(93,26)(95,24)

\drawline(95,16)(93,14)
\drawline(93,14)(93,1)\put(79,5){$\scriptstyle{i\!-\!1}$}
\drawline(93,1)(95,-1)

\drawline(95,-19)(93,-21)
\drawline(93,-21)(93,-34)\put(73,-30){$\scriptstyle{x_2-i}$}
\drawline(93,-34)(95,-36)

\put(135,-3){$\displaystyle{
= \sum_{i=1}^{x_2}\frac{q^{i-1}
(1-q^2)a(zq^{m-1})
\prod_{n=x_1+1}^{x_1+i-1}b(zq^{-m-1+2n})}
{\prod_{n=x_1+1}^{x_1+i}a(zq^{-m-1+2n})}}$.}

\end{picture}
\end{equation}
The red and blue edges are assigned with the local states $1$ and $2$,
respectively.
The incoming state (left column) represents 
$v_2^{\otimes x_2}\otimes v_1^{\otimes x_1}$.
The factor $q^{i-1}$ accounts for the effect of 
rearranging the outgoing state into the base form by using the relation
$v_1\otimes v_2 \equiv qv_2\otimes v_1 \mod A$ as
\begin{equation*}
v_2^{\otimes x_2-i}\otimes v_1\otimes v_2^{\otimes i-1}
\otimes v_1^{\otimes x_1}
\equiv q^{i-1}v_2^{\otimes y_2}\otimes v_1^{\otimes y_1}
\in V_m,
\end{equation*}
where $y=(y_1,y_2)=(x_1+1,x_2-1)$ for $(j,k)=(1,2)$.

One can fuse $R^{(m,1)}(z)$ further 
along the other component of the tensor product 
in a completely parallel fashion.
The composition  
\begin{equation}\label{rcom2}
R^{(m,1)}_{0,n}(zq^{n-1})\cdots 
R^{(m,1)}_{0,2}(zq^{-n+3})R^{(m,1)}_{0,1}(zq^{-n+1})
\in {\rm End}(V_m \otimes V_1^{\otimes n})
\end{equation}
can be restricted to $V_m \otimes V_n$.
The result yields the quantum $R$ matrix 
$R^{(m,n)}(z) \in {\rm End}(V_m \otimes V_n)$.
The $R$ matrices so obtained again satisfy the Yang-Baxter equation
in ${\rm End}(V_l \otimes V_m \otimes V_n)$:
\begin{equation}\label{ybe2}
R^{(m,n)}_{23}(z')R^{(l,n)}_{13}(z)R^{(l,m)}_{12}(z/z')
= R^{(l,m)}_{12}(z/z')R^{(l,n)}_{13}(z)R^{(m,n)}_{23}(z').
\end{equation}
It is depicted as (\ref{ybe}) with the three lines to be interpreted 
as representing $V_l, V_m$ and $V_n$.

The quantum $R$ matrix $R^{(m,n)}(z)$ gives rise to  
a fusion vertex model on a planar square lattice by the 
same rule as diagrams (\ref{re}) and (\ref{r1m}).
The local variables on the horizontal and vertical edges are
taken from $V_m$ and $V_n$, respectively.

\subsection{Transfer matrices}\label{ss:tm}

Here we use the additive spectral parameter $u$ as well as 
the multiplicative one $z$.
They are related as $z=q^u$.
We introduce the row to row transfer matrix 
\begin{equation}\label{trme}
\begin{picture}(100,60)(-5,-7)
\put(-73,40){$\displaystyle{T_m(u) =\mathrm{Tr}_{V_m}
\left(R^{(m, s_N)}_{0,N}(z/w_N)\cdots 
R^{(m, s_1)}_{0,1}(z/w_1)\right)}$}

\put(-43,8){$= \displaystyle{\sum_{x \in V_m}}$}
\put(-8,8){$x$}\put(91,8){$x$.}
\put(0,10){\line(1,0){88}}
\put(12,20){\line(0,-1){20}}
\drawline(12,7)(15,10)\put(15,3){$\scriptstyle{z/w_1}$}

\put(37,20){\line(0,-1){20}}

\put(43.5,12){$\cdots$}

\put(65,20){\line(0,-1){20}}
\drawline(65,7)(68,10)\put(68,3){$\scriptstyle{z/w_N}$}
\end{picture}
\end{equation}
The horizontal line is associated with $V_m$ 
which is called the auxiliary space.
The trace over it corresponds to the periodic boundary condition.
There are $N$ vertical lines corresponding to 
$V_{s_1}\otimes \cdots \otimes V_{s_N}$ which is called the quantum space.
The $T_m(u)$ is a linear operator acting on the quantum space.
The data $s_i, w_i$ represent the inhomogeneity 
in the spins and coupling constants.

The first consequence of the Yang-Baxter equation (\ref{ybe2}) is 
the commutativity of the transfer matrices
acting on the common quantum space 
(common $s_i$ and $w_i$ in the present context)
\begin{equation}\label{cmut}
[T_m(u), T_n(v)]=0.
\end{equation}

Let us take $s_i=1$ for all $i$ for simplicity and 
demonstrate the functional relation 
\begin{equation}\label{shos}
\begin{split}
&T_1(u+1)T_1(u-1) = T_0(u)T_2(u) + g_1(u)\mathrm{id},\\
&T_0(u) = \prod_{i=1}^Na(z_i/q),\quad
g_1(u) = \prod_{i=1}^Na(z_iq)b(z_i/q),
\end{split}
\end{equation}
where $z_i = z/w_i$.
This corresponds to the T-system for $A_1$ 
(\ref{ta1}) with $m=1$ modified by a model dependent 
factors $T_0(u)$ and $g_1(u)$.
Consider the diagram for $T_1(u+1)T_1(u-1)$ corresponding to 
the matrix element for the transition
$v_{\alpha_1}\otimes \cdots \otimes v_{\alpha_N}
\mapsto 
v_{\beta_1}\otimes \cdots \otimes v_{\beta_N}$:
\begin{equation}\label{ttdi}
\begin{picture}(60,60)(0,-27)

\put(-46,0){$\displaystyle{\sum_{k,l=1,2}}$}

\put(-8,7){$k$}    \put(88,7){$k$}

\put(-7,-14){$l$} \put(88,-14){$l$}

\put(8,24){$\alpha_1$}\put(60,24){$\alpha_N$}

\put(0,10){\line(1,0){82}}
\put(12,20){\line(0,-1){20}}
\drawline(12,7)(15,10)\put(14,3){$\scriptstyle{z_1\!/\!q}$}

\put(38,12){$\cdots$}

\put(65,20){\line(0,-1){20}}
\drawline(65,7)(68,10)\put(67,3){$\scriptstyle{z_N\!/\!q}$}

\multiput(0,-20)(0,0){1}{
\put(0,10){\line(1,0){82}}
\put(12,20){\line(0,-1){20}}
\drawline(12,7)(15,10)\put(14,3){$\scriptstyle{z_1q}$}

\put(38,12){$\cdots$}

\put(65,20){\line(0,-1){20}}
\drawline(65,7)(68,10)\put(67,3){$\scriptstyle{z_Nq}$}
}

\put(8,-30){$\beta_1$}\put(61,-30){$\beta_N$}

\end{picture}
\end{equation}
Given $\alpha_i, \beta_i$, 
the sum over $k,l$ is regarded as the trace of an operator acting  
on the auxiliary space $V_1\otimes V_1$ horizontally.
The space  $V_1\otimes V_1$ possesses the invariant subspace
${\rm Im }\,{\check R}(q^{-2}) = \C(v_1\otimes v_2-q v_2\otimes v_1)$ 
which propagates to the right owing to the Yang-Baxter equation (\ref{ybe}).
In fact, the following identity can be checked directly.
\begin{equation}\label{lemv}
\begin{picture}(90,65)(70,-5)

\put(-51,33){$2$}
\put(-51,13){$1$}

\put(44,33){$1$}
\put(44,13){$2$}
\multiput(-30,25)(95,0){2}{

\put(-3,26){$\alpha$}

\put(-15,10){\line(1,0){30}}

\put(18,7){$k$}\put(18,-13){$l$}

\drawline(0,7)(3,10)
\put(3,4){$\scriptstyle{z\!/\!q}$}
\put(-15,-10){\line(1,0){30}}
\drawline(0,-13)(3,-10)
\put(3,-16){$\scriptstyle{zq}$}
\put(0,-22){\line(0,1){44}}

\put(-3,-31){$\beta$}
}

\put(9,21){$-$}

\put(27,21){$q \times$}

\put(100,21)
{$= \delta_{\alpha \beta} a(zq)b(z/q)\times
\begin{cases}
1 & (k,l)=(2,1),\\
-q & (k,l) = (1,2),\\
0 & \text{otherwise}.
\end{cases}$}
\end{picture}
\end{equation}
Thus ${\rm Im }\,{\check R}(q^{-2})$ 
contributes to $\mathrm{Tr}_{V_1 \otimes V_1}$ (\ref{ttdi}) as 
$\prod_{i=1}^N\delta_{\alpha_i, \beta_i} a(z_iq)b(z_i/q)$,
giving the second term in the RHS of (\ref{shos}).
The other contribution to the trace is from 
$(V_1\otimes V_1)/{\rm Im }\,{\check R}(q^{-2})=V_2$.
This is equal to $T_0(u)T_2(u)$ by the definition,
where the factor $T_0(u)$ 
is due to the denominator in (\ref{rcomp}) with $m=2$.
In this way one observes that the exact sequence 
\begin{equation}\label{esv}
0 \rightarrow {\rm Im }\,{\check R}(q^{-2}) 
\rightarrow V_1 \otimes V_1 \rightarrow V_2 \rightarrow 0
\end{equation}
plays a key role in deriving (\ref{shos}).

In Section \ref{ss:fdr}, we will introduce the 
Kirillov-Reshetikhin module $W^{(a)}_m(u)$ for general quantum affine algebra
$U_q(\hat{\mathfrak g})$.
The case ${\mathfrak g}=A_1$ relevant here, denoted by 
$W_m(u) = W^{(1)}_m(u)$, 
will be described explicitly in Section \ref{ss:rf}. 
In such a formalism, one endows each line in the diagrams like 
(\ref{ttdi})--(\ref{lemv}) with a spectral parameter $z=q^u$
which corresponds to a 
Kirillov-Reshetikhin module $W_m(u)$.
The $R$ matrix  
$R^{(m,n)}(z)\in \mathrm{End}(V_m\otimes V_n)$
is actually to be understood as
$R^{(m,n)}(z_1/z_2)\in \mathrm{End}(W_m(u_1)\otimes W_n(u_2))$
with $z_i = q^{u_i}$.
Up to an overall scalar, it is characterized by the intertwining property 
$\Delta(g)PR^{(m,n)}(z_1/z_2) =
PR^{(m,n)}(z_1/z_2) \Delta (g)$ 
where $g$ is any element from $U_q(A^{(1)}_1)$ and 
$\Delta$ is the coproduct (\ref{cpr})  \cite{J1}.
Accordingly, we say that the transfer matrix $T_m(u)$ (\ref{trme})
has the auxiliary space $W_m(u)$ and 
acts on the quantum space 
$W_{s_1}(v_1)\otimes \cdots \otimes W_{s_N}(v_N)$
with $w_i= q^{v_i}$.

The exact sequence (\ref{esv}) will also be refined into the one 
among tensor product of Kirillov-Reshetikhin modules.
See (\ref{wes2}).
The T-system relation 
$T_m(u+1)T_m(u-1)=T_{m+1}(u)T_{m-1}(u) + g_m(u) \mathrm{id}$
for general $m$ follows from Theorem \ref{th:es} with $n=j=m$. 
An additional feature here is that one actually needs to consider  
the central extension of $U_q(A^{(1)}_1)$
to properly cope with the factor $g_m(u)$.
We refer to \cite[section 2.2]{KNS2} for this point.
See also \cite{RW}.

To summarize, the Kirillov-Reshetikhin module of the 
quantum affine algebra and 
their exact sequence form the representation theoretical background 
for the $R$ matrix, fusion procedure and the T-system among 
commuting family of transfer matrices.

\subsection{Restricted solid-on-solid (RSOS) models 
and fusion}\label{ss:rsos}

Besides vertex models, 
there is another class of solvable lattice models called
Interaction Round Face (IRF or simply face) models \cite{Ba3}. 
The relation of the two classes of models
has been studied from various viewpoints \cite{Ba2,Pas,DJO,Fe,JKOS}.
Here we recall the 8 vertex solid-on-solid (8VSOS) model \cite{ABF}.
It is the fundamental example 
associated with $U_q(A^{(1)}_1)$ at $q$ a root of unity
and serves as the prototype of  
restricted solid-on-solid (RSOS) models.
It generalizes to $U_q(\hat{\mathfrak g})$ for any ${\mathfrak g}$
in principle.
We illustrate the fusion procedure \cite{DJKMO1} and 
the derivation of the simplest case of the 
T-system for the commuting transfer matrices \cite{BP,BR1}.
The contents are parallel with the 6 vertex model 
discussed in the previous subsection.
For simplicity we concentrate on 
the critical case\footnote{The RSOS models allow
elliptic Boltzmann weights in general. The critical case means the 
trigonometric case of them.
The fusion procedure and the T-system 
are equally valid in the elliptic case as well.}. 

Consider the two dimensional square lattice, where each site is 
assigned with a local state belonging to $\Z$.
On the two local states $a, b$ on neighboring sites,
the condition $|a-b|=1$ is imposed.
With the allowed configuration round a face, 
the following Boltzmann weights are assigned \cite{ABF}.
\begin{equation}\label{abf}
\begin{split}
W\!\left(\left. \begin{matrix}
a &a\!\mp\!1\\ a\!\pm\!1&a
\end{matrix}\right|u\right)
&=\frac{[2+u]_{q^{1/2}}}{[2]_{q^{1/2}}},\quad
W\!\left(\left. \begin{matrix}
a\!\pm\!1 & a \\ a & a\!\pm\!1
\end{matrix}\right|u\right)
=\frac{[2\xi+2a\mp u]_{q^{1/2}}}{[2\xi+2a]_{q^{1/2}}},\\
W\!\left(\left. \begin{matrix}
a\!\pm\!1 & a \\ a & a\!\mp\!1
\end{matrix}\right|u\right)
& =\frac{[2\xi+2a\pm 2]_{q^{1/2}}[u]_{q^{1/2}}}
{[2\xi+2a]_{q^{1/2}}[2]_{q^{1/2}}},
\end{split}
\end{equation}
where $u$ is the spectral parameter, $q$ and 
$\xi$ are generic constants which will be 
specialized when considering the restriction in Section \ref{ss:rest}.
The function $[u]_{q^{1/2}}$ is given by replacing 
$q\rightarrow  q^{1/2}$ in  
\begin{equation}\label{qnb}
[u]_q = \frac{q^u-q^{-u}}{q-q^{-1}}.
\end{equation}
The Boltzmann weights (\ref{abf}) are depicted as
\begin{equation}
\begin{picture}(50,30)(-17,7)

\put(-8,20){$b$}\put(23,21){$c$}
\put(-8,-5){$a$}\put(23,-6){$d$}
\put(0,20){\line(1,0){20}}
\put(0,0){\line(0,1){20}}\put(20,0){\line(0,1){20}}
\put(0,0){\line(1,0){20}}
\drawline(16,0)(20,4)\put(7,8){$u$}

\end{picture}
= W\!\left(\left. \begin{matrix}
b & c\\ a & d
\end{matrix}\right|u\right).
\end{equation}
It satisfies the (generalized) star-triangle relation \cite{Ba3} 
which plays the role of the Yang-Baxter equation in face models:
\begin{equation}\label{str}
\begin{split}
&\sum_g
W\!\left(\left. \begin{matrix}f & g\\ a & b \end{matrix}\right|u\right)
W\!\left(\left. \begin{matrix}e & d\\ f & g \end{matrix}\right|v\right)
W\!\left(\left. \begin{matrix}g & d\\ b & c \end{matrix}\right|u-v\right)\\
&=\sum_g
W\!\left(\left. \begin{matrix}e & d\\ g & c \end{matrix}\right|u\right)
W\!\left(\left. \begin{matrix}g & c\\ a & b \end{matrix}\right|v\right)
W\!\left(\left. \begin{matrix}f & e\\ a & g \end{matrix}\right|u-v\right).
\end{split}
\end{equation}
The sum over $g$ consists of at most two terms in each side because
of the neighboring condition, e.g. $|f-g|=|b-g|=|d-g|=1$ 
for the LHS.
We depict (\ref{str}) as
\begin{equation}\label{str2}
\begin{picture}(200,75)(-30,-37)

\drawline(0,0)(13,22.5)
\drawline(0,0)(-26,0)
\drawline(0,0)(13,-22.5)

\drawline(-6,0)(3.5,6)\drawline(23.4,-3.8)(23.4,3.8)
\drawline(8,-22.5)(11,-19)

\put(-2.2,-2.2){$\bullet$}

\put(-10,10){$\scriptstyle{v}$}
\put(5,-2){$\scriptstyle{u-v}$}
\put(-10,-15){$\scriptstyle{u}$}

\put(-16,26){$e$}\put(11,26){$d$}
\put(-34,-3){$f$}\put(30,-3){$c$}
\put(-16,-32){$a$}\put(11,-32){$b$}

\put(84,26){$e$}\put(111,26){$d$}
\put(66,-3){$f$}\put(130,-3){$c$}
\put(84,-32){$a$}\put(111,-32){$b$}

\drawline(-13,22.5)(13,22.5)
\drawline(-26,0)(-13,22.5)\drawline(13,22.5)(26,0)
\drawline(-26,0)(-13,-22.5)\drawline(13,-22.5)(26,0)
\drawline(-13,-22.5)(13,-22.5)

\drawline(87,22.5)(113,22.5)
\drawline(74,0)(87,22.5)\drawline(113,22.5)(126,0)
\drawline(74,0)(87,-22.5)\drawline(113,-22.5)(126,0)
\drawline(87,-22.5)(113,-22.5)

\put(48,-2){$=$}

\drawline(100,0)(87,22.5)
\drawline(100,0)(126,0)
\drawline(100,0)(87,-22.5)

\drawline(97.1,-4.6)(97.1,4.6)
\drawline(121,0)(124,3.5)
\drawline(108,-22.5)(115.5,-18)

\put(98,-2){$\bullet$}

\put(105,10){$\scriptstyle{u}$}
\put(77,-2){$\scriptstyle{u-v}$}
\put(106,-15){$\scriptstyle{v}$}

\end{picture}
\end{equation}
where $\bullet$ stands for the sum over the local state.
The faces drawn together are to be understood as the product of the
attached Boltzmann weights. 

One can apply the fusion procedure to the 8VSOS model \cite{DJKMO1}.
Note the properties
\begin{equation}\label{kmi}
\begin{picture}(360,50)(-10,-22)

\put(8,-2){$\scriptstyle{-2}$}
\put(0,0){\line(1,1){15}}\put(15,15){\line(1,-1){15}}
\put(0,0){\line(1,-1){15}}\put(15,-15){\line(1,1){15}}
\drawline(26,-4)(26,4)\put(28,-3){$\bullet = 0\;\;\therefore$}

\put(74,-2){$\scriptstyle{a+1}$}\put(115,-2){$-$}
\put(128,-2){$\scriptstyle{a-1}$}

\multiput(90,0)(55,0){2}{
\put(-5,22){$\scriptstyle{a}$}\put(-5,-25){$\scriptstyle{a}$}
\put(18,-3){$\bullet$}
\put(3,7){$\scriptstyle{u-1}$}\put(3,-12){$\scriptstyle{u+1}$}
\put(0,20){\line(1,0){20}}
\put(0,0){\line(1,0){20}}\drawline(16,0)(20,4)\drawline(16,-20)(20,-16)
\put(0,-20){\line(1,0){20}}
\put(0,-20){\line(0,1){40}}\put(20,-20){\line(0,1){40}}

}

\multiput(220,0)(0,0){1}{
\put(-47,-2){$\propto$}
\put(-17,18){$\scriptstyle{a}$}\put(-17,-21){$\scriptstyle{a}$}
\multiput(-30,0)(0,0){1}{
\put(8,-2){$\scriptstyle{-2}$}
\put(0,0){\line(1,1){15}}\put(15,15){\line(1,-1){15}}
\put(0,0){\line(1,-1){15}}\put(15,-15){\line(1,1){15}}
\drawline(26,-4)(26,4)}
\put(-2,-3){$\bullet$}\put(18,-3){$\bullet$}

\put(-5,22){$\scriptstyle{a}$}\put(-5,-25){$\scriptstyle{a}$}
\put(0,20){\line(1,0){20}}
\put(3,7){$\scriptstyle{u-1}$}\put(3,-12){$\scriptstyle{u+1}$}
\put(0,0){\line(1,0){20}}\drawline(16,0)(20,4)\drawline(16,-20)(20,-16)
\put(0,-20){\line(1,0){20}}
\put(0,-20){\line(0,1){40}}\put(20,-20){\line(0,1){40}}
}

\multiput(260,0)(0,0){1}{
\put(-14,-2){$=$}
\put(-5,22){$\scriptstyle{a}$}\put(-5,-25){$\scriptstyle{a}$}
\put(0,20){\line(1,0){20}}
\put(0,0){\line(1,0){20}}\drawline(16,0)(20,4)\drawline(16,-20)(20,-16)
\put(0,-20){\line(1,0){20}}
\put(3,-12){$\scriptstyle{u-1}$}\put(3,7){$\scriptstyle{u+1}$}
\put(0,-20){\line(0,1){40}}\put(20,-20){\line(0,1){40}}
\put(18,-3){$\bullet$}\put(48,-3){$\bullet$}
\multiput(20,0)(0,0){1}{
\put(8,-2){$\scriptstyle{-2}$}
\put(0,0){\line(1,1){15}}\put(15,15){\line(1,-1){15}}
\put(0,0){\line(1,-1){15}}\put(15,-15){\line(1,1){15}}
\drawline(26,-4)(26,4)}
\put(55,-2){$=0$}
}

\end{picture}
\end{equation}
where the second equality from the right is due to the star-triangle relation.
This implies that for $m=2$, the quantity 
\begin{equation}\label{wm1}
\begin{picture}(40,98)(0,-6)

\multiput(0,0)(0,22){5}{
\put(0,0){\line(1,0){22}}
}

\multiput(22,22)(0,22){3}{
\put(-3,-3){$\bullet$}
}

\put(23,89){$\scriptstyle{c}$}
\put(-6,89){$\scriptstyle{b}$}

\multiput(0,0)(0,22){4}{
\put(17.3,0){\line(1,1){5}}}

\put(-11,64){$\scriptstyle{\alpha_{1}}$}
\put(-11,43){$\scriptstyle{\alpha_{2}}$}
\put(-8,27){$\vdots$}\put(9,27){$\vdots$}
\put(-22,19){$\scriptstyle{\alpha_{m-1}}$}

\put(-6,-5){$\scriptstyle{a}$}
\put(23,-5){$\scriptstyle{d}$}

\put(2,78){$\scriptstyle{u-m}$}\put(8,72){$\scriptstyle{+1}$}
\put(2,55){$\scriptstyle{u-m}$}\put(8,49){$\scriptstyle{+3}$}
\put(2,10){$\scriptstyle{u+m}$}\put(8,4){$\scriptstyle{-1}$}

\put(0,0){\line(0,1){88}}
\put(22,0){\line(0,1){88}}
\end{picture}
\end{equation}
is independent of $\alpha_1,\ldots, \alpha_{m-1}$ 
as long as they are chosen so that $|\alpha_i-\alpha_{i+1}|=1$
($\alpha_0 = b,\, \alpha_m = a$).
The independence for general $m$ can be shown similarly.
Thus (\ref{wm1}) only depends on the local states $a,b,c,d$ 
on the corners.
We define the fused Boltzmann weight 
$W_{m,1}\!\left(\left. \begin{matrix}b & c\\ a & d 
\end{matrix}\right|u\right)$
to be (\ref{wm1}) divided by 
$\prod_{j=1}^{m-1}
\frac{[u+m+1-2j]_{q^{1/2}}}{[2]_{q^{1/2}}}$.
By induction on $m$, the following formulas are easily established
($W_{1,1}=W$).
\begin{equation}\label{wme}
\begin{split}
W_{m,1}\!\left(\left. \begin{matrix}b \;\;\;b\mp 1\\ a \;\;\; a\mp 1 
\end{matrix}\right|u\right) = &
\frac{\left[2\xi+a+b\pm m\right]_{q^{1/2}}
\left[1\pm(a-b)+u\right]_{q^{1/2}}}
{[2]_{q^{1/2}}[2\xi+2a]_{q^{1/2}}},\\
W_{m,1}\!\left(\left. \begin{matrix}b \;\;\; b\pm 1\\ a \;\;\; a\mp 1 
\end{matrix}\right|u\right) = &
\frac{\left[m\pm(a-b)\right]_{q^{1/2}}
\left[2\xi+a+b\pm 1\pm u\right]_{q^{1/2}}}
{[2]_{q^{1/2}}[2\xi+2a]_{q^{1/2}}}.
\end{split}
\end{equation}
One can fuse them further in the horizontal direction.
A similar argument shows that the quantity  
\begin{equation}\label{wmn}
\begin{picture}(200,60)(-50,-10)

\put(18,48){$\scriptstyle{\beta_1}$}
\put(39,48){$\scriptstyle{\beta_2}$}\put(49,48){$\cdots$}
\put(63,48){$\scriptstyle{\beta_{n-1}}$}

\put(-4,47){$\scriptstyle{b}$}\put(90,47){$\scriptstyle{c}$}
\put(-5,-6){$\scriptstyle{a}$}\put(91,-6){$\scriptstyle{d}$}

\put(49,18){$\cdots$}

\multiput(0,0)(22,0){4}{
\put(0,44){\line(1,0){22}}
\put(0,0){\line(0,1){44}}
\put(22,0){\line(0,1){44}}
\put(0,0){\line(1,0){22}}
\drawline(16,0)(22,6)
}

\put(2,20){$\scriptstyle{u-n}$}\put(8,13){$\scriptstyle{+1}$}
\put(24,20){$\scriptstyle{u-n}$}\put(30,13){$\scriptstyle{+3}$}
\put(68,20){$\scriptstyle{u+n}$}\put(74,13){$\scriptstyle{-1}$}

\put(19,-2){$\bullet$}\put(41,-2){$\bullet$}
\put(63,-2){$\bullet$}

\end{picture}
\end{equation}
is independent of $\beta_1,\ldots, \beta_{n-1}$
as long as 
$|\beta_i-\beta_{i+1}|=1$ ($\beta_0 = b,\, \beta_n = c$).
Here each rectangle stands for the weight 
$W_{m,1}$ (\ref{wme}) with the specified spectral parameters.
Thus we define $W_{m,n}\!\left(\left. 
\begin{matrix}b & c \\ a & d \end{matrix}\right|u\right)$ to be (\ref{wmn}).
By construction, it is zero unless
\begin{equation}\label{ad1}
b-a, c-d \in \{-m, -m+2,\ldots, m\},
\quad
c-b, d-a \in \{-n, -n+2,\ldots, n\}.
\end{equation}
The the star-triangle relation (\ref{str})
is generalized to 
\begin{equation}\label{strg}
\begin{split}
&\sum_g
W_{l,n}\left(\left. \begin{matrix}f & g\\ a & b \end{matrix}\right|u\right)
W_{m,n}\left(\left. \begin{matrix}e & d\\ f & g \end{matrix}\right|v\right)
W_{l,m}\left(\left. \begin{matrix}g & d\\ b & c \end{matrix}\right|u-v\right)\\
&=\sum_g
W_{l,n}\left(\left. \begin{matrix}e & d\\ g & c \end{matrix}\right|u\right)
W_{m,n}\left(\left. \begin{matrix}g & c\\ a & b \end{matrix}\right|v\right)
W_{l,m}\left(\left. \begin{matrix}f & e\\ a & g \end{matrix}\right|u-v\right).
\end{split}
\end{equation}

\subsection{Relation to vertex models}\label{ss:fvc}
The trigonometric face models under consideration are related to 
the 6 vertex model and its fusion in Section \ref{ss:vm} \cite{Pas}.
Let us explain it along the simplest cases (\ref{abf}) and (\ref{eqa:r}).
Let $a\in \Z_{\ge 1}$ and $V_{a-1}$ 
be the spin $\frac{a-1}{2}$ representation of 
$U_q(A_1)$ in Section \ref{ss:vm}\footnote{
Actually, $V_{a-1}$ can be 
the Verma module with the highest 
weight vector $v^{a-1}_1$ such that 
$k_1v^{a-1}_1=q^{a-1}v^{a-1}_1$ for generic $a$.}.
We use coproduct (\ref{cpr}) and the concrete form (\ref{hyog1}).
In the irreducible decomposition
$V_{a-1}\otimes V_1= \bigoplus_{b=a\pm 1} V_{b-1}$,
the highest weight vector $v_{a,b} \in V_{b-1}$ 
is given by 
$v_{a,a+1} = v^{a-1}_1\otimes v^1_1$ and 
$v_{a,a-1} = v^{a-1}_1\otimes v^1_2 -
q^{a-1}v^{a-1}_2\otimes v^1_1$.
Repeating this once more, one gets the 
highest weight vectors $v_{a,b,c}$ in the irreducible 
component $V_{c-1}$ in the decomposition of 
$V_{a-1}\otimes V_1\otimes V_1$
labeled with $a,b,c$ such that  $|a-b|=|b-c|=1$.
Explicitly, they read
\begin{equation}
\begin{split}
v_{a,a+1,a+2} &= v^{a-1}_1\otimes v^1_1\otimes v^1_1,\\
v_{a,a-1,a} & = [a-1]_q(v^{a-1}_1\otimes v^1_2\otimes v^1_1
-q^{a-1}v^{a-1}_2\otimes v^1_1\otimes v^1_1),\\
v_{a,a+1,a} &= [a]_qv^{a-1}_1\otimes v^1_1\otimes v^1_2
- q^{a-1}[a-1]_qv^{a-1}_2\otimes v^1_1\otimes v^1_1
-q^av^{a-1}_1\otimes v^1_2\otimes v^1_1,\\
v_{a,a-1,a-2} &=v^{a-1}_1\otimes v^1_2\otimes v^1_2
-q^{a-1}v^{a-1}_2\otimes v^1_1\otimes v^1_2
-q^{a-2}v^{a-1}_2\otimes v^1_2\otimes v^1_1\\
&\quad+q^{2a-4}v^{a-1}_3\otimes v^1_1\otimes v^1_1.
\end{split}
\end{equation}
Now consider the operator $1\otimes \check{R}(z)$ 
acting on $V_{a-1}\otimes V_1\otimes V_1$.
Since it commutes with $U_q(A_1)$, the images of the 
highest weight vectors are again highest.
The face Boltzmann weights can be extracted from the matrix elements
between those highest weight vectors as
\begin{equation}\label{par}
\left(1\otimes \check{R}(q^{u})\right)v_{a,b,c}
= -(q-q^{-1})q^{1+\frac{u}{2}}\sum_d 
W\!\left(\left. \begin{matrix}b & c\\ a & d
\end{matrix}\right|u\right)v_{a,d,c}.
\end{equation}
Here $\xi=0$ in the RHS and 
the sum is over $d$ such that $|a-d|=|d-c|=1$.
A similar relation holds also between the fusion models.
 
Conversely, one can deduce the 
$R$ matrix from the face Boltzmann weights as a limit
where the site variables or effectively $\xi$ tends to infinity.  
For instance, (\ref{wjk}) is obtained from (\ref{wme}) as
\begin{align}
&-(q-q^{-1})q^{\frac{m+1+u}{2}}
\frac{(a,b)_m(b,c)_1}{(d,c)_m(a,d)_1}
\lim_{q^{\xi}\rightarrow 0}
W_{m,1}\!\left(\left. \begin{matrix}b & c\\ a & d
\end{matrix}\right|u\right) =
\begin{picture}(35,22)(-17,-3)
\put(-2,14){$j$}\put(-17,-3){$x$}\put(3,-7){$z$}
\put(12,-3){$y$,}\put(-2,-18){$k$}\drawline(3,0)(0,-3)
\put(-10,0.2){\line(1,0){20}}
\put(-10,0){\line(1,0){20}}
\put(-10,-0.2){\line(1,0){20}}
\put(0,-10){\line(0,1){20}}
\end{picture}\label{wr}\\
&(a,b)_m = q^{\frac{1}{8}(a-b)^2+\frac{1}{4}m(a+b)},
\quad z = q^{u},\\
&x=(x_1,x_2)= \left(\frac{m-a+b}{2},\frac{m+a-b}{2}\right),
\quad
j=\frac{3+b-c}{2},\; k=\frac{3+a-d}{2}.
\end{align}
The factor on the LHS of (\ref{wr}) does not
spoil the star-triangle relation. 

\subsection{Restriction}\label{ss:rest}
The (fusion) face models constructed thus far possess local states ranging over the 
infinite set $\Z$ and are called unrestricted.
To obtain a model with finitely many local states,  we make {\em restriction}.
We introduce the integer $\ell \in \Z_{\ge 2}$ called {\em level},
and specialize the parameters as follows:
\begin{equation}\label{bd}
\xi=0,\quad q=\exp\left(\frac{\pi\sqrt{-1}}{\ell+2}\right),\quad
[u]_{q^{1/2}}= \frac{\sin\frac{\pi u}{2(\ell+2)}}
{\sin\frac{\pi}{2(\ell+2)}}.
\end{equation}
We further set  
$W_{m,n}\!\left(\left. \begin{matrix}b & c\\ a & d \end{matrix}\right|u\right)
=0$ unless the pairs $(a,b), (d,c)$ (resp. $(a,d), (b,c)$) are $m$-admissible 
(resp. $n$-admissible).
We say that a pair $(a,b)$ is $m$-admissible if 
\begin{align}
b-a &\in \{-m, -m+2,\ldots, m\}, \label{adm1}\\
a+b &\in \{m+2,m+4, \ldots, 2\ell+2-m\}.\label{adm2}
\end{align}
Notice that the admissibility forces $a,b \in \{1,2,\ldots, \ell+1\}$.
The resulting 
$W_{m,n}\!\left(\left. 
\begin{matrix}b & c\\ a & d \end{matrix}\right|u\right)$
with $a,b,c,d \in \{1,2,\ldots, \ell+1\}$
is called the restricted Boltzmann weight.
One may wonder if $[0]_{q^{1/2}}=[2\ell+4]_{q^{1/2}}=0$ may cause 
a divergence somewhere in the construction.
However it has been proved \cite{DJKMO1} that the restricted 
Boltzmann weights are well-defined and 
satisfy the star-triangle relation (\ref{strg}) 
among themselves\footnote{
Actually the statement holds for appropriately symmetrized $W_{m,n}$.
See \cite[section 2.2]{DJKMO1}.}.
In this way one obtains the 
level $\ell$ RSOS model whose local states belong to 
$\{1,2,\ldots, \ell+1\}$ and 
the fusion degree specified by $m$ and $n$. 

Let us comment on the admissibility condition 
among which the first one (\ref{adm1}) already appeared in (\ref{ad1}).
When $\ell\rightarrow \infty$, the admissibility reduces to  
the Clebsch-Gordan rule:
\begin{equation}\label{cgr}
V_{a-1}\otimes V_m 
= \bigoplus_{b-1=|a-1-m|,\ldots, a+m-3,a+m-1}V_{b-1}.
\end{equation}
The RHS contains precisely those $b$ such that $(a,b)$ is $m$-admissible
at $\ell=\infty$.
For $\ell$ finite, the necessity of $a+b\le 2\ell+2-m$
can be seen for example in the first Boltzmann weight in (\ref{wme}).
Under the specialization (\ref{bd}),
it contains the factor 
$\sin\left(\frac{\pi(a+b+m)}{2\ell+4}\right)$ in the numerator.
Thus the ``next"' $b$ for which $a+b=2\ell+4-m$ ``can not be reached".
Such a truncation is also observed at the 
level of characters associated with (\ref{cgr}).
Denoting 
the $q$-dimension of $V_{a-1}$ at root of unity 
by $\dim_q\!V_{a-1}
= \sin\left(\frac{\pi a}{\ell+2}\right)/
\sin\left(\frac{\pi}{\ell+2}\right)$, we have
\begin{equation}\label{mad}
(\dim_q\!V_{m})(\dim_q\!V_{a-1})
= \sum_{b: (a,b) \text{ is $m$-admissible}}
\dim_q\!V_{b-1}.
\end{equation}
This truncated decomposition is also known as the fusion rule 
in the SU(2) level $\ell$ WZW conformal field theory \cite{Ver}.

Finally we remark that given $\ell$, one can not fuse too much.
In fact, (\ref{adm1}) and (\ref{adm2}) fix
the admissible pairs to 
$\{(a,a)\mid 1 \le a \le \ell+1\}$ at $m=0$ 
and to $\{(a,\ell+2-a)\mid 1 \le a \le \ell+1 \}$ at $m=\ell$.
They lead to completely frozen models.
Nontrivial situations correspond to the
fusion degrees in the range $1\le m \le \ell-1$.
This is an origin of the truncation condition in the 
restricted T-system (Section \ref{ss:res1}) for ${\mathfrak g}=A_1$.

\subsection{Transfer matrices}\label{ss:tmrsos}
We consider the row to row transfer matrix $T_m(u)$ 
with periodic boundary condition whose elements 
$T_m(u)_{a_1,\ldots, a_N}^{b_1,\ldots, b_N}$ are given by
\begin{equation*}
W_{m,s_1}\!\left(\left. \begin{matrix}b_1\;\;b_2\\ a_1\;\; a_2
\end{matrix}\right|u-v_1\!\right)\cdots
W_{m,s_{N-1}}\!\left(\left. 
\begin{matrix}b_{N-1}\;\;b_N\\ a_{N-1}\;\; a_N
\end{matrix}\right|u-v_{N-1}\!\right)
W_{m,s_{N}}\!\left(\left. 
\begin{matrix}b_{N}\;\; b_1\\ a_{N} \;\; a_1
\end{matrix}\right|u-v_N\!\right).
\end{equation*}
No sum is involved. It is depicted as 
\begin{equation}\label{tuba}
\begin{picture}(200,45)(-48,-20)

\put(-80,0)
{$\displaystyle{T_m(u)_{a_1,\ldots, a_N}^{b_1,\ldots, b_N} = }$}

\put(-4,18){$\scriptstyle{b_1}$}
\put(21,18){$\scriptstyle{b_2}$}
\put(43,18){$\scriptstyle{b_{N-1}}$}
\put(72,18){$\scriptstyle{b_N}$}
\put(98,18){$\scriptstyle{b_1}$}

\put(5,0){$\scriptstyle{u\!-\!v_1}$}
\put(32,0){$\cdots$}
\put(51.7,0){$\scriptstyle{u\!-\!v_{\!N\!-\!1}}$}
\put(80,0){$\scriptstyle{u\!-\!v_N}$}

\put(-4,-16){$\scriptstyle{a_1}$}
\put(21,-16){$\scriptstyle{a_2}$}
\put(43,-16){$\scriptstyle{a_{N-1}}$}
\put(72,-16){$\scriptstyle{a_N}$}
\put(98,-16){$\scriptstyle{a_1}$}

\put(0,-10){\line(0,1){25}}
\put(25,-10){\line(0,1){25}}
\put(50,-10){\line(0,1){25}}
\put(75,-10){\line(0,1){25}}
\put(100,-10){\line(0,1){25}}
\put(0,15){\line(1,0){100}}
\put(0,-10){\line(1,0){100}}

\multiput(25,-10)(25,0){4}{\drawline(-5,0)(0,5)}

\end{picture}
\end{equation}
Here $(a_i,a_{i+1}), (b_i,b_{i+1})$ are 
$s_i$-admissible ($a_{N+1}=a_1, b_{N+1}=b_1)$
and $(a_i,b_i)$ is $m$-admissible for all $i$.
The inhomogeneity $s_i, v_i$ 
in fusion degrees and coupling constants are fixed and suppressed in the notation.
The $T_m(u)$ is zero unless the parity condition
$\sum_{i=1}^Ns_i \equiv 0\mod 2$ is satisfied. 
The star-triangle relation (\ref{strg}) implies the commutativity \cite{Ba3}
\begin{equation}\label{cmut2}
[T_m(u), T_n(v)]=0.
\end{equation}

Let us take $s_1=1$ for all $i$ for simplicity and demonstrate 
the functional relation
\begin{equation}\label{shos2}
\begin{split}
&T_1(u+1)T_1(u-1) = T_0(u)T_2(u) + g_1(u)\mathrm{id},\\
&T_0(u) = \prod_{i=1}^N\frac{[u_i+1]_{q^{1/2}}}{[2]_{q^{1/2}}},\quad 
g_1(u) = \prod_{i=1}^N\frac{[u_i+3]_{q^{1/2}}[u_i-1]_{q^{1/2}}}
{[2]_{q^{1/2}}^2},
\end{split}
\end{equation}
where $u_i = u-v_i$.
We first consider the case $a_N=b_N=a$.
Set 
\begin{equation}\label{Lde}
\begin{picture}(150,68)(-30,-8)

\put(-39,23){$\displaystyle{{\mathcal L}_{c,d}} = $}
\put(-7,23){$\scriptstyle{c}$}\put(103,23){$\scriptstyle{d}$}

\put(-4,54){$\scriptstyle{a}$}
\put(22,54){$\scriptstyle{b_1}$}
\put(47,54){$\scriptstyle{b_2}$}
\put(70,54){$\scriptstyle{b_{N-1}}$}
\put(97,54){$\scriptstyle{b_N}$}

\put(0,50){\line(1,0){100}}
\put(0,25){\line(1,0){100}}
\put(0,0){\line(1,0){100}}

\put(0,0){\line(0,0){50}}
\put(25,0){\line(0,1){50}}
\put(50,0){\line(0,1){50}}
\put(75,0){\line(0,1){50}}
\put(100,0){\line(0,1){50}}

\put(3,35){$\scriptstyle{u_N-1}$}
\put(28,35){$\scriptstyle{u_1-1}$}
\put(56,35){$\cdots$}
\put(76,35){$\scriptstyle{u_{N\!-\!1}\!-\!1}$}

\put(3,10){$\scriptstyle{u_N+1}$}
\put(28,10){$\scriptstyle{u_1+1}$}
\put(56,10){$\cdots$}
\put(76,10){$\scriptstyle{u_{N\!-\!1}\!+\!1}$}

\put(22.5,22.5){$\bullet$}\put(47.5,22.5){$\bullet$}
\put(72.5,22.5){$\bullet$}

\put(-4,-7){$\scriptstyle{a}$}
\put(22,-7){$\scriptstyle{a_1}$}
\put(47,-7){$\scriptstyle{a_2}$}
\put(70,-7){$\scriptstyle{a_{N-1}}$}
\put(97,-7){$\scriptstyle{a_N}$}

\multiput(25,25)(25,0){4}{\drawline(-5,0)(0,5)}
\multiput(25,0)(25,0){4}{\drawline(-5,0)(0,5)}

\end{picture}
\end{equation}
where each face stands for $W=W_{1,1}$.
To the difference ${\mathcal L}_{a+1,d}-{\mathcal L}_{a-1,d}$,
one can apply the same trick as (\ref{kmi}).
In particular, the repeated use of the 
star-triangle relation and the 
property $W\!\left(\left. \begin{matrix}b\;\;c\\ a\;\; d
\end{matrix}\right|-2\!\right)\propto \delta_{a c}$
tells that it vanishes unless $a_i=b_i$ for all $i$.
Then the induction on $N$ leads to the identity 
\begin{equation}\label{lele} 
{\mathcal L}_{a+1,d}-{\mathcal L}_{a-1,d}
= \frac{[2a]_{q^{1/2}}}{[2a_N]_{q^{1/2}}}
\prod_{i=1}^N\left(\delta_{a_i,b_i}
\frac{[u_i+3]_{q^{1/2}}[u_i-1]_{q^{1/2}}}{[2]_{q^{1/2}}^2}\right)
\times \begin{cases}
1 & d=a_N+1,\\
-1 & d=a_N-1.
\end{cases}
\end{equation}
Now we are ready to 
evaluate the matrix elements of $T_1(u+1)T_1(u-1)$.
When $a_N=b_N=a$, we have 
\begin{align*}
\left(T_1(u+1)T_1(u-1)\right)^{b_1,\ldots, b_N}_{a_1,\ldots, a_N}
&={\mathcal L}_{a-1,a-1}+ {\mathcal L}_{a+1,a+1}\\
&={\mathcal L}_{a-1,a-1}+{\mathcal L}_{a-1,a+1}
+{\mathcal L}_{a+1,a+1}-{\mathcal L}_{a-1,a+1}
\end{align*}
The first two terms yield 
$T_0(u)T_2(u)^{b_1,\ldots, b_N}_{a_1,\ldots, a_N}$
by the definition (\ref{wm1}).
The other two terms are equal to 
$\left(g_1(u)\mathrm{id}\right)^{b_1,\ldots, b_N}_{a_1,\ldots, a_N}$
due to (\ref{lele}) with $a=a_N$.
When $a_N=b_N\pm 2$,
one can more easily check (\ref{shos2}) since 
$g_1(u)\mathrm{id}$ does not contribute.

\subsection{\mathversion{bold}Vertex and RSOS models for general 
${\mathfrak g}$}\label{ss:vrg}

We include a formal and partly conjectural 
description of solvable vertex and RSOS models 
and their T-system for general ${\mathfrak g}$.
We will use the terminology 
introduced in later sections.
(Therefore this technical section 
may better be skipped on the first reading.)

Let $W^{(a)}_m(u)$ be the Kirillov-Reshetikhin module (Section \ref{ss:fdr}), 
where $a \in I$ (set of vertices on the Dynkin diagram
of ${\mathfrak g}$) 
and $m \in \Z_{\ge 1}$.
It is an irreducible finite dimensional representation 
of untwisted quantum affine algebra $U_q=U_q(\hat{\mathfrak g})$.
Up to an overall scalar, there is the unique element, the $R$ matrix,
$R \in \mathrm{End}(W^{(a)}_m(u_1)\otimes W^{(b)}_n(u_2))$
characterized by the intertwining property 
$\Delta(U_q)PR = PR \Delta(U_q)$, where $P$ is the transposition.
It can in principle be constructed concretely 
by solving this linear equation, or by the fusion of the simpler cases
$m=n=1$ (cf. Theorem \ref{th:FR}) 
or by taking the image of the universal $R$.
Let us denote the resulting $R$ matrix by  
$R^{(a,m;  b,n)}(z_1/z_2)$, 
where $z_i= q^{tu_i}$, $t$ is defined by (\ref{eq:t1})
and the dependence through $z_1/z_2$ is due to the general theory.
\begin{equation}
\begin{picture}(250,55)(-160,-19)

\put(-145,-2){$R^{(a,m;  b,n)}(z_1/z_2)=$}

\put(-18,22){$\scriptstyle{W^{(b)}_n(u_2)}$}
\put(-54,-2){$\scriptstyle{W^{(a)}_m(u_1)}$}

\put(-17,0){\line(1,0){35}}
\put(0,17){\line(0,-1){35}}

\drawline(0,-5)(5,0)

\put(4,-9){$\scriptstyle{z_1/z_2}$}

\end{picture}
\end{equation}

As in (\ref{trme}), 
one introduces the row to row transfer matrix with 
the auxiliary space $W^{(a)}_m(u)$ by ($z=q^{tu}$)
\begin{equation}\label{tamg}
T^{(a)}_m(u)= \mathrm{Tr}_{W^{(a)}_m(u)}
\left(R^{(a,m ;  r_N,s_N)}_{0,N}(z/w_N)\cdots
R^{(a,m ;  r_1,s_1)}_{0,1}(z/w_1)\right),
\end{equation}
which acts on the quantum space
$W^{(r_1)}_{s_1}(v_1)\otimes \cdots \otimes W^{(r_N)}_{s_N}(v_N)$
with $w_i = q^{tv_i}$.
They are all commutative, i.e.
$[T^{(a)}_m(u), T^{(b)}_n(v)]=0$ thanks to the  Yang-Baxter relation. 
It is a corollary of
the exact sequence underlying Theorem \ref{th:nh} 
and the argument on the central extension (cf. \cite[section 2.2]{KNS2})
that $T^{(a)}_m(u)$ satisfies the unrestricted T-system for 
${\mathfrak g}$ (\ref{tga}) with some scalars 
$T^{(a)}_0(u)$ and $g^{(a)}_m(u)$ appropriately chosen 
depending on the normalization of $T^{(a)}_m(u)$.

Let $\ell \in \Z_{\ge 2}$.
{}From the $R$ matrix one can in principle construct the face 
Boltzmann weights for level $\ell$ $U_q(\hat{\mathfrak g})$ 
RSOS model 
at $q=\exp\left(\frac{\pi\sqrt{-1}}{t(\ell+h^\vee)}\right)$\footnote{
Actually any primitive $2t(\ell+h^\vee)$ th root of unity. 
$h^\vee$ is the dual Coxeter number of ${\mathfrak g}$ (\ref{hhd}).}.
Let us introduce 
\begin{equation}\label{PP}
P_+=\Z_{\ge 0}\,\omega_1 \oplus 
\cdots \oplus \Z_{\ge 0}\,\omega_r,\quad
P_\ell=\{ \lambda \in P_+\mid 
(\lambda | \text{maximal root})\le \ell \},
\end{equation}
where $\omega_a$ is a fundamental weight of ${\mathfrak g}$ 
(Section \ref{ss:utw}).
$P_\ell$ is the classical projection of the set of 
level $\ell$ dominant integral weights of the affine Lie algebra
$\hat{\mathfrak g}$ at level $\ell$ \cite{Ka}.
For $\lambda \in P_+$, 
let $V_\lambda$ be the irreducible $U_q(\mathfrak{g})$-module 
with highest weight $\lambda$.
Let $\mathrm{res} \,W^{(a)}_m$ be the 
(not necessarily irreducible) $U_q(\mathfrak{g})$-module
obtained by restricting the $U_q(\hat{\mathfrak g})$-module 
$W^{(a)}_m(u)$. 
It is independent of $u$. See the text around (\ref{rew}).
When $q$ is not a root of unity,
one has the irreducible decomposition
\begin{equation}\label{yada}
V_\lambda \otimes \mathrm{res} \,W^{(a)}_m \otimes \mathrm{res} \,
W^{(b)}_n =
\bigoplus_{\mu \in P_+}\Omega(\lambda)_{\mu} \otimes V_\mu,
\end{equation}
where $\Omega(\lambda)_{\mu}$ is the space of highest weight vectors
of weight $\mu$.
Since ${\check R}(z) = PR^{(a,m; b,n)}(z)$ 
commutes with $U_q({\mathfrak g})$, the space $\Omega(\lambda)_{\mu}$
is invariant under $\mathrm{id}\otimes {\check R}(z)$.
Thus its matrix elements yield the Boltzmann weights 
of unrestricted SOS model as in (\ref{par}).
The star-triangle relation for them follows from this construction.

To make the restriction, we consider the case
$q = \exp\left(\frac{\pi\sqrt{-1}}{t(\ell+h^\vee)}\right)$, where
the decomposition (\ref{yada}) no longer holds \cite{Lu, RA}.
However, based on the observation for ${\mathfrak g} = A_1$ \cite{Pas}, 
we conjecture that if $\lambda$ is taken from $P_\ell$ 
and $m \le t_a\ell, n \le t_b\ell$, 
the quotient of the RHS of (\ref{yada}) by the 
type I modules  \cite{PasS, Ke}\footnote{Indecomposable modules 
with $\dim_q = 0$. See (\ref{qdi}). } 
reduces the sum $\mu \in P_+$ to $\mu \in P_\ell$, and  
$\mathrm{id}\otimes {\check R}(z)$ remains well defined on it.
Then the RSOS Boltzmann weights are defined as
the matrix elements of $\mathrm{id}\otimes {\check R}(z)$
on the quotient space, and satisfy the star-triangle relation.

The RSOS model so constructed has the fluctuating variables 
on edges as well as sites in general (cf. \cite[Fig.1]{JKMO}).
\begin{equation}\label{jkmo}
\begin{picture}(150,65)(0,-12)

\put(-6,26){$\scriptstyle{\mu}$}
\put(26,26){$\scriptstyle{\nu}$}
\put(-6,-6){$\scriptstyle{\lambda}$}
\put(26,-6){$\scriptstyle{\kappa}$}

\put(14,27){$\scriptstyle{\beta}$}
\put(-7,15){$\scriptstyle{\gamma}$}\put(27,14){$\scriptstyle{\delta}$}
\put(14,-6){$\scriptstyle{\alpha}$}


\put(-39,10){$\scriptstyle{W^{(a)}_m}$}
\multiput(-18,12)(2,0){30}{\put(0,0){\line(1,0){1}}}

\put(7,47){$\scriptstyle{W^{(b)}_n}$}

\multiput(12,42)(0,-2){30}{\put(0,0){\line(0,1){1}}}

\put(0,0){\line(1,0){24}}
\put(0,0){\line(0,1){24}}
\put(24,0){\line(0,1){24}}
\put(0,24){\line(1,0){24}}

\drawline(19,0)(24,5)

\put(112,28)
{$\displaystyle{\beta \in \Omega^{(b,n)}_{\mu \nu}}$}

\put(70,10)
{$\displaystyle{\gamma \in \Omega^{(a,m)}_{\lambda \mu}}$}

\put(156,10)
{$\displaystyle{\delta \in \Omega^{(a,m)}_{\kappa \nu}}$}

\put(112,-8)
{$\displaystyle{\alpha \in \Omega^{(b,n)}_{\lambda \kappa}}$}

\end{picture}
\end{equation}
The site variables belong to $P_\ell$.
In fact for ${\mathfrak g}=A_1$, one may 
regard the set of site variables 
$\{1,2,\ldots, \ell+1\}$ as 
$P_\ell = \{0,\omega_1,\ldots, \ell\omega_1\}$.
To describe the edge variables, we consider the decomposition
$V_{\lambda}\otimes \mathrm{res} \,W^{(a)}_m=
 \bigoplus_{\mu \in P_+} \overline{\Omega}^{(a,m)}_{\lambda \mu}
\otimes V_{\mu}$ at generic $q$.
When $q = \exp\left(\frac{\pi\sqrt{-1}}{t(\ell+h^\vee)}\right)$,
we need to take the quotient of
the RHS by the type I modules,  
and this induces the quotient 
$\Omega^{(a,m)}_{\lambda \mu}$ 
of  $\overline{\Omega}^{(a,m)}_{\lambda \mu}$.
The edge variable associated to 
$W^{(a)}_m$ belongs to the space
$\Omega^{(a,m)}_{\lambda \mu}$.
We set ${\mathcal A}^{(a,m)}_{\lambda \mu} = 
\dim \Omega^{(a,m)}_{\lambda \mu}$ and 
say that an (ordered) pair of site variables 
$(\lambda,\mu) \in P_\ell \times P_\ell$ 
is admissible under $W^{(a)}_m$ 
if ${\mathcal A}^{(a,m)}_{\lambda \mu}\ge 1$\footnote{
The type $A_r$ is bit special in that  
${\mathcal A}^{(a,m)}_{\lambda \mu} \in \{0,1\}$ holds
for any $(a,m)$ and $\lambda, \mu$, 
hence effectively no edge variable exists.
However, the situation 
${\mathcal A}^{(a,m)}_{\lambda \mu}\ge 2$ still happens 
for the fusion types more general than those specified by 
Kirillov-Reshetikhin modules \cite{JKMO}.}.
The matrix 
${\mathcal A}^{(a,m)} = 
({\mathcal A}^{(a,m)}_{\lambda \mu})_{\lambda, \mu \in P_\ell}$ 
is called the admissibility matrix of $W^{(a)}_m$.

Let us formulate the row to row transfer matrix $T^{(a)}_m(u)$ 
that corresponds to the dual of the one for the vertex model (\ref{tamg}).
It acts on the space of paths
\begin{align}
\EuScript{H}(N)&= \bigoplus_{\lambda_i \in P_\ell}
\Omega^{(r_1,s_1)}_{\lambda_1 \lambda_2}\otimes \cdots 
\otimes \Omega^{(r_N,s_N)}_{\lambda_N \lambda_1},\label{sop1}\\
\dim \EuScript{H}(N) 
&= \mathrm{Tr}\left({\mathcal A}^{(r_1,s_1)}\cdots 
{\mathcal A}^{(r_N,s_N)}\right).\label{sop2}
\end{align}
The matrix elements are depicted as follows
($u_i = u-v_i, \lambda_i = \lambda_{i+N},
\mu_i = \mu_{i+N}$):
\begin{equation}\label{tame}
\begin{picture}(260,60)(-150,-20)
\put(-173,12){$\displaystyle{T^{(a)}_m
(u)_{\lambda_1,\alpha_1,\lambda_2, \alpha_2, \ldots, \lambda_N,\alpha_N}
^{\mu_1,\beta_1, \mu_2, \beta_2, 
\ldots, \mu_N,\beta_N}=
\sum_{\gamma \in 
\Omega^{(a,m)}_{\lambda_1 \mu_1}}}$}

\put(0,0){\line(1,0){115}}
\put(0,25){\line(1,0){115}}

\put(0,0){\line(0,1){25}}
\put(25,0){\line(0,1){25}}
\put(50,0){\line(0,1){25}}
\put(90,0){\line(0,1){25}}
\put(115,0){\line(0,1){25}}

\drawline(20,0)(25,5)
\drawline(45,0)(50,5)
\drawline(110,0)(115,5)

\put(-8,11){$\scriptstyle{\gamma}$}
\put(118.3,11){$\scriptstyle{\gamma}$}

\put(23,10){$\bullet$}\put(48,10){$\bullet$}
\put(88,10){$\bullet$}

\put(-5,29){$\scriptstyle{\mu_1}$}
\put(9,28){$\scriptstyle{\beta_1}$}
\put(22,29){$\scriptstyle{\mu_2}$}
\put(34,28){$\scriptstyle{\beta_2}$}
\put(45,29){$\scriptstyle{\mu_3}$}

\put(87,29){$\scriptstyle{\mu_N}$}
\put(100,28){$\scriptstyle{\beta_N}$}
\put(115,29){$\scriptstyle{\mu_1}$}

\put(8,10){$u_1$}\put(33.5,10){$u_2$}
\put(64,10){$\cdots$}
\put(97,10){$u_N$}

\put(-5,-8){$\scriptstyle{\lambda_1}$}
\put(9,-7.5){$\scriptstyle{\alpha_1}$}
\put(22,-8){$\scriptstyle{\lambda_2}$}
\put(34,-7.5){$\scriptstyle{\alpha_2}$}
\put(45,-8){$\scriptstyle{\lambda_3}$}

\put(87,-8){$\scriptstyle{\lambda_N}$}
\put(100,-7.5){$\scriptstyle{\alpha_N}$}
\put(115,-8){$\scriptstyle{\lambda_1}$.}
\end{picture}
\end{equation}
Here the symbols $\alpha_i$ and $\beta_i$ denote a  
basis of $\Omega^{(r_i,s_i)}_{\lambda_i \lambda_{i+1}}$
and $\Omega^{(r_i,s_i)}_{\mu_i \mu_{i+1}}$, respectively.
The pairs $(\lambda_i,\lambda_{i+1})$ and 
$(\mu_i, \mu_{i+1})$ are both admissible under 
$W^{(r_i)}_{s_i}$, whereas
$(\lambda_i, \mu_i)$ is so under $W^{(a)}_m$.
The RHS stands for 
the product of the $N$ 
Boltzmann weights attached to the elementary squares summed 
over the states on the vertical edges 
accommodating $\Omega^{(a,m)}_{\lambda_i \mu_i}$
for $i=1,\ldots, N$.
As for the weights, 
$\lambda_{i+1}-\lambda_i \equiv 
\mu_{i+1}-\mu_i \equiv s_i\omega_{r_i}$
mod the root lattice; therefore,
the $T^{(a)}_m(u)$ under consideration is vanishing unless
\begin{equation}\label{na}
\sum_{i=1}^Ns_i\left(C^{-1}\right)_{a\, r_i} \in \Z\;\;
\text{ for all } a \in I,
\end{equation}
where $C$ is the Cartan matrix of ${\mathfrak g}$ 
(Section \ref{ss:utw}).
Due to the star-triangle relation (including sums over edge variables),
the commutativity $[T^{(a)}_m(u), T^{(b)}_n(v)]=0$ holds.
We conjecture that $T^{(a)}_m(u)$ 
satisfies the level $\ell$ restricted T-system for  
${\mathfrak g}$ of the form (\ref{tga}) 
with some scalars 
$T^{(a)}_0(u)$ and $g^{(a)}_m(u)$ appropriately chosen 
depending on the normalization.
In particular, this implies that the $|P_\ell|$ by $|P_\ell|$ matrices 
${\mathcal A}^{(a,m)}$ with $a\in I, 0 \le m \le t_a\ell$ are commutative and
satisfy the level $\ell$ restricted Q-system (cf. Section \ref{ss:rwq})
with the boundary condition
\begin{equation}\label{AA}
{\mathcal A}^{(a,1)}=
1,\quad {\mathcal A}^{(a,t_a\ell+1)}=0\footnote{This 
leads to $\prod_{b\in I}({\mathcal A}^{(b,t_b\ell)})^{C_{a b}} = 1$ 
for any $a \in I$, 
which is a weaker constraint than ${\mathcal A}^{(a,t_a\ell)}=1$
employed in the definition of the level $\ell$ restricted Q-system 
in Section \ref{ss:rwq}. }.
\end{equation}

Let $\dim_qV_\lambda$ be the $q$-dimension of $V_\lambda$ at 
$q = \exp\left(\frac{\pi\sqrt{-1}}{t(\ell+h^\vee)}\right)$ 
defined in (\ref{qdi}).
We set  $Q^{(a)}_m=\dim_q {\rm res}\, W^{(a)}_m$, 
which supposedly
satisfies the level $\ell$ restricted Q-system (\ref{ss:rwq})
(Conjecture \ref{conj:rq}).
Now the generalization of (\ref{mad}) is given as
\begin{equation}
Q^{(a)}_m\dim_qV_{\lambda} 
= \sum_{\mu \in P_\ell}
{\mathcal A}^{(a,m)}_{\lambda \mu}
\dim_qV_{\mu}\quad (\lambda \in P_\ell).
\end{equation}
Since $\dim_qV_{\lambda} >0$ for any $\lambda \in P_\ell$,
the Perron-Frobenius theorem tells that $Q^{(a)}_m$ is the 
largest eigenvalue of the admissibility matrix ${\mathcal A}^{(a,m)}$.
Therefore in the homogeneous case where
$(r_i,s_i) = (p,s)$ for all $i$,
we find from (\ref{sop2}) that
\begin{equation}\label{ehq}
\lim_{N\rightarrow \infty}
\left(\dim \EuScript{H}(N)\right)^{1/N}
= Q^{(p)}_s.
\end{equation}
This property will be re-derived in the TBA analysis in (\ref{hteq2}).

In general, the Boltzmann weights (\ref{jkmo}) are expressed in terms of 
the function $[u]_{q^{t/2}} \propto 
\sin\left(\frac{\pi u}{2(\ell+h^\vee)}\right)$.
($t$ is defined in (\ref{eq:t1}).)
This is indeed the case for $A_1$ as in (\ref{abf}) and 
in the other known examples.
It is also consistent with the Bethe equation (\ref{be}). 
Consequently, the transfer matrix with an appropriate normalization
possesses the periodicity
\begin{equation}\label{pe1}
T^{(a)}_m(u+ 2(\ell+h^\vee)) = T^{(a)}_m(u).
\end{equation}
We will see in Theorem \ref{t:thm:period2} that the 
level $\ell$ restricted T-system in Section \ref{ss:res1}\footnote{
In this case, the normalization is 
$T^{(a)}_0(u) = T^{(a)}_{t_a\ell}(u)=1$.} alone compels this property.

\subsection{Bibliographical notes}
The integrability of the 6 vertex model (\ref{6v}) (first solved 
in \cite{Lie,Su1}) has been formulated in terms of 
the Yang-Baxter equation and commuting transfer matrices in 
\cite{Ba3}.
Solutions of the Yang-Baxter equation
that have been known by 1980 are surveyed in \cite{KSk}
from the perspective of the quantum inverse scattering method.
Subsequent generalizations of trigonometric 
vertex models to type $A$ \cite{BDV, Ch, PS} and 
many other ${\mathfrak g}$ \cite{J2,Baz} have been assembled in 
the reprint volume \cite{J3}.
The fusion of vertex models is formulated in \cite{KRS}.
See also \cite{SAA}.
The idea of utilizing the functional relations of transfer matrices 
goes back to Baxter \cite{Ba1,Ba3}.
Some simplest examples of the T-system 
have been obtained for the XXZ chain \cite{KR1}, 
the $O(n)$-symmetry models \cite{R1} and vertex models
associated with some other ${\mathfrak g}$ \cite{R2}.

With regard to the RSOS models, 
the 8VSOS model is the fundamental example
containing the Ising and (generalized) hard hexagon models
as the level $\ell=2,3$ cases, respectively.
The one point function \cite{ABF} 
essentially gives rise to the character of the Virasoro minimal series, 
and this fact inspired intensive studies on 
the relations with conformal field theory and 
representation theory of quantum affine algebras.
In the terminology in Section \ref{ss:vrg},
the 8VSOS model corresponds to the level 
$\ell$ RSOS model for  
${\mathfrak g} = A_1$ with fusion type $W^{(1)}_1$
(both on the horizontal and vertical edges).

Beyond the $A_1$ case, concrete constructions of
RSOS models for untwisted affine Lie algebra 
$\hat{\mathfrak g}$ 
have been done for 
non exceptional series ${\mathfrak g} = A_r, B_r, C_r, D_r$ 
\cite{JMO1, JMO2} 
associated with $W^{(1)}_1$ (``vector representation") 
and  ${\mathfrak g} = G_2$ \cite{KS3} 
with $W^{(2)}_1$.
The fusion of RSOS models have been worked out 
explicitly only for type $A$ \cite{DJKMO1,JKMO}.
One of the earliest examples of the T-system for RSOS models (except the Ising)
is \cite{BP} for the generalized hard hexagon model.
It was systematized to the general level restricted T-system for $A_1$ 
in \cite{KP2}. See also \cite{BR1} 
where the relation of the form ``$T_mT_1 = T_{m-1}+ T_{m+1}$" was given. 
In \cite{BR2}, the Jacobi-Trudi type functional relations 
(cf. Theorem \ref{th:jta} and \ref{th:jta2}) 
were given for the fusion RSOS models of type $A_r$.
The T-system for $A_r$ is extracted from them in \cite{KNS2}, 
where the extension to all ${\mathfrak g}$ was proposed
based on the connection to the Y-system and the Q-system.
Finally, one can construct the quantum field theory analog of the 
commuting transfer matrices that act on Virasoro Fock spaces and 
satisfy the T-system. 
See \cite{BLZ1} 
for the original construction for $\mathfrak{g}=A_1$
and \cite{Ru} for a recent application.

\section{T-system in quantum group theory}\label{s:qg}

\subsection{Quantum affine algebra}\label{ss:qaa}
For simplicity we concentrate on the untwisted quantum affine algebra
$U_q(\hat{\mathfrak g})$ until Section \ref{ss:tsq}.
We assume that 
$q\in \mathbb{C}^{\times}$ is not a root of unity
and set $q=e^\hbar$; therefore, 
the domain $U$ of the spectral parameter $u$ should be understood as 
$U=\C_{t\hbar}$. See Section \ref{ss:utw}.
We set $\hat{I} = \{0\} \sqcup I$ and 
let $\hat{C} = (\hat{C}_{ij})_{i,j \in \hat{I}}$
be the Cartan matrix of 
the affine Kac-Moody algebra $\hat{\mathfrak{g}}$ \cite{Ka}.
For $i, j \in I$, one has $\hat{C}_{ij} = C_{ij}$ where the latter is an 
element of the Cartan matrix $C$ of $\mathfrak{g}$. 
By definition, the (untwisted) quantum affine algebra 
$U_q(\hat{\mathfrak{g}})$ \cite{D1,J1} is 
the associative algebra over $\mathbb{C}$ with generators 
$x^{\pm}_i, k^{\pm 1}_i, \,(i \in \hat{I})$ and the relations:
\begin{equation}\label{djr}
\begin{split}
&k_i k^{-1}_i=k^{-1}_ik_i=1,\; \;\; k_ik_j = k_jk_i,\\
&k_i x^{\pm}_j k^{-1}_i=q_i^{\pm \hat{C}_{ij}}x^{\pm}_j,
\quad [x^+_i, x^-_j]=\delta_{ij}
\frac{k_i-k^{-1}_i}{q_i-q^{-1}_i},\\
&\sum_{\nu=0}^{1-\hat{C}_{ij}}(-1)^\nu
\left[{1-\hat{C}_{ij} \atop \nu}\right]_{q_i}\!
(x^\pm_i)^{1-\hat{C}_{ij}-\nu}x^{\pm}_j (x^{\pm}_i)^\nu = 0\;\;
(i\neq j).
\end{split}
\end{equation}
Here $q_0 = q$ and $q_i = q^{t/t_i}$
for $i \in I$.  
For the notations $t$ and $t_i$, see (\ref{eq:t1}).
Furthermore, for $0 \le n \le m$,
\begin{equation}\label{qb}
\left[{m \atop n}\right]_{q}
= \frac{[m]_q!}{[n]_q! [m-n]_q!},\quad
[m]_q!=[1]_q[2]_q\cdots [m]_q.
\end{equation} 
See (\ref{qnb}) for the definition of $[m]_q$.
The algebra $U_q(\hat{\mathfrak{g}})$ is denoted by 
$U'_q(\hat{\mathfrak{g}})$ in some 
literature indicating that the analog 
of the derivation operator in $\hat{\mathfrak{g}}$ has not been included.
There are $2^{r+1}$ algebra automorphisms of 
$U_q(\hat{\mathfrak{g}})$ given on generators by
\begin{equation}\label{twisting}
k_i \mapsto \sigma_i k_i, \;\;
x^+_i \mapsto \sigma_i x^+_i,\;\;
x^-_i \mapsto x^-_i
\end{equation}
for any set of signs $\sigma_0,\ldots, \sigma_r  \in \{\pm 1\}$. 
Obviously, $U_q(\hat{\mathfrak{g}})$ contains 
$U_q(\mathfrak{g})$ as a subalgebra.

There is another realization of $U_q(\hat{\mathfrak{g}})$ 
called the Drinfeld new realization \cite{D2,Bec}.
Namely,  $U_q(\hat{\mathfrak{g}})$  is isomorphic to the 
algebra with generators 
$x_{i,n}^{\pm}$ ($i\in I$, $n\in\Z$),
$k_i^{\pm 1}$ ($i\in I$), $h_{i,n}$ ($i\in I$, $n\in
\Z\backslash\{0\}$) and central elements $c^{\pm 1/2}$, with the
following relations:
\begin{equation}\label{dr}
\begin{split}
&k_ik_j = k_jk_i,\quad  k_ih_{j,n} =h_{j,n}k_i,\quad
k_ix^\pm_{j,n}k_i^{-1} = q_i^{\pm C_{ij}}x_{j,n}^{\pm},\\
&[h_{i,n} , x_{j,m}^{\pm}] = \pm \frac{1}{n} [n C_{ij}]_{q_i} 
c^{\mp{|n|/2}}x_{j,n+m}^{\pm},\quad
[h_{i,n},h_{j,m}]
=\delta_{n,-m} \frac{1}{n} [n C_{ij}]_{q_i} \frac{c^n -
c^{-n}}{q_j-q^{-1}_j},\\ 
&x_{i,n+1}^{\pm}x_{j,m}^{\pm}
-q_i^{\pm C_{ij}}x_{j,m}^{\pm}x_{i,n+1}^{\pm} 
=q_i^{\pm C_{ij}}x_{i,n}^{\pm}x_{j,m+1}^{\pm}
-x_{j,m+1}^{\pm}x_{i,n}^{\pm},\\ 
&[x_{i,n}^+ , x_{j,m}^-]=\delta_{ij}  \frac{
c^{(n-m)/2}\phi_{i,n+m}^+ - c^{-(n-m)/2} \phi_{i,n+m}^-}{q_i -
q_i^{-1}},\\
&\sum_{\pi\in\Sigma_s}\sum_{k=0}^s(-1)^k
\left[{s \atop k}\right]_{q_i} x_{i, n_{\pi(1)}}^{\pm}\ldots
x_{i,n_{\pi(k)}}^{\pm}  x_{j,m}^{\pm} x_{i,
n_{\pi(k+1)}}^{\pm}
\ldots x_{i,n_{\pi(s)}}^{\pm} =0,\ \ i\ne j
\end{split}
\end{equation}
for all sequences of integers $n_1,\ldots,n_s$, where $s=1-C_{ij}$,
$\Sigma_s$ is the symmetric group on $s$ letters, and
$\phi_{i,n}^{\pm}$'s are determined by the formal power series
\begin{equation}
\sum_{n=0}^{\infty}\phi_{i,\pm n}^{\pm}\zeta^{\pm n} = k_i^{\pm
1} \exp\left(\pm(q_i-q^{-1}_i)\sum_{m=1}^{\infty}h_{i,\pm m} 
\zeta^{\pm m}\right).
\end{equation}
In the two realizations (\ref{djr}) and (\ref{dr}),
the symbol $k^{\pm 1}_i \,(i \in I)$ stands for the same generator
under the isomorphism.
$U_q(\hat{\mathfrak g})$ admits a Hopf algebra structure \cite{D1,J1}.

\subsection{Finite dimensional representations}\label{ss:fdr}
A representation $W$ of $U_q(\hat{\mathfrak g})$ is called  {\em type 1}
if the generators $k_0, k_1, \ldots, k_r$ act semi simply on $W$ with 
eigenvalues in $q^\Z$ and $c^{1/2}$ in (\ref{dr}) acts as 1 on $W$.
A vector $v \in W$ is called a highest weight vector if
\begin{equation}\label{hwg}
x_{i,n}^+ \cdot v=0,\quad \quad \phi_{i,n}^\pm \cdot v =
\psi_{i,n}^\pm v,\quad \quad c^{1/2} v =v,
\end{equation}
for some complex numbers $\psi_{i,n}^{\pm}$. 
A type 1 representation $W$ is called a highest weight representation if 
$W=U_q(\hat{\mathfrak g})\cdot v$
for some highest weight vector $v$. 

\begin{theorem}[\cite{CP2,CP3}]\label{th:cp}
(1) Every finite dimensional irreducible
representation of $U_q(\hat{\mathfrak g})$ 
can be obtained from a type 1 representation
by a twisting with an automorphism (\ref{twisting}).

(2) Every finite dimensional irreducible representation of 
$U_q(\hat{\mathfrak g})$ of type 1 is a highest weight representation.

(3) A type 1 highest weight representation with the highest weight vector 
$v$ in (\ref{hwg}) is finite dimensional if and only if 
there exist polynomials $\EuScript{P}_a(\zeta) \in \C[\zeta]\,
(a \in I)$ such that $\EuScript{P}_a(0)=1$ and 
\begin{equation}\label{pwd}
\sum_{n \ge 0}\psi^\pm_{a,\pm n}\zeta^{\pm n}
=q_a^{\deg \EuScript{P}_a}
\frac{\EuScript{P}_a(\zeta q_a^{-1})}{\EuScript{P}_a(\zeta q_a)}
\in \C[[\zeta^{\pm 1}]].
\end{equation}
\end{theorem} 

The polynomials $\EuScript{P}_a(\zeta)$ are called Drinfeld polynomials
after the analogous classification theorem by Drinfeld 
for Yangians \cite{D2}.

The Kirillov-Reshetikhin module $W^{(a)}_m(u)$ 
$(a \in I, m \in \Z_{\ge 1}, u \in \C_{t\hbar})$ 
is the irreducible finite dimensional representation 
of $U_q(\hat{\mathfrak g})$ that corresponds to the 
Drinfeld polynomial
\begin{equation}\label{dpd}
\EuScript{P}_b(\zeta) = \begin{cases}
\prod_{s=1}^m(1-\zeta q^{tu}q_a^{m+1-2s})
& \text{if } b=a,\\
1 & \text{otherwise}.
\end{cases}
\end{equation}
This $W^{(a)}_m(u)$ is equal to 
$W^{(a)}_{m, q^{tu}q_i^{-m+1}}$ in \cite{N3, Her1}.
In particular, 
$W^{(1)}_1(u), \ldots, W^{(r)}_1(u)$ are
called fundamental representations.

\subsection{\mathversion{bold}Example}\label{ss:rf}

Consider the simplest example  $U_q=U_q(A^{(1)}_{1})$.
In realization (\ref{djr}),  $\hat{I} = \{0,1\}$ and 
the Cartan matrix is
$\hat{C}= \begin{pmatrix}2 & -2\\ -2 & 2\end{pmatrix}$.
The coproduct is given by
\begin{equation}\label{cpr}
\Delta x^+_i = x^+_i\otimes 1 + k_i\otimes x^+_i,\quad
\Delta x^-_i = x^-_i\otimes k^{-1}_i + 1 \otimes x^-_i,\quad
\Delta k^{\pm 1}_i = k^{\pm 1}_i  \otimes  k^{\pm 1}_i.
\end{equation} 
For $m \in \Z_{\ge 0}$,
let $W_m(u)=W^{(1)}_m(u)$ be the Kirillov-Reshetikhin module.
Plainly, it is the $m+1$ dimensional (i.e. spin $\frac{m}{2}$)
irreducible representation 
$W_m(u)= \C v^m_1 \oplus \cdots \oplus \C v^m_{m+1}$ given by 
$(z=q^u)$
\begin{align}
x^-_1v^m_j &= [m+1-j]v^m_{j+1},\quad
x^+_1v^m_j = [j-1]v^m_{j-1},\quad
k_1^{\pm 1} v^m_j = q^{\pm(m+2-2j)}v^m_j,\label{hyog1}\\
x^+_0v^m_j &= z[m+1-j]v^m_{j+1},\quad
x^-_0v^m_j = z^{-1}[j-1]v^m_{j-1},\quad
k_0^{\pm 1} v^m_j = q^{\mp(m+2-2j)}v^m_j,\label{hyog2}
\end{align} 
where $[j]=[j]_q = \frac{q^j-q^{-j}}{q-q^{-1}}$ 
as in (\ref{qnb}).
In the Drinfeld new realization (\ref{dr}), 
the highest weight vector is identified with $v^m_1$ and 
the eigenvalues in (\ref{hwg}) read
\begin{equation*}
\psi^\pm_{1,{\pm n}} = \begin{cases}
q^{\pm m} & n = 0,\\
\pm (q^m-q^{-m})(zq^m)^{\pm n} & n\ge 1.
\end{cases}
\end{equation*}
The relation (\ref{pwd}) holds with 
the Drinfeld polynomial 
\begin{equation*}
\EuScript{P}_1(\zeta) = (1-\zeta q^{u-m+1})(1-\zeta q^{u-m+3})
\cdots (1-\zeta q^{u+m-1})
\end{equation*}
in agreement with (\ref{dpd}).

The exact sequence (\ref{esv}) is refined along the definitions here.
The vectors $v_i \in V_1$ and $x=(x_1,x_2) \in V_m$ in Section \ref{ss:vm} 
are to be identified with 
$v^1_i$ and $v^m_{x_2+1}$ in (\ref{hyog1})--(\ref{hyog2}), respectively.
We introduce the base of $W_1(u)\otimes W_1(v)$ as
\begin{equation}
\begin{split}
{\bf u}_1 &= v^1_1\otimes v^1_1,\\
{\bf u}_2 &= \frac{1}{[2]}\Delta(x^-_1){\bf u}_1 
= \frac{v^1_1\otimes v^1_2 
+ q^{-1}v^1_2\otimes v^1_1}{[2]},\qquad
{\bf u'}_1 = v^1_1\otimes v^1_2-qv^1_2\otimes v^1_1,\\ 
{\bf u}_3 &= \Delta(x^-_1){\bf u}_2 = v^1_2\otimes v^1_2.
\end{split}
\end{equation}
Under the action of $x^\pm_1, k^{\pm 1}_1$, the set of vectors
$\{{\bf u}_1,{\bf u}_2,{\bf u}_3\}$ 
and $\{{\bf u'}_1\}$ behave as the triplet and the singlet 
representations as usual.
On the other hand, with regard to $x^\pm_0$,
they are mixed as follows. ($x=q^u, y=q^v$)
\begin{equation}\label{tir}
\begin{picture}(280,120)(-20,40)

\put(184,145){\color{red}$\Delta(x^+_0)\,: $}
\put(224,147){\color{red}\vector(1,0){20}}

\put(184,125){\color{blue}$\Delta(x^-_0)\,: $}
\put(224,127){\color{blue}\vector(1,0){20}}

\put(50,150){${\bf u}_1$}

\put(52,111){\color{blue}\vector(0,1){31}}
\put(58,141){\color{red}\vector(0,-1){31}}

\put(-10,126){\color{blue}$\scriptstyle{(x^{-1}+y^{-1})[2]^{-1}}$}
\put(63,126){\color{red}$\scriptstyle{x+y}$}

\put(66,152){\color{red}\vector(2,-1){80}}
\put(145,107){\color{blue}\vector(-2,1){80}}

\put(80,110){\color{blue}
$\scriptstyle{y^{-1}-q^2x^{-1}}$}

\put(95,143)
{\color{red}$\scriptstyle{(yq^{-2}-x)[2]^{-1}}$}

\put(50,100){${\bf u}_2$} \put(150,100){${\bf u'}_1$}

\put(62,81){\color{red}$\scriptstyle{(x+y)[2]^{-1}}$}
\put(11,81){\color{blue}$\scriptstyle{x^{-1}+y^{-1}}$}

\put(66,53){\color{blue}\vector(2,1){80}}
\put(145,98){\color{red}\vector(-2,-1){80}}

\put(109,95){\color{red}$\scriptstyle{x-q^2y}$}
\put(120,69){\color{blue}
$\scriptstyle{(q^{-2}x^{-1}-y^{-1})[2]^{-1}}$}

\put(52,61){\color{blue}\vector(0,1){31}}
\put(58,91){\color{red}\vector(0,-1){31}}

\put(50,50){${\bf u}_3$}

\end{picture}
\end{equation}
The diagram means
$\Delta(x^+_0){\bf u}_1 = (x+y){\bf u}_2
+ \frac{yq^{-2}-x}{[2]}{\bf u'}_1$ for instance.
{}From (\ref{tir}),  we find that $W_1(u)\otimes W_1(v)$ is irreducible 
if and only if $\frac{x}{y}\neq q^{\pm 2}$, namely
$u-v \neq \pm 2$.
In the reducible cases, (\ref{tir}) looks as
\begin{equation}\label{tir2}
\begin{picture}(300,150)(-70,15)

\put(184,145){\color{red}$\Delta(x^+_0)\,: $}
\put(224,147){\color{red}\vector(1,0){20}}

\put(184,125){\color{blue}$\Delta(x^-_0)\,: $}
\put(224,127){\color{blue}\vector(1,0){20}}

\multiput(-75,0)(0,0){1}{
\put(50,150){${\bf u}_1$}

\put(52,111){\color{blue}\vector(0,1){31}}
\put(58,141){\color{red}\vector(0,-1){31}}

\put(37,122){\color{blue}$\scriptstyle{z^{-1}}$}
\put(61,122){\color{red}$\scriptstyle{z[2]}$}

\put(100,110){\color{blue}\vector(-1,1){37}}

\put(50,100){${\bf u}_2$} \put(100,100){${\bf u'}_1$}

\put(62,80){\color{red}$\scriptstyle{z}$}
\put(27,80){\color{blue}$\scriptstyle{z^{-1}[2]}$}

\put(102,92){\color{red}\vector(-1,-1){37}}

\put(52,61){\color{blue}\vector(0,1){31}}
\put(58,91){\color{red}\vector(0,-1){31}}

\put(50,50){${\bf u}_3$}

\put(35,28){(i) $z:=qx=q^{-1}y$}
}

\multiput(50,0)(0,0){1}{
\put(50,150){${\bf u}_1$}

\put(52,111){\color{blue}\vector(0,1){31}}
\put(58,141){\color{red}\vector(0,-1){31}}

\put(37,122){\color{blue}$\scriptstyle{z^{-1}}$}
\put(61,122){\color{red}$\scriptstyle{z[2]}$}

\put(63,147){\color{red}\vector(1,-1){37}}

\put(50,100){${\bf u}_2$} \put(100,100){${\bf u'}_1$}

\put(62,80){\color{red}$\scriptstyle{z}$}
\put(27,80){\color{blue}$\scriptstyle{z^{-1}[2]}$}

\put(65,55){\color{blue}\vector(1,1){37}}

\put(52,61){\color{blue}\vector(0,1){31}}
\put(58,91){\color{red}\vector(0,-1){31}}

\put(50,50){${\bf u}_3$}

\put(35,28){(ii) $z:=q^{-1}x=qy$.}
}

\end{picture}
\end{equation}
In the both cases, $W_1(u)\otimes W_1(v)$ is indecomposable
and the subspace $\C{\bf u}_1\oplus \C{\bf u}_2 \oplus \C{\bf u}_3$
becomes isomorphic to $W_2(\frac{u+v}{2})$
corresponding to the multiplicative spectral parameter $z$.
The difference is that $W_2(\frac{u+v}{2})$ 
is the irreducible submodule in the case of (i)
while it is the irreducible quotient for (ii).
Denoting the trivial one dimensional module $\C{\bf u'}_1$ by $W_0$, 
we thus get the exact sequences of $U_q$-modules:
\begin{align}
{\rm (i) }\;\;&0 \rightarrow W_2(u) \rightarrow 
W_1(u-1)\otimes W_1(u+1) \rightarrow
W_0 \rightarrow 0,\label{wes1}\\
{\rm (ii) }\;\;&0 \rightarrow W_0 \rightarrow 
W_1(u+1)\otimes W_1(u-1) \rightarrow
W_2(u) \rightarrow 0.\label{wes2}
\end{align}
The general case, which was first worked out in the context of Yangian, 
is summarized in  
\begin{theorem}[\cite{CP0}]\label{th:es}
$W_m(u)\otimes W_n(v)$ is reducible if and only if 
$|u-v| = m+n-2j+2$ for some $1 \le j \le \min(m,n)$.
In these case, the following exact sequences are valid:
\begin{equation*}
\begin{split}
 0 &\rightarrow 
W_{j-1}(u+m-j+1)\otimes W_{m+n -j+1}(v-m+j-1) \rightarrow
W_m(u)\otimes W_{n}(v)\\
& \rightarrow 
W_{m-j}(u-j)\otimes W_{n-j}(v+j) \rightarrow 0
\end{split}
\end{equation*}
for $v-u = m+n - 2j + 2$.
\begin{equation}\label{wes4}
\begin{split}
0 &\rightarrow 
W_{m-j}(u+j)\otimes W_{n-j}(v-j) \rightarrow 
W_m(u)\otimes W_{n}(v)\\
&\rightarrow W_{j-1}(u-m+j-1)\otimes
W_{m+n -j+1}(v+m-j+1) \rightarrow 0
\end{split}
\end{equation}
for $u-v = m+n - 2j + 2$.
\end{theorem}

\subsection{\mathversion{bold}$q$-characters}\label{ss:qc}
Let $\mathrm{Rep}\, U_q(\hat{\mathfrak{g}})$ be
the Grothendieck ring of the category of the
type 1 finite dimensional $U_q(\hat{\mathfrak{g}})$-modules.
Such a module $W$ allows the direct sum decomposition
\begin{equation*}
W = \bigoplus_{\gamma=(
\gamma^\pm_{a,\pm n})_{i\in I, n \ge 0}}W_\gamma,\quad
W_\gamma = \{v \in W\mid
\exists p \ge 0, \forall a \in I, n\ge 0, 
(\phi^{\pm}_{a,\pm n}-\gamma^\pm_{a,\pm n})^p v=0\}.
\end{equation*}
It can be shown \cite{FR2} that the generating function
of the (generalized) eigenvalues is expressed as
\begin{equation}\label{Raz}
\sum_{n>0} \gamma^\pm_{a,\pm n} \zeta^{\pm n}
= q_a^{\deg R^+_a - \deg R^-_a} \frac{R^+_a(\zeta q_a^{-1})
R^-_a(\zeta q_a)}{R^+_a(\zeta q_a) R^-_a(\zeta q_a^{-1})}
\in \C[[\zeta^{\pm 1}]]
\end{equation}
in terms of some polynomials $R^\pm_a(\zeta)$ in 
$\zeta$ with constant term $1$. 

Let $\Z[Y^{\pm 1}_{a, z}]_{a\in I, z \in \C^{\times}}$ be the 
ring of integer coefficient Laurent polynomials 
in infinitely many algebraically independent variables 
$\{Y_{a,z}\mid a\in I, z \in \C^\times\}$\footnote{The
variable $Y_{a,z}$ is unrelated to the Y of Y-systems.}.
The Frenkel-Reshetikhin $q$-character 
$\chi_q$ is the injective ring morphism
\begin{equation}\label{rmo}
\chi_q  :  \mathrm{Rep}\, U_q(\hat{\mathfrak{g}}) 
\rightarrow \Z[Y^{\pm 1}_{a, z}]_{a\in I, z \in \C^{\times}},\qquad
\chi_q(W) = \sum_\gamma \dim(W_\gamma)m_\gamma,
\end{equation}
where the monomial $m_\gamma$ is specified from 
$R^{\pm}_a(\zeta)$ (\ref{Raz}) by
\begin{equation}\label{mpq}
m_\gamma = \prod_{a\in I, z\in \C^\times}
Y^{r^+_{a,z}-r^-_{a,z}}_{a,z},
\quad
R^{\pm}_a(\zeta) = \prod_{z \in \C^\times}
(1-\zeta z)^{r^{\pm}_{a,z}}.
\end{equation}

Suppose that $W$ is the irreducible representation with 
Drinfeld polynomials 
$\EuScript{P}_a(\zeta) = \prod_{s=1}^{m_a}(1-\zeta z^{(a)}_s)$.
Comparing (\ref{pwd}) with (\ref{Raz}) and (\ref{mpq}),
one finds that its $q$-character $\chi_q(W)$ contains 
the monomial 
$\prod_{a=1}^r\prod_{s=1}^{m_a}Y_{a,z^{(a)}_s}$
corresponding to the highest weight vector.
Such a monomial is called a highest weight monomial.
Thus, in particular, the $q$-character of the Kirillov-Reshetikhin module 
$W^{(a)}_m(u)$ is a Laurent polynomial 
containing the highest weight monomial as
\begin{equation}\label{tamy}
\chi_q(W^{(a)}_m(u)) = \prod_{s=1}^m
Y_{a, zq_a^{m+1-2s}} + \cdots,
\end{equation}
where we have set $z = q^{tu}$. 
The case $m=1$ is called the fundamental $q$-character.
For an analogous treatment of the Yangians, see \cite{Kin}.

Define $\mathrm{Ch}\, U_q(\hat{\mathfrak{g}})$ 
to be the image $\mathrm{Im}\, \chi_q$
and call it the $q$-character ring 
of $U_q(\hat{\mathfrak{g}})$. 
By the definition, $\mathrm{Ch}\, U_q(\hat{\mathfrak{g}})$
is an integral domain and a commutative ring 
isomorphic to $\mathrm{Rep}\, U_q(\hat{\mathfrak{g}})$.
The following fact is well known.
\begin{theorem}[\cite{FR2}, Corollary\ 2]\label{th:FR}
The ring $\mathrm{Ch}\, U_q(\hat{\mathfrak{g}})$
is freely generated by the fundamental $q$-characters
$ \chi_q(W^{(a)}_1(u))$ ($a\in I,u\in U$).
\end{theorem}

\begin{example}\label{ex:qc1}
For ${\mathfrak g} = A_1$,  
the $q$-character of the Kirillov-Reshetikhin module 
$W^{(1)}_m(u)$ is given by ($z = q^{u}$, $Y_z=Y_{1,z}$)
\begin{align*}
\chi_q(W^{(1)}_1(u))&= Y_z + Y^{-1}_{zq^{2}},\\
\chi_q(W^{(1)}_2(u)) &= 
Y_{zq^{-1}}Y_{zq} + Y_{zq^{-1}}Y^{-1}_{zq^3}
+Y^{-1}_{zq}Y^{-1}_{zq^3},\quad\text{and, in general},\\
\chi_q(W^{(1)}_m(u)) &
= \sum_{j=0}^m\prod_{i=1}^{m-j}Y_{zq^{-m-1+2i}}
\prod_{k=1}^jY^{-1}_{zq^{m+3-2k}}.
\end{align*}

\end{example}

\begin{example}\label{ex:qc2}
We write down the fundamental $q$-characters 
$\chi_q(W^{(a)}_1(u))$ for
${\mathfrak g}$ with rank 2 ($z = q^{tu}$).
\begin{align*}
A_2: \chi_q(W^{(1)}_1(u)) &= 
Y_{1,z} + Y^{-1}_{1,zq^{2}}Y_{2,zq} + Y^{-1}_{2,zq^3},\\
\chi_q(W^{(2)}_1(u)) &= 
Y_{2,z} + Y_{1,zq}Y^{-1}_{2,zq^{2}} + Y^{-1}_{1,zq^3},\\
B_2:  \chi_q(W^{(1)}_1(u)) &= 
Y_{1,z} + Y^{-1}_{1,zq^4}Y_{2,zq}Y_{2,zq^3} 
+Y_{2,zq}Y^{-1}_{2,zq^5}+
Y_{1,zq^2}Y^{-1}_{2,zq^3}Y^{-1}_{2,zq^5}+Y^{-1}_{1,zq^6},\\
\chi_q(W^{(2)}_1(u)) &= 
Y_{2,z} + Y_{1,zq}Y^{-1}_{2,zq^2} + Y^{-1}_{1, zq^5}Y_{2,zq^4}
+Y^{-1}_{2, zq^6},\\
C_2: \chi_q(W^{(a)}_1(u)) &= (\chi_q(W^{(3-a)}_1(u)) 
\text{ for $B_2$})|_{Y_{1,z}\leftrightarrow Y_{2,z}}\;\;(a=1,2),\\
G_2:   \chi_q(W^{(1)}_1(u)) &= 
Y_{1, z}+ Y_{2, zq} Y_{2, zq^3} Y_{2, zq^5} Y^{-1}_{1, zq^6}+ 
 Y_{2, zq} Y_{2, zq^3} Y^{-1}_{2, zq^7}\\
&+ 
 Y_{1, zq^4} Y_{2, zq} Y^{-1}_{2, zq^5} Y^{-1}_{2, zq^7}+ 
 Y_{1, zq^2} Y_{1, zq^4} Y^{-1}_{2, zq^3} Y^{-1}_{2, zq^5} 
Y^{-1}_{2, zq^7}\\
&+ 
 Y_{2, zq} Y_{2, zq^9} Y^{-1}_{1, zq^{10}}+ 
 Y_{1, zq^2} Y_{2, zq^9} Y^{-1}_{1, zq^{10}} 
Y^{-1}_{2, zq^3}+ 
 Y_{2, zq} Y^{-1}_{2, zq^{11}}\\
&+ Y_{1, zq^4} Y^{-1}_{1, zq^{8}}+
 Y_{2, zq^5} Y_{2, zq^7} Y_{2, zq^9} Y^{-1}_{1, zq^8} 
Y^{-1}_{1, zq^{10}}+ 
 Y_{1, zq^2} Y^{-1}_{2, zq^3} Y^{-1}_{2, zq^{11}}\\
&+ 
 Y_{2, zq^5} Y_{2, zq^7} Y^{-1}_{1, zq^8} Y^{-1}_{2, zq^{11}}+ 
 Y_{2, zq^5} Y^{-1}_{2, zq^9} Y^{-1}_{2, zq^{11}}\\
&+ 
 Y_{1, zq^6} Y^{-1}_{2, zq^7} Y^{-1}_{2, zq^9} 
Y^{-1}_{2, zq^{11}}+ Y^{-1}_{1, zq^{12}},
\\
\chi_q(W^{(2)}_1(u)) &= Y_{2,z} + Y_{1, zq}Y^{-1}_{2, zq^2}
+Y^{-1}_{1,zq^7}Y_{2,zq^4}Y_{2,zq^6}
+Y_{2,zq^4}Y^{-1}_{2,zq^8}\\
&+Y_{1,zq^5}Y^{-1}_{2,zq^6}Y^{-1}_{2,zq^8}
+Y^{-1}_{1,zq^{11}}Y_{2,zq^{10}} + Y^{-1}_{2,zq^{12}}.
\end{align*}
\end{example}

More examples will be given in Sections \ref{ss:A}--\ref{ss:D}.

Any finite dimensional $U_q(\hat{\mathfrak{g}})$-module $W$ defines a  
representation of the subalgebra $U_q(\mathfrak{g})$,
which we denote by ${\rm res}\,W$.
The (usual) character $\chi$ of the latter lives in 
$\Z[y^{\pm 1}_a]_{a\in I}$ with $y_a=e^{\omega_a}$
with $\omega_a$ being a fundamental weight.
The $q$-character is a deformation of the character by $z$ in that 
\begin{equation}\label{rew}
{\rm res }\, \chi_q(W) =\chi({\rm res}\,W),
\end{equation}
where ${\rm res}$ on the LHS is to be understood as
\begin{equation}\label{res}
\begin{split}
{\rm res }\,:\, \Z[Y^{\pm 1}_{a,z}]_{a\in I, z \in \C^\times}
&\rightarrow
\Z[y^{\pm 1}_a]_{a\in I}\\
Y_{a,z}&\mapsto y_a.
\end{split}
\end{equation}
Note that ${\rm res}\,W$ is not necessarily an irreducible 
$U_q(\mathfrak{g})$-module even if 
$W$ is so as a $U_q(\hat{\mathfrak{g}})$-module.
Therefore the irreducible $q$-character $\chi_q(W^{(a)}_m(u))$
does not restrict to an irreducible character in general.
In fact in Example \ref{ex:qc2}, one observes
\begin{equation}\label{g2e}
{\rm res }\,\chi_q(W^{(1)}_1(u)) 
= \begin{cases}
\chi(V_{\omega_1}) + \chi(V_0) 
&\text{if $\mathfrak{g}=G_2$ and $a=1$},\\
\chi(V_{\omega_a})
&\text{otherwise},
\end{cases}
\end{equation}
where $V_\lambda$ denotes the irreducible 
$U_q(\mathfrak{g})$-module with highest weight $\lambda$. 
The algebra $\mathfrak{g}=A_r$ is exceptional in that 
${\rm res }\,\chi_q(W^{(a)}_m(u)) = \chi(V_{m\omega_a})$
holds for all $a$ and $m$. 
See (\ref{rea}) and (\ref{nca}).
A systematic treatment of such decompositions is related to the 
{\em Kirillov-Reshetikhin conjecture} 
which has been fully solved by now. 
See Section \ref{s:q}, especially Section \ref{ss:eqc}.

For $a\in I$ and $z \in \C^\times$, set 
\begin{equation}\label{Ade}
A_{a,z} = Y_{a,zq^{-1}_a}Y_{a, zq_a}
\prod_{b: C_{ba}=-1}Y^{-1}_{b,z}
\prod_{b: C_{ba}=-2}Y^{-1}_{b,zq^{-1}}Y^{-1}_{b,zq}
\prod_{b: C_{ba}=-3}Y^{-1}_{b,zq^{-2}}
Y^{-1}_{b,z}Y^{-1}_{b,zq^{2}}.
\end{equation}
By the definition, one has ${\rm res}\,A_{a,z} = 
\prod_{b\in I}y_b^{C_{b a}}=e^{\alpha_a}$
with $\alpha_a$ being a simple root.

Let $S_a \,(a \in I)$ be the screening operator \cite{FR2}.
Namely, $S_a$ sends 
$\Z[Y^{\pm 1}_{a,z}]_{a\in I, z \in \C^\times}$
to the extended ring adjoined with the extra symbols 
$S_{a,z}$ with $a \in I, z \in \C^\times$.
The action is given by 
\begin{equation}\label{sac1}
S_a\cdot Y_{b,z} = \delta_{ab}Y_{a,z}S_{a,z}
\end{equation}
and the Leibniz rule
$S_a\cdot(YZ)=(S_a\cdot Y)Z + Y(S_a\cdot Z)$.
Thus for example, 
$S_a\cdot Y^{-1}_{b, z} = -\delta_{ab}Y^{-1}_{a, z}S_{a,z}$.
The symbol $S_{a,z}$ is assumed to obey the relation 
\begin{equation}\label{sac2}
S_{a, zq_a^2}=A_{a, zq_a}S_{a, z}
\end{equation}
in the extended ring.

\begin{theorem}[\cite{FR2,FM}]\label{th:frm}
(1) The $q$-character of an irreducible finite dimensional 
$U_q(\hat{\mathfrak g})$-module $W$ has the form
$\chi_q(W) = m_+(1+\sum_p M_p)$, where 
$m_+$ is the highest weight monomial and 
each $M_p$ is a monomial in $A^{-1}_{a, z}$,
$a \in I, z \in \C^{\times}$, (i.e. it does not contain 
any positive power factors of $A_{a,z}$). 

(2) The image $\mathrm{Im}\, \chi_q (\simeq 
\mathrm{Ch}\, U_q(\hat{\mathfrak{g}}))$ of the
$q$-character morphism (\ref{rmo})
is equal to $\bigcap_{a=1}^r {\rm Ker}\,S_a$.
\end{theorem}
The assertion (1) is a natural analog of its undeformed version
${\rm res}\chi_q(W) \in 
{\rm res}(m_+)(1+\sum_\alpha \Z_{\ge 0}\, e^{-\alpha})$,
where ${\rm res}(m_+) = e^{\text{highest weight}}$ and 
the $\alpha$-sum runs over 
$\Z_{\ge 0}\alpha_1\oplus \cdots \oplus
\Z_{\ge 0}\alpha_r\setminus \{0\}$.

The assertion (2) has a background in the characterization of 
the (deformed) $W$-algebra as the intersection of the kernel of  
screening operators \cite{FR2}.

\begin{example}\label{ex:fmc}
Let us illustrate Theorem \ref{th:frm} along ${\mathfrak g}=A_2$. 
The definition (\ref{Ade}) reads 
\begin{equation*}
A_{1,z} = Y_{1,zq^{-1}}Y_{1,zq}Y^{-1}_{2,z},\quad
A_{2,z} = Y_{2,zq^{-1}}Y_{2,zq}Y^{-1}_{1,z}.
\end{equation*}
Take $\chi_q=\chi_q(W^{(1)}_1(u))
= Y_{1,z} + Y^{-1}_{1,zq^{2}}Y_{2,zq} + Y^{-1}_{2,zq^3}$ 
for $A_2$ in Example \ref{ex:qc2}.
The highest weight monomial is $Y_{1,z}$.
$\chi_q$ is expressed as 
\begin{equation}
\chi_q=Y_{1,q}(1+A^{-1}_{1, zq}+
A^{-1}_{1, zq}A^{-1}_{2, zq^2})
\end{equation}
in agreement with (1).
With regard to (2), let us check that $\chi_q$ belongs to 
${\rm Ker }S_1\bigcap {\rm Ker }S_2$.
\begin{align*}
S_1\cdot \chi_q &= Y_{1,z}S_{1,z}-Y^{-1}_{1,zq^2}Y_{2,zq}S_{1,zq^2}
= Y_{1,z}S_{1,z}-Y^{-1}_{1,zq^2}Y_{2,zq}A_{1,zq}S_{1,z}
=0,\\
S_2\cdot \chi_q &= Y^{-1}_{1,zq^2}Y_{2,zq}S_{2,zq}
-Y^{-1}_{2,zq^3}S_{2,zq^3}
=Y^{-1}_{1,zq^2}Y_{2,zq}S_{2,zq}
-Y^{-1}_{2,zq^3}A_{2,zq^2}S_{2,zq}=0.
\end{align*}
\end{example}

\subsection{\mathversion{bold}T-system 
and $q$-characters}\label{ss:tsq}
We continue to set $u \in U=\mathbb{C}_{t\hbar}$ in this subsection.
The following is the fundamental result that relates 
the Kirillov-Reshetikhin modules with the T-system.
\begin{theorem}[\cite{N3, Her1}]\label{th:nh}
For any ${\mathfrak g}$, 
$T^{(a)}_m(u) = \chi_q(W^{(a)}_m(u))$
satisfies the unrestricted T-system for ${\mathfrak g}$.
\end{theorem}

In fact, the exact sequence corresponding to the ${\mathfrak g}$-version 
of $j=n=m$ case of (\ref{wes4}) has been obtained.
It is an elementary exercise to check that the
$q$-characters for ${\mathfrak g} = A_1$ 
in Example \ref{ex:qc1} satisfy the T-system (\ref{ta1}).

Theorem \ref{th:nh} leads to a description of the ring 
${\rm Rep} \,U_q(\hat{\mathfrak g}) \simeq
\mathrm{Ch}\, U_q(\hat{\mathfrak{g}})$ 
by the $q$-characters of the Kirillov-Reshetikhin modules and 
the unrestricted T-system, which we shall now explain.
Let $T=\{T^{(a)}_m(u) \mid a\in I, 
m\in \Z_{\ge 1}, u\in U \}$ 
denote the family of variables.
Let $\EuScript{T}({\mathfrak g})$ be the ring with generators 
$T^{(a)}_m(u)^{\pm 1}$ with the relations given by the T-system
for ${\mathfrak g}$.
Define $\EuScript{T}^{\circ}({\mathfrak g})$ to be the subring 
of $\EuScript{T}({\mathfrak g})$ generated by $T$.

\begin{theorem}[\cite{IIKNS}]\label{th:28}
The ring
 $\EuScript{T}^{\circ}({\mathfrak g})$
is isomorphic to $\mathrm{Rep}\, U_q(\hat{\mathfrak{g}})$
by the correspondence $T^{(a)}_m(u)\mapsto W^{(a)}_m(u)$.
\end{theorem}

\subsection{\mathversion{bold}T-system for quantum affinizations of
quantum Kac-Moody algebras}\label{ss:qaq}
The T-systems have been generalized by Hernandez \cite{Her3}
to the quantum affinizations
of a wide class of quantum Kac--Moody algebras
studied in \cite{D2,VV,Jin,Mi,N1,Her2}.
The most distinct feature compared from the setting so far is 
that the category
${\rm Rep}\, U_q(\hat{\mathfrak g})$ and the tensor product 
$\otimes$ need to be replaced by 
${\rm Mod}(U_q(\hat{\mathfrak g}))$ consisting of 
not necessarily finite dimensional modules and 
the fusion product $\ast_f$, respectively.
Nevertheless, with an appropriate definition of the 
Kirillov-Reshetikhin modules and their $q$-characters, 
the latter satisfy the (generalized) T-system \cite{Her3}.
  
Here we only give the definition of 
the quantum affinization of quantum Kac-Moody algebras and 
write down the T-system, leaving many details 
to \cite{Her3}. 
Instead, 
we include the explicit form of the corresponding Y-system \cite{KNS4}
on which our presentation is mainly based.

We begin by resetting the definitions and notations such as
$C, t, q_i, {\mathfrak g}, \hat{\mathfrak g}, 
U_q({\mathfrak g})$ and $U_q(\hat{\mathfrak g})$ 
introduced so far\footnote{This reset is only for the current subsection. }.
Let $I=\{1,\dots,r\}$
and let $C=(C_{ij})_{i,j \in I}$ be a
{\em generalized Cartan matrix\/} in \cite{Ka};
namely, it satisfies
$C_{ij}\in \mathbb{Z}$, $C_{ii}=2$, $C_{ij}\leq 0$ for
any $i\neq j$, and $C_{ij}=0$ if and only if $C_{ji}=0$.
We assume that $C$ is {\em symmetrizable\/},
i.e. there is a diagonal matrix $D=\mathrm{diag}
(d_1,\dots,d_r)$ with $d_i\in \mathbb{Z}_{\ge 1}$
such that $B=DC$ is symmetric.
We assume that there is no common divisor
for $d_1,\dots,d_r$ except for 1.

Let $(\mathfrak{h},\Pi,\Pi^{\vee})$ be a realization
of the Cartan matrix $C$ \cite{Ka};
namely, $\mathfrak{h}$ is a $(2r-\mathrm{rank}\,C)$ dimensional
$\mathbb{Q}$-vector space,
and $\Pi=\{\alpha_1,\dots,\alpha_r\}\subset \mathfrak{h}^*$,
$\Pi^{\vee}=\{\alpha_1^{\vee},\dots,\alpha_r^{\vee}\}
\subset \mathfrak{h}$
such that $\alpha_j(\alpha_i^{\vee})=C_{ij}$.
Let $q\in \mathbb{C}^{\times}$ be not a root of unity.
We set $q_i = q^{d_i}$ ($i\in I$) and use the symbols 
defined in (\ref{qb}).
Let $U_q({\mathfrak g})$ be 
the quantum Kac-Moody algebra \cite{D1,J1},
which is a $q$-analog of the 
Kac-Moody algebra ${\mathfrak g}$ associated with $C$
\cite{Ka}.

The {\em quantum affinization\/} (without central elements)
of the quantum Kac--Moody algebra $U_q(\mathfrak{g})$,
denoted by $U_q(\hat{\mathfrak{g}})$,
is the $\C$-algebra with generators
$x^{\pm}_{i,n}$ ($i\in I$, $n\in \mathbb{Z}$),
$k_h$ ($h\in \mathfrak{h}$),
$h_{i,n}$ ($i\in I$, $n\in \mathbb{Z}\setminus\{0\}$)
and the following relations:
\begin{gather}
k_h k_{h'} = k_{h+h'},\qquad
k_0=1,\qquad
k_h \phi^{\pm}_i(z) = \phi^{\pm}_i(z)k_h,\nonumber\\
k_h x^{\pm}_i(z) = q^{\pm \alpha_i(h)}x^{\pm}_i(z)k_h,\nonumber\\
\phi^{+}_i(z)x^{\pm}_j(w)=
\frac{q^{\pm B_{ij}}w-z}{w-q^{\pm B_{ij}}z}
x^{\pm}_j(w)\phi^+_i(z),\nonumber\\
\phi^{-}_i(z)x^{\pm}_j(w)=
\frac{q^{\pm B_{ij}}w-z}{w-q^{\pm B_{ij}}z}
x^{\pm}_j(w)\phi^-_i(z),\nonumber\\
x^+_i(z)x^-_j(w)-x^-_j(w)x^+_i(z)=
\frac{\delta_{ij}}{q_i-q_i^{-1}}
\left(
\delta\left(\frac{w}{z}\right)
\phi^+_i(w)
-
\delta\left(\frac{z}{w}\right)
\phi^-_i(z)
\right),\nonumber\\
(w-q^{\pm B_{ij}}z)x^{\pm}_i(z)x^{\pm}_j(w)=
(q^{\pm B_{ij}}w-z)
x^{\pm}_j(w)x^{\pm}_i(z),\nonumber\\
\sum_{\pi\in \Sigma}
\sum_{k=1}^{1-C_{ij}}
(-1)^k
{1-C_{ij} \brack{k}}_{q_i}
x^{\pm}_i(w_{\pi(1)})\cdots
x^{\pm}_i(w_{\pi(k)})
x^{\pm}_j(z)\nonumber\\
\qquad{}\times
x^{\pm}_i(w_{\pi(k+1)})\cdots
x^{\pm}_i(w_{\pi(1-C_{ij})})=0
\qquad (i\neq j).\label{eq:serre1}
\end{gather}
In (\ref{eq:serre1}) $\Sigma$ is the symmetric group
for the set $\{1,\dots,1-C_{ij}\}$.
We have also used the following formal series:
\begin{equation*}
x^{\pm}_i(z) =\sum_{n\in \mathbb{Z}} x^{\pm}_{i,n} z^n,\quad
\phi^{\pm}_i(z) =
k_{\pm d_i\alpha_i^{\vee}}
\exp \left(
\pm \big(q-q^{-1}\big)\sum_{n\geq 1} h_{i,\pm n} z^{\pm n}
\right).
\end{equation*}
and the formal delta function
$\delta(z)= \sum\limits_{n\in \mathbb{Z}} z^n$.

When $C$ is of finite type,
the above $U_q(\hat{\mathfrak{g}})$ is
called  an {\em $($untwisted$)$ quantum affine algebra
$($without central elements$)$} or {\em quantum loop algebra\/};
it is isomorphic to a subquotient 
of the previously introduced one (\ref{dr}) by 
the ideal generated by $c^{\pm 1/2}-1$ \cite{D2,Bec}.
When $C$ is of affine type,
the quantum Kac-Moody algebra $U_q({\mathfrak g})$ 
is the one in (\ref{djr}).
Its quantum affinization 
$U_q(\hat{\mathfrak{g}})$
 is called a {\em quantum toroidal algebra\/}
(without central elements).
In general, if $C$ is not of finite type,
$U_q(\hat{\mathfrak{g}})$ is no longer isomorphic to a subquotient of
any quantum Kac--Moody algebra
and has no Hopf algebra structure.

From now on we shall exclusively consider 
a symmetrizable generalized Cartan matrix $C$
satisfying the following condition due to Hernandez
\cite{Her3}:
\begin{gather}
\label{eq:Ccond1}
\mbox{If $C_{ij}< -1$, then $d_i=-C_{ji}=1$,}
\end{gather}
where $D=\mathrm{diag}(d_1,\dots,d_r)$
is the diagonal matrix symmetrizing $C$.
We say that a
generalized Cartan matrix $C$ is {\em tamely laced\/} if
it is symmetrizable and satisfies the condition~\eqref{eq:Ccond1}.
A generalized Cartan matrix $C$ is {\em simply laced\/} if
$C_{ij}=0$ or $-1$ for any $i\neq j$.
If $C$ is simply laced, then it is symmetric,
$d_a=1$ for any $a\in I$,
and it is tamely laced.

With a tamely laced generalized Cartan matrix $C$,
we  associate a {\em Dynkin diagram\/} in the standard way \cite{Ka}:
For any pair $i\neq j \in I$ with $C_{ij}<0$,
the vertices $i$ and $j$ are connected by
 $\max\{|C_{ij}|,|C_{ji}|\}$ lines,
and the lines are equipped with an arrow from $j$ to $i$
 if $C_{ij}<-1$.
Note that the condition \eqref{eq:Ccond1} means

\begin{itemize}\itemsep=0pt
\item[(i)] the vertices $i$ and $j$ are not connected
if $d_i,d_j>1$ and $d_i\neq d_j$,

\item[(ii)]
the vertices $i$ and $j$ are connected by $d_i$ lines with
 an arrow from $i$ to $j$
or not connected if $d_i>1$ and $d_j=1$,

\item[(iii)]
the vertices $i$ and $j$ are connected by a single line
or not connected if $d_i=d_j$.
\end{itemize}

\begin{example}
$(1)$ Any Cartan matrix of finite or affine type is
tamely laced except for types~$A^{(1)}_1$ and $A^{(2)}_{2\ell}$.

$(2)$ The following generalized Cartan matrix $C$
is tamely laced:
\begin{gather*}
C=
\begin{pmatrix}
2 & -1 & 0 & 0\\
-3& 2 & -2 & -2\\
0 & -1 & 2 & -1\\
0 & -1 & -1 & 2\\
\end{pmatrix},
\qquad
D=
\begin{pmatrix}
3 & 0 & 0 & 0\\
0& 1 & 0 & 0\\
0 & 0 & 2 & 0\\
0 & 0 & 0 & 2\\
\end{pmatrix}.
\end{gather*}
The corresponding Dynkin diagram is
\begin{figure}[h]
\begin{picture}(100,20)(0,13)

\put(-5,20){\circle{6}}\put(-8,7){1}
\put(20,20){\circle{6}}\put(17,7){2}
\put(42.5,31.5){\circle{6}}\put(48,29){3}
\put(42.5,9.5){\circle{6}}\put(48,3){4}

\put(-2.2,22){\line(1,0){19.4}}
\put(-1.8,20){\line(1,0){18.3}}
\put(-2.2,18){\line(1,0){19.4}}

\put(22.2,22.5){\line(2,1){17}}
\put(23.3,20.5){\line(2,1){17}}

\put(23.3,19.5){\line(2,-1){17}}
\put(22,17.5){\line(2,-1){17.3}}

\put(42.6,12.4){\line(0,1){15.9}}

\put(9,20){\line(-1,1){5}}
\put(9,20){\line(-1,-1){5}}

\put(30,25){\line(1,2){3}}
\put(30,25){\line(1,0){6.5}}

\put(30,14.9){\line(1,0){6.5}}
\put(30,14.9){\line(1,-2){3}}

\end{picture}
\end{figure}
\end{example}

Define the integer $t$ by
\begin{gather*}
t=\mathrm{lcm}(d_1,\dots,d_r).
\end{gather*}
For $a,b\in I$, we write $a\sim b$ if
$C_{ab}<0$, i.e. $a$ and $b$ are adjacent
in the corresponding Dynkin diagram.
Let $U$ be either $\frac{1}{t}\mathbb{Z}$, 
the complex plane $\mathbb{C}$, or
the cylinder $\mathbb{C}_{\xi}:= \mathbb{C}/
(2\pi \sqrt{-1}/\xi) \mathbb{Z}$ for some
 $\xi\in \mathbb{C}
\setminus 2\pi \sqrt{-1}\mathbb{Q}$,
depending on the situation under consideration.

For a tamely laced generalized Cartan matrix $C$,
the unrestricted T-system associated with $C$
\cite{Her3} is the following relations among the commuting variables
$\{T^{(a)}_m(u) \mid a\in I, m\in \Z_{\ge 1}, u\in U \}$:
\begin{gather}
\label{eq:T1}
T^{(a)}_m\left(u-\frac{d_a}{t}\right)
T^{(a)}_m\left(u+\frac{d_a}{t}\right)
 =
T^{(a)}_{m-1}(u)T^{(a)}_{m+1}(u)
+
\prod_{b: b\sim a}
T^{(b)}_{\frac{d_a}{d_b}m}(u)
\qquad\mbox{if \ \ $d_a>1$},
\\
\label{eq:T2}
T^{(a)}_m\left(u-\frac{d_a}{t}\right)
T^{(a)}_m\left(u+\frac{d_a}{t}\right)
 =
T^{(a)}_{m-1}(u)T^{(a)}_{m+1}(u)
+
\prod_{b: b\sim a}
S^{(b)}_{m}(u)
\qquad\mbox{if \ \ $d_a=1$},
\end{gather}
where
$T^{(a)}_0 (u)= 1$ if they occur
on the RHS in the relations.
The symbol $S^{(b)}_{m}(u)$ is defined by
\begin{gather}
\label{eq:Sbm}
S^{(b)}_{m}(u)=
{\prod_{k=1}^{d_b}}T^{(b)}_{1+E\left[\frac{m-k}{d_b}\right]}
\left(u+\frac{1}{t}\left(2k-1-m+E\left[\frac{m-k}{d_b}\right]d_b\right)
\right),
\end{gather}
and $E[x]$ ($x\in \mathbb{Q}$) denotes the largest integer
not exceeding $x$.

Explicitly, $S^{(b)}_m(u)$ is written as follows:
For $0\leq j < d_b$,
\begin{gather*}
S^{(b)}_{d_bm+j}(u)
=
\left\{
\prod_{k=1}^j
 T^{(b)}_{m+1}
\left(u+\frac{1}{t}(j+1-2k)\right)
\right\}\!
\left\{
\prod_{k=1}^{d_b-j}
 T^{(b)}_{m}\left(u+\frac{1}{t}(d_b-j+1-2k)\right)
\right\}\!.\!\!\!
\end{gather*}
For example, for $d_b=1$,
\begin{gather*}
S^{(b)}_{m}(u)=T^{(b)}_m(u),
\end{gather*}
for $d_b=2$,
\begin{gather*}
S^{(b)}_{2m}(u)
 =
T^{(b)}_{m}\left(u- \frac{1}{t}\right)
T^{(b)}_{m}\left(u+\frac{1}{t}\right),\\
S^{(b)}_{2m+1}(u)
 =
T^{(b)}_{m+1}(u)
T^{(b)}_{m}(u),
\end{gather*}
for $d_b=3$,
\begin{gather*}
S^{(b)}_{3m}(u)
 =
T^{(b)}_{m}\left(u- \frac{2}{t}\right)
T^{(b)}_{m}(u)
T^{(b)}_{m}\left(u+\frac{2}{t}\right),\\
S^{(b)}_{3m+1}(u)
 =
T^{(b)}_{m+1}(u)
T^{(b)}_{m}\left(u- \frac{1}{t}\right)
T^{(b)}_{m}\left(u+\frac{1}{t}\right),\\
S^{(b)}_{3m+2}(u)
 =
T^{(b)}_{m+1}\left(u- \frac{1}{t}\right)
T^{(b)}_{m+1}\left(u+ \frac{1}{t}\right)
T^{(b)}_{m}(u),
\end{gather*}
and so on.
The second terms on the RHS of
\eqref{eq:T1} and \eqref{eq:T2} can be written in a unified way
as follows \cite{Her3}:
\begin{gather*}
\prod_{b:b\sim a}
\prod_{k=1}^{-C_{ab}}
T^{(b)}_{-C_{ba}+E\left[\frac{d_a(m-k)}{d_b}\right]}
\left(u+\frac{d_b}{t}\left(\frac{-2k+1}{C_{ab}}-C_{ba}
+E\left[\frac{d_a(m-k)}{d_b}\right]-1\right)
-\frac{d_am}{t}
\right).
\end{gather*}

When $C$ is of finite type ${\mathfrak g}$, the above T-system coincides with 
the one for $U_q(\hat{\mathfrak g})$ in Section \ref{ss:utw}.
For $C$ of affine type, it was also studied by \cite{T1}
as a discrete Toda field equation. 

Let us proceed to the Y-system.
For a tamely laced generalized Cartan matrix $C$,
the unrestricted Y-system associated with~$C$
is the following relations among the commuting variables
$\{Y^{(a)}_m(u) \mid a\in I, m\in \Z_{\ge 1}, u\in U \}$,
where
$Y^{(a)}_0 (u)^{-1}= 0$ if they occur
on the RHS in the relations:
\begin{gather}
\label{eq:Y1}
Y^{(a)}_m\left(u-\frac{d_a}{t}\right)
Y^{(a)}_m\left(u+\frac{d_a}{t}\right)
=
\frac{
{ \prod\limits_{b:b\sim a}}
Z^{(b)}_{\frac{d_a}{d_b},m}(u)
}
{
(1+Y^{(a)}_{m-1}(u)^{-1})(1+Y^{(a)}_{m+1}(u)^{-1})}
\qquad\mbox{if \ \ $d_a>1$},
\\
\label{eq:Y2}
Y^{(a)}_m\left(u-\frac{d_a}{t}\right)
Y^{(a)}_m\left(u+\frac{d_a}{t}\right)
 =
\frac{
{\prod\limits_{b:b\sim a}}
\big(1+Y^{(b)}_{\frac{m}{d_b}}(u)\big)
}
{
(1+Y^{(a)}_{m-1}(u)^{-1})(1+Y^{(a)}_{m+1}(u)^{-1})}
\qquad\mbox{if \ \ $d_a=1$},
\end{gather}
where for $p\in \Z_{\ge 1}$
\begin{gather*}
Z^{(b)}_{p,m}(u)=
{\prod_{j=-p+1}^{p-1}}
\left\{
{\prod_{k=1}^{p-|j|}}
\left(
1+Y^{(b)}_{pm+j}\left(
u+\frac{1}{t}(p-|j|+1-2k)
\right)
\right)
\right\},
\end{gather*}
and $Y^{(b)}_{\frac{m}{d_b}}(u)=0$ in \eqref{eq:Y2}
if $\frac{m}{d_b}\not\in \Z_{\ge 1}$.

The Y-systems here are formally in the same form
as (\ref{yade})--(\ref{yg})
for the quantum affine algebras.
However, $p$ in $Z^{(b)}_{p,m}(u)$
here may be greater than 3.
On the RHS of \eqref{eq:Y1},
$\frac{d_a}{d_b}$ is either 1 or $d_a$
due to \eqref{eq:Ccond1}.
The term
 $Z^{(b)}_{p,m}(u)$ is written more explicitly as follows:
for $p=1$,
\begin{gather*}
Z^{(b)}_{1,m}(u)
=1+Y^{(b)}_{m}(u),
\end{gather*}
for $p=2$,
\begin{gather*}
Z^{(b)}_{2,m}(u)
=\big(1+Y^{(b)}_{2m-1}(u)\big)
\left(1+Y^{(b)}_{2m}\left(u-\frac{1}{t}\right)\right)
\left(1+Y^{(b)}_{2m}\left(u+\frac{1}{t}\right)\right)
\big(1+Y^{(b)}_{2m+1}(u)\big),
\end{gather*}
for $p=3$,
\begin{gather*}
Z^{(b)}_{3,m}(u)
=
\big(1+Y^{(b)}_{3m-2}(u)\big)
\left(1+Y^{(b)}_{3m-1}\left(u-\frac{1}{t}\right)\right)
\left(1+Y^{(b)}_{3m-1}\left(u+\frac{1}{t}\right)\right)\\
\phantom{Z^{(b)}_{3,m}(u)=}{}
\times \left(1+Y^{(b)}_{3m}\left(u-\frac{2}{t}\right)\right)
\big(1+Y^{(b)}_{3m}(u)\big)
\left(1+Y^{(b)}_{3m}\left(u+\frac{2}{t}\right)\right)\\
\phantom{Z^{(b)}_{3,m}(u)=}{}
\times \left(1+Y^{(b)}_{3m+1}\left(u-\frac{1}{t}\right)\right)
\left(1+Y^{(b)}_{3m+1}\left(u+\frac{1}{t}\right)\right)
\big(1+Y^{(b)}_{3m+2}(u)\big),
\end{gather*}
and so on.
There are $p^2$ factors in $Z^{(b)}_{p,m}(u)$.

The T and Y-systems in this subsection satisfy 
formally the same relations as those explained in 
Section \ref{ss:rty}.
Their restricted versions have also been 
formulated in \cite{KNS4}.

\subsection{Bibliographical notes}

The origin of the Kirillov-Reshetikhin modules (they are named so in  
\cite[Definition 1.1]{KN3})
goes back to \cite{KR3}, where the spectral parameter dependence
was not considered.
The idea of treating them as one family of 
$Y(\mathfrak{g})$ or $U_q(\hat{\mathfrak{g}})$
modules with spectral parameter
satisfying the  T-system in the Grothendieck ring
was initiated by \cite{KNS2},
where the identification by Drinfeld polynomials was also given
in the context of Yangian based on the result of \cite{CP0}.
Meanwhile, the representation theory of
finite dimensional $U_q(\hat{\mathfrak{g}})$ modules was
pushed forward by \cite{CP1,CP2},
where the Kirillov-Reshetikhin modules
were characterized and studied as {\em minimal affinizations\/} of  
$U_q(\hat{\mathfrak{g}})$ modules \cite{C1, CP4, CP5, C2, CH}.

The relation between the Kirillov-Reshetikhin modules and T-systems
became transparent after the introduction of $q$-character by \cite{FR2}.
The case of Yangian goes back to \cite{Kin}.
Theorem \ref{th:nh} is  due to \cite{N3} 
for simply laced ${\mathfrak g}$ and \cite{Her1} for general 
${\mathfrak g}$.
Under certain circumstances, 
there are algorithms to compute $q$-characters \cite{FM}
or its further generalization called $t$-analog 
of $q$-characters $\chi_{q,t}$ \cite{N2,N4} for ADE case.
In particular, $\chi_{q,t}$ of all the fundamental representations  
has been produced \cite{N4}, among which 
the $E_8$ case requires a supercomputer.

The T-systems for the quantum affinizations of 
quantum Kac--Moody algebras
in Section \ref{ss:qaq} are due to \cite{Her3}.
The corresponding Y-system and formulation by 
cluster algebra are given in \cite{KNS4}.

\section{Formulation by cluster algebras}\label{s:ca} 

\subsection{Dilogarithm identities in conformal field theory}
\label{t:ss:di}

Let $L(x)$ be the {\em Rogers dilogarithm function\/}
\cite{Le,Zag}
\begin{align}
\label{rodi}
L(x)=-\frac{1}{2}\int_{0}^x 
\left\{ \frac{\ln(1-y)}{y}+
\frac{\ln y}{1-y}
\right\} dy
\quad (0\leq x\leq 1).
\end{align}
It is well known that the following properties hold
$(0\leq x,y\leq 1)$.
\begin{gather}
\label{t:eq:L1}
L(0)=0,
\quad L(1)=\frac{\pi^2}{6},\\
\label{t:eq:L2}
\quad L(x) + L(1-x)=\frac{\pi^2}{6},\\
\label{t:eq:L3}
\quad L(x) + L(y)+ L(1-xy)+
L\left( \frac{1-x}{1-xy}\right)
+L\left( \frac{1-y}{1-xy}\right)
=\frac{\pi^2}{2}.
\end{gather}

In the series of works by
Bazhanov, Kirillov, and Reshetikhin
\cite{KR1,BR1,KR3,Kir3,BR2}, they reached a remarkable
conjecture on identities expressing the central charges
of conformal field theories
in terms of $L(x)$,
and partly established it.

In what follows, 
$\mathfrak{g}$ denotes any one of 
the simple Lie algebras $A_r, B_r, \ldots, G_2$ as in the previous sections.
In Section \ref{ss:res1}, we defined 
the level $\ell$ restricted Y-system for $\mathfrak{g}$ for 
$\ell \in \Z_{\ge 1}$.
Let us introduce the system of relations
for the variable $\{Y^{(a)}_m \mid a\in I, 1\leq
m \leq t_a\ell-1 \}$ 
obtained from the level $\ell$ restricted Y-system
by setting $Y^{(a)}_m(u)=Y^{(a)}_m$ 
dropping the dependence on the spectral parameter $u$.
We call it the {\em level $\ell$ restricted constant Y-system}.

\begin{theorem}[{\cite{NK,IIKKN1}}]
\label{t:thm:unique}
There exists a unique solution of the 
level $\ell$ restricted constant Y-system for 
${\mathfrak g}$  
satisfying $Y^{(a)}_m \in \R_{>0}$ for all $a\in I, 1\leq
m \leq t_a\ell-1$.
\end{theorem}

Theorem \ref{t:thm:unique}
 was proved by \cite{NK} for simply laced case,
and extended  to nonsimply laced case
by \cite{IIKKN1} using the same method.
For more information on the constant Y-system, 
see Section \ref{ss:cy} and \ref{ss:qru}.

The following theorem  was
originally conjectured by \cite{KR3} and \cite{BR2} for simply laced case,
and conjectured by \cite{Kir3} and properly corrected by
\cite{Ku} for nonsimply laced case.

\begin{theorem}[{Dilogarithm identities \cite{Kir3,Na1,IIKKN1,IIKKN2}}]
\label{t:thm:DI}
Suppose that a family of positive real numbers
$\{Y^{(a)}_m \mid a\in I, 1\leq m \leq t_a\ell-1\}$
satisfy the level $\ell$ constant Y-system for $\mathfrak{g}$.
Then, the following identities hold:
\begin{align}
\label{t:eq:DI}
\frac{6}{\pi^2}\sum_{a\in I}
\sum_{m=1}^{t_a\ell-1}
L\left(\frac{Y^{(a)}_m}{1+Y^{(a)}_m}\right)
=
\frac{\ell \dim \mathfrak{g}}{\ell+h^{\vee}} -
\mathrm{rank}\, \mathfrak{g},
\end{align}
where $h^{\vee}$ is the dual Coxeter number of $\mathfrak{g}$ (\ref{hhd}).
\end{theorem}

The rational number of the first term on the RHS
of \eqref{t:eq:DI} is the central charge of the {\em Wess-Zumino-Witten
 conformal field theory\/} associated with ${\mathfrak g}$ with level $\ell$
\cite{KZ, GW}.
The rational number on the RHS of \eqref{t:eq:DI} itself
is also the central charge of the {\em parafermion conformal
field theory\/} associated with ${\mathfrak g}$ with level $\ell$ \cite{ZF,Gep}.
The identity \eqref{t:eq:DI} is crucial  to establish the connection between
conformal field theories and various types of
non conformal integrable models in various limits
(cf. Section \ref{ss:lcs}).

\begin{example}
[\cite{KR1}]
\label{t:ex:A1}
 Consider the case $\mathfrak{g}=A_1$ and any $\ell$,
which is equivalent to the case
$\mathfrak{g}=A_{\ell-1}$ and $\ell=2$ by the {\em level-rank duality}.
Then, one has the solution
\begin{align}
Y^{(1)}_m = \frac{\sin^2\frac{\pi}{\ell+2}}
    {\sin\frac{m\pi}{\ell+2}
\sin\frac{(m+2)\pi}{\ell+2}},
\end{align}
and the corresponding identity  \eqref{t:eq:DI} reads 
\begin{align}
\label{t:eq:DIex}
\frac{6}{\pi^2}
\sum_{m=1}^{\ell-1}
L\left(
\frac{
\sin^2 \frac{\pi}{\ell+2}
}
{
\sin^2\frac{(m+1)\pi}{\ell+2}
}
\right)
=
\frac{3\ell}{2+\ell} - 1.
\end{align}
This identity has been known and studied by
various authors in various points of view.
See \cite{Kir4,Nah} and reference therein.
In particular,
the identity is derived  \cite{NRT, Te} from
the following $q$-series expression
\cite{LP}  for the parafermion conformal
character (``string function" in \cite{Ka} multiplied with 
Dedekind's eta function):
\begin{align}
\sum_{n_1,\dots,n_{\ell-1}=0}^{\infty}
\displaystyle
q^{\sum_{k,m=1}^{\ell-1} n_k n_m (\min (k,m) -\frac{km}{\ell})}
 \displaystyle \prod_{m=1}^{\ell-1} (q)_{n_m}^{-1},
\quad
(q)_k:=\prod_{j=1}^k (1-q^j),
\end{align}
where the sum is under the constraint
$ \sum_{m=1}^{\ell-1} m n_m 
\equiv 0
\ \mathrm{mod}\ 2\ell$.
In fact, a crude estimate by a saddle point method tells that as $q\rightarrow 1$,
this series diverges as  
${\rm const}\cdot(\overline{q})^{-c/24}$ 
where $c$ is the LHS of (\ref{t:eq:DIex}) and 
$\overline{q}\rightarrow 0$ is the modular 
conjugate specified by $(\ln q)(\ln \overline{q})=4\pi^2$. 
Comparing this with the known asymptotics of the string function \cite{Ka}
yields (\ref{t:eq:DIex}).
For general ${\mathfrak g}$, see around (\ref{qkns}). 
\end{example}

For $\mathfrak{g}=A_r$, Kirillov \cite{Kir3} gave
the explicit expression of the solution (cf. Example \ref{ex:rqy}),
 and  proved the corresponding
identity \eqref{t:eq:DI} by the analytic method,
but an extension of the proof to the other cases seemed difficult.

In the 1990s, people pursued  a proof through
lifting the dilogarithm identities
to the Rogers-Ramanujan type identities as
Example \ref{t:ex:A1} (e.g.
 \cite{KNS1,KKMM1,KKMM2,FSt,Geo}).
This created a new subject called the {\em Fermionic
formula} of conformal characters and their variants,
which turned out to be a rich subject itself,
and it has been 
intensively studied to this day by its own right.
See (ii) in Section \ref{ss:xm}.
In spite of this successful development,
the original problem of proving the
dilogarithm identities \eqref{t:eq:DI} itself did not
make much progress.

The scene changed after the introduction of
a new class of commutative algebras
called {\em cluster algebras\/} by Fomin-Zelevinsky
\cite{FZ1} around 2000,
which we explain in this section.

\subsection{Cluster algebras with coefficients}
\label{t:subsec:cluster}

Here
we recall the definition
of the cluster algebras with
coefficients
and some of their basic properties,
following the convention in \cite{FZ4}
with slight change of notations and terminology.
See \cite{FZ4} for more detail and information.

Fix an arbitrary  semifield $\mathbb{P}$,
i.e. an abelian multiplicative group
endowed with a binary
operation of addition $\oplus$ which is commutative,
associative, and distributive with respect to the
multiplication \cite{HW}.
Let $\mathbb{Q}\mathbb{P}$
denote the quotient field of the group ring $\mathbb{Z}\mathbb{P}$
of $\mathbb{P}$.
Let $I$ be a finite set\footnote{
This $I$ does not necessarily correspond to 
the $I$ in Section \ref{ss:utw} for the index set of 
Dynkin diagrams.}, and let $B=(B_{ij})_{i,j\in I}$ be a
 skew
symmetrizable (integer) matrix; namely, there is a diagonal positive integer
matrix $D$ such that
${}^{t}(DB)=-DB$.
Let $x=(x_i)_{i\in I}$
be an $I$-tuple of formal variables,
and let $y=(y_i)_{i\in I}$ be an $I$-tuple of elements in $\mathbb{P}$.
For the triplet $(B,x,y)$, called the {\em initial seed},
the  {\em cluster algebra $\mathcal{A}(B,x,y)$ with
coefficients in $\mathbb{P}$} is defined as follows.

Let $(B',x',y')$ be a triplet consisting of
skew symmetrizable matrix $B'$,
an $I$-tuple $x'=(x'_i)_{i\in I}$ with
 $x'_i\in \mathbb{Q}\mathbb{P}(x)$,
and 
an $I$-tuple $y'=(y'_i)_{i\in I}$ with $y'_i\in \mathbb{P}$.
For each $k\in I$, we define another triplet
$(B'',x'',y'')=\mu_k(B',x',y')$, called the {\em mutation
of $(B',x',y')$ at $k$}, as follows.

{\it  (i) Mutations of the matrix.}
\begin{align}
\label{t:eq:Bmut}
B''_{ij}=
\begin{cases}
-B'_{ij}& \mbox{$i=k$ or $j=k$},\\
B'_{ij}+\frac{1}{2}
(|B'_{ik}|B'_{kj} + B'_{ik}|B'_{kj}|)
&\mbox{otherwise}.
\end{cases}
\end{align}

{\it (ii)  Exchange relation of the coefficient tuple.}
\begin{align}
\label{t:eq:coef}
y''_i =
\begin{cases}
\displaystyle
{y'_k}{}^{-1}&i=k,\\
\displaystyle
y'_i \frac{1}{(1\oplus {y'_k}^{-1})^{B'_{ki}}}&
i\neq k,\ B'_{ki}\geq 0,\\
y'_i (1\oplus y'_k)^{-B'_{ki}}&
i\neq k,\ B'_{ki}\leq 0.\\
\end{cases}
\end{align}

{\it (iii)   Exchange relation of the cluster.}
\begin{align}
\label{t:eq:clust}
x''_i =
\begin{cases}
\displaystyle
\frac{y'_k
\prod_{j: B'_{jk}>0} {x'_j}^{B'_{jk}}
+
\prod_{j: B'_{jk}<0} {x'_j}^{-B'_{jk}}
}{(1\oplus y'_k)x'_k}
&
i= k.\\
{x'_i}&i\neq k,\\
\end{cases}
\end{align}
It is easy to see that  $\mu_k$ is an involution,
namely, $\mu_k(B'',x'',y'')=(B',x',y')$.
Now, starting from the initial seed
$(B,x,y)$, iterate mutations and collect all the
resulted triplets $(B',x',y')$.
We call $(B',x',y')$ the {\em seeds\/},
$y'$ and $y'_i$ a {\em coefficient tuple} and
a {\em coefficient\/},
$x'$  and $x'_i$, a {\em cluster\/} and
a {\em cluster variable}, respectively.
The {\em cluster algebra $\mathcal{A}(B,x,y)$ with
coefficients in $\mathbb{P}$} is the
$\mathbb{Z}\mathbb{P}$-subalgebra of the
rational function field $\mathbb{Q}\mathbb{P}(x)$
generated by all the cluster variables.

It is standard to identify
a {\em skew-symmetric} (integer) matrix $B=(B_{ij})_{i,j\in I}$
with a {\em quiver $Q$
without loops or 2-cycles}.
The set of the vertices of $Q$ is given by $I$,
and we put $B_{ij}$ arrows from $i$ to $j$ 
if $B_{ij}>0$.
The mutation $Q''=\mu_k(Q')$ of a quiver $Q'$ is given by the following
rule:
For each pair of an incoming arrow $i\rightarrow k$
and an outgoing arrow $k\rightarrow j$ in $Q'$,
add a new arrow $i\rightarrow j$.
Then, remove a maximal set of pairwise disjoint 2-cycles.
Finally, reverse all arrows incident with $k$.

Let  $\mathbb{P}_{\mathrm{univ}}(y)$ 
be the {\em universal semifield\/} of
the $I$-tuple of generators $y=(y_i)_{i\in I}$, namely,
the semifield consisting of 
the {\em subtraction-free\/} rational functions of formal
variables $y$ with
usual multiplication and addition in the rational function
 field $\mathbb{Q}(y)$.
We write $\oplus$ in $\mathbb{P}_{\mathrm{univ}}(y)$ as $+$
for simplicity.

{}{\em From now on,
unless otherwise mentioned,
we set the semifield $\mathbb{P}$ for $\mathcal{A}(B,x,y)$
to be $\mathbb{P}_{\mathrm{univ}}(y)$,
where $y$ is the coefficient tuple in the initial seed $(B,x,y)$.}

Let $\mathbb{P}_{\mathrm{trop}}(y)$ 
 be the {\em tropical semifield\/} 
of $y=(y_i)_{i\in I}$, which
is the abelian multiplicative group freely generated by
$y$ endowed with the addition $\oplus$
\begin{align}
\label{t:eq:trop}
\prod_i y_i^{a_i}\oplus
\prod_i y_i^{b_i}
=
\prod_i y_i^{\min(a_i,b_i)}.
\end{align}
There is a canonical surjective semifield homomorphism
$\pi_{\mathbf{T}}$ (the {\em tropical evaluation})
from $\mathbb{P}_{\mathrm{univ}}(y)$
to $\mathbb{P}_{\mathrm{trop}}(y)$ defined by $\pi_{\mathbf{T}}(y)= y$.
For any coefficient $y'_i$ of $\mathcal{A}(B,x,y)$,
let us write $[y'_i]_{\mathbf{T}}:= \pi_{\mathbf{T}}(y'_i)$ for simplicity.
We call $[y'_i]_{\mathbf{T}}$'s the {\em tropical coefficients\/}
(the {\em principal coefficients\/} in \cite{FZ4}).
They satisfy the exchange relation \eqref{t:eq:coef}
by replacing $y'_i$ with $[y'_i]_{\mathbf{T}}$ 
with $\oplus$ being the addition in \eqref{t:eq:trop}.
We also extend this homomorphism to
the homomorphism of fields
$\pi_{\mathbf{T}}:(\mathbb{Q}\mathbb{P}_{\mathrm{univ}}(y))(x)
\rightarrow 
(\mathbb{Q}\mathbb{P}_{\mathrm{trop}}(y))(x)$.

To each seed $(B',x',y')$ of $\mathcal{A}(B,x,y)$
we attach the {\em $F$-polynomials\/} $F'_i(y)\in
 \mathbb{Q}(y)$ ($i\in I$)
by the specialization of $[x'_i]_{\mathbf{T}}$
at $x_j=1$ ($j\in I$).
It is, in fact, a polynomial in $y$ with integer coefficients
due to the Laurent phenomenon \cite[Proposition 3.6]{FZ4}.
For definiteness, let us take  $I=\{1,\dots,n\}$.
Then,
$x'$ and $y'$ have the following factorized expressions
\cite[Proposition 3.13, Corollary 6.3]{FZ4}
by the $F$-polynomials.
\begin{align}
\label{t:eq:gF}
x'_i &=
\left(
\prod_{j=1}^n x_j^{g'_{ji}}
\right)
\frac{
F'_i(\hat{y}_1, \dots,\hat{y}_n)
}
{
F'_i(y_1, \dots,y_n)
},
\quad
\hat{y}_i=y_i\prod_{j=1}^n x_j^{B_{ji}},
\\
\label{t:eq:Yfact}
y'_i&=
[y'_i]_{\mathbf{T}}
\prod_{j=1}^n F'_j(y_1,\dots,y_n)^{B'_{ji}}.
\end{align}
The integer vector $\mathbf{g}'_i=(g'_{1i},\dots,g'_{ni})$ ($i=1,\dots,n$)
uniquely determined by \eqref{t:eq:gF}
for each $x'_i$ is called the {\em $g$-vector}
for $x'_i$.

Let $\mathbf{i}=(i_1,\dots,i_r)$ be an $I$-sequence,
namely, $i_1,\dots,i_r\in I$.
We define the {\em composite mutation\/} $\mu_{\mathbf{i}}$
by $\mu_{\mathbf{i}}=\mu_{i_r}
\cdots \mu_{i_2} \mu_{i_1}$, where the product means
the composition.

\begin{lemma}
\label{t:lem:order}
Let $B=(B_{ij})_{i,j\in I}$ be a skew symmetrizable matrix
and let $\mathbf{i}=(i_1,\dots,i_r)$ be an $I$-sequence.
Suppose that $B_{{i_a}{i_b}}=0$ for any $1\leq a,b \leq r$.
Then, the following facts hold.
\par
(a) For any permutation $\sigma$ of
$\{1,\dots, r\}$, we have
\begin{align}
\mu_{{\mathbf{i}}}(B,x,y)=
\mu_{(i_{\sigma(1)},\dots,i_{\sigma(r)})}(B,x,y).
\end{align}
\par
(b)
Let $B'=\mu_{\mathbf{i}}(B)$.
Then, $B'_{{i_a}{i_b}}=0$ holds for any $1\leq a,b \leq r$.
\par
(c)
Let  $(B',x',y')=\mu_{\mathbf{i}}(B,x,y)$.
Then, $(B,x,y)=\mu_{\mathbf{i}}(B',x',y')$.
\end{lemma}

\subsection{T and Y-systems in cluster algebras}
\label{t:subsec:TY1}

All the T and Y-systems in Sections \ref{ss:utw}--\ref{ss:res2} 
are regarded as relations among a cluster
among cluster variables and coefficients in certain
cluster algebras $\mathcal{A}(B,x,y)$.

Let us mention  two big advantages of cluster algebra formulation.

\begin{itemize}
\item[(a)]
The T and Y-systems are integrated
in one algebra $\mathcal{A}(B,x,y)$, and
commonly controlled by 
$F$-polynomials (together with tropical coefficients
and $g$-vectors) through
the formulas \eqref{t:eq:gF} and
\eqref{t:eq:Yfact}.
This fact may be hardly realized just by treating
the T and Y-systems only.

\item[(b)]
The cluster algebra $\mathcal{A}(B,x,y)$ itself is further
controlled by the {\em (generalized) cluster category\/}
developed 
by \cite{CC,BMRRT,Kel1,A,Pal2,Pal1}.
\end{itemize}

Here we concentrate on an example 
of level $4$ restricted T and Y-systems for $A_4$
to present a basic idea.
Let $Q$ be the following quiver with index set 
$\mathcal{I}=\{1,2,3,4\}\times
\{1,2,3\}$. Note that we also attached the property $+/-$ to
each vertex.
\begin{align}
\label{t:eq:qA}
\raisebox{-40pt}{
\begin{picture}(90,92)(0,-17)
\put(90,30){\circle{6}}
\put(0,0){\circle{6}}
\put(30,0){\circle{6}}
\put(60,0){\circle{6}}
\put(90,0){\circle{6}}
\put(0,30){\circle{6}}
\put(30,30){\circle{6}}
\put(60,30){\circle{6}}
\put(0,60){\circle{6}}
\put(30,60){\circle{6}}
\put(60,60){\circle{6}}
\put(90,60){\circle{6}}
\put(0,3){\vector(0,1){24}}
\put(0,57){\vector(0,-1){24}}
\put(30,27){\vector(0,-1){24}}
\put(30,33){\vector(0,1){24}}
\put(60,3){\vector(0,1){24}}
\put(60,57){\vector(0,-1){24}}
\put(90,27){\vector(0,-1){24}}
\put(90,33){\vector(0,1){24}}
\put(27,0){\vector(-1,0){24}}
\put(33,0){\vector(1,0){24}}
\put(87,0){\vector(-1,0){24}}
\put(3,30){\vector(1,0){24}}
\put(57,30){\vector(-1,0){24}}
\put(63,30){\vector(1,0){24}}
\put(27,60){\vector(-1,0){24}}
\put(33,60){\vector(1,0){24}}
\put(87,60){\vector(-1,0){24}}
\put(-12,3){\small $+$}
\put(18,3){\small $-$}
\put(48,3){\small $+$}
\put(78,3){\small $-$}
\put(-12,33){\small $-$}
\put(18,33){\small $+$}
\put(48,33){\small $-$}
\put(78,33){\small $+$}
\put(-12,63){\small $+$}
\put(18,63){\small $-$}
\put(48,63){\small $+$}
\put(78,63){\small $-$}
\put(-25,-12){\small $(1,1)$}
\put(5,-12){\small $(2,1)$}
\put(35,-12){\small $(3,1)$}
\put(65,-12){\small $(4,1)$}
\put(-25,18){\small $(1,2)$}
\put(5,18){\small $(2,2)$}
\put(35,18){\small $(3,2)$}
\put(65,18){\small $(4,2)$}
\put(-25,48){\small $(1,3)$}
\put(5,48){\small $(2,3)$}
\put(35,48){\small $(3,3)$}
\put(65,48){\small $(4,3)$}
\end{picture}
}
\end{align}
Below we identify $Q$ with the corresponding 
skew symmetric matrix $B$ as described in Section \ref{t:subsec:cluster}.

Let
$\mathbf{i}_+$ (resp.  $\mathbf{i}_-$)
be a sequence of all the distinct elements of
$\mathcal{I}$ with property $+$ (resp. $-$),
where the order of the sequence is chosen arbitrarily
thank to Lemma \ref{t:lem:order}.
Then, the quiver $Q$ has the following periodicity
under the sequences of mutation $\mathbf{i}_+$ and $\mathbf{i}_-$:
\begin{align}
Q
\
\mathop{\longleftrightarrow}^{\mu_{\mathbf{i}_+}}
\
 Q^{\mathrm{op}}
\
\mathop{\longleftrightarrow}^{\mu_{\mathbf{i}_-}}
\
Q,
\end{align}
where $Q^{\mathrm{op}}$ is the {\em opposite quiver\/} of $Q$,
namely, the quiver obtained from $Q$ by inverting
all the arrows.

Now we set $(Q(0),x(0),y(0)):=(Q,x,y)$ (the initial seed
of $\mathcal{A}(Q,x,y)$)
and consider the corresponding infinite
sequence of mutations of {\em seeds\/}
\begin{align}
\label{t:eq:seedmutseq}
\begin{split}
\cdots
\mathop{\longleftrightarrow}^{\mu_{\mathbf{i}_+}}
\
(Q(-1),x(-1),&y(-1))
\
\mathop{\longleftrightarrow}^{\mu_{\mathbf{i}_-}}
\
(Q(0),x(0),y(0))
\
\mathop{\longleftrightarrow}^{\mu_{\mathbf{i}_+}}
\\
&(Q(1),x(1),y(1))
\
\mathop{\longleftrightarrow}^{\mu_{\mathbf{i}_-}}
\
(Q(2),x(2),y(2))
\
\mathop{\longleftrightarrow}^{\mu_{\mathbf{i}_+}}
\
\cdots,
\end{split}
\\
Q(u)&=
\begin{cases}
Q& \mbox{$u$ is even}\\
Q^{\mathrm{op}}& \mbox{$u$ is odd},\\
\end{cases}
\end{align}
thereby introducing a family of clusters $x(u)$ ($u\in \mathbb{Z}$)
and coefficients tuples $y(u)$ ($u\in \mathbb{Z}$).

For $((i,i'),u)\in \mathcal{I} \times \mathbb{Z}$,
 we write $((i,i'),u):\mathbf{p}_+$
if $i+i'+u$ is even,
or equivalently,
if $u$ is even and $(i,i')$ has the property $+$,
or $u$ is odd and $(i,i')$ has the property $-$.
Plainly speaking, $((i,i'),u):\mathbf{p}_+$ 
 is a {\em forward mutation point\/}
in \eqref{t:eq:seedmutseq}.

For $((i,i'),u)\in \mathcal{I} \times \mathbb{Z}$, we set
 $((i,i'),u):\tilde{\mathbf{p}}_+$
if $((i,i'),u+1):\mathbf{p}_+$.
Consequently, we have
\begin{align}
\textstyle
((i,i'),u):\tilde{\mathbf{p}}_+
\Longleftrightarrow 
((i,i'),u\pm1):\mathbf{p}_+.
\end{align}

First, we explain how the Y-system appears in cluster algebra.
The sequence of mutations \eqref{t:eq:seedmutseq} gives
various relations among coefficients $y_{i,i'}(u)$ ($((i,i'),u)\in \mathcal{I} 
\times \mathbb{Z}$) by the exchange relation
\eqref{t:eq:coef}.
Then,
all these coefficients are 
 products of the
``generating" coefficients $y_{i,i'}(u)$ and $1+y_{i,i'}(u)$
 ($((i,i'),u):\mathbf{p}_+$).
Furthermore, these generating coefficients
obey some relations,
which are the Y-system.

Let us write down the relations explicitly.
Take $((i,i'),u):\mathbf{p}_+$ and consider the mutation
at $((i,i'),u)$,
where $y_{i,i'}(u)$ is exchanged to
$y_{i,i'}(u+1)=y_{i,i'}(u)^{-1}$,
by \eqref{t:eq:coef}.
In the next step going from $Q(u+1)$ to $Q(u+2)$, 
the (forward) mutation points are those satisfying $((j,j'),u+1):\mathbf{p}_+$.
Therefore the above $y_{i,i'}(u+1)$ gets 
multiplied by factors $(1+y_{j,j'}(u+1))$ 
if the quiver $Q(u+1)$ has an arrow from $(i,i')$ to $(j,j')$,
and $(1+y_{j,j'}(u+1)^{-1})^{-1}$ 
if the quiver $Q(u+1)$ has an arrow from $(j,j')$ to $(i,i')$.
The result coincides with the coefficient $y_{i,i'}(u+2)$.
In summary, we have the following relations:
For  $((i,i'),u):\mathbf{p}_+$,
\begin{align}
\label{t:eq:yi'}
y_{i,i'}(u)y_{i,i'}(u+2)
&= \frac{
(1+y_{i-1,i'}(u+1))(1+y_{i+1,i'}(u+1))
}
{
(1+y_{i,i'-1}(u+1)^{-1})(1+y_{i,i'+1}(u+1)^{-1})
},
\end{align}
where $y_{0,i'}(u+1)=y_{5,i'}(u+1)=0$
and $y_{i,0}(u+1)^{-1}=y_{i,4}(u+1)^{-1}=0$ on the RHS.
Or, equivalently, for  $((i,i'),u):\tilde{\mathbf{p}}_+$,
\begin{align}
\label{t:eq:yi}
\textstyle
y_{i,i'}(u-1)y_{i,i'}(u+1)
&= \frac{
(1+y_{i-1,i'}(u))(1+y_{i+1,i'}(u))
}
{
(1+y_{i,i'-1}(u)^{-1})(1+y_{i,i'+1}(u)^{-1})
}.
\end{align}
This certainly agrees with the level $4$ restricted 
Y-system for $A_4$
under the identification of $y_{i,i'}(u)$ with $Y^{(i)}_{i'}(u)$.

Next, we explain how the T-system appears in cluster algebra.
The sequence of mutations \eqref{t:eq:seedmutseq} gives
various relations among cluster variables $x_{i,i'}(u)$ ($((i,i'),u)\in \mathcal{I} 
\times \mathbb{Z}$) by the exchange relation
\eqref{t:eq:clust}.
All these coefficients are represented 
by the
``generating" cluster variables $x_{i,i'}(u)$ ($((i,i'),u):\mathbf{p}_+$).
Furthermore, these generating cluster variables
obey some relations,
which are the T-system.

Let us write down the relations explicitly.
Take $((i,i'),u):\mathbf{p}_+$ and consider the mutation
at $((i,i'),u)$.
Then, by \eqref{t:eq:clust} and
the fact that $((i\pm1,i'),u)$ and $((i, i'\pm1),u)$
are {\em not\/} forward mutation points, we have
\begin{align}
\label{t:eq:xi'}
\begin{split}
x_{i,i'}(u)x_{i,i'}(u+2)
&=
\frac{y_{i,i'}(u)}{1+y_{i,i'}(u)}
 x_{i-1,i'}(u+1)x_{i+1,i'}(u+1)\\
&\quad +
\frac{1}{1+y_{i,i'}(u)}
 x_{i,i'-1}(u+1)x_{i,i'+1}(u+1),
\end{split}
\end{align}
where $x_{0,i'}(u+1)=x_{5,i'}(u+1)=
x_{i,0}(u+1)=x_{i,4}(u+1)=1$ on the RHS.
By introducing the ``shifted cluster variables"
$\tilde{x}_i(u):=x_i(u+1)$ for $((i,i'),u):\tilde{\mathbf{p}}_+$,
these relations can be written  in a more ``balanced" form
and become parallel to \eqref{t:eq:yi}
as follows:
For $((i,i'),u):\mathbf{p}_+$,
\begin{align}
\label{t:eq:xi}
\begin{split}
\textstyle
\tilde{x}_{i,i'}(u-1)\tilde{x}_{i,i'}(u+1)
&=
\frac{y_{i,i'}(u)}{1+y_{i,i'}(u)}
\tilde{x}_{i-1,i'}(u)\tilde{x}_{i+1,i'}(u)\\
&\quad
+
\frac{1}{1+y_{i,i'}(u)}
\tilde{x}_{i,i'-1}(u)\tilde{x}_{i,i'+1}(u).
\end{split}
\end{align}

Let $\mathcal{A}(B,x)$ be the cluster algebra with
trivial coefficients with initial seed $(B,x)$.
Namely, we set every coefficient to be $1$
in the trivial semifield $\mathbf{1}=\{1\}$.
Let $\pi_{\mathbf{1}}:\mathbb{P}_{\mathrm{univ}}(y)
\rightarrow \mathbf{1}$ be the projection.
Let $[x_i(u)]_{\mathbf{1}}$ be the image
of $x_i(u)$ by the algebra homomorphism
$\mathcal{A}(B,x,y)\rightarrow \mathcal{A}(B,x)$
induced from $\pi_{\mathbf{1}}$.
By the specialization of \eqref{t:eq:xi},
we have
\begin{align}
\label{t:eq:ti}
\textstyle
[\tilde{x}_{i,i'}(u-1)]_{\mathbf{1}}
[\tilde{x}_{i,i'}(u+1)]_{\mathbf{1}}
&=
 [\tilde{x}_{i-1,i'}(u)]_{\mathbf{1}}
  [\tilde{x}_{i+1,i'}(u)]_{\mathbf{1}}
+
 [\tilde{x}_{i,i'-1}(u)]_{\mathbf{1}}
  [\tilde{x}_{i,i'+1}(u)]_{\mathbf{1}}.
\end{align}
This certainly agrees with the level $4$ restricted T-system for 
$A_4$ under the identification of 
$[\tilde{x}_{i,i'}(u)]_{\mathbf{1}}$ with $T^{(i)}_{i'}(u)$.

For $\mathfrak{g}$ simply laced,
the quiver relevant to the level $\ell$ restricted T and Y-systems 
is drawn similarly to (\ref{t:eq:qA}) on the vertex set 
${\mathcal I} = \{\text{nodes of the Dynkin diagram}\}
\times \{1,2,\ldots, \ell-1\}$.
For $\mathfrak{g}$ nonsimply laced,
it is slightly more involved \cite{IIKKN1,IIKKN2}.
Here we only give examples for 
$B_3$ with level 2 (left) and level 3 (right).
\begin{equation*}
\begin{picture}(310,70)(-70,-35)
\put(-100,0){
\put(30,0){\circle{5}}
\put(60,0){\circle{5}}
\put(90,0){\circle{5}}
\put(120,0){\circle{5}}
\put(150,0){\circle{5}}
\put(90,15){\circle{5}}
\put(57,0){\vector(-1,0){24}}
\put(87,0){\vector(-1,0){24}}
\put(93,0){\vector(1,0){24}}
\put(147,0){\vector(-1,0){24}}
\put(90,-15){\circle{5}}
\put(90,12){\vector(0,-1){9}}
\put(90,-12){\vector(0,1){9}}
\put(63,-2){\vector(2,-1){24}}
\put(63,2){\vector(2,1){24}}}

\put(80,-15){
\multiput(0,0)(0,30){2}{
\put(30,0){\circle{5}}
\put(60,0){\circle{5}}
\put(90,0){\circle{5}}
\put(120,0){\circle{5}}
\put(150,0){\circle{5}}
\put(90,15){\circle{5}}}

\put(33,30){\vector(1,0){24}}
\put(87,30){\vector(-1,0){24}}
\put(93,30){\vector(1,0){24}}
\put(123,30){\vector(1,0){24}}

\put(57,0){\vector(-1,0){24}}
\put(87,0){\vector(-1,0){24}}
\put(93,0){\vector(1,0){24}}
\put(147,0){\vector(-1,0){24}}

\put(90,-15){\circle{5}}
\multiput(0,0)(0,30){2}{
\put(90,12){\vector(0,-1){9}}
\put(90,-12){\vector(0,1){9}}}

\put(30,3){\vector(0,1){24}}
\put(60,27){\vector(0,-1){24}}
\put(120,3){\vector(0,1){24}}
\put(150,27){\vector(0,-1){24}}

\put(117,28){\vector(-2,-1){24}}
\put(117,32){\vector(-2,1){24}}

\put(63,-2){\vector(2,-1){24}}
\put(63,2){\vector(2,1){24}}

}

\end{picture}
\end{equation*}

\begin{remark}\label{re:ty}
Once we realize that  the T and Y-systems are integrated in a single  
cluster algebra
with coefficients as above,
the relation between T and Y-systems in Theorem \ref{th:ty1}
becomes an immediate consequence of  a more general relation
between cluster variables and coefficients in 
\cite[Prop.~3.9] {FZ4},
where \eqref{ty1} is a special case of
\cite[eq.~(3.7)]{FZ4}
with the specialization of the base semifield $\mathbb{P}$ therein
to the trivial semifield.
See also \cite[Prop.~5 .11]{Na2} for the relation between more  
general T and Y-systems.
\end{remark}

\subsection{Application to periodicity and dilogarithm identities}

As remarkable applications of the cluster algebra formulation,
one can prove the periodicities of T and Y-systems
and dilogarithm identities \eqref{t:eq:DI}.

The following periodicity property was originally
conjectured for
type $A_1$ by \cite{Z1},
for simply laced case by Ravanini-Tateo-Valleriani \cite{RTV},
and 
for nonsimply laced case
by Kuniba-Nakanishi-Suzuki \cite{KNS2}.

\begin{theorem}[{Periodicity 
\cite{GT2,FS,FZ3, Sz,V,Kel1,Kel2,  IIKNS,IIKKN1,
IIKKN2}}]
\label{t:thm:period1}
For any family of variables
 $\{Y^{(a)}_m (u)\mid a\in I, 1\leq m \leq t_a\ell-1,
u\in \mathbb{Z}\}$
satisfying the level $\ell$ restricted Y-system for $\mathfrak{g}$,
one has the periodicity
\begin{align}
Y^{(a)}_{m}(u+2(h^{\vee}+\ell))=Y^{(a)}_{m}(u).
\end{align}
\end{theorem}

To prove Theorem \ref{t:thm:period1} in full generality,
the use of the categorification 
of the cluster algebra by the cluster category by \cite{Pal2, Pal1} is essential.

Since the T-system is integrated in the same cluster algebra,
one can simultaneously prove  the periodicity of T-system 
as well, which was overlooked in the literature until recently
\cite{CGT, IIKNS}.

\begin{theorem}[{Periodicity \cite{FZ4,Hen,
 V, Kel1,IIKNS,IIKKN1,IIKKN2}}]
\label{t:thm:period2}
For any family of variables
 $\{T^{(a)}_m (u)\mid a\in I, 1\leq m \leq t_a\ell-1,
u\in \mathbb{Z}\}$
satisfying the level $\ell$ restricted T-system for $\mathfrak{g}$,
one has the periodicity
\begin{align}
T^{(a)}_{m}(u+2(h^{\vee}+\ell))=T^{(a)}_{m}(u).
\end{align}
\end{theorem}

Closely related to the periodicity of Y-systems,
the following (significant) functional generalization
of the dilogarithm identities \eqref{t:eq:DI}
was originally
conjectured for simply laced case by Gliozzi-Tateo \cite{GT1}.

\begin{theorem}[{Functional dilogarithm identities 
\cite{GT2, FS,Cha, Na1, IIKKN1,IIKKN2}}]
\label{t:thm:DI2}
Suppose that 
a family of positive real numbers
 $\{Y^{(a)}_m(u)\mid a\in I, 1\leq m \leq t_a\ell-1,
u\in \mathbb{Z}\}$
satisfy the level $\ell$ restricted Y-system for $\mathfrak{g}$.
Then, the following identities hold:
\begin{align}\label{t:eq:DI2}
\frac{6}{\pi^2}\sum_{a\in I}\sum_{m=1}^{t_a\ell-1}
\sum_{u=0}^{2(h^{\vee}+\ell)-1}
L\left(
\frac{Y^{(a)}_m(u)}{1+Y^{(a)}_m(u)}
\right)
&=
2(\ell h - h^{\vee})\mathrm{rank}\,\mathfrak{g},
\end{align}
where $h$ is the Coxeter number of ${\mathfrak g}$ (\ref{hhd}).
\end{theorem}

\begin{example}[\cite{GT1}]
(i) In the simplest case, type $A_1$,
the identity \eqref{t:eq:DI2} is equivalent to
\eqref{t:eq:L2}.
\par
(ii) In the next simplest case, type  $A_2$,
the identity \eqref{t:eq:DI2} is equivalent to
the  5-term relation \eqref{t:eq:L3}.
\end{example}

Theorem \ref{t:thm:DI2}
implies Theorem \ref{t:thm:DI};
namely, take a  {\em constant solution} $Y^{(a)}_m=Y^{(a)}_m(u)$
of the Y-system with respect to the spectral parameter $u$.
Or equivalently, take a solution to the constant Y-system 
in Section \ref{ss:cy}.
Then, one obtains  \eqref{t:eq:DI} from \eqref{t:eq:DI2}.

See Section \ref{t:subsec:cbib} for more precise
account of contributions
to Theorems \ref{t:thm:period1},
\ref{t:thm:period2}, and \ref{t:thm:DI2}.

\subsection{Bibliographical notes}
\label{t:subsec:cbib}

The cluster algebraic formulation
of Y-systems was given for the simply laced case with
level 2 by \cite{FZ3},
for the simply laced case with general level by
\cite{Kel1},
for the nonsimply laced case by
\cite{IIKKN1,IIKKN2},
and for the quantum affinizations of
the tamely laced quantum Kac-Moody algebras
by \cite{KNS4,Na3}.
The recognition of T-systems in the cluster algebras
was made a little later than Y-systems
in \cite{DK2,IIKNS,HL},
though the simply laced case with level 2 clearly
appeared in \cite{FZ2}.
The formulation here is due to
\cite{KNS4,IIKKN1,IIKKN2}.
See  \cite{Na2} for a further generalization of T and Y-systems
in view of cluster algebras.

Theorem \ref{t:thm:period1} was proved
for type $A_r$ with level 2
by \cite{GT2,FS},
for the simply laced case with level 2
by \cite{FZ2},
for type $A_r$ with general level
by \cite{Sz} and \cite{V},
for the simply laced case with general level
by \cite{Kel1,Kel2},
and for all the cases with unified method by \cite{IIKKN1,IIKKN2}.

Theorem \ref{t:thm:period2} was proved
for the simply laced case with level 2
by \cite{FZ2},
for type $A_r$ with general level
by \cite{Hen} and \cite{V},
for the simply laced case with general level
by \cite{Kel1} and \cite{IIKNS},
for type $C_r$ with general level
by \cite{IIKNS},
and for all the cases with a unified method by \cite{IIKKN1,IIKKN2}.
Actually in \cite{IIKNS,IIKKN1,IIKKN2}, 
refinements of Theorem \ref{t:thm:period1} and \ref{t:thm:period2} 
have been obtained concerning the property under 
the half shift $u \rightarrow u+ h^\vee+\ell$. 

Theorem \ref{t:thm:DI2} was proved
for type $A_r$ with level 2
by \cite{GT2,FS},
for the simply laced case with level 2
by \cite{Cha},
for the simply laced case with general level
by \cite{Na1}
and
for the nonsimply laced case by \cite{IIKKN1,IIKKN2}.
See  \cite{Na2} for a further generalization of dilogarithm identities
in view of cluster algebras.

There is a dilogarithm conjecture that generalizes (\ref{t:eq:DI})
involving $-24\times$(scaling dimensions) in addition to 
the central charge on the RHS. 
See  \cite{KN1} and \cite[appendix D]{KNS3}.
Some of them has been proved in \cite[section 1.3, 1.4]{Kir4}.

\section{Jacobi-Trudi type formula}\label{s:jt}

\subsection{\mathversion{bold}Introduction: Type $A_r$}\label{ss:jta}
In this section we exclusively consider unrestricted T-systems.
By Theorem \ref{th:FR}, we know that 
$T^{(a)}_m(u)$ is expressible as a polynomial
in the fundamental ones $T^{(1)}_1(v), \ldots, T^{(r)}_1(v)$
with various $v$.
Such formulas can be derived directly.
Consider for instance the unrestricted T-system for $A_2$:
\begin{equation*}
\begin{split}
T^{(1)}_m(u-1)T^{(1)}_m(u+1)
&=T^{(1)}_{m-1}(u)T^{(1)}_{m+1}(u)+T^{(2)}_{m}(u),\\
T^{(2)}_m(u-1)T^{(2)}_m(u+1)
&=T^{(2)}_{m-1}(u)T^{(2)}_{m+1}(u)+T^{(1)}_{m}(u).
\end{split}
\end{equation*}
Setting $m=1,2$ and noting $T^{(1)}_0(u) = T^{(2)}_0(u)=1$, 
one gets
\begin{align*}
T^{(1)}_2(u) &= T^{(1)}_1(u-1)T^{(1)}_1(u+1)-T^{(2)}_1(u),\\
T^{(2)}_2(u) &= T^{(2)}_1(u-1)T^{(2)}_1(u+1)-T^{(1)}_1(u),\\
T^{(1)}_3(u) 
&= T^{(1)}_1(u-2)T^{(1)}_1(u)T^{(1)}_1(u+2)
-T^{(1)}_1(u-2)T^{(2)}_1(u+1)\\
&\quad-T^{(1)}_1(u+2)T^{(2)}_1(u-1)+1.
\end{align*}
The formulas generated in this manner are systematized in 
a determinant form:
\begin{equation*}
\begin{split}
T^{(1)}_2(u) 
&= \begin{vmatrix}
T^{(1)}_1(u-1) & T^{(2)}_1(u)\\
1 & T^{(1)}_1(u+1)
\end{vmatrix},\quad
T^{(2)}_2(u)=\begin{vmatrix}
T^{(2)}_1(u-1) & 1\\
T^{(1)}_1(u) & T^{(2)}_1(u+1)
\end{vmatrix},\\
T^{(1)}_3(u) 
&= \begin{vmatrix}
T^{(1)}_1(u-2) & T^{(2)}_1(u-1) & 1 \\
1 & T^{(1)}_1(u) & T^{(2)}_1(u+1)\\
0 & 1 & T^{(1)}_1(u+2)
\end{vmatrix}.
\end{split}
\end{equation*}
Proceeding similarly, one gets
\begin{theorem}[\cite{BR2}]\label{th:jta}
For the unrestricted T-system for $A_r$,
the following formula is valid:
\begin{align}
T^{(a)}_m(u) 
= \det(T^{(a-i+j)}_1(u+i+j-m-1))_{1\le i,j \le m}, \label{jta1}
\end{align}
where $T^{(a)}_1(u)=0$ unless $0 \le a \le r+1$,
and $T^{(0)}_1(u)=T^{(r+1)}_1(u)=1$.
\end{theorem}

The proof reduces to the Jacobi identity among the determinants
\begin{equation}\label{dd}
D[{\textstyle {m+1 \atop m+1}}]
D[{\textstyle {1 \atop 1}}]
=D[{\textstyle {1, m+1 \atop 1, m+1}}]D 
+D[{\textstyle {1 \atop m+1}}]
D[{\textstyle {m+1 \atop 1}}],
\end{equation}
where $D[{i_1,i_2,\ldots \atop j_1,j_2,\ldots}]$ is the minor of $D$ 
removing $i_k$'s rows and $j_k$'s columns.

Alternatively, one can also solve the T-system to express 
everything by $T^{(1)}_k(v)$ with various $v$ and $k$.
By the same method as before, one can easily systematize such 
formulas and establish 
\begin{theorem}[\cite{BR2}]\label{th:jta2}
For the unrestricted T-system for $A_r$ (\ref{ta})
without assuming $T^{(r+1)}_m(u)=1$, 
the following formula is valid:
\begin{equation}\label{jta2}
T^{(a)}_m(u) 
= \det(T^{(1)}_{m-i+j}(u+i+j-a-1))_{1\le i,j \le a}\quad
(1 \le a \le r+1),
\end{equation}
where $T^{(1)}_0(u)=1$ and $T^{(1)}_m(u)=0$ for $m<0$.
\end{theorem}

The formulas (\ref{jta1}) and (\ref{jta2}) are 
quantum analog of the Jacobi-Trudi formula
for Schur functions \cite{Ma1}.

In the remainder of this section, we present the Jacobi-Trudi type formulas 
analogous to (\ref{jta1}) for the T-systems for  
$B_r, C_r$ and $D_r$.
The result involves not only determinants but also 
Pfaffians for $T^{(r)}_m(u)$ in $C_r$ and 
$T^{(r-1)}_m(u)$ and $T^{(r)}_m(u)$ in $D_r$.

\subsection{\mathversion{bold}Type $B_r$}\label{ss:jtb}

For any $k \in {\bf C}$, set 
\begin{equation}\label{xak}
x^a_k = \begin{cases}
T^{(a)}_1(u+k) & 1 \le a \le r,\\ 
1& a=0. 
\end{cases}
\end{equation}
We introduce the infinite dimensional matrices
${\mathcal T} = ({\mathcal T}_{i j} ) _{i,j \in \Z}$
and
${\mathcal E} = ({\mathcal E}_{i j} ) _{i,j \in \Z}$
as follows.
\begin{align}
{\mathcal T}_{i j} 
&= \begin{cases}
x^{{j-i \over 2}+1}_{{i+j\over 2}-1} &
\text{if  $i \in 2\Z+1$ 
and ${i-j \over 2} \in \{1,0,\ldots,2-r\}$},\\
-x^{{i-j \over 2}+2r-2}_{{i+j \over 2}-1} &
\text{if $i \in 2\Z+1$ and
${i-j \over 2} \in
\{1-r,-r,\ldots,2-2r\}$},\\ 
-x^r_{r+i-{5 \over 2}} & 
\text{if $i \in 2\Z$ and $j=i+2r-3$},\\
0 & \text{otherwise}.
\end{cases}\\
{\mathcal E}_{i j} 
&= \begin{cases}
\pm 1 & \text{if $i = j - 1 \pm 1$ and $i \in 2\Z$},\\
x^r_{i-1} & \text{if $i = j - 1$ and $i \in 2\Z + 1$},\\
0 & \text{otherwise}.
\end{cases}
\end{align}
For instance for $B_3$, they read
\begin{align}
({\mathcal T}_{i j})_{i,j \ge 1}&=
\begin{pmatrix}    x^1_0&     0&     x^2_1&      0&   
             -x^2_2&     0&   -x^1_3 &      0&    -1&       \\
                  0&     0&         0&      0&    -x^3_{5/2}&
                  0&     0&         0&     0&               \\
                  1&     0& x^1_2&      0&         x^2_3 &  0&
               -x^2_4&   0&  -x^1_5  &\cdots                \\
                   0& 0& 0& 0& 0& 0& -x^3_{9/2}& 0& 0&      \\
                   0& 0& 1& 0& x^1_4& 0& x^2_5& 0& -x^2_6&  \\
                    & &   &  &  \vdots &&&&& \ddots
\end{pmatrix},\label{t7a}\\
({\mathcal E}_{i j})_{i,j \ge 1}&=
\begin{pmatrix}0& x^3_0&  0&     0&     0&     0&     0&   \\
          0&     1&  0&    -1&     0&     0&     0&   \\
          0&     0&  0& x^3_2&     0&     0&     0&   \\
          0&     0&  0&     1&     0&    -1&     0&  \cdots \\
          0&     0&  0&     0&     0& x^3_4&     0&   \\
          0&     0&  0&     0&     0&     1&     0&   \\
           &      &   &    \vdots&  &      &      &\ddots 
\end{pmatrix}.
\end{align}
Let ${\mathcal T}\vert_{u \rightarrow u + s}$ be 
the overall shift of the lower index
$x^a_k \rightarrow x^a_{k+s}$ 
in ${\mathcal T}$ in accordance with (\ref{xak}).
As is evident from this example,
the quantity $x^a_k$ 
is contained in ${\mathcal T}\vert_{u \rightarrow u + s}$
at most once as its matrix element
for any $1 \le a \le r$ and $k$.
For example, the shift $s = 1$ is needed to
accommodate $x^1_1$ as the (1,1) element of 
${\mathcal T}\vert_{u \rightarrow u + s}$.
In view of this, we employ the notation
${\mathcal T}_m(i,j,\pm x^a_k)$ to mean the 
$m$ by $m$ sub-matrix of 
${\mathcal T}\vert_{u \rightarrow u + s}$, where
$s$ is chosen so that its $(i,j)$ element becomes  
exactly $\pm x^a_k$.
For example in (\ref{t7a}),
\begin{alignat*}{2}
{\mathcal T}_3(1,1,x^1_0) &= \begin{pmatrix}
x^1_0 & 0 & x^2_1 \\
0     & 0 &   0 \\
1     & 0 &  x^1_2
\end{pmatrix},&
\quad 
{\mathcal T}_3(1,1,x^1_1) &= \begin{pmatrix}
x^1_1 & 0 & x^2_2 \\
0     & 0 &   0 \\
1     & 0 &  x^1_3
\end{pmatrix}, \\
{\mathcal T}_2(1,2,-x^3_{5/2}) &= \begin{pmatrix}
0 & -x^3_{5/2} \\
0 & x^2_3 
\end{pmatrix},&
{\mathcal T}_2(1,2,-x^3_2) &= \begin{pmatrix}
0 & -x^3_2 \\
0 & x^2_{5/2}
\end{pmatrix}.
\end{alignat*}
We also use the similar notation 
${\mathcal E}_m(i,j,\pm x^r_k)$.
Now the result for $B_r$ is stated as
\begin{theorem}[\cite{KNH}]\label{th:jtb}
For unrestricted T-system for $B_r$, 
the following formula is valid:
\begin{align*}
T^{(a)}_m(u) &= \det\bigl(
{\mathcal T}_{2m-1}(1,1,x^a_{-m+1}) +
{\mathcal E}_{2m-1}(1,2,x^r_{-m+r-a+{1\over 2}})\bigr)  \quad
(1 \le a < r), \\
T^{(r)}_m(u) &= (-1)^{m(m-1)/2} \det\bigl(
{\mathcal T}_m(1,2,-x^{r-1}_{-{m\over 2}+1}) +
{\mathcal E}_m(1,1,x^r_{-{m\over 2}+{1\over 2}})\bigr).
\end{align*}
\end{theorem}

\subsection{\mathversion{bold}Type $C_r$}\label{ss:jtc}

Here we introduce the infinite dimensional 
matrix ${\mathcal T}$ by
\begin{equation}
{\mathcal T}_{i j} = \begin{cases}
x^{j-i+1}_{{i+j\over 2}-1} & \text{if $i - j \in \{1,0, \ldots, 1-r\}$},\\
- x^{i-j+2r+1}_{{i+j\over 2}-1} & 
\text{if $i - j \in \{-1-r,-2-r, \ldots, -1-2r\}$},\\ 
0 & \text{otherwise}.
\end{cases}
\end{equation}
For instance, for $C_2$, it reads
\begin{equation*}
({\mathcal T}_{i j})_{i,j \ge 1} = \begin{pmatrix}
x^1_0 &  x^2_{1/2} & 0 & -x^2_{3/2} & -x^1_2 & -1 & 0 & 0 &    \\
1 &  x^1_1  &  x^2_{3/2} & 0 & -x^2_{5/2} & -x^1_3 & -1 & 0 & \cdots \\
0 & 1 &  x^1_2  &  x^2_{5/2} & 0 & -x^2_{7/2} & -x^1_4 & -1 &  \\
0 & 0 & 1 & x^1_3 & x^2_{7/2} & 0 & -x^2_{9/2} & -x^1_5 & \\
&&& & \vdots &&&& \ddots
\end{pmatrix}.
\end{equation*}
We keep the notation (\ref{xak}) and 
${\mathcal T}_m(i,j,\pm x^a_k)$ ($1 \le a \le r$) 
as in Section \ref{ss:jtb}.
Note that ${\mathcal T}_m(1,2,-x^r_k)$ is an anti-symmetric matrix
for any $m$.
\begin{theorem}[\cite{KNH}]\label{th:jtc}
For unrestricted T-system for $C_r$, 
the following formula is valid:
\begin{align}
T^{(a)}_m(u) &= \det
{\mathcal T}_m(1,1,x^a_{-{m\over 2}+{1\over 2}})
\quad (1 \le a < r),\label{jtc1}\\
T^{(r)}_m(u) &= (-1)^m {\rm  pf }\,
{\mathcal T}_{2m}(1,2,-x^r_{-m+1}).\label{jtc2}
\end{align}
\end{theorem}

As an additional result, we have the following relations.
\begin{align}
T^{(r)}_m(u-\hf)T^{(r)}_m(u+\hf) &=
\det{\mathcal T}_{2m}(1,1,x^r_{-m+{1\over 2}}),\label{jtc3}\\
T^{(r)}_m(u) T^{(r)}_{m+1}(u) &=
\det {\mathcal T}_{2m+1}(1,1,x^r_{-m}).\label{jtc4}
\end{align}

If one extends the definition of $x^a_k$ (\ref{xak})  
by  $x^a_k + x^{2r+2-a}_k = 0$ in accordance with (\ref{ttr}),
then (\ref{jtc1}) is identical with the result (\ref{jta1}) 
for $A_{2r+1}$.

As remarked in the end of Section \ref{ss:utw},
the T-systems for $B_2$ and $C_2$ are equivalent 
by the interchange $T^{(1)}_m(u) \leftrightarrow T^{(2)}_m(u)$.
Therefore Theorems \ref{th:jtb} and \ref{th:jtc} 
supply these T-systems with two kinds of Jacobi-Trudi type formulas.

\subsection{\mathversion{bold}Type $D_r$}\label{ss:jtd}

Here we define the infinite dimensional matrices
${\mathcal T}$ and ${\mathcal E}$ by
\begin{align}
{\mathcal T}_{i j} &= \begin{cases}
x^{{j-i \over 2}+1}_{{i+j\over 2}-1} &
\text{if $i \in 2\Z+1$ and 
${i-j\over 2} \in \{1,0,\ldots,3-r\}$},\\
-x^{r-1}_{i+j-1\over 2}& 
\text{if $i \in 2\Z+1$ and 
${i-j\over 2}={5\over 2}-r$},\\
-x^{r}_{i+j-3\over 2}&
\text{if $i \in 2\Z+1$ and 
${i-j\over 2}={3\over 2}-r$},\\
-x^{{i-j\over 2}+2r-3}_{{i+j\over 2}-1}&
\text{if $i \in 2\Z+1$ and 
${i-j\over 2} \in \{1-r,-r,\ldots,3-2r\}$},\\
0& \text{otherwise}.
\end{cases}\\
{\mathcal E}_{i j} &= \begin{cases}
\pm 1 & \text{if $i=j-2 \pm 2$ and $i \in 2\Z$},\\
x^{r-1}_i & \text{if $i=j-3$ and $i \in 2\Z$},\\
x^{r}_{i-2}& \text{if $i=j-1$ and $i \in 2\Z$},\\
0& \text{otherwise}. 
\end{cases}
\end{align}
For instance for $D_4$, they read
\begin{align*}
\setcounter{MaxMatrixCols}{12}
({\mathcal T}_{i j})_{i,j \ge 1} 
&=  \begin{pmatrix} 
x^1_0 & 0 & x^2_1 & -x^{3}_2 & 0 & 
-x^{4}_2 & -x^{2}_3 & 0 & -x^{1}_4 & 0 & -1& \\
0& 0& 0& 0& 0& 0& 0& 0& 0& 0& 0&\cdots \\
1& 0& x^1_2& 0&x^{2}_3&  -x^{3}_4&  
0&  -x^{4}_4& -x^{2}_5& 0&-x^1_{6}& \\
0& 0& 0& 0& 0& 0& 0& 0& 0& 0& 0&  \\
& &  &  &  & &  \vdots & & & & & \ddots 
\end{pmatrix},\\
({\mathcal E}_{i j})_{i,j \ge 1} &= 
\begin{pmatrix}
0&  0&  0& 0&  0& 0&  0&  0& 0&   \\
0&  1& x^4_{0}&  0& x^{3}_{2}&  -1& 0& 
0&  0&   \ldots \\
0& 0& 0& 0&  0& 0& 0&0&  0& \\
0&  0&  0& 1& x^4_{2}& 0& x^{3}_{4}&  -1&  0& \\
& &  & & & \vdots & & & &
\end{pmatrix}. 
\end{align*}
We keep the notations (\ref{xak}), 
${\mathcal T}_m(i,j,\pm x^a_k)$ ($1 \le a \le r-2$) 
and ${\mathcal T}_m(i,j,-x^a_k), 
{\mathcal E}_m(i,j,x^a_k)$ ($a = r-1, r$) 
as in Section \ref{ss:jtb}.

\begin{theorem}[\cite{KNH}]\label{th:jtd}
For unrestricted T-system for $D_r$, 
the following formula is valid:
\begin{align}
&T^{(a)}_m(u) = \det\bigl(
{\mathcal T}_{2m-1}(1,1,x^a_{-m+1}) +
{\mathcal E}_{2m-1}(2,3,x^r_{-m-r+a+4}) \bigr)\;\;
(1 \le a \le r-2),\label{d6a}\\
&T^{(r-1)}_m(u) = {\rm pf} \bigl(
{\mathcal T}_{2m}(2,1,-x^{r-1}_{-m+1})+
{\mathcal E}_{2m}(1,2,x^{r-1}_{-m+1}) \bigr),\label{d6b}\\
&T^{(r)}_m(u) = (-1)^m {\rm pf} \bigl(
{\mathcal T}_{2m}(1,2,-x^r_{-m+1})+
{\mathcal E}_{2m}(2,1,x^r_{-m+1}) \bigr).\label{d6c}
\end{align}
\end{theorem} 

The matrices in (\ref{d6b}) and (\ref{d6c}) 
are indeed anti-symmetric.
The following relations also hold.
\begin{align*}
&T^{(r-1)}_m(u)T^{(r)}_m(u) = (-1)^m
\det \bigl( {\mathcal T}_{2m}(1,1,-x^{r-1}_{-m+1})+
{\mathcal E}_{2m}(2,2,x^r_{-m+1}) \bigr),
\\
&T^{(r-1)}_m(u+1)T^{(r)}_m(u-1) = (-1)^m
\det \bigl( {\mathcal T}_{2m}(1,1,-x^{r}_{-m})+
{\mathcal E}_{2m}(2,2,x^{r-1}_{-m+2}) \bigr),
\\
&T^{(r-1)}_{m+1}(u)T^{(r)}_m(u-1) = (-1)^{m+1}
\det \bigl( {\mathcal T}_{2m+1}(1,1,-x^{r-1}_{-m})+
{\mathcal E}_{2m+1}(2,2,x^r_{-m}) \bigr),
\\
&T^{(r-1)}_m(u+1)T^{(r)}_{m+1}(u) = (-1)^m
\det \bigl( {\mathcal T}_{2m+1}(2,1,x^{r-2}_{-m+1})+
{\mathcal E}_{2m+1}(1,1,x^r_{-m}) \bigr).
\end{align*}

Theorems \ref{th:jtb}--\ref{th:jtd} can only be proved 
by using (\ref{dd}) and the fact $({\rm pf})^2 = \det$.

\subsection{\mathversion{bold}Another 
Jacobi-Trudi type formula for $B_r$}\label{ss:jtb2}

For $B_r$ and $D_r$, a variant of the Jacobi-Trudi  type 
formula is known which has a quite similar structure to the $A_r$ case.
Compared with the rather sparse matrices  
${\mathcal T}$ and ${\mathcal E}$,
the relevant matrices are dense and involve 
some auxiliary variables.
Here we present the result for $B_r$.
The $D_r$ case is similar although slightly more involved.

Given $T^{(1)}_1(u),\ldots, T^{(r)}_1(u)$,  
we introduce the auxiliary variable $T^a(u)$ for all $a \in \Z$ by 
\begin{align}
T^a(u) &= \begin{cases}
0 & a < 0,\\
1 & a=0,\\
T^{(a)}_1(u) & 1 \le a \le r-1,
\end{cases}\label{tsp1}\\
T^a(u)+T^{2r-1-a}(u) &
= T^{(r)}_1(u-r+a+\textstyle\frac{1}{2})
T^{(r)}_1(u+r-a-\textstyle\frac{1}{2})\quad \text{for all } a \in \Z.
\label{tsp2}
\end{align}

Recall that $t_a=1$ for $a\neq r$ and $t_r=2$ for $B_r$
according to (\ref{eq:t1}).

\begin{theorem}[\cite{KOS}]\label{th:jtb2}
For unrestricted T-system for $B_r$, 
the following formula is valid:
\begin{align}
&T^{(a)}_{t_am}(u) = \det (T^{a+i-j}(u+i+j-m-1))_{1\le i,j\le m}
\quad (1 \le a \le r),\label{jtb3}\\
&T^{(r)}_{2m+1}(u) \nonumber\\
&=
\begin{small}
\begin{vmatrix}
T^{(r)}_1(u-m) & T^{r-1}(u-m+\frac{1}{2}) & 
T^{r-2}(u-m+\frac{3}{2}) & \cdots & T^{r-m}(u-\frac{1}{2})\\
T^{(r)}_1(u-m+2) & T^{r}(u-m+\frac{3}{2}) & 
T^{r-1}(u-m+\frac{5}{2}) & \cdots & T^{r-m+1}(u+\frac{1}{2})\\
\vdots & \vdots & \vdots & \ddots & \vdots\\
T^{(r)}_1(u+m) & T^{r+m-1}(u+\frac{1}{2}) & 
T^{r+m-2}(u+\frac{3}{2}) & \cdots & T^r(u+m-\frac{1}{2})
\end{vmatrix},
\end{small}\label{jtb4}
\end{align}
where the matrix (\ref{jtb4}) is of size $m+1$, 
its $(i\!+\!1,1)$ element is $T^{(r)}_1(u-m+2i)$ 
and the rest has the same pattern as (\ref{jtb3})
for $T^{(r)}_{2m+2}(u-\hf)$.
\end{theorem}

\subsection{Bibliographical notes}
The formulas (\ref{jta1})--(\ref{jta2}) for $A_r$ in Theorem \ref{th:jta} 
first appeared  in \cite{BR2} before the T-system was formulated.
There,  transfer matrices more general than $T^{(a)}_m(u)$ were considered.
Theorems \ref{th:jtb}--\ref{th:jtd} 
supplemented the determinant conjectures in \cite{KNS2}
with Pfaffians.
A result for $D_r$ analogous to Theorem \ref{th:jtb2} is available in \cite{TK}.

\section{Tableau sum formula}\label{s:tab}

\subsection{\mathversion{bold}Type $A_r$.}\label{ss:A}

Let $\fbox{$1$}_{\,u}, \ldots, \fbox{$r\!+\!1$}_{\, u}$ 
be variables depending on $u$.  
If we set $T^{(1)}_1(u) = \sum_{a=1}^{r+1} \fbox{$a$}_u$,
then
\begin{equation}\label{ehb}
T^{(1)}_1(u-1)T^{(1)}_1(u+1) = \sum_{a\le b} 
 \fbox{$a$}_{u-1}\fbox{$b$}_{u+1}
+\sum_{a>b} \;{\fbox{$b$}_{u+1}\atop \fbox{$a$}_{u-1}},
\end{equation}
where the both arrays of the boxes stand for the product. 
Comparing this with the T-system relation 
$T^{(1)}_1(u-1)T^{(1)}_1(u+1)
= T^{(1)}_2(u) + T^{(2)}_1(u)$,
one may identify $T^{(1)}_2(u)$ and $T^{(2)}_1(u)$ individually 
with the two terms in (\ref{ehb}),
and try to further establish similar formulas for higher $T^{(a)}_m(u)$.
Such a procedure leads to a solution of the T-system expressed 
as a sum of tableaux.
In fact, if one forgets the spectral parameter $u$ in (\ref{ehb}),
it can be viewed as the identity among Schur functions 
corresponding to the irreducible decomposition of the 
$A_r$-modules:
\begin{equation}\label{dst}
\begin{picture}(130,25)(0,-5)

\drawline(0,0)(0,12)
\drawline(0,12)(12,12)
\drawline(12,0)(12,12)
\drawline(0,0)(12,0)

\drawline(30,0)(30,12)
\drawline(30,12)(42,12)
\drawline(42,0)(42,12)
\drawline(30,0)(42,0)

\put(17,3){$\otimes$}
\put(50,3){$=$}

\drawline(62,0)(62,12)\drawline(74,0)(74,12)
\drawline(62,12)(86,12)
\drawline(86,0)(86,12)
\drawline(62,0)(86,0)

\put(92,3){$\oplus$}

\drawline(105,-6)(105,18)
\drawline(105,18)(117,18)
\drawline(117,-6)(117,18)
\drawline(105,6)(117,6)
\drawline(105,-6)(117,-6)

\end{picture}
\end{equation}
In this sense the result presented in what follows for $A_r$
is a deformation of the classical tableau sum formula
for the Schur functions \cite{Ma1}.

Consider the Young diagram $(m^a)$ 
of $a\times m$ rectangular shape.
Let $\text{Tab}(m^a)$ be the set of semistandard tableaux
on $(m^a)$ with numbers $\{1,2,\ldots, r\!+\!1\}$.
The inscribed numbers are strictly increasing to the bottom
and non-decreasing to the right.
For example when $r=2$, 
\begin{equation*}
\begin{picture}(200,45)

\put(-10,37){$\text{Tab}(2)=\{$}
\multiput(43.7,35)(30,0){6}{
\put(0,10){\line(1,0){20}}
\put(0,0){\line(1,0){20}}
\put(0,0){\line(0,1){10}}
\put(10,0){\line(0,1){10}}
\put(20,0){\line(0,1){10}}
}

\put(218,37){$\}$,}

\multiput(65,36)(30,0){5}{,}

\put(-14,7){$\text{Tab}(2^2)
=\left\{\phantom{\frac{P^P}{P_P}}\right.$}
\multiput(43.7,0)(30,0){6}{
\put(0,20){\line(1,0){20}}
\put(0,10){\line(1,0){20}}
\put(0,0){\line(1,0){20}}
\put(0,0){\line(0,1){20}}
\put(10,0){\line(0,1){20}}
\put(20,0){\line(0,1){20}}}

\put(46,36.9){$\textstyle{1}$}
\put(56,36.9){$\textstyle{1}$}
\put(76,36.9){$\textstyle{1}$}
\put(86,36.9){$\textstyle{2}$}
\put(106,36.9){$\textstyle{1}$}
\put(116,36.9){$\textstyle{3}$}
\put(136,36.9){$\textstyle{2}$}
\put(146,36.9){$\textstyle{2}$}
\put(166,36.9){$\textstyle{2}$}
\put(176,36.9){$\textstyle{3}$}
\put(196,36.9){$\textstyle{3}$}
\put(206,36.9){$\textstyle{3}$}

\put(202,7){$\left.\phantom{\frac{P^P}{P_P}}\right\}$.}

\multiput(65,1)(30,0){5}{,}

\put(46,11.9){$\textstyle{1}$}
\put(56,11.9){$\textstyle{1}$}
\put(76,11.9){$\textstyle{1}$}
\put(86,11.9){$\textstyle{1}$}
\put(106,11.9){$\textstyle{1}$}
\put(116,11.9){$\textstyle{1}$}
\put(136,11.9){$\textstyle{1}$}
\put(146,11.9){$\textstyle{2}$}
\put(166,11.9){$\textstyle{1}$}
\put(176,11.9){$\textstyle{2}$}
\put(196,11.9){$\textstyle{2}$}
\put(206,11.9){$\textstyle{2}$}

\put(46,1.9){$\textstyle{2}$}
\put(56,1.9){$\textstyle{2}$}
\put(76,1.9){$\textstyle{2}$}
\put(86,1.9){$\textstyle{3}$}
\put(106,1.9){$\textstyle{3}$}
\put(116,1.9){$\textstyle{3}$}
\put(136,1.9){$\textstyle{2}$}
\put(146,1.9){$\textstyle{3}$}
\put(166,1.9){$\textstyle{3}$}
\put(176,1.9){$\textstyle{3}$}
\put(196,1.9){$\textstyle{3}$}
\put(206,1.9){$\textstyle{3}$}

\end{picture}
\end{equation*}
Note that $\text{Tab}(m^a)$ is empty for $a>r+1$.
We define
\begin{equation}\label{tu}
T_u= \prod_{i=1}^a\prod_{j=1}^m
\fbox{$t_{ij}$}_{\,u+a-m-2i+2j}
\qquad\text{for}\;
T =(t_{ij}) \in \text{Tab}(m^a),
\end{equation}
where $t_{ij}$ denotes the entry of the box in the  
$i$th row and the $j$th column from the top left.

\begin{theorem}\label{th:taba}
\begin{equation}\label{atab}
T^{(a)}_m(u) = \sum_{T \in \mathrm{Tab}(m^a)} T_u
\quad (1\le a \le r+1)
\end{equation}
is a solution of the T-system for $A_r$ (\ref{ta}).
\end{theorem}

We note that $T^{(r+1)}_m(u)$ here is not just $1$ but
non trivially chosen as (\ref{atab}) 
as opposed to the original definition of the T-system.
However, $\text{Tab}(m^{r+1})$ consists of a unique tableau; 
therefore, (\ref{atab}) 
states that $T^{(r+1)}_m(u)$ is a monomial: 
\begin{equation}
T^{(r+1)}_m(u) = \prod_{j=1}^mT^{(r+1)}_1(u-m-1+2j),\quad
T^{(r+1)}_1(u)=\prod_{i=1}^{r+1}
\fbox{$i$}_{\, u+r+2-2i}.
\end{equation}
Thus the situation $T^{(r+1)}_m(u)=1$ can be restored  
if the variables 
$\fbox{$1$}_{\,u}, \ldots, \fbox{$r\!+\!1$}_{\, u}$
are chosen so as to satisfy the simple relation $T^{(r+1)}_1(u)=1$.
Theorem \ref{th:taba} yields the $q$-characters 
by the special choice
\begin{equation}\label{zaa}
\fbox{$a$}_u = 
z_a(u):=Y^{-1}_{a-1, q^{u+a}}Y_{a, q^{u+a-1}}
\qquad (Y_{0, q^u}=Y_{r+1, q^u}=1),
\end{equation}
which indeed satisfies the condition $T^{(r+1)}_m(u)=1$.
The restriction (\ref{rew})-(\ref{res}) 
of the resulting $q$-character 
$T^{(a)}_m(u)=\chi_q(W^{(a)}_m(u))$ is given by
\begin{equation}\label{rea}
{\rm res}\, T^{(a)}_m(u) = \chi (V_{m\omega_a})
\end{equation}
in the notation of (\ref{g2e}) since 
the $a\times m$ rectangle Young diagram corresponds to the  
highest weight $m\omega_a$.

In the rest of this section we shall present the 
tableau sum formulas for ${\mathfrak g}= B_r, C_r, D_r$ 
along the context of 
the $q$-characters $T^{(a)}_m(u) = \chi_q(W^{(a)}_m(u))$.
The contents cover all the fundamental ones 
$T^{(1)}_1(u), \ldots, T^{(r)}_1(u)$,
which is enough in principle to determine all the higher ones 
$T^{(a)}_m(u)$ due to Theorem \ref{th:FR}.
Some $T^{(a)}_m(u)$ 
allowing a relatively simple description will also be included. 

\subsection{\mathversion{bold}Type $B_r$}\label{ss:B}
Let us introduce the index set and a total order on it as
\begin{equation}\label{Jb}
J=\{1,2\ldots, r, 0, \overline{r},\ldots, \overline{2},\overline{1}\},\quad
1 \prec \cdots \prec r \prec 0 \prec 
 \overline{r} \prec \cdots  \prec \overline{1}.
\end{equation}
We introduce the variables corresponding to single box tableaux.
\begin{equation}\label{zab}
\begin{aligned}
& z_a(u)= Y_{a, q^{2u+2a-2}}Y_{a-1, q^{2u+2a}}^{-1}
\qquad (1\le a \le r-1),\\
& z_r(u)= Y_{r, q^{2u+2r-3}}Y_{r, q^{2u+2r-1}}
Y_{r-1, q^{2u+2r}}^{-1},\\
& z_0(u) = Y_{r, q^{2u+2r-1}} Y_{r, q^{2u+2r-3}} 
Y^{-1}_{r, q^{2u+2r+1}} Y^{-1}_{r, q^{2u+2r-1}},\\
& z_{\overline{r}}(u) = Y_{r-1,q^{2u+2r-2}}Y_{r, q^{2u+2r-1}}^{-1}
Y_{r, q^{2u+2r+1}}^{-1},\\
& z_{\overline{a}}(u)= Y_{a-1, q^{2u+4r-2a-2}}
Y_{a, q^{2u+4r-2a}}^{-1}
\qquad (1\le a \le r-1),
\end{aligned}
\end{equation}
where $Y_{0, q^k}=1$.
($z_0(u)$ in p1427 of \cite{KOSY} contains a misprint.) 
Consider the Young diagram $(m^a)$ 
of $a\times m$ rectangular shape.
Let $\text{Tab}(B_r, m^a)$ be the set of tableaux
on $(m^a)$ with entries from $J$.
The letter $t_{i,j}\in J$ inscribed on the $i$th row and the $j$th column
from the top left corner should satisfy the 
following conditions for any adjacent pair:
\begin{equation}
\begin{split}
&t_{i,j}\preceq t_{i,j+1}\; \;\text{and}\; \; (t_{i,j}, t_{i,j+1})\neq (0, 0),\\
&t_{i,j} \prec t_{i+1,j} \;\; \; \text{or} \; \;\;(t_{i,j}, t_{i+1, j})=(0, 0).
\end{split}
\end{equation}
Given a tableau $T=(t_{i,j}) \in \text{Tab}(B_r, m^a)$ we set 
\begin{equation}
T_u = \prod_{i=1}^a\prod_{j=1}^m z_{t_{i,j}}(u+a-m+2i+2j).
\end{equation}
This is an analog of the $A_r$ case (\ref{tu}).
\begin{theorem}[\cite{KOS,Her1}]\label{th:tabb}
The $q$-character $T^{(a)}_{t_am}(u) = \chi_q(W^{(a)}_{t_am}(u))$
is given by
\begin{equation}\label{btab1}
T^{(a)}_{t_am}(u) = \sum_{T \in \text{Tab}(B_r, m^a)}T_u\qquad
(1 \le a \le r).
\end{equation}
\end{theorem}

Recall that $t_a$ (\ref{eq:t1}) is $1$ except $t_r=2$ for $B_r$.
The formula (\ref{btab1}) is related to (\ref{jtb3}) in a parallel way with the 
$A_r$ case explained in the previous subsection.
A similar result is available for the remaining case $T^{(r)}_{2m+1}(u)$
based on (\ref{jtb4}) \cite{KOS}.
Theorem \ref{th:tabb} follows by combining the facts that 
the RHS and the $T^{(r)}_{2m+1}(u)$ 
satisfy the T-system \cite{KOS},
$q$-characters also satisfy the T-system \cite{Her1},
and the $T^{(a)}_m(u)$ is uniquely determined by the T-system 
and $T^{(a)}_1(u) \, (a \in I)$.  
See also \cite{NN3}.

Here we only give the formula for $T^{(r)}_1(u)$.  
It is known that the $U_q(B^{(1)}_r)$-module $W^{(r)}_1(u)$ is
isomorphic as a $U_q(B_r)$-module to the spin representation of the latter. 
Its weights are multiplicity-free and naturally labeled with the arrays
$(\sigma_1,\ldots, \sigma_r) \in \{\pm 1\}^r$.
Accordingly we introduce 
\begin{align}
&(\sigma_1,\ldots, \sigma_r)_u = 
\prod_{a=1}^r
\left(Y_{a, q^{2u+2r-1-\rho_a}}
\right)^{\frac{1}{2}(\sigma_a-\sigma_{a+1})},\\
&\rho_a = 2(\sigma_1+\cdots  + \sigma_{a-1})
+\frac{\sigma_a-\sigma_{a+1}}{t_a},\quad
\sigma_{r+1}=-\sigma_r.
\end{align}
Then we have
\begin{equation}
T^{(r)}_1(u) = \sum_{\sigma_1,\ldots, \sigma_r=\pm 1}
(\sigma_1,\ldots, \sigma_r)_u.
\end{equation}
For $r=2$, 
$T^{(2)}_1(u) = \chi_q(W^{(2)}_1(u))$ 
has been written down in Example \ref{ex:qc2}.

\subsection{\mathversion{bold}Type $C_r$}\label{ss:C}

Let us introduce the index set and a total order on it as
\begin{equation}\label{Jc}
J=\{1,2\ldots, r, \overline{r},\ldots, \overline{2},\overline{1}\},\quad
1 \prec \cdots \prec r \prec 
 \overline{r} \prec \cdots  \prec \overline{1}.
\end{equation}
For $1 \le a \le r$ we set
\begin{equation}\label{zac}
\begin{aligned}
& z_a(u)= Y_{a, q^{2u+a-1}}Y_{a-1, q^{2u+a}}^{-1},\\
& z_{\overline{a}}(u)= Y_{a-1, q^{2u+2r-a+2}}
Y_{a, q^{2u+2r-a+3}}^{-1},
\end{aligned}
\end{equation}
where $Y_{0, q^k}=1$.
Here we present the tableau sum formulas for 
$T^{(1)}_m(u)$ and $T^{(a)}_1(u)$.
Consider the Young diagram $(m)$ with length $m$ one row shape.
Let $\text{Tab}(C_r, (m))$ be the set of tableaux on it 
with entries from $J$ having the following form:
\begin{equation}\label{cm0}
\begin{picture}(300,45)(-60,-20)

\drawline(-3,-3)(-3,11)
\drawline(-3,11)(190,11)
\drawline(-3,-3)(190,-3)
\drawline(190,-3)(190,11)

\put(11,-3){\line(0,1){14}}
\put(41,-3){\line(0,1){14}}
\put(57,-3){\line(0,1){14}}
\put(71,-3){\line(0,1){14}}
\put(85,-3){\line(0,1){14}}
\put(105,-3){\line(0,1){14}}
\put(119,-3){\line(0,1){14}}
\put(133,-3){\line(0,1){14}}

\put(148,-3){\line(0,1){14}}
\put(175,-3){\line(0,1){14}}

\put(58,6)
{$\overbrace{\phantom{XXXXXXXX}}^{2n}$}

\put(1,0){$i_1$}
\put(20,0){$\cdots$}
\put(45,0){$i_k$}

\put(61,0){$\overline{r}$}\put(76,0){$r$}

\put(89,0){$\cdots$}
\put(109,0){$\overline{r}$}\put(123,0){$r$}

\put(137,1){$\overline{j}_l$}
\put(155,0){$\cdots$}
\put(178,1){$\overline{j}_1$}

\put(0,-20){$1 \preceq i_1 \preceq \cdots \preceq i_k \preceq r,
\quad
\overline{r} \preceq \overline{j}_l \preceq 
\cdots \preceq \overline{j}_1
\preceq \overline{1}$.}
\end{picture}
\end{equation}
Here $k, l$ and $n$ are any nonnegative integers satisfying 
$k+2n+l=m$.
Let those tableaux be denoted simply 
by the array of entries as $(i_1,\ldots, \overline{j}_1)
\in J^m$.
Then we have
\begin{equation}\label{cm1}
T^{(1)}_m(u) = \sum_{(i_1,\ldots, i_m) 
\in \text{Tab}(C_r, (m))} 
\prod_{k=1}^m z_{i_k}\bigl(u+\frac{2k-m-1}{2}\bigr).
\end{equation}

Consider the Young diagram $(1^a)$ with length $a$ one column shape.
Let $\text{Tab}(C_r, (1^a))$ be the set of tableaux on it 
with entries from $J$. 
The letter $i_k \in J$ inscribed on the $k$th row from the top
should satisfy the conditions:
\begin{equation}\label{ccba}
\begin{split}
&i_1 \prec \cdots \prec i_a,\\
&r+k-l \ge c \;\text{ for any $k, l, c$ such that }
\; i_k = c,\; i_l = \overline{c}.
\end{split}
\end{equation}
Denote such a tableau by the array $(i_1,\ldots, i_a) \in J^a$.
Then we have
\begin{equation}\label{cta1}
T^{(a)}_1(u) = \sum_{(i_1,\ldots, i_a) 
\in \text{Tab}(C_r, (1^a))} 
\prod_{k=1}^a z_{i_k}\bigl(
u+\frac{a+1-2k}{2}\bigr) \qquad(1\le a \le r).
\end{equation}

We note that $T^{(1)}_m(u)$ and $T^{(a)}_1(u)$ are 
the simplest cases in that the tableau rules can actually be   
described just by arrays without introducing a tableau.

\subsection{\mathversion{bold}Type $D_r$}\label{ss:D}
Here we treat $T^{(1)}_m(u)$ and the fundamental 
$q$-characters $T^{(a)}_1(u)$.
Let us introduce the index set and a partial order on it as
\begin{equation}\label{Jd}
J=\{1,2\ldots, r, \overline{r},\ldots, \overline{2},\overline{1}\},\quad
1 \prec \cdots \prec r-1 \prec 
{r \atop \overline{r}}
\prec\overline{r-1} \prec \cdots  \prec \overline{1},
\end{equation}
where no order is assumed between $r$ and $\overline{r}$.
For $i \in J$, define $z_i(u)$ by 
\begin{equation}\label{zad}
\begin{aligned}
& z_a(u)= Y_{a, q^{u+a-1}}Y_{a-1, q^{u+a}}^{-1}
\qquad (1 \le a \le r-2),\\
&z_{r-1}(u) 
= Y_{r-1, q^{u+r-2}}Y_{r, q^{u+r-2}}Y_{r-2,q^{u+r-1}}^{-1},\\
&z_r(u) = Y_{r, q^{u+r-2}}Y_{r-1,q^{u+r}}^{-1},\\
&z_{\overline{r}}(u) = Y_{r-1,q^{u+r-2}}Y_{r,q^{u+r}}^{-1},\\
&z_{\overline{r-1}}(u) = Y_{r-2,q^{u+r-1}}
Y_{r-1,q^{u+r}}^{-1}Y_{r,q^{u+r}}^{-1},\\
& z_{\overline{a}}(u)= Y_{a-1, q^{u+2r-a-2}}
Y_{a, q^{u+2r-a-1}}^{-1}
\qquad (1 \le a \le r-2),
\end{aligned}
\end{equation}
where $Y_{0, q^k}=1$.

Let $\text{Tab}(D_r, (m))$ be the set of one row tableaux
$(i_1,\ldots, i_m) \in J^m$ obeying the condition:
\begin{equation}
\begin{split}
&i_1\prec \cdots \prec i_m,\\
&r \;\text{ and }\; \overline{r} \; \text{ do not appear simultaneously}.
\end{split}
\end{equation}
Then we have
\begin{equation}\label{dtab1}
T^{(1)}_m(u) = \sum_{(i_1,\ldots, i_m) 
\in \text{Tab}(D_r, (m))} 
\prod_{k=1}^m z_{i_k}(u+2k-m-1).
\end{equation}

For $1\le a \le r-2$, 
let $\text{Tab}(D_r, (1^a))$ be the set of one column tableaux
$(i_1,\ldots, i_a) \in J^a$ obeying the condition:
\begin{equation}
i_k\prec i_{k+1}\; \text{ or }\; 
(i_k, i_{k+1}) = (r, \overline{r}) 
\; \text{ or }\; 
(i_k, i_{k+1}) = (\overline{r}, r) 
\;\text{ for  } 1 \le k \le a-1.
\end{equation}
Then we have
\begin{equation}\label{dtab2}
T^{(a)}_1(u) = \sum_{(i_1,\ldots, i_a) 
\in \text{Tab}(D_r, (1^a))} 
\prod_{k=1}^a z_{i_k}(u+a+1-2k) \qquad(1\le a \le r-2).
\end{equation}

It is known that the $U_q(D^{(1)}_r)$-modules 
$W^{(r-1)}_1(u)$ and $W^{(r)}_1(u)$ are
isomorphic as $U_q(D_r)$-modules to the spin representations of the latter. 
Their weights are multiplicity-free and naturally labeled with the arrays
$(\sigma_1,\ldots, \sigma_r) \in \{\pm 1\}^r$.
Accordingly we introduce 
\begin{align}
&(\sigma_1,\ldots, \sigma_r)_u = 
\left(Y_{r, q^{u+r-1-\rho_r}}
\right)^{\frac{1}{2}(\sigma_r+\sigma_{r-1})}
\prod_{a=1}^{r-1}
\left(Y_{a, q^{u+r-1-\rho_a}}
\right)^{\frac{1}{2}(\sigma_a-\sigma_{a+1})},\\
&\rho_a 
= \begin{cases}
\sigma_1+\cdots  + \sigma_{a-1}
+\frac{\sigma_a-\sigma_{a+1}}{2} &1 \le a \le r-1,\\
\sigma_1+\cdots  + \sigma_{r-2}
+\frac{\sigma_r+\sigma_{r-1}}{2}& a=r.
\end{cases}
\end{align}
It follows that 
\begin{equation}
(\sigma_1,\ldots, \sigma_{r-1},-\sigma_r)_u 
= (\sigma_1,\ldots, \sigma_r)_u|_{Y_{r,q^k} \leftrightarrow Y_{r-1,q^k}}.
\end{equation}
We have
\begin{equation}
\begin{split}
T^{(r-1)}_1(u) 
= \sum_{\begin{subarray}{c} \sigma_1,\ldots, \sigma_r=\pm 1\\ 
\sigma_1 \cdots \sigma_r=-1\end{subarray}}
(\sigma_1,\ldots, \sigma_r)_u,
\quad
T^{(r)}_1(u) 
= \sum_{\begin{subarray}{c} \sigma_1,\ldots, \sigma_r=\pm 1\\ 
\sigma_1 \cdots \sigma_r=1\end{subarray}}
(\sigma_1,\ldots, \sigma_r)_u.
\end{split}
\end{equation}

\subsection{Bibliographical notes}
Tableau sums in Theorems \ref{th:taba} and \ref{th:tabb} were 
respectively given in 
\cite{BR2} and \cite{KOS} in the context of 
analytic Bethe ansatz for more 
general skew shape Young diagrams.
A uniform proof of the 
equality between the Jacobi-Trudi type determinant and 
the tableau sum is available in \cite{NN1}.
For type $A_r$,  see also \cite{NNSY} for an account 
from the viewpoint of Macdonald's ninth variation of Schur functions \cite{Ma2}.
The tableau sums in Sections \ref{ss:C} and \ref{ss:D}  
first appeared in the analytic Bethe ansatz \cite{KS1}.
The sums of the same structure are used in 
the deformed $W$-algebras \cite{FR1}. 
Tableau constructions of higher $T^{(a)}_m(u)$ for $C_r$ and 
$D_r$, which are significantly more involved than $A_r$ and 
$B_r$, have been achieved in \cite{NN2, NN3}. 
In this section we have only treated the untwisted case $U_q(\hat{\mathfrak g})$.
For tableau sum formulas for T-systems in twisted case,
see \cite{KS2, Her4} and reference therein.

\section{Analytic Bethe ansatz}\label{s:aba}

Let $T^{(a)}_m(u)$ be the commuting transfer matrix
of a solvable lattice model in the sense of Section \ref{s:ctm}.
There is an empirical method called {\it analytic Bethe ansatz} to produce 
eigenvalues of $T^{(a)}_m(u)$ in many cases.
Those eigenvalue formulas possess a specific ``dressed vacuum form"
which necessarily satisfy the T-system in Remark \ref{re:g}
with a nontrivial $g^{(a)}_m(u)$.
Here we consider the Bethe equation and dressed vacuum forms 
for general ${\mathfrak g}$ and $T^{(a)}_m(u)$, 
and reformulate the conventional analytic Bethe ansatz 
via its connection with $q$-characters.

\subsection{\mathversion{bold}$A_1$ case}\label{ss:aba1}
Consider the 6 vertex model (\ref{6v}).
Here we employ the normalization 
\begin{equation}\label{6vr}
\begin{picture}(330,62)(0,-3)
\multiput(14.5,38)(55,0){6}{
\put(-12,0){\line(1,0){24}}
\put(0,-10){\line(0,1){20}}
}

\put(12,51){1}
\put(-5,34){1}
\put(29,34){1}
\put(12,16){1}

\multiput(5,0)(0,0){1}{
\put(62,51){2}
\put(45,34){2}
\put(79,34){2}
\put(62,16){2}}

\multiput(10,0)(0,0){1}{
\put(112,51){1}
\put(95,34){2}
\put(129,34){2}
\put(112,16){1}}

\multiput(15,0)(0,0){1}{
\put(162,51){2}
\put(145,34){1}
\put(179,34){1}
\put(162,16){2}}

\multiput(20,0)(0,0){1}{
\put(212,51){2}
\put(195,34){1}
\put(229,34){2}
\put(212,16){1}}

\multiput(25,0)(0,0){1}{
\put(262,51){1}
\put(245,34){2}
\put(279,34){1}
\put(262,16){2}}

\put(0,0){$[2+u]_{q^{1/2}}$}
\put(55,0){$[2+u]_{q^{1/2}}$}
\put(117,0){$[u]_{q^{1/2}}$}
\put(171,0){$[u]_{q^{1/2}}$}

\put(215,0){$z^{1/2}[2]_{q^{1/2}}$}

\put(272,0){$z^{-1/2}[2]_{q^{1/2}}$,}
\end{picture}
\end{equation}
which is obtained by dividing (\ref{6v}) by 
$(zq)^{1/2}(1-q)$ and setting $z=q^u$.
For the definition of the symbol $[u]_q$, see (\ref{qnb}).
Let $T_1(u)$ be the transfer matrix (\ref{trme}) with 
$m=1$ and $w_j=q^{v_j}$.
Its eigenvalue (denoted by the same symbol) is given by \cite{Ba3}
\begin{align}
T_1(u) &= \fbox{$1$}_{\,u} + \fbox{$2$}_{\,u},
\label{1+2}\\
\fbox{$1$}_{\,u} &= \phi(u+2)\frac{Q(u-1)}{Q(u+1)},\quad
\fbox{$2$}_{\,u} = \phi(u)\frac{Q(u+3)}{Q(u+1)}.
\label{12}
\end{align}
Here $\phi(u)=\prod_{j=1}^N[u-v_j]_{q^{1/2}}$ and 
$Q(u)=Q_1(u)$ is called Baxter's $Q$-function 
$Q(u) = \prod_{j=1}^n[u-u_j]_{q^{1/2}}$ with 
$u_1, \ldots, u_n$ determined from the Bethe equation
\begin{equation}\label{be1}
-\frac{\phi(u_j+1)}{\phi(u_j-1)}
=\frac{Q(u_j+2)}{Q(u_j-2)}\qquad (1\le j \le n).
\end{equation}
Here, $n$ is the number of down spins preserved 
under $T_1(u)$.
The factors $\phi(u+2)$ and $\phi(u)$ in (\ref{12}) are 
called vacuum parts in the sense that they are already present 
in the vacuum sector $n=0$ where $Q(u)=1$.
In fact, the vector 
$1 1 \ldots 1$ is obviously 
the unique eigenvector with the vacuum eigenvalue:
\begin{equation} \label{R+R}
\prod_{j=1}^N[u-v_j+2]_{q^{1/2}} 
+ \prod_{j=1}^N[u-v_j]_{q^{1/2}} = \phi(u+2) + \phi(u).
\end{equation}
The factors involving $Q$-functions in (\ref{12}) are called dress parts,
and the eigenvalue formula of the form 
(\ref{1+2})--(\ref{12}) is called a dressed vacuum form.
The vacuum part is non-universal in that 
it is directly affected by the normalization of the Boltzmann weights
(relevant $R$ matrix) 
and also depends on the quantum space data such as 
inhomogeneity $\{v_j\}$ entering $\phi(u)$.
On the other hand, the dress part encodes the
structure of the auxiliary space  essentially as we will see below.
 
The dressed vacuum form 
has an apparent pole at $u=-1+u_j$ because of $Q(u_j)=0$.
The Bethe equation (\ref{be1}) tells that it is actually 
spurious provided that $u_j$ is distinct from the other roots.
This is compatible with the property that eigenvalues of the 
transfer matrix are regular functions of $u$ if the local Boltzmann 
weights are so. 

The analytic Bethe ansatz is a hypothesis that one can reverse these
arguments to reproduce the eigenvalue formula 
from its characteristic properties bypassing the 
construction of eigenvectors.
One starts with the ansatz dressed vacuum form with the 
prescribed vacuum part
\begin{equation}
T_1(u) = \phi(u+2)\frac{Q(u+a)}{Q(u+b)} 
+ \phi(u)\frac{Q(u+c)}{Q(u+d)}.
\end{equation}
Then $a,b,c,d$ are determined by demanding that  
the pole-freeness is formally guaranteed 
by the Bethe equation (\ref{be1}) 
which one somehow admits from the onset.
In the present example, 
this certainly fixes $a,b,c,d$ uniquely as in (\ref{12}).
Further supplementary conditions 
may also be taken into account 
such as asymptotic behavior as $|u| \rightarrow \infty$ 
and the symmetry under complex conjugation, etc.
It is not known whether such a procedure 
indeed leads to the unique and correct eigenvalue formula in general.
Instead we shall propose in Section \ref{ss:dvfq} 
a constructive way of producing the dressed vacuum form
for general $U_q(\hat{\mathfrak g})$ by utilizing $q$-characters.

In the remainder of this subsection, 
we illustrate the simplest solution 
of the T-system for $A_1$ in the dressed vacuum form.
Although the result is obtainable by specializing the tableau sum 
formula (\ref{tu}), we re-derive it here for later convenience.
For simplicity $T^{(1)}_m(u)$ will be denoted by $T_m(u)$. 
Then the product of (\ref{1+2}) is written as 
\begin{align*}
T_1(u-1)T_1(u+1) 
= \fbox{1}_{u-1} \fbox{1}_{u+1} 
+\fbox{1}_{u-1} \fbox{2}_{u+1} 
+ \fbox{2}_{u-1} \fbox{2}_{u+1} 
+\;{\fbox{1}_{u+1}\atop \fbox{2}_{u-1}}.
\end{align*}
By (\ref{12}),  the last term becomes
$\phi(u-1)\phi(u+3)$, which is independent of $Q(u)$.
Identifying the other three terms with $T_2(u)$, one has  
\begin{equation*}
T_1(u-1)T_1(u+1) = T_2(u) + \phi(u-1)\phi(u+3),
\end{equation*}
which is an affinization of the identity
$(\text{doublet})^{\otimes 2} = 
(\text{triplet}) \oplus (\text{singlet})$ depicted as (\ref{dst}).
It is easy to systematize this calculation to show that
\begin{equation}\label{tmu}
T_m(u) = \sum_{1\le i_1\le \cdots \le i_m\le 2}
\fbox{$i_1$}_{\, u-m+1} 
\fbox{$i_2$}_{\,u-m+3}\cdots \fbox{$i_m$}_{\,u+m-1}
\end{equation}
is a solution of the unrestricted T-system for $A_1$
on the eigenvalues:
\begin{align}
T_m(u-1)T_m(u+1) &= T_{m-1}(u)T_{m+1}(u) + g_m(u),
\label{tA1}\\
g_m(u) &= \prod_{k=0}^{m-1}\phi(u+2k-m)\phi(u+4+2k-m).
\label{gm}
\end{align}
Explicitly, (\ref{tmu}) reads as
\begin{align}\label{tmexp}
T_m(u) = 
\Bigl(\prod_{k=1}^{m-1}\phi(u+m+1-2k)\Bigr)
\sum_{j=0}^m
{Q(u-m)Q(u+m+2)\phi(u+m+1-2j)\over Q(u+m-2j)Q(u+m+2-2j)}.
\end{align}
The summands in (\ref{tmu}) are naturally labeled with the 
semistandard tableaux of length $m$ row shape $(m)$ 
on numbers $\{1,2\}$.
Note that 
\begin{equation}\label{gggg}
g_m(u-1)g_m(u+1) = g_{m-1}(u)g_{m+1}(u)
\end{equation}
is satisfied with $g_0(u) = 1$.
Although the explicit form (\ref{tmexp}) is not 
particularly more 
illuminating than (\ref{tmu}), one can easily 
check that it is formally pole-free in the same manner as before
thanks to the Bethe equation (\ref{be1}).
Another way of seeing this is of course by the Jacobi-Trudi type 
formula (\ref{jta1}) with $r=1$ modified as
$T^{(2)}_1(u) = g_1(u)$, e.g.
\begin{equation*}
T_3(u) 
= \begin{vmatrix}
T_1(u-2) & g_1(u-1) & 0 \\
1 & T_1(u) & g_1(u+1)\\
0 & 1 & T_1(u+2)
\end{vmatrix}.
\end{equation*}
Thus the pole-freeness of $T_m(u)$ 
is an obvious corollary of that for $T_1(u)$.

\subsection{\mathversion{bold}Dressed vacuum form 
and $q$-characters}\label{ss:dvfq}
The analytic Bethe ansatz is extended to the general 
$U_q(\hat{\mathfrak g})$ and further sharpened by a connection with the 
theory of $q$-characters.
First we make a motive observation on the simplest example.
Recall the $q$-character of $W^{(1)}_1(u)$, 
the ``spin $\frac{1}{2}$ representation" of  $U_q(A^{(1)}_1)$  in 
Example \ref{ex:qc1}:
\begin{equation}\label{qbak}
\chi_q(W^{(1)}_1(u))= Y_z + Y^{-1}_{zq^{2}}\qquad
(z=q^u).
\end{equation} 
On the other hand, the dressed vacuum form (\ref{1+2})--(\ref{12})
of the 6-vertex model transfer matrix reads
\begin{equation}\label{dvb}
T^{(1)}_1(u) = \phi(u+2)\frac{Q(u-1)}{Q(u+1)}+
\phi(u)\frac{Q(u+3)}{Q(u+1)}.
\end{equation}
Upon substitution 
\begin{equation*}
Y_{q^u} \rightarrow \frac{\eta(u-1)Q(u-1)}{\eta(u+1)Q(u+1)},
\end{equation*}
the $q$-character (\ref{qbak}) becomes
\begin{equation*}
\frac{\eta(u-1)}{\eta(u+1)}\frac{Q(u-1)}{Q(u+1)}+
\frac{\eta(u+3)}{\eta(u+1)}\frac{Q(u+3)}{Q(u+1)}.
\end{equation*}
Thus the above substitution with the following 
overall renormalization
\begin{equation*}
\phi(u+2)\frac{\eta(u+1)}{\eta(u-1)}
\chi_q(W^{(1)}_1(u))
= \phi(u+2)\frac{Q(u-1)}{Q(u+1)}+
\phi(u+2)\frac{\eta(u+3)}{\eta(u-1)}\frac{Q(u+3)}{Q(u+1)}
\end{equation*}
reproduces the dressed vacuum form (\ref{dvb}) if 
$\eta(u)$ is assumed to obey the difference equation
\begin{equation}\label{ede}
\frac{\phi(u+1)}{\phi(u-1)} = \frac{\eta(u-2)}{\eta(u+2)}.
\end{equation} 
Note that this equation has the form of the Bethe equation (\ref{be1}):
\begin{equation*} 
-\frac{\phi(u_j+1)}{\phi(u_j-1)}
=\frac{Q(u_j+2)}{Q(u_j-2)}
\end{equation*}
without the sign factor, and 
$Q$ and $u_j$ being replaced by $\eta^{-1}$ and $u$, respectively.
The same feature will be adopted in (\ref{ebe}).
The connection of  (\ref{qbak}) and (\ref{dvb}) 
originates in the fact that 
the former is the $q$-character of $W^{(1)}_1(u)$ which is the auxiliary space of 
the transfer matrix relevant to the latter.

Now we generalize these observations to $U_q(\hat{\mathfrak g})$.
Consider the trigonometric vertex model associated 
with $U_q(\hat{\mathfrak g})$ under the periodic boundary condition.
Let $T^{(a)}_m(u)$ be the transfer matrix (\ref{tamg})
with the auxiliary space $W^{(a)}_m(u)$ and the quantum space 
$W^{(r_1)}_{s_1}(v_1)\otimes \cdots \otimes W^{(r_N)}_{s_N}(v_N)$: 
\begin{equation}\label{tamb}
\begin{split}
T^{(a)}_m(u) &= \mathrm{Tr}_{W^{(a)}_m(u)}
\left(R^{(a,m ;  r_N,s_N)}_{0,N}(z/w_N)\cdots
R^{(a,m ;  r_1,s_1)}_{0,1}(z/w_1)\right)\\
&\in {\rm End}(
W^{(r_1)}_{s_1}(v_1)\otimes \cdots \otimes W^{(r_N)}_{s_N}(v_N)),
\end{split}
\end{equation}
where $z=q^{tu}, w_i = q^{tv_i}$.
Due to the Yang-Baxter equation, they are commutative, i.e.  
$[T^{(a)}_m(u), T^{(b)}_n(v)]=0$.
The problem is to find their joint spectrum.

Let us construct 
a relevant dressed vacuum form $\Lambda^{(a)}_m(u)$ 
for $T^{(a)}_m(u)$.
In the following, a simple identity  
\begin{equation}\label{sf}
A_{a, z}|_{
Y_{c, z}\rightarrow 
\frac{f_c(u-1/t_c)}{f_c(u+1/t_c)}}
= \prod_{b=1}^r
\frac{f_b(u-(\alpha_a|\alpha_b))}{f_b(u+(\alpha_a|\alpha_b))}
\quad (z=q^{tu})
\end{equation}
for any functions $f_1, \ldots, f_r$ will be useful.
See (\ref{eq:t1}) and (\ref{Ade}) for the definitions of $t_a, t$ and $A_{a,z}$.
First we introduce an ``unnormalized" dressed vacuum form: 
\begin{equation}\label{Lam}
\tilde{\Lambda}^{(a)}_m(u) =
\chi_q(W^{(a)}_m(u))\;\;\text{with substitution }\;
Y_{c, q^{tv}}\rightarrow 
\frac{\eta_c(v-\frac{1}{t_c})}{\eta_c(v+\frac{1}{t_c})}
\frac{Q_c(v-\frac{1}{t_c})}{Q_c(v+\frac{1}{t_c})}.
\end{equation}
Let ${\tilde A}_{c,q^{tv}}$ be the result of the same substitution 
into $A_{c,q^{tv}}$.
By the definition we have
\begin{align}
\tilde{\Lambda}^{(a)}_m(u) 
&=\frac{\eta_a(u-\frac{m}{t_a})}{\eta_a(u+\frac{m}{t_a})}
\frac{Q_a(u-\frac{m}{t_a})}{Q_a(u+\frac{m}{t_a})}
\left(1 + \sum_{c,v}
\text{monomial in ${\tilde A}^{-1}_{c,q^{tv}}$}\right).
\end{align}
Here the factor $(\eta_aQ_a)/(\eta_aQ_a)$ 
is the top term specified by (\ref{tamy}) and (\ref{Lam}).
The appearance of ${\tilde A}^{-1}_{c,q^{tv}}$ is due to 
Theorem \ref{th:frm} (1).
As for the functions $\eta_1,\ldots, \eta_r$, 
we postulate, as the generalization of (\ref{ede}),  
the following difference equation
\begin{equation}\label{ebe}
\prod^N_{\begin{subarray}{c} k=1\\ r_k = a \end{subarray}}
\frac{\bigl[u-v_k+\frac{s_k}{t_a}\bigr]_{q^{t/2}}}
{\bigl[u-v_k-\frac{s_k}{t_a}\bigr]_{q^{t/2}}}
= \prod_{b=1}^r
\frac{\eta_b(u-(\alpha_a|\alpha_b))}
{\eta_b(u+(\alpha_a|\alpha_b))}\quad(1\le a \le r),
\end{equation}
where $[u]_p$ is defined in $(\ref{qnb})$.
Then using (\ref{sf}) and (\ref{ebe}) we find
\begin{equation}\label{ac}
\begin{split}
{\tilde A}_{a, q^{tu}}&=\prod_{b=1}^r
\frac{\eta_b(u-(\alpha_a|\alpha_b))Q_b(u-(\alpha_a|\alpha_b))}
{\eta_b(u+(\alpha_a|\alpha_b))Q_b(u+(\alpha_a|\alpha_b))}\\
&=\prod^N_{\begin{subarray}{c} k=1\\ r_k = a \end{subarray}}
\frac{\bigl[u-v_k+\frac{s_k}{t_a}\bigr]_{q^{t/2}}}
{\bigl[u-v_k-\frac{s_k}{t_a}\bigr]_{q^{t/2}}}
\cdot
\prod_{b=1}^r
\frac{Q_b(u-(\alpha_a|\alpha_b))}{Q_b(u+(\alpha_a|\alpha_b))}.
\end{split}
\end{equation}

Next we adjust the overall normalization.
Consider the $R$ matrix on
$W^{(a)}_m(u)\otimes W^{(b)}_s(v)$ and 
write its unique diagonal matrix element between the 
tensor product of the highest weight vectors as
$\phi^{(a, b)}_{m, s}(u-v)$. 
Namely, 
\begin{equation}\label{batsu}
\begin{picture}(460,34)(-10,-16)

\put(0,-3){$\phi^{(a, b)}_{m, s}(u-v)=$ 
Boltzmann weight of the vertex}

\multiput(252,0)(0,0){1}{
\put(-10,0){\line(1,0){20}}
\put(0,-10){\line(0,1){20}}

\put(-33,-3){$m\omega_a$}\put(16,-3){$m\omega_a$.}
\put(-6,-18){$s\omega_b$}\put(-6,14){$s\omega_b$}
}

\end{picture} 
\end{equation}
Now we define the normalized dressed vacuum form by
\begin{align}
\Lambda^{(a)}_m(u) &= 
\left(\prod_{k=1}^N
\phi^{(a, r_k)}_{m, s_k}(u-v_k)\right)
\frac{\eta_a(u+\frac{m}{t_a})}{\eta_a(u-\frac{m}{t_a})}
\tilde{\Lambda}^{(a)}_m(u)\nonumber\\
&=\left(\prod_{k=1}^N
\phi^{(a, r_k)}_{m, s_k}(u-v_k)\right)
\frac{Q_a(u-\frac{m}{t_a})}{Q_a(u+\frac{m}{t_a})}
\left(1 + \sum_{c,v}
\text{monomial in ${\tilde A}^{-1}_{c,q^{tv}}$}\right). \label{dvff}
\end{align}
Besides the (in principle) known Boltzmann weights 
$\phi^{(a,b)}_{m,k}$,
this only contains the $Q$-functions $Q_1,\ldots, Q_r$ 
and the LHS of (\ref{ebe}).

Recall that the transfer matrices preserve 
the subspaces (sectors) of the quantum space specified by the weight. 
Let us parameterize the weight by the nonnegative integers $n_1, \ldots, n_r$ as
\begin{equation}\label{swm}
\sum_{k=1}^Ns_k\omega_{r_k}-\sum_{a=1}^rn_a\alpha_a,
\end{equation}
where $\omega_1, \ldots, \omega_r$ denote 
the fundamental weights of ${\mathfrak g}$ (\ref{aaw}).
Given $n_a$, we set
\begin{equation}\label{qa}
Q_a(u) = \prod_{j=1}^{n_a}[u-u^{(a)}_j]_{q^{t/2}}
\end{equation}
by introducing the unknowns 
$\{u^{(a)}_j| 1\le a \le r, 1 \le j \le n_a\}$.

\begin{conjecture}\label{conj:aba}
Let $T^{(a)}_m(u)$ (\ref{tamb}) be the transfer matrix 
normalized as (\ref{batsu}). 
Then its eigenvalues in the sector (\ref{swm}) 
are given by the dressed vacuum form $\Lambda^{(a)}_m(u)$
(\ref{dvff}), (\ref{qa}) 
with the numbers $\{u^{(a)}_j| 1\le a \le r, 1 \le j \le n_a\}$
satisfying the Bethe equation:
\begin{equation}\label{be}
\prod^N_{\begin{subarray}{c} k=1\\ r_k = a \end{subarray}}
\frac{\bigl[u^{(a)}_j-v_k+\frac{s_k}{t_a}\bigr]_{q^{t/2}}}
{\bigl[u^{(a)}_j-v_k-\frac{s_k}{t_a}\bigr]_{q^{t/2}}}
= -\prod_{b=1}^r
\frac{Q_b(u^{(a)}_j+(\alpha_a|\alpha_b))}
{Q_b(u^{(a)}_j-(\alpha_a|\alpha_b))}.
\end{equation}
\end{conjecture}

Practically the results in Section \ref{s:tab} serve as a large input to 
the prescription (\ref{Lam}) to produce $\Lambda^{(a)}_m(u)$. 
The functions $Q_a(u)$ are called the (generalized) Baxter $Q$-functions.
In view of Theorem \ref{th:frm} (2), 
we expect that their zeros, if in a generic position, do not cause a pole in   
$\Lambda^{(a)}_m(u)$ due to the Bethe equation.

Let $\EuScript{P}_a(\zeta)$ be 
the product of the $a$th Drinfeld polynomial (\ref{dpd})
for each component in the quantum space
$W^{(r_1)}_{s_1}(v_1)
\otimes \cdots \otimes W^{(r_N)}_{s_N}(v_N)$:
\begin{align}
\EuScript{P}_a(\zeta) = 
\prod_{\begin{subarray}{c} k=1\\ r_k = a \end{subarray}}^N\prod_{i=1}^{s_k}
(1-\zeta q^{t(v_k+(s_k+1-2i)/t_a)}),
\quad \deg \EuScript{P}_a = 
\sum_{\begin{subarray}{c} k=1\\ r_k = a \end{subarray}}^Ns_k.
\end{align}
We remark that the LHS of (\ref{ebe}) is expressed as
\begin{equation}
\prod^N_{\begin{subarray}{c} k=1\\ r_k = a \end{subarray}}
\frac{\bigl[u-v_k+\frac{s_k}{t_a}\bigr]_{q^{t/2}}}
{\bigl[u-v_k-\frac{s_k}{t_a}\bigr]_{q^{t/2}}}
= q_a^{\deg \EuScript{P}_a}
\frac{\EuScript{P}_a(\zeta q^{-1}_a)}{\EuScript{P}_a(\zeta q_a)}
\qquad (\zeta = q^{-tu}),
\end{equation}
which further becomes the LHS of the Bethe equation (\ref{be})
by the specialization $u=u^{(a)}_j$.
This has formally the same form 
as (\ref{pwd}).
Note however that the quantum space 
$W^{(r_1)}_{s_1}(v_1)
\otimes \cdots \otimes W^{(r_N)}_{s_N}(v_N)$
under consideration is not necessarily irreducible in general, 
and the above $\EuScript{P}_a(\zeta)$ is 
the $a$th Drinfeld polynomial of its irreducible quotient 
containing the tensor product of the highest weight vectors. 

By the construction (\ref{Lam}) and Theorem \ref{th:nh}, 
the unnormalized dressed vacuum form 
$\tilde{\Lambda}^{(a)}_m(u)$ satisfies the 
unrestricted T-system for ${\mathfrak g}$. 
It follows that the normalized one
$T^{(a)}_m(u)=\Lambda^{(a)}_m(u)$ (\ref{dvff}) 
satisfies the modified T-system containing an extra factor $g^{(a)}_m(u)$ as
(\ref{tga}):
\begin{equation*}
\textstyle
T^{(a)}_m(u-\frac{1}{t_a})T^{(a)}_m(u+\frac{1}{t_a})
=T^{(a)}_{m-1}(u)T^{(a)}_{m+1}(u)+g^{(a)}_m(u)M^{(a)}_m(u),
\end{equation*}
where the original T-system corresponds to $g^{(a)}_m(u)=1$ as in (\ref{ttm}).
The scalar factor $g^{(a)}_m(u)$ has the properties:

(i)Apart from $(a,m,u)$, 
it only depends on the quantum space data

\hspace{0.3cm}
$W^{(r_1)}_{s_1}(v_1)
\otimes \cdots \otimes W^{(r_N)}_{s_N}(v_N)$.

(ii) It satisfies relation (\ref{gae}):
\begin{equation*}
\textstyle
g^{(a)}_m(u-\frac{1}{t_a})g^{(a)}_m(u+\frac{1}{t_a})
=g^{(a)}_{m-1}(u)g^{(a)}_{m+1}(u).
\end{equation*}
In fact this has been encountered for ${\mathfrak g}=A_1$ in (\ref{gggg}).
To derive these properties, note that 
the fusion construction implies that the diagonal element of the $R$-matrix 
(\ref{batsu}) is factorized as 
$\phi^{(a,b)}_{m,s}(u)=\prod_{i=1}^m
\phi^{(a,b)}_{1,s}(u+(m+1-2i)/t_a)$.
Thus the first relation in (\ref{dvff}) is written as
\begin{align}
\tilde{\Lambda}^{(a)}_m(u) &=
\Lambda^{(a)}_m(u)
\prod_{i=1}^m\gamma_a\Bigl(u+\frac{m+1-2i}{t_a}\Bigr),\label{lala}\\
\gamma_a(u) &=
\frac{\eta_a(u-\frac{1}{t_a})}{\eta_a(u+\frac{1}{t_a})}
\prod_{k=1}^N\phi^{(a,r_k)}_{1,s_k}(u-v_k)^{-1}.\label{gep}
\end{align}
In view of (\ref{lala}),  replace $T^{(a)}_m(u)$ in the original T-system with 
$T^{(a)}_m(u)\prod_{i=1}^m\gamma_a(u+(m+1-2i)/t_a)$.
After removing the common factor,  
the result is indeed reduced to the form (\ref{tga}) with 
\begin{align}
g^{(a)}_m(u) &= \prod_{i=1}^mg^{(a)}_1\Bigl(u+\frac{m+1-2i}{t_a}\Bigr),
\label{gm1}\\
g^{(a)}_1(u) &= A^{-1}_{a,z}|_{Y_{c,z}\rightarrow \gamma_c(u)}
\quad(z = q^{tu}).\label{ggp}
\end{align}
The property (ii) directly follows from (\ref{gm1})
without using the concrete form of $g^{(a)}_1(u)$.
The property (i) is essentially due to the remark after (\ref{dvff}).
In fact, it is attributed to $g^{(a)}_1(u)$ (\ref{ggp}).
With regard to $\gamma_c(u)$ therein, 
$\phi^{(c, r_k)}_{1,s_k}(u-v_k)$ 
in (\ref{gep}) depends on the quantum space data only,
and so does the contribution from $\eta_c$
because of (\ref{sf}) and (\ref{ebe}).

\begin{remark}\label{re:jiba}
The transfer matrix (\ref{tamb}) can be 
generalized by the ``magnetic field" as
$T^{(a)}_m(u) =
\mathrm{Tr}_{W^{(a)}_m(u)}
(e^{\mathcal H}
R^{(a,m ;  r_N,s_N)}_{0,N}(z/w_N)\cdots
R^{(a,m ;  r_1,s_1)}_{0,1}(z/w_1))$
without spoiling the commutativity and the T-system.
Here ${\mathcal H}$ is any element in 
the Cartan subalgebra of $U_q({\mathfrak g})$ acting on the auxiliary space.
The dressed vacuum form for such $T^{(a)}_m(u)$ is obtained by modifying  
the substitution (\ref{Lam}) into
$Y_{c, q^{tv}} \rightarrow 
e^{\omega_c({\mathcal H})}
\frac{\eta_c(v-\frac{1}{t_c})Q_c(v-\frac{1}{t_c})}
{\eta_c(v+\frac{1}{t_c})Q_c(v+\frac{1}{t_c})}$.
Accordingly $\tilde{A}_{a,q^{tu}}$ (\ref{ac}) 
and the LHS of the Bethe equation (\ref{be})
get multiplied by the extra factor $e^{\alpha_a({\mathcal H})}$.
\end{remark}

\subsection{RSOS models}\label{ss:dvfrs}
We consider 
the spectrum of the transfer matrix 
$T^{(a)}_m(u)\, (1 \le m \le t_a\ell)$ 
(\ref{tame}) for the trigonometric level $\ell$ RSOS models
sketched in Section \ref{ss:vrg}. 
($T^{(a)}_{t_a\ell}(u)$ corresponds to a frozen model.)
Conjecturally, it is covered by 
the dressed vacuum form in Remark \ref{re:jiba} 
specialized along (i)--(iii) in what follows.

(i) The parameter $q$ entering through $[u]_{q^{t/2}}$ is set 
$q=\exp\left(\frac{\pi\sqrt{-1}}{t(\ell+h^\vee)}\right)$,
where $h^\vee$ is the dual Coxeter number of $\mathfrak{g}$ (\ref{hhd}).

(ii) The integers $n_1,\ldots, n_r$ entering (\ref{qa}) 
are fixed by demanding (\ref{swm}) be $0$,
which is possible thanks to (\ref{na}). 

(iii) The magnetic field is taken so that
$\omega_c({\mathcal H}) 
= \frac{2\pi\sqrt{-1}(\omega_c|\Lambda+\rho)}{\ell+h^\vee}$, 
where $\rho = \sum_{a \in I}\omega_a$ and 
$\Lambda$ is an element from $P_\ell$ (\ref{PP}).

Introduce the specialized $q$-character 
$Q^{(a)}_m(\Lambda) 
:= \chi_q(W^{(a)}_{t_a\ell}(u))|_{Y_{c,q^{tv}}\rightarrow
e^{\omega_c({\mathcal H})}}$, where $\Lambda$-dependence 
enters through the above ${\mathcal H}$.
Then according to the conjecture in \cite[(A.8)-(A.9)]{KNS2},
the relation
$\prod_{b \in I}Q^{(b)}_{t_b\ell}(\Lambda)^{C_{a b}} = 1$
holds.
The quantity $Q^{(a)}_m=\dim_q\mathrm{res}\, W^{(a)}_{t_a\ell}$ 
in Section \ref{ss:qru} is equal to 
$Q^{(a)}_m(0)$ in the notation here.
The above relation is a generalization of 
$Q^{(a)}_{t_a\ell}(0)=1$ in 
Section \ref{ss:qru}.

\subsection{Bibliographical notes}
The analytic Bethe ansatz was proposed in \cite{R1}
by extracting the idea from Baxter's solution of the 8-vertex model
\cite{Ba1}.
It was applied systematically in \cite{R2, KS1, KOS}
to a wide class of solvable vertex models. 
Formulation of the Bethe equation by root system goes back, 
for instance, to \cite{OW,R2}.
A relation between dressed vacuum forms and $q$-characters
similar to Section \ref{ss:dvfq} 
has also been argued in \cite[section 6]{FR2}.

\section{Wronskian type (Casoratian) formula}\label{s:tq}

Here we present the solution of the T-system for $A_r$ and $C_r$
in terms of Casoratian (difference analog of Wronskian).
It is most naturally done by introducing 
a difference analog of $L$-operators in soliton theory.
It also provides a Casoratian 
interpretation and generalization of the Baxter $Q$-functions. 
Our description is along the context of $q$-characters; hence, 
the identification of the variables 
\begin{equation}\label{YQ}
Y_{a,q^{tu}} = \frac{Q_a(u-\frac{1}{t_a})}{Q_a(u+\frac{1}{t_a})}
\end{equation}
is assumed. See (\ref{Lam}).
($t, t_a$ are defined in (\ref{eq:t1}).)
Resulting formulas can suitably be modified to 
fit transfer matrices with specific normalizations according to 
the argument in Section \ref{ss:dvfq}. 
We will also give analogous $L$-operators for 
$B_r, D_r$ and $sl(r|s)$.

\subsection{\mathversion{bold}Difference $L$ operators}\label{ss:dl}
We treat the $A_r$ case first as an illustration.
Let $D=e^{2\partial_u}$ be the shift operator
$Df(u) = f(u+2)D$.
Using $z_a(u)$ (\ref{zaa}), we introduce the 
difference $L$ operator: 
\begin{equation}\label{ldef1}
L(u) = (1-z_{r+1}(u)D)\cdots (1-z_2(u)D)
(1-z_1(u)D).
\end{equation}
Expanding the product, one identifies the
coefficients with $m=1$ case of (\ref{atab}) to find
\begin{equation}\label{ldef2}
L(u) = \sum_{a=0}^{r+1}(-1)^aT^{(a)}_1(u+a-1)D^a,
\end{equation}
where $T^{(0)}_1= T^{(r+1)}_1=1$.
Thus $L(u)$ is a generating function of the
fundamental $q$-characters
$T^{(a)}_1(u) = \chi_q(W^{(a)}_1(u))$.

Define the action of the screening operator $S_a$
(\ref{sac1}) on difference operators by 
$S_a\cdot (\sum_i f_i(u)D^i) = \sum_i(S_a\cdot f_i(u))D^i$.
Let us calculate $S_a\cdot L(u)$ by using the factorized form (\ref{ldef1}).
According to the rule (\ref{sac1}),  
$S_a$ acts non trivially only on the variable $Y_{a, z}$.
{}From (\ref{zaa}), it is contained only in 
$z_a(u)$ and $z_{a+1}(u)$.
The action on this part is calculated as
\begin{align*}
&S_a\cdot(1-z_{a+1}(u)D)(1-z_{a}(u)D)\\
&= S_a\cdot(
1-Y_{a,q^{u+a+1}}^{-1}Y_{a+1, q^{u+a}}D
-Y_{a-1,q^{u+a}}^{-1}Y_{a, q^{u+a-1}}D
+Y_{a-1,q^{u+a+2}}^{-1}Y_{a+1, q^{u+a}}D^2)\\
&=S_{a,q^{u+a+1}}Y_{a,q^{u+a+1}}^{-1}Y_{a+1, q^{u+a}}D
-S_{a,q^{u+a-1}}Y_{a-1,q^{u+a}}^{-1}Y_{a, q^{u+a-1}}D=0,
\end{align*}
where the last equality is due to (\ref{sac2}) and (\ref{Ade}):
\begin{align*}
S_{a,q^{u+a+1}} = A_{a,q^{u+a}}S_{a,q^{u+a-1}}
= Y_{a,q^{u+a-1}}Y_{a,q^{u+a+1}}
Y_{a-1,q^{u+a}}^{-1}Y_{a+1,q^{u+a}}^{-1}S_{a,q^{u+a-1}}.
\end{align*}
In this way one gets
\begin{equation}\label{Lz}
S_a\cdot L(u) = 0\quad (1 \le a \le r).
\end{equation}
In view of (\ref{ldef2}), this offers a simple way of checking 
$T^{(a)}_1(u) \in \bigcap_{b=1}^r{\rm Ker} S_b$ 
in agreement with Theorem \ref{th:frm} (2).
When $r=1$, the change of variables 
from $\{z_a(u)\}$ to $\{ T^{(a)}_1(u)\}$  is a 
difference analog of the Miura transformation
$q=q(u) \rightarrow f=f(u)=q^2-\partial_u q$ by
\begin{equation*}
(\partial_u-q)(\partial_u + q) = \partial^2_u -f.  
\end{equation*} 
With regard to the inverse
\begin{align*}
L(u)^{-1} = (1-z_1(u)D)^{-1}(1-z_2(u)D)^{-1}\cdots 
(1-z_{r+1}(u)D)^{-1},
\end{align*}
the simple expansion formula
\begin{equation}\label{ldef3}
L(u)^{-1} = \sum_{m\ge 0}T^{(1)}_m(u+m-1)D^m
\end{equation}
holds due to (\ref{atab}),  
confirming similarly that $T^{(1)}_m(u) 
\in \bigcap_{b=1}^r{\rm Ker} S_b$.
The product of (\ref{ldef2}) and (\ref{ldef3}) leads to the 
two types of TT-relations:
\begin{align*}
&\sum_{0 \le a \le \min(r+1,m)}(-1)^aT^{(a)}_1(u+a)
T^{(1)}_{m-a}(u+m+a)=\delta_{m 0},\\
&\sum_{0 \le a \le \min(r+1,m)}(-1)^aT^{(a)}_1(u+m-a)
T^{(1)}_{m-a}(u-a) = \delta_{m 0}
\end{align*}
for $m \ge 0$.

\subsection{Casoratian formula}\label{ss:wr}
Consider the linear difference equation on $w(u)$
\begin{equation}\label{Lw}
L(u)w(u) = 0.
\end{equation}
This is of order $r+1$ with respect to $D$.
Letting $\{w_1(u), \ldots, w_{r+1}(u)\}$ be a basis of the solution,
we denote the Casoratian by
\begin{equation}\label{iku}
C_u[i_1, \ldots, i_k] = 
\det\begin{pmatrix}
w_1(u+i_1) & \cdots & w_1(u+i_k) \\
\vdots & & \vdots \\
w_k(u+i_1) & \cdots & w_k(u+i_k)
\end{pmatrix}
\end{equation}
for $1 \le k \le r+1$. 
Thus for example 
$C_{u+2}[i_1,\ldots, i_k]=C_u[i_1+2,\ldots, i_k+2]$.
By using (\ref{ldef2}), 
the relations $L(u)w_k(u)=0$ with $k=1,\ldots, r+1$ 
are expressed in the matrix form:
\begin{equation*}
\begin{pmatrix}
w_1(u)\\
w_2(u)\\
\vdots \\
w_{r+1}(u)
\end{pmatrix}
=\begin{pmatrix}
w_1(u+2)\;\; w_1(u+4) & \cdots & w_1(u+2r+2) \\
w_2(u+2) \;\; w_2(u+4) & \cdots & w_2(u+2r+2) \\
\vdots & & \vdots \\
w_{r+1}(u+2) \;\;w_{r+1}(u+4) & \cdots & w_{r+1}(u+2r+2) 
\end{pmatrix}
\begin{pmatrix}
T^{(1)}_1(u)\\
(-1)T^{(2)}_1(u+1)\\
\vdots \\
(-1)^rT^{(r+1)}_1(u+r)
\end{pmatrix},
\end{equation*}
where $T^{(r+1)}_1(u)=1$ in our normalization here ($q$-characters)
as noted under (\ref{zaa}).
By Cramer's formula, we have
\begin{equation}\label{taw}
T^{(a)}_1(u+a-1) = 
\frac{C_u[0,\ldots, 2a-2, 2a+2,\ldots, 2r+2]}
{C_u[2,\ldots, 2r+2]}\quad(0 \le a \le r+1),
\end{equation}
where $\ldots$ signifies that the omitted arrays are consecutive 
with difference $2$.
The relation $L(u)w_k(u)=0$ means that 
$w_k(u+2r+2) = (-1)^rw_k(u) 
+ \text{terms involving}\;w_k(u+2), \ldots, w_k(u+2r)$. 
It follows the periodicity
\begin{equation}\label{qco}
C_u[0,2,\ldots, 2r] = C_{u+2}[0,2,\ldots, 2r].
\end{equation}
Its actual value becomes important in physical applications, and 
the resulting relation on $C_u[0,2,\ldots, 2r]$ is called the 
quantum Wronskian condition. 
See for example \cite{BLZ2,DDT3}.

The solution to the T-system for $A_r$ 
that matches (\ref{taw}) is given by
\begin{equation}\label{tamw}
T^{(a)}_m(u+a+m-2) =
\frac{C_u[0,\ldots, 2a-2, 2a+2m,\ldots, 2r+2m]}
{C_u[0,\ldots, 2r]}\quad(0 \le a \le r+1).
\end{equation} 
This satisfies the boundary conditions 
$T^{(0)}_m(u)=T_0^{(a)}(u)=1$ and 
$T^{(a)}_{-1}(u) = 0$.
In fact, if (\ref{tamw}) is substituted into (\ref{ta}),
the denominator can be removed as an overall factor owing to (\ref{qco}).
Then (\ref{ta}) is identified with a simplest Pl\"ucker relation
\begin{equation}
\xi^{(a)}_m(u)\xi^{(a)}_m(u+2) -
\xi^{(a)}_{m+1}(u)\xi^{(a)}_{m-1}(u+2) 
- \xi^{(a+1)}_{m}(u)\xi^{(a-1)}_{m}(u+2)=0
\end{equation}
among the determinant 
$\xi^{(a)}_m(u) = C_u[0,\ldots, 2a-2, 2a+2m,\ldots, 2r+2m]$.

The Casoratian formula (\ref{tamw}) 
is a Yang-Baxterization ($u$-dependent generalization)
of the Weyl character formula.
To see this, recall the restriction map 
${\rm res}$ (\ref{res}).
{}From (\ref{zaa}) we have ${\rm res }\,(z_a(u)) = x_a$, 
where the latter is defined by 
$x_a=y_a/y_{a-1} = e^{\omega_a-\omega_{a-1}}$
with $\omega_0 = \omega_{r+1}=0$.
We extend ${\rm res}$  naturally to 
the difference $L$ operator and the wave functions as
\begin{align}
{\rm res }\, L(u) = (1-x_{r+1}D)\cdots (1-x_1D),
\quad {\rm res }\, (w_i(u)) = x_i^{-u/2}.
\end{align}
The latter is certainly annihilated by the former.
By using $x_1\cdots x_{r+1}=1$, it is straightforward to see that 
the restriction of (\ref{tamw}) becomes
\begin{equation}
{\rm res} \left(\frac{C_u[0,\ldots, 2a-2, 2a+2m,\ldots, 2r+2m]}
{C_u[0,\ldots, 2r]}\right)
=\frac{\det(x_i^{\lambda_j+r+1-j})_{1\le i,j \le r+1}}
{\det(x_i^{r+1-j})_{1\le i,j \le r+1}},
\end{equation} 
where $(\lambda_j)$ corresponds to the 
$a \times m$ rectangular Young diagram, namely, 
$\lambda_j =m$ if $1 \le j \le a$ and $\lambda_j = 0$ otherwise. 
The RHS is the Weyl character formula of  
the Schur function for $(\lambda_j)$ as is well known.

The Casoratian formula here and the 
tableau sum formula (Section \ref{ss:A}) 
are connected by the following general fact.

\begin{proposition}[\cite{NNSY}]\label{pr:nnsy}
Let $C_u[i_1,\ldots, i_k]$ be as in (\ref{iku}).
($L(u)w_j(u)=0$ is not assumed.)
Given even integers  $0=i_0 <i_1< \cdots < i_{N-1}$, 
let $\mu= (\mu_j)$ be the Young diagram with 
depth less than $N$ specified by $\mu_j = \frac{i_{N-j}}{2}+j - N$.
Take any $d \ge\mu_1$. Then
\begin{align*}
\frac{C_u[0,i_1, i_2, \cdots, i_{N-1}]}
     {C_{u+2d}[0,2,\ldots, 2N-2]}
&= \sum_{\mathcal{T}}\prod_{(\alpha,\beta) \in (d^N)/\mu}
{\tilde x}_{\mathcal{T}(\alpha,\beta)}
(u+2\alpha+2\beta-4),
\end{align*}
where 
$\tilde{x}_j(u) = \frac{C_u[0,2,\ldots,2j-2]C_u[4,6,\ldots,2j]}
{C_u[2,4,\ldots,2j]C_u[2,4,\ldots,2j-2]}$ and 
the sum $\sum_\mathcal{T}$ extends over the semistandard tableaux 
on the skew Young diagram $(d^N)/\mu$ \cite{Ma1}
on letters $\{1,\ldots, N\}$.
$\mathcal{T}(\alpha,\beta)$ denotes the entry of $\mathcal{T}$ at the 
$\alpha$th row and the $\beta$th column {}from the bottom left corner.
\end{proposition}

According to Proposition \ref{pr:nnsy},
the RHS of (\ref{tamw}) equals the sum over 
semistandard tableaux on $a\times m$ Young diagram
on letters $\{1,\ldots, r+1\}$.
The building block of the tableau variable $\tilde{x}_j(u)$
is the principal minors of the Casoratian (quantum Wronskian) 
$C_u[0,2,\ldots, 2r]$.
Combined with (\ref{Lw}),
they are identified with the Baxter $Q$-functions as 
we will see in the next subsection.

\subsection{\mathversion{bold}$Q$-functions}\label{ss:Q}

{}From the full $L$ operator (\ref{ldef1}),  we extract the partial ones by
\begin{equation}\label{Lj}
L_j(u) = (1-z_j(u)D)\cdots (1-z_2(u)D)(1-z_1(u)D)
\quad (1 \le j \le r+1).
\end{equation}
The original one corresponds to $L_{r+1}(u)$.
By the definition we have
\begin{equation}
{\rm Ker}\, L_1(u) \subset {\rm Ker}\, L_2(u) \subset
\cdots \subset {\rm Ker}\, L_{r+1}(u).
\end{equation}
Choose the basis 
of ${\rm Ker}\, L_{j}(u)$ according to this flag structure as
\begin{equation}\label{wf}
\{w_1(u)\} \subset \{w_1(u), w_2(u) \} \subset \cdots
\subset  \{w_1(u), \ldots, w_{r+1}(u)\}.
\end{equation}
As the simplest example, $w_1(u) \in {\rm Ker}\, L_1(u)$ is the condition
$0=(1-z_1(u)D)w_1(u)$.
In view of (\ref{zaa}) and (\ref{iku}), 
this is the $j=1$ case of 
\begin{equation}\label{ydw}
\bigl(1-Y_{j,q^{u+j-1}}D\bigr)C_u[0,\ldots, 2j-2] =0
\quad(1 \le j \le r).
\end{equation}
To derive this,  note that a direct calculation using (\ref{zaa}) leads to
\begin{equation*}
L_j(u) = 1+(-1)^jY_{j,q^{u+j-1}}D^j + \text{terms involving}\,
D, \ldots, D^{j-1}.
\end{equation*}
Therefore $L_j(u)w_k(u)=0$ $(1 \le k \le j)$ implies
\begin{equation*}
Y_{j,q^{u+j-1}}w_k(u+2j) = (-1)^{j-1}w_k(u) + 
\sum_{l=1}^{j-1}c_{j,l}(u)w_k(u+2l),
\end{equation*}
where $c_{j,l}(u)$ is independent of $k$.
The second term in (\ref{ydw}) is equal to \\ 
$Y_{j,q^{u+j-1}}C_u[2,\ldots, 2j-2,2j]$.
Applying the above relation to the last column of this,
we find the result is equal to $C_u[0,\ldots, 2j-2]$,  
hence (\ref{ydw}).

If we express the variable 
$Y_{a,q^u}$ in $q$-characters in terms of $Q$-functions as in 
(\ref{YQ}), 
the solution of the first order difference equation (\ref{ydw}) is given by
\begin{equation}\label{wq}
C_u[0,\ldots, 2j-2]= \sigma_j(u)Q_j(u+j-2)\quad (1\le j \le r),
\end{equation}
where $\sigma_j(u)$ is 
any variable satisfying $\sigma_j(u+2)=\sigma_j(u)$.
In this way, the $Q$-functions are identified with the principal minors of 
the Casoratian $C_u[0,\ldots, 2r]_u$
made of the wave functions $\{w_i(u)\}$ especially chosen along the 
scheme (\ref{wf}).
The simplest case $j=1$ of (\ref{wq}) is
$w_1(u) = \sigma_1(u)Q_1(u-1)$.
Thus $L(u)w_1(u)=0$ is rephrased as 
\begin{equation}
\sum_{a=0}^{r+1}(-1)^aT^{(a)}_1(u+a)Q_1(u+2a) = 0,
\end{equation}
which is an example of TQ-relations.

\subsection{B\"acklund transformations}\label{ss:bt}
Here we remove the boundary condition 
$T^{(a)}_0(u) = T^{(0)}_m(u)=1$ 
and redefine  $T^{(a)}_m(u)$ in (\ref{tamw}) and 
$Q_j(u)$ in (\ref{wq}) as
\begin{align}
&T^{(a)}_m(u+a+m-2) =C_u[0,\ldots, 2a-2, 2a+2m,\ldots, 2r+2m],
\label{twd}\\
&Q_a(u+a-1)=C_u[0,\ldots, 2a-2]. \label{qaw}
\end{align}
These functions are special cases of more general ones: 
\begin{align}
&T^{(s,a)}_m(u+a+m-2) \nonumber\\
&= 
\begin{small}
\begin{vmatrix}
w_1(u) & \cdots & 
w_1(u+2a-2)  \;\;\;\; w_1(u+2a+2m)& \cdots & w_1(u+2s+2m)\\
\vdots & && & \vdots\\
w_{s+1}(u) & \cdots & 
w_{s+1}(u+2a-2) \;\; w_{s+1}(u+2a+2m)& \cdots & w_{s+1}(u+2s+2m)
\end{vmatrix},
\end{small}\nonumber\\
&Q_{\{i_1,\ldots, i_a\}}(u+a-1) = 
\begin{small}
\begin{vmatrix}
w_{i_1}(u) & \cdots & w_{i_1}(u+2a-2)\\
\vdots & & \vdots \\
w_{i_a}(u) & \cdots & w_{i_a}(u+2a-2)
\end{vmatrix}
\end{small},\label{qg}
\end{align}
where $\cdots$ in determinants signify that $u$ increases by 2.
$T^{(s,a)}_m(u)$ is defined for $0 \le a \le s+1, 0 \le s \le r$ and 
$m \ge 0$.
The set $\{i_1,\ldots, i_a\}$ is any subset of $\{1,\ldots, r+1\}$.
By the definition, $T^{(r,a)}_m(u) = T^{(a)}_m(u)$ 
and $Q_{\{1,\ldots, a\}}(u) = Q_a(u)$.
These functions obey various relations 
as the consequence of identities among determinants.
Let us mention a few of them that have analogy with soliton theory.

The symmetric group ${\mathfrak S}_{r+1}$ acts on the basis 
$w_1(u),\ldots, w_{r+1}(u)$ as their permutations keeping 
$L(u)$ invariant.
This can be viewed as B\"acklund transformations generating the 
functions $Q_{\{i_1,\ldots, i_a\}}$ from $Q_1,\ldots, Q_{r+1}$.
Its generator,  the transposition $s_a$ of $w_a(u)$ and $w_{a+1}(u)$,
acts trivially as $s_a(Q_b) = Q_b$ for $a>b$ and similarly as 
$s_a(Q_b) = -Q_b$ for $a<b$.
The nontrivial case 
$s_a(Q_a) = Q_{\{1, \ldots, a-1, a+1\}}$ satisfies the QQ-relation:
\begin{equation}\label{qq}
D(Q_a)s_a(Q_a)- Q_a Ds_a(Q_a) + D(Q_{a-1})Q_{a+1}=0,
\end{equation}
where the first term denotes $Q_a(u+2)s_a(Q_a)(u)$ for instance.
This is derived by applying the Jacobi identity (\ref{dd}) to the 
$a, a+1$ rows and $1, a+1$ columns for the determinant of $Q_{a+1}$.

With regard to $T^{(s,a)}_m(u)$, it is the T-function for $A_s(\subset A_r)$.
Writing $T^{(s,a)}_m(u)$ and $T^{(s-1,a)}_m(u)$ simply as 
$T^{(a)}_m(u)$ and ${\tilde T}^{(a)}_m(u)$, respectively,  one can derive 
\begin{equation}
\begin{split}
T^{(a)}_m(u){\tilde T}^{(a-1)}_m(u-1) &=
T^{(a-1)}_m(u-1){\tilde T}^{(a)}_m(u) 
+T^{(a)}_{m-1}(u-1){\tilde T}^{(a-1)}_{m+1}(u) ,\\
T^{(a)}_{m+1}(u-1){\tilde T}^{(a)}_m(u) &=
T^{(a)}_{m}(u){\tilde T}^{(a)}_{m+1}(u)
+T^{(a+1)}_{m}(u-1){\tilde T}^{(a-1)}_{m+1}(u)
\end{split}
\end{equation}
from the Pl\"ucker relation.
This is a B\"acklund transformation between T-functions
associated with $A_s$ and $A_{s-1}$.
The T-system for $T^{(a)}_m(u)$ arises as a compatibility of the two
linear equations on ${\tilde T}^{(a)}_m(u)$ \cite{KLWZ}.
For more examples, see \cite{SS, KV, T6, KSZ, Heg} and references therein.
It is an open problem to construct such a Lax representation of the
T-system for general ${\mathfrak g}$.

\subsection{\mathversion{bold}Type $C_r$}\label{ss:Lc}
Let $D$ be the difference operator $Df(u) = f(u+1)D$. 
We use the variable $z_a(u) \, (a \in J)$ (\ref{zac}) which are related to the 
$Q$-functions by (\ref{YQ}).
We also introduce the variables $x_1(u),\ldots, x_{2r+2}(u)$ by
\begin{equation}\label{zxc}
\begin{split}
x_a(u) &= z_a(u),\quad x_{2r+3-a}(u) = z_{\overline{a}}(u)
\;\;(1 \le a \le r),\\
x_{r+1}(u) &= -x_{r+2}(u) = 
\frac{Q_r(u+\frac{r-1}{2})Q_r(u+\frac{r+3}{2})}
{Q_r(u+\frac{r+1}{2})^2}.
\end{split}
\end{equation}
Note that 
$x_{r+1}(u)$ and $x_{r+2}(u)$ are {\em not} contained in 
$\Z[Y^{\pm}_{a,z}]_{a\in I, z \in \C^\times}$.
With the notation
\begin{equation}
\prod_{1 \le i \le k}^{\longrightarrow}X_i = X_1X_2\cdots X_k,
\qquad
\prod_{1 \le i \le k}^{\longleftarrow}X_i = X_k\cdots X_2X_1,
\end{equation}
the difference $L$-operator is 
\begin{equation}\label{Lc}
L(u) = \prod_{1 \le a \le r}^{\longrightarrow}
(1-z_{\overline{a}}(u)D)\cdot
(1-z_{\overline{r}}(u)z_r(u+1)D^2)\cdot
\prod_{1 \le a \le r}^{\longleftarrow}(1-z_a(u)D).
\end{equation}
One can easily check $S_a\cdot L(u)=0$ as in type $A$.
The middle quadratic operator can be factorized as
\begin{align*}
1-Y_{r,q^{2u+r+1}}Y_{r,q^{2u+r+3}}^{-1}D^2
&= 1-\frac{Q_r(u+\frac{r+5}{2})Q_r(u+\frac{r-1}{2})}
{Q_r(u+\frac{r+1}{2})Q_r(u+\frac{r+3}{2})}D^2\\
&=(1\pm x_{r+2}(u)D)(1\pm x_{r+1}(u)D).
\end{align*}
Thus (\ref{Lc}) is expressed as 
\begin{equation}\label{Lxd}
L(u) = \prod_{1 \le i \le 2r+2}^{\longleftarrow}(1-x_i(u)D),
\end{equation}
which resembles curiously the $A_{2r+1}$ case rather than $A_{2r-1}$. 
The operator $L(u)$ generates each fundamental $q$-character ``twice".
\begin{theorem}[\cite{KOSY}]\label{th:kosy1}
\begin{equation*}
L(u) = \sum_{a=0}^r(-1)T^{(a)}_1(u+\frac{a-1}{2})D^a
- \sum_{a=r+2}^{2r+2}(-1)^aT^{(2r+2-a)}_1(u+\frac{a-1}{2})D^a,
\end{equation*}
where $T^{(0)}_1=1$.
\end{theorem}
{}From Theorem \ref{th:kosy1} and (\ref{Lxd}), 
we obtain another tableau sum formula for the 
fundamental $q$-characters:
\begin{equation}
T^{(a)}_1(u+\frac{a-1}{2}) = \sum_{1\le i_1\le \cdots \le i_a\le 2r+2}
\prod_{k=1}^ax_{i_k}(u+a-k)\quad(1 \le a \le r).
\end{equation} 
Although this is formally the same form as $A_{2r+1}$ case (\ref{atab}), 
the variable $x_{r+2}(u)$ (\ref{zxc}) is ``negative" here.
It is highly nontrivial that the cancellation due to the sign yields 
the previous formula (\ref{cta1})
described by the rule (\ref{ccba}), 
which constitutes a substantial part of the proof of Theorem \ref{th:kosy1}.
On the other hand it is easy to see 
\begin{equation}\label{lic}
L(u)^{-1}=\sum_{m\ge 0}T^{(1)}_m(u+\frac{m-1}{2})D^m
\end{equation}
from (\ref{cm1}), (\ref{cm0}) and (\ref{Lc}).

The rest of this subsection will be brief as
the content is more or less parallel with $A_{2r+1}$ case.
We formally extend the fundamental $q$-characters 
$T^{(a)}_1(u)$ to $1 \le a \le 2r+2$ by
\begin{equation}\label{ttr}
T^{(a)}_1(u) + T^{(2r+2-a)}_1(u)=0 \quad (0 \le a \le 2r+2).
\end{equation}
Then Theorem \ref{th:kosy1} is rephrased as 
\begin{equation}
L(u) = \sum_{a=0}^{2r+2}(-1)^aT^{(a)}_1(u+\frac{a-1}{2})D^a.
\end{equation}
We consider the difference equation $L(u)w(u)=0$ and 
a basis of the solution $\{w_1(u),\ldots, w_{2r+2}(u)\}$.
With the same notation $C_u[i_1,\ldots, i_k]$ as (\ref{iku}),
we have the Casoratian formula
\begin{equation}\label{tawc}
T^{(a)}_1(u+\frac{a-1}{2}) = 
\frac{C_u[0,\ldots, a-1, a+1,\ldots, 2r+2]}
{C_u[1,\ldots, 2r+2]}\quad(0 \le a \le 2r+2),
\end{equation}
where $\ldots$ signifies that the omitted arrays are consecutive 
with difference $1$.
The denominator possesses the periodicity
\begin{equation}\label{qcoc}
C_u[0,1,\ldots, 2r+1] =- C_{u+1}[0,1,\ldots, 2r+1],
\end{equation}
which is a $C_r$ analog of the quantum Wronskian condition.

Set 
\begin{equation}
\begin{split}
\xi^{(a)}_m(u) &= C_u[0,\ldots, a-1, a+m,\ldots, 2r+1+m],\\
\xi(u) &= C_u[0,1,\ldots, 2r+1].
\end{split}
\end{equation}

The solution of the unrestricted T-system for $C_r$ that matches 
(\ref{tawc}) is given by 
\begin{theorem}[\cite{KOSY}]\label{th:kosy2}
The following is a solution of the T-system for $C_r$.
\begin{align*}
&T^{(a)}_m(u+\frac{a+m-2}{2})
=(-1)^{m-1}  \frac{\xi^{(a)}_m (u) }{\xi(u+1)}
\quad (1 \le a \le r-1),\\
&T^{(r)}_m(u+\frac{r+2m-1}{2}) T^{(r)}_m(u+\frac{r+2m-3}{2})
=\frac{\xi^{(r)}_{2m} (u) }{\xi(u)},\\
&T^{(r)}_m(u+\frac{r+2m-1}{2}) T^{(r)}_{m+1}(u+\frac{r+2m-1}{2})
=\frac{\xi^{(r)}_{2m+1} (u) }{\xi(u+1)}, \\
&T^{(r)}_m(u+\frac{r+2m-1}{2})^2 =
\frac{\xi^{(r+1)}_{2m} (u) }{\xi(u)}.
\end{align*}
\end{theorem}
As for the first three, there is 
an alternative expression derived by using the
identity 
$\xi^{(a)}_m(u) = (-1)^{a+m+r+1}\xi^{(2r+2-a)}_m(u+a-r-1)$.
See Proposition 4.3 in \cite{KOSY} for details. 

\subsection{\mathversion{bold}Type $B_r$ and $D_r$}\label{ss:Lbd}
Here we only give the $L$-operators and their expansions. 
Let $D$ be the difference operator $Df(u) = f(u+2)D$.
We use the variables $z_a(u)$ for $B_r$ (\ref{zab}) 
and $D_r$ (\ref{zad}) which are related 
to the $Q$-function by (\ref{YQ}).
The difference $L$-operators are 
\begin{align}
B_r:\;\;L(u) &= \prod_{1 \le a \le r}^{\longrightarrow}
(1-z_{\overline{a}}(u)D)\cdot
(1+z_0(u)D)^{-1}\cdot
\prod_{1 \le a \le r}^{\longleftarrow}(1-z_a(u)D),\label{Lb}\\
D_r:\;\;L(u) &= \prod_{1 \le a \le r}^{\longrightarrow}
(1-z_{\overline{a}}(u)D)\cdot
(1-z_r(u)z_{\overline{r}}(u+2)D^2)^{-1}\cdot
\prod_{1 \le a \le r}^{\longleftarrow}(1-z_a(u)D).\label{Ld}
\end{align}
One can check $S_a\cdot L(u)=0$ by 
expanding the middle factor into a power series in $D$.
Introduce the expansion coefficients of $L(u)$ as
\begin{align}
L(u) = \sum_{a \ge 0} (-1)^a T^a(u+a-1)D^{a},\quad
L(u)^{-1} = \sum_{m \ge 0}  T_m(u+m-1)D^{m}.
\end{align}
They are related to the previous tableau constructions as follows:
\begin{align*}
T_m(u) &= T^{(1)}_m(u) \;\;
\text{(\ref{btab1})  for $B_r$ and  (\ref{dtab1}) for $D_r$},\\
T^a(u) &= T^{(a)}_{t_a}(u) \;\;
\text{(\ref{btab1})  for $B_r, 1 \le a\le r$ 
and  (\ref{dtab2}) for $D_r, 1 \le a\le r-2$}.
\end{align*}
With the convention $T^a(u)=0$ for $a<0$, 
the coefficient 
$T^a(u)$ beyond these upper bound is characterized by the 
following relations with the $q$-characters of spin representations:
\begin{align}
B_r:\;\;T^a(u) + T^{h^\vee-a}(u)&=T^{(r)}_1(u+\frac{h^\vee}{2}-a)
T^{(r)}_1(u-\frac{h^\vee}{2}+a),\label{btt}\\
D_r:\;\;T^a(u)+T^{h^\vee-a}(u) 
&=T^{(r)}_1(u+\frac{h^\vee}{2}-a)T^{(r-\delta)}_1(u-\frac{h^\vee}{2}+a)
\nonumber\\
&+
T^{(r-1)}_1(u+\frac{h^\vee}{2}-a)T^{(r-1+\delta)}_1(u-\frac{h^\vee}{2}+a).
\label{dtt}
\end{align}
Here $a \in \Z$ is arbitrary and 
$\delta=0$ if $a\equiv r \mod 2$ and $\delta=1$ otherwise.
$h^\vee$ is the dual Coxeter number (\ref{hhd}), i.e. 
$h^\vee=2r-1$ for $B_r$ and $h^\vee=2r-2$ for $D_r$.
In particular, one has $T^{r-1}(u) = T^{(r)}_1(u)T^{(r-1)}_1(u)$
for $D_r$.

\subsection{\mathversion{bold}Type $sl(r|s)$}\label{ss:supdvf} 

There are two kinds of roots, odd and even 
for the graded algebra $sl(r|s)$.
The choice of simple roots is not unique.
The most standard one is called distinguished, where
all roots but $\alpha_r$ is even.  
Here we follow \cite{T2} and 
set ${\rm I}=\{1,\cdots, r+s\}={\rm I}_1\cup {\rm I}_2$,
${\rm I}_1=\{1,2,\ldots, r\}$, 
${\rm I}_2=\{r+1,r+2,\ldots, r+s \}$, and 
assign the grading $p_a$  by 
$p_a=1\,(-1)$ if  $ a \in  {\rm I}_1 \,({\rm I}_2)$.
The Cartan matrix is expressed by the grading as
\begin{equation*}
(\alpha_k|\alpha_j) 
=(p_k+p_{k+1})\delta_{kj}-p_{k+1} \delta_{k+1, j}-p_k \delta_{k, j+1}.
\end{equation*}
Now the analog of (\ref{zaa}) is 
\begin{equation*}
z_a(u)=Y_{a-1, q^{u+s_a}}^{-p_a} Y_{a, q^{u+s_{a-1}}}^{p_a}
\quad (a \in {\rm I}),
\end{equation*}
where $s_a=\sum_{j=1}^a p_j$ and $Y_{0,q^u}=Y_{r+s,q^u}=1$.
Let $D$ be the difference operator $Df(u) = f(u+2)D$. 
Then the analog of (\ref{ldef2}) and (\ref{ldef3}) are given as
\begin{align*}
&(1+z_{r+s}(u)D)^{p_{r+s}}\cdots
(1+z_{1}(u)D)^{p_1} 
= \sum_{a=0}^{\infty} T^{(a)}_1(u+a-1)D^a,  \\
&(1-z_{1}(u)D)^{-p_1}\cdots
(1-z_{r+s}(u)D)^{-p_{r+s}}
= \sum_{m=0}^{\infty}T^{(1)}_m (u+m-1)D^m. 
\end{align*}

\begin{example}\label{jex:dvf}
$\phantom{}$

\noindent
$sl(2|1)$, $p_1=p_2=-p_3=1$.
\begin{align*}
T^{(1)}_1(u) &= 
Y_{1,z} + Y_{1,z q^2}^{-1} Y_{2, z q} - Y_{2, z q},   \\
T^{(2)}_1(u) &=  
Y_{2,z} -  Y_{1, z q}Y_{2,z} 
-Y_{1, z q^3}^{-1}Y_{2, z} Y_{2, z q^2}+ Y_{2, z }Y_{2, z q^2}, \\
T^{(3)}_1(u) &= -Y_{2,z q^{-1}} Y_{2,z q} 
+ Y_{1, z q^2}Y_{2,z q^{-1}} Y_{2,z q}
+Y_{1, z q^4}^{-1} Y_{2, zq^{-1} }Y_{2, z q}Y_{2, z q^3} \\
&- Y_{2, zq^{-1}}Y_{2, z q}Y_{2, z q^3}.
\end{align*}

\noindent
$sl(2|1)$, $p_1=-p_2=p_3=1$.
\begin{align*}
T^{(1)}_1(u) &= Y_{1,z} - Y_{1,z} Y_{2, z q}^{-1} + Y_{2, z q}^{-1}, \\
T^{(2)}_1(u) &=  - Y_{1,zq^{-1}}Y_{1,zq}   Y_{2,z}^{-1}
 + Y_{1, z q} Y_{2,z}^{-1} + Y_{1, z q^{-1}} Y_{1, z q} Y_{2, z q^2}^{-1}  Y_{2, z}^{-1}  -Y_{1,z q}  Y_{2, z q^2}^{-1}  Y_{2, z}^{-1},  \\
T^{(3)}_1(u)&=Y_{1,zq^{-2}} Y_{1,z}
Y_{1,zq^{2}}Y_{2,zq^{-1}}^{-1}Y_{2,zq}^{-1}
-Y_{1,z} Y_{1,zq^{2}}  Y_{2,zq^{-1}}^{-1} Y_{2,zq}^{-1},  \\
&-Y_{1,zq^{-2}} Y_{1,z} Y_{1,zq^{2}}Y_{2,zq^{-1}}^{-1}
Y_{2,zq}^{-1} Y_{2,zq^3}^{-1}
+  Y_{1,z}Y_{1,zq^{2}}Y_{2,zq^{-1}}^{-1}Y_{2,zq}^{-1} Y_{2,zq^3}^{-1}.
\end{align*}
\end{example}

For the formulas for general case, see \cite{KV, T6}.

\subsection{Bibliographical notes}
The Casoratian solution (\ref{tamw}) for $A_r$ 
has been known in various contexts.
For the T-system of transfer matrices, 
a slightly more general solution than (\ref{twd}) 
was given in eq.(2.25) in \cite{KLWZ}
containing $2r+2$ arbitrary functions.
It does not satisfy 
the natural boundary condition 
$T^{(a)}_{-1}(u)=0$ for fusion transfer matrices in general.
As usual, such a ``Dirichlet" condition halves the arbitrary functions
to $w_1(u),\ldots, w_{r+1}(u)$,  which brings one back to (\ref{twd}).
Casoratian solutions are known also for the 
restricted T-systems for  
$A_r$ \cite{V} and $C_r$ \cite{IIKNS}.

The $L$-operator for type $A$ has been studied from the viewpoint of 
difference analog of Drinfeld-Sokolov reduction \cite{FRS}.
The concrete forms for type $BCD$ and their application to $q$-characters 
were given in \cite{KOSY}.
Analogous difference $L$-operators for all the
twisted cases except $E^{(2)}_6$ have been constructed in \cite{T5}.
The results (\ref{btt}) and (\ref{dtt}) 
are taken from Theorem 2.3 in \cite{KOS}
and Proposition 2.3 in \cite{TK}, respectively.

\section{T-system in ODE}\label{s:ODE}

T-system appears also in the connection problem of 
1D Schr{\"o}dinger equation,
which is a typical example of the 
ODE (ordinary differential equations)/IM (integrable models) 
correspondence.
As a comprehensible review on the ODE/IM correspondence 
is already available in \cite{DDT3}, 
we only discuss the issue briefly in view of T-system. 
Wronskians appear naturally in the context of ODE.
They will be shown to coincide with the analogous object,
the Casoratian (\ref{iku}) 
in the difference equation in Section \ref{s:tq}.
 
 \subsection{Generalized Stokes multipliers 
- the 2nd order case} \label{ss:ODE2}
As the simplest example, 
we consider the 1D Schr{\"o}dinger equation 
on the real axis with a potential term:
\begin{equation}
 \Bigl( -\frac{d^2}{dx^2} + x^{2 M} \Bigr) \psi(x)  = E \psi(x),
\label{j:Schr}
\end{equation}
where  $M \in \Z_{>0}$.
The boundary condition $\psi(\pm \infty)=0$ is imposed. 
We find it convenient to extend $x$ into the complex plane\footnote{ 
For a general reference to ODE in the complex domain, we recommend \cite{Si}.
}.

Since the Schr{\"o}dinger equation has the irregular singularity at $\infty$, 
we expect a sudden change of 
$\psi(x)$ when crossing a border line  of sectors defined below.
This is called the Stokes phenomenon. 
The change  is characterized by the Stokes multiplier $\tau_1$.
Below we will introduce a set of generalized 
Stokes multipliers $\{\tau_j\}_{j=1}^{2M}$
and show that they satisfy the level $2M$ restricted T-system for $A_1$.

First, let ${\mathcal S}_j$ be a sector in the complex plane defined by
$$
{\mathcal S}_j=\Bigl\{ x \big\vert \;
\Big\vert{\rm arg}\, x - \frac{j \pi}{M+1} \Big\vert 
< \frac{\pi}{2M+2}
\Bigr\}.
$$
The sector ${\mathcal S}_0$ thus includes the positive real axis.
We then introduce a solution $\phi(x, E)$ to (\ref{j:Schr}) 
which decays exponentially 
as $x$ tends to $\infty$  inside ${\mathcal S}_0$ as
\begin{equation}
\phi(x,E) \sim \frac{ x^{-M/2} }{\sqrt{2i}}
\exp\Bigl(-\frac{ x^{M+1} }{M+1}\Bigr), 
\qquad x \in {\mathcal S}_0.
\label{j:asym}   
\end{equation}
This is referred to as the subdominant solution.
There should be another solution to (\ref{j:Schr})
which diverges exponentially in 
${\mathcal S}_0$ as $x$ tends to $\infty$. 
We call it  dominant.
It is also represented by $\phi$.
To see this, note the invariance of (\ref{j:Schr}) 
under the simultaneous transformations
$x \rightarrow q^{-1} x$ and $E \rightarrow  E q^2$,
where $q=\exp(\frac{\pi i}{M+1})$.  
We call this ``discrete rotational symmetry".
We thus introduce  $y_j= q^{j/2} \phi(q^{-j} x,  q^{2 j} E)$
so that $y_0=\phi$. 
The above observation tells that 
any $y_j$ is  a solution to  (\ref{j:Schr}).  
Moreover, we can show that the pair 
$(y_j, y_{j+1})$ forms the fundamental system of solutions (FSS) 
in ${\mathcal S}_j$.
This is easily seen by introducing 
the Wronskian matrix $\Phi_j$ and the Wronskian $W[y_i, y_j]$:
$$
\Phi_j=
\begin{pmatrix}
y_j &  y_{j+1}\\
\partial y_j& \partial y_{j+1}
\end{pmatrix},
\qquad
W[y_i, y_j]={\rm det}\, 
\begin{pmatrix}
y_i &  y_j\\
\partial y_i& \partial y_j
\end{pmatrix}.
$$
By using the asymptotic form (\ref{j:asym}),  
one can check $W[y_j, y_{j+1}]=1$, hence 
the pair $(y_j, y_{j+1})$ is independent.
Thus, $y_0$  (equals to $\phi$) is the subdominant 
solution in ${\mathcal S}_0$, while  $y_1$ is a dominant one.

We are interested in the relation among FSS in different sectors.
Let us start from  ${\mathcal S}_0$ and   ${\mathcal S}_1$.
Obviously $y_2$ must be represented 
by the linear combination of $y_0$ and $y_1$
as $y_2=a_0 y_0 + a_1 y_1$. 
As $W[y_j, y_{j+1}] =1$ for any $j$,  we find $a_0=-1$.
The coefficient $a_1$ can be regarded as a function of $E$ 
and we write it as $\tau_1(E)=a_1$,
which is referred to as the Stokes multiplier.
The result can be neatly represented in the matrix form
$$
\Phi_0=\Phi_1 {\mathcal M}_{1,0}, \qquad
{\mathcal M}_{1,0}=
\begin{pmatrix}
  \tau_1(E) & 1 \\
   -1& 0  \\
\end{pmatrix}.
$$
The general adjacent FSS $\Phi_{j}$ and  $\Phi_{j+1}$ are connected 
by $ \Phi_{j}=\Phi_{j+1}{\mathcal M}_{j+1, j}$, 
and the ``discrete rotational symmetry" leads to
${\mathcal M}_{j+1, j}
= {\mathcal M}_{1,0}|_{E \rightarrow E q^{2j}}$.
We introduce the matrix connecting well separated sectors
\begin{equation}
\Phi_0=\Phi_j {\mathcal M}_{j,0}.
\label{j:Mj0def}
\end{equation}
By the definition, the recursion relation
\begin{equation}\label{recurM}
{\mathcal M}_{j,0} ={\mathcal M}_{j,1} {\mathcal M}_{1,0}
\end{equation}
holds. 
The solution to this takes the form  
\begin{equation}
{\mathcal M}_{j,0} =
\begin{pmatrix}
         \tau_{j}(E)& \tau_{j-1}(Eq^2) \\
          - \tau_{j-1}(E)& -\tau_{j-2}(Eq^2)
\end{pmatrix}.
\label{j:Mj0sol}
\end{equation}
Here $\tau_j$ is the function 
uniquely determined from $\tau_1$ and the recursion relation
\begin{equation}
\tau_{j}(q^{2}E)\tau_1(E)\!=\!\tau_{j+1}(E)\!+\!\tau_{j-1}(q^{4}E)
\label{recursiontau}
\end{equation}
with $\tau_0(E)=1$.
We set $\tau_{-1}(E)=0$ so that 
this holds also at $j=0$. 
In addition we have $\tau_{2M}(E)=1, \tau_{2M+1}(E)=0$ as 
after $360^\circ$ rotation, FSS must come coincide with the original
one times $(-1)$.
(cf. \cite[(21.31)]{Si}.)
We call $\tau_j\, (j \ge 2)$  generalized Stokes multipliers.
The generalized Stokes multipliers satisfy the relation 
\begin{equation}
\tau_{j}(E) \tau_j(Eq^2) = \tau_{j-1}(E q^2) \tau_{j+1}(E) + 1.
\label{j:tauT}
\end{equation}
This is equivalent to $\det {\mathcal M}_{j,0}=1$.
It is shown either by (\ref{j:Mj0def}) 
or by induction on $j$ using (\ref{recursiontau}).
See also the discussion in Section \ref{ss:WCdual}.
Setting
$$
T_j(u) = \tau_j (Eq^{-j-1}), \qquad 
\text{where}\;\; E=\exp\Bigl(\frac{\pi iu}{M+1}\Bigr),
$$
we therefore have
\begin{proposition}
$\{T_j(u)\}$ satisfy the level $2M$ restricted T-system for $A_1$
\begin{equation}
T_j(u+1) T_j(u-1) = T_{j-1}(u) T_{j+1}(u)+1  \qquad (j=1,\cdots, 2M),
\label{j:Tode}
\end{equation}
where  $ T_{0}(u)=1$ and $T_{2M+1}(u)=0$.
\end{proposition}

\begin{example}
By (\ref{j:Mj0def}), (\ref{j:Mj0sol}) 
and $\det {\mathcal M}_{j,0}=1$, one has
\begin{equation*}
\tau_j(E) =W[y_0, y_{j+1}],
\end{equation*}
where the RHS is independent of $x$.
The consistency of   
$\tau_{2M}=1$ and $\tau_{2M+1}=0$ with 
$y_{2M+1}=-y_{-1}$ and 
$y_{2M+2}=-y_0$ is reconfirmed.
Relation (\ref{j:tauT}) is also re-derived 
from the simple identity among Wronskians
$[y_\alpha, y_\beta][y_\gamma, y_\delta] 
=[y_\alpha, y_\gamma][y_\beta,y_\delta] 
+ [y_\alpha, y_\delta][y_{\gamma},y_\beta]$ 
by the specialization  $\alpha=0, \beta=j+1, \gamma=1, \delta=j+2$.
Note $W[y_k,y_{k+1}]=1$ for any $k$.
\end{example}

\subsection{Higher order ODE}
\label{ss:slnODE}

One can extend the observation on the second order ODE to 
higher order case corresponding to 
$\mathfrak{g}=A_r$ \cite{DDT1, Suz2, BHK, FL}.
Consider a natural generalization of  (\ref{j:Schr}):
\begin{equation}
(-1)^r \frac{d^{r+1} y}{dx^{r+1}} + x^{\ell} y =Ey  =\lambda^{r+1} y.
\label{NthODE}
\end{equation}
Let $q={\rm e}^{i\theta}$ with $\theta=\frac{2\pi}{\ell+r+1}$.
The sector ${\mathcal S}_k$ is now defined by
$|{\rm arg}\, x -k \theta  | \le \frac{\theta}{2}$.
We pay attention to the solution $\phi(x,  \lambda)$ 
in ${\mathcal S}_0$ which decays most rapidly as 
$x \rightarrow \infty$ as
$$
\phi(x,  \lambda) \sim  C x^{-r\ell /(2r+2)}
\exp\Bigl(- \frac{x^{\nu}}{\nu}\Bigr),
\qquad
\nu= \frac{\ell+r+1}{r+1}.
$$
The normalization factor $C$ will be determined later. 
As in the 2nd order ODE case, 
(\ref{NthODE}) is invariant 
under $x \rightarrow  x q^{-1}, E \rightarrow  E q^{r+1}$.
Thus in terms of  $\lambda$,  
$y_k=q^{rk/2} \phi(x q^{-k}, \lambda q^k)$  
is also a solution to  (\ref{NthODE}) for any $k \in \Z$.

The FSS in ${\mathcal S}_k$ consists of $(y_{k}, \cdots, y_{k+r})$. 
It is convenient to
introduce a Wronskian matrix
\begin{equation*}
\Phi_k=
\begin{pmatrix}
y_k            & y_{k+1}          & \cdots  & y_{k+r}\\
\vdots          &                    &        &\vdots \\
\partial^r y_k & \partial^r y_{k+1}& \cdots&\partial^r y_{k+r}
\end{pmatrix}.
\end{equation*}
We write the determinant of a slightly more 
general matrix (for $m\le r$) as 
\begin{equation}\label{j:W}
W[y_{i_0}, y_{i_1}, \cdots \,  ,y_{i_m}]={\rm det}
\begin{pmatrix}
y_{i_0}& y_{i_1} &  \cdots & y_{i_m}\\
\vdots  &          &          &  \vdots \\
\partial^{m} y_{i_0}&\partial^{m} y_{i_1}&
\cdots&   \partial^{m}y_{i_m}
\end{pmatrix}.
\end{equation}
Due to (\ref{NthODE}), the Wronskians ($m=r$ cases) 
are independent of $x$.
In particular, the normalization constant 
$C$ can be fixed 
so that  ${\rm det}\, \Phi_k=W[y_{k},\cdots ,y_{k+r}]=1$ 
for any $k$.
We introduce the connection matrix ${\mathcal M}_{k+1,k}$ by
\begin{equation}
\Phi_{k}= \Phi_{k+1}{\mathcal M}_{k+1,k}.
\label{j:defcM}
\end{equation}
It has the form 
\begin{equation*}
{\mathcal M}_{k+1,k} =
\begin{pmatrix}
\tau^{(1)}_1(\lambda q^k)  & 1 & 0&  0&  \cdots& 0\\
\tau^{(2)}_1(\lambda q^k) & 0 & 1&  0&  \cdots& 0\\
      \vdots   &    &    &  &          & \vdots \\
\tau^{(r)}_1(\lambda q^k) & 0& 0&  0&  \cdots & 1 \\
\tau^{(r+1)}_1(\lambda q^k)& 0& 0&  0& \cdots & 0
\end{pmatrix}.
\end{equation*}
By using  Cramer's formula, 
$\tau_1^{(a)}(\lambda q^k)$ is expressed as the Wronskian
$$
\tau_1^{(a)}(\lambda q^k) 
=W[y_{k+1},\cdots, y_{k+a-1}, y_k, y_{k+a+1},\cdots, y_{k+r+1}].
$$
Especially, one finds $\tau_1^{(r+1)}(\lambda q^k) =(-1)^r$.
We further introduce the generalized Stokes multipliers
$\tau_m^{(a)}(\lambda)$ for $m \ge 2$ by
\begin{equation}\label{j:tam}
\tau^{(a)}_m (\lambda)= W[
y_1, y_2, \cdots  y_{a-1},y_0 , y_{a+m},y_{a+m+1}\cdots y_{r+m} ].
\end{equation}
Note that $m$ does not extend to infinity.
Due to $y_{r+1+\ell}=(-)^r y_{0}$, one has  
$\tau^{(a)}_{\ell+1}(\lambda)=0$.  
This causes a truncation analogous to the 
level restriction in quantum group at root of unity.
It is elementary to prove
\begin{proposition}\label{TsystemStokes}
The generalized Stokes multipliers $\tau_m^{(a)}(\lambda)$ 
satisfy the level $\ell$ restricted T-system for $A_r$
\begin{equation*}
\tau^{(a)}_m(\lambda) \tau^{(a)}_m(\lambda q)
= \tau^{(a)}_{m+1}(\lambda) \tau^{(a)}_{m-1}(\lambda q)+
 \tau^{(a+1)}_m(\lambda) \tau^{(a-1)}_m(\lambda q)
\qquad (1\le a \le r),
\end{equation*}
where the boundary conditions are modified as
$\tau^{(0)}_m(\lambda)=1, 
\tau^{(r+1)}_m(\lambda) = (-1)^{r}$ 
and $\tau^{(a)}_0(\lambda) = (-1)^{a-1}$.
\end{proposition}

\begin{remark}
One might expect that $\tau_m^{(a)}(\lambda)$ 
may appear in the generalized connection matrix 
${\mathcal M}_{k+m,k}$ connecting  $\Phi_k$ and 
$\Phi_{k+m}\, (m \ge 2)$. 
This is not the case.  
As the Schur functions, one can define generalized 
Stokes multipliers associated with 
(skew) Young tableaux of a general shape.  
Entries of ${\mathcal M}_{k+m,k}$ 
are generally identified with such objects.
Especially the $(a,1)$ component of ${\mathcal M}_{k+m,k}$ 
corresponds to the Young tableau of the hook shape
of width $m$ and height $a$. 
\end{remark}

\subsection{Wronskian-Casoratian duality}
\label{ss:WCdual}

The $(i+1,1)$ element from 
the matrix relation (\ref{j:defcM}) with $k=0$ reads
$\partial^i y_0 = \tau^{(1)}_1(\lambda)\partial^i y_1+\cdots +
\tau^{(r+1)}_1(\lambda)\partial^i y_{r+1}$.
Remember that $y_k = q^{rk/2}\phi(xq^{-k},\lambda q^k)$
involves $x$ but $\tau^{(a)}_1(\lambda)$ does not.
Thus one obtains an $x$-independent relation by setting $x=0$ as
\begin{equation}\label{pty}
\partial^i y_0|_{x=0} 
= \tau^{(1)}_1(\lambda)\partial^i y_1|_{x=0}+\cdots +
\tau^{(r+1)}_1(\lambda)\partial^i y_{r+1}|_{x=0}
\quad (0 \le i \le r).
\end{equation} 
In view of $y_k=q^{rk/2} \phi(x q^{-k}, \lambda q^k)$,
this has the same form as 
the difference equation (TQ-relation) 
(\ref{Lw}) with (\ref{ldef2}):
\begin{equation}\label{Lww}
w(u) - T^{(1)}_1(u)w(u+2) + 
\cdots + (-1)^{r+1}T^{(r+1)}_1(u+r)w(u+2r+2)=0.
\end{equation}
In fact, under the formal (ODE/IM) correspondence between 
the Stokes multipliers and the transfer matrix eigenvalues
\begin{equation}\label{fid}
\tau^{(a)}_1(\lambda)  = (-1)^{a-1}T^{(a)}_1(u+a-1)\quad 
(1 \le a \le r+1),
\end{equation}
the identification $w(u+2j) = \partial^iy_j|_{x=0}$ 
provides a solution to (\ref{Lww}) for any $0 \le i \le r$.
The variables $u$ and $\lambda$ are 
related so that the shift $u \rightarrow u+2$ corresponds to 
$\lambda \rightarrow \lambda q$.
Now we are entitled to substitute 
\begin{equation}\label{wuy}
w_i(u+2j) = \partial^{i-1}y_j|_{x=0}\qquad  
(1\le i \le r+1)
\end{equation}
into the Casoratian $C_u$ (\ref{iku}).
The result is the equality 
\begin{equation}\label{cw}
 W[y_{i_1},\ldots, y_{i_k}]|_{x=0} = C_u[2i_1,\ldots, 2i_k],
\end{equation}
which we call the Wronskian-Casoratian duality.
One can remove ``$|_{x=0}$"  when $k=r+1$.
Remember that in Section \ref{ss:dl}--\ref{ss:Q},
a variety of generalizations of $T^{(a)}_1$ are 
expressed in terms of Casoratians $C_u$.
The relations (\ref{fid}) and (\ref{cw}) enable us 
to import those results 
to establish a number of Wronskian formulas for 
the generalized Stokes multipliers.
For example, the formula (\ref{tamw}) leads to 
(\ref{j:tam}).

The Wronskian-Casoratian duality further provides 
the Stokes multipliers with dressed vacuum forms 
like the ones for $A_r$ in Section \ref{s:aba}.
Recall that Proposition \ref{pr:nnsy} expresses the  
Casoratians as the sums over semistandard tableaux like 
(skew) Schur functions.
The variables attached to tableau letters are ratio of 
the principal minors of $C_u[0,2,\ldots,2r]$, namely
$Q_a(u+a-1) = C_u[0,\ldots, 2a-2]$ (\ref{qaw}), 
which are called Baxter's Q-functions. 
Via the Wronskian-Casoratian duality, this is translated to
a dressed vacuum form for Stokes multipliers. 
The tableau variables are ratio of 
$W[y_{k+1},y_{k+2},\ldots, y_{k+a}]|_{x=0}$, 
which are to be identified with Baxter's Q-functions 
$Q_a(\lambda q^{a+k})$ in the present context.

As explained in Section \ref{ss:bt} for Casoratians, the solutions 
$w_1,\ldots, w_{r+1}$ to (\ref{Lww}) may be renumbered 
arbitrarily, and this freedom generates B\"acklund transformations
among Q-functions.
Even more generally, one may consider 
arbitrary linear combinations of 
(\ref{pty}) instead of (\ref{wuy}) as
\begin{equation}\label{wuy2}
w_i(u+2j) = \sum_{n=0}^{r}A_{in}\partial^{n}y_j|_{x=0}\qquad  
(1\le i \le r+1),
\end{equation}
where $(A_{in})_{1\le i \le r+1, 0 \le n\le r}$ is 
any invertible matrix.  
In the Wronskian language, this corresponds to 
identifying $Q_a(\lambda q^{a+k})$ with
\begin{equation*}
\sum_{0\le n_1<\cdots < n_a\le r}
\det (A_{i, n_j})_{1 \le i, j \le a}\det
\begin{pmatrix}
\partial^{n_1}y_{k+1}& 
\partial^{n_1}y_{k+2}&  \cdots & 
\partial^{n_1}y_{k+a}\\
\vdots  &          &          &  \vdots \\
\partial^{n_a}y_{k+1}& 
\partial^{n_a}y_{k+2}&  \cdots & 
\partial^{n_a}y_{k+a}
\end{pmatrix}
\end{equation*}
evaluated at $x=0$.
In this way the same Stokes multiplier acquires
a variety of representations.

We note that in the simple cases like $\tau^{(1)}_1(\lambda)$, 
the recursion relation (see for example \cite{DDT1,Suz2})
\begin{equation}
\frac{[y_0, y_2, \cdots y_m] }{ [ y_1, \cdots ,y_m  ]}=
\frac{[y_0, y_2, \cdots y_{m-1}] }{ [ y_1, \cdots ,y_{m-1} ]}+
\frac{[y_0, y_1, \cdots y_{m-1}][y_2, \cdots, y_m] }
  { [ y_1, \cdots ,y_m ][ y_1, \cdots ,y_{m-1} ]}
\end{equation}
is handy to derive the dressed vacuum forms 
without recourse to 
Proposition \ref{pr:nnsy} and 
the Wronskian-Casoratian duality (\ref{cw}).

\subsection{Bibliographical notes}\label{ss:bode}

The functional relations have appeared in
ODE in the context of asymptotic analysis \cite{Si} or
of complex WKB method \cite{Vor}.
The connection to integrable models 
has been realized in \cite{DT1} and
the machineries of the latter 
have been applied since then \cite{Suz1, BLZ3, DT2}.
The connection not only provides the information on Stokes multipliers but
also solves the spectral problem of ODE.  
With an assumption on analyticity, one can transform
(\ref{j:Tode}) to the thermodynamic Bethe ansatz equation 
that describes a conformal field theory (CFT) in the ground state.
It provides a quantitative tool to 
obtain the eigenvalues of (\ref{j:Schr}).
A more direct relation can be established between the 
spectral determinant associated to ODE and the 
vacuum expectation value of the
Baxter's $Q$ operator in CFT \cite{BLZ3,DT2}. 

It is tempting to consider Schr{\"o}dinger operators with 
more general polynomial potentials.
Although we can argue the algebraic part in an almost same manner, 
the problem with the analyticity defies most attempts up to now.
The case with $V(x)=\alpha x^{M-1}+x^{2M}$ 
is exceptionally treated nicely \cite{Suz3}.
The underlying model seems to possess $gl(2|1)$ symmetry. 
The fundamental reason why this symmetry
appears remains to be clarified. 
This case seems interesting in its relation to 
${\mathcal PT}$ symmetric quantum systems \cite{BB} and
spontaneous breakdown of the symmetry \cite{DDT2}. 
The integro-differential systems corresponding 
to non exceptional classical Lie algebras in the similar sense 
are proposed in \cite{DDMST}.

The role played by the excited states of CFT is studied in \cite{BLZ4}.
The corresponding Schr{\"o}dinger operators 
with potentials possessing singularities are identified.   
A further argument from the viewpoint of the Langlands correspondence is
given in \cite{FeFr}.  

In general, CFTs are realized as scaling limits of lattice models. 
Then one may wonder if there exists an ODE which corresponds
to a lattice model on a finite system.
This is investigated in \cite{FSZ, DST} for particular cases.
As for generalizations related to 
massive deformations of CFT, see \cite{BM, LZ}. 

\section{Applications in gauge/string theories}\label{s:ags}

The AdS/CFT correspondence 
is a huge subject in theoretical and mathematical physics.
Here we pick just two topics rather briefly, 
planar AdS/CFT spectrum (Section \ref{ss:pacs}--\ref{ss:baba})
and area of minimal surface in AdS (Section \ref{ss:msa}--\ref{ss:af}),
from the gauge and the string theory sides, respectively.
These subjects have been growing rapidly during
the last couple of years where 
some specific T and Y-systems have found notable applications.

\subsection{Planar AdS/CFT spectrum}\label{ss:pacs}

Recall the AdS$_5$/CFT$_4$ correspondence between
the type IIB superstring on the curved space time 
$\mathrm{AdS}_5\times \mathrm{S}^5$
and the large $N$ 
conformal ${\mathcal N}=4$ super Yang-Mills (SYM) gauge theory 
in four dimensions on the boundary of $\mathrm{AdS}_5$
\cite{Mal, GKP, Wi}.
The correspondence implies that the energies of 
specific string states 
should coincide with anomalous scaling dimensions of 
local gauge invariant operators in the SYM.
We call the sought common spectrum the {\em planar AdS/CFT spectrum}.

To be concrete, let us consider simplest examples from
the SYM side,  
linear combinations of single trace scalar operators without derivatives
\begin{equation}\label{sitr}
\textstyle{\sum_{i_1\ldots i_L}} 
c^{i_1\ldots i_L}\mathrm{Tr}\; \Phi_{i_1}\cdots \Phi_{i_L},
\end{equation}
where $\Phi_i\,(i=1,\ldots,6)$ denote 
the six scalar fields of ${\mathcal N}=4$ SYM
in the adjoint representation of $SU(N)$. 
They contain important examples like chiral primary and 
BMN operators \cite{BMN} as special cases
and form an interesting sector that are mixed 
only among themselves at one-loop renormalization.
In fact, the last property reduces the one-loop calculation of 
scaling dimensions of (\ref{sitr}) to the 
diagonalization of the Wilson matrix 
$\left(\frac{\partial \ln Z}{\partial \ln \Lambda}\right)$ 
consisting of the wave function renormalization factors $Z=(Z_{i j})$,
where $\Lambda$ is the UV cutoff.
This problem turns out rather remarkably identical 
with a periodic spin chain of length $L$ 
associated with rational $R$ matrix for $SO(6)$.
Thus in the large $L$ limit, 
one can evaluate, for example,
the largest possible scaling dimension of (\ref{sitr}) 
by the Bethe ansatz as \cite{MZ1}
\begin{equation*}
L + \frac{\lambda L}{8\pi^2}
\bigl(\frac{\pi}{2}+\ln 2\bigr) + {\mathcal O}(\lambda^2),
\end{equation*}
where $\lambda=g_{\rm YM}^2N$ is the 't Hooft coupling.
One sees how the bare dimension $L$ (1st term) acquires 
the anomalous correction.

Although this is a one-loop perturbative approximation to the   
planar AdS/CFT spectrum in a very limited sector,
the connection to the Bethe ansatz is a signal of 
the integrability of the full problem.
In fact, this theme has been explored 
both from the gauge and string theory perspectives extensively by 
an enormous amount of works.
We do not intend to cover them here but refer to the 
literatures that will be cited in the next subsection and 
\cite{BMN, MZ1, BSt1, BPR, AFS, KLOV, KMMZ, 
BDKM, HLo, Ja, BKM, Dor, Bel, BJ} for example and references therein.
See also \cite{Lip, FK} for earlier observations before AdS/CFT.

\subsection{\mathversion{bold}T and Y-system 
for AdS$_5$/CFT$_4$}\label{ss:tyads}

The planar AdS/CFT spectrum is accessible from the gauge theory side
via an integrable long range quantum spin chain 
with $PSU(2,2|4)$ symmetry \cite{BSt2}.
This is actually so at least asymptotically if the relevant quantum numbers 
like the bare scaling dimension are large enough. 
In the language of spin chains, such situations correspond to 
the thermodynamic limit where ``impurities" 
(Bethe roots) are kept dilute.

Complementally,  the {\em exact} spectrum including 
``finite size effects" may be encoded in 
some T and Y-systems together 
with an appropriate, albeit highly elaborate, analyticity input\footnote{
Such features are illustrated along 
the elementary example of the XXZ chain 
in Section \ref{s:app}.}.
A candidate for such a Y-system has been proposed in 
\cite{BFT1, GKKV, AF3} based on the ground state TBA equation 
associated with the asymptotic Bethe ansatz (ABA) 
equation \cite{St, BSt2, BES}
in the mirror form \cite{AF2}.

The underlying symmetry of the ABA equation is 
$PSU(2,2|4)$ \cite{BSt2}.
Reflecting this fact, the Y-system in question
contains two copies of the Y-systems for the subgroup
$SU(2|2)$\footnote{
It is essentially the Y-system for $U_q(sl(2|2))$ 
in Section \ref{ss:sup}.}  
denoted by $SU(2|2)_L$ and $SU(2|2)_R$.
Apparently it takes the same form as type $A$ case: 
\begin{equation}\label{ygkv}
\frac{Y_{a,s}(u-\frac{i}{2})Y_{a,s}(u+\frac{i}{2})}
{Y_{a+1,s}(u)Y_{a-1,s}(u)} = 
\frac{(1+Y_{a,s+1}(u))(1+Y_{a,s-1}(u))}
{(1+Y_{a+1,s}(u))(1+Y_{a-1,s}(u))}.
\end{equation}
A peculiarity here is that $Y_{a,s}(u)$ is defined for those $(a,s)$ 
that correspond to the black nodes in 
the following T-shaped fat hook:
\begin{equation}\label{brf}
\begin{picture}(240,122)(-120,-27)
\unitlength=0.3mm
\thicklines

\put(1,93){\vector(0,1){15}}
\put(6,105){$a$}

\put(-25,60){\color{red}\line(-1,0){10}}
\put(25,60){\color{red}\line(1,0){10}}
\put(-25,40){\color{red}\line(-1,0){10}}
\put(25,40){\color{red}\line(1,0){10}}

\multiput(0,0)(80,0){2}{
\multiput(-40,20)(0,20){3}{
\put(0,5){\color{red}\line(0,1){10}}}}

\put(-40,60){\color{red}\circle{10}}\put(40,60){\color{red}\circle{10}}
\put(-40,40){\color{red}\circle{10}}\put(40,40){\color{red}\circle{10}}

\multiput(0,0)(140,0){2}{
\multiput(-40,20)(-20,0){3}{
\put(-5,0){\color{red}\line(-1,0){10}}}}

\put(-80,20){\color{red}\circle{10}}\put(-60,20){\color{red}\circle{10}}
\put(60,20){\color{red}\circle{10}}\put(80,20){\color{red}\circle{10}}

\multiput(-80,0)(140,0){2}{
\put(0,5){\color{red}\line(0,1){10}}
\put(20,5){\color{red}\line(0,1){10}}}

\put(-41,79){\color{red}$\vdots$}
\put(-21,79){$\vdots$}
\put(-1,79){$\vdots$}
\put(19,79){$\vdots$}
\put(39,79){\color{red}$\vdots$}

\put(25,20){\line(1,0){10}}\put(-40,20){\circle{10}}
\put(40,20){\circle{10}}\put(-25,20){\line(-1,0){10}}

\multiput(0,0)(0,20){4}{
\put(-20,0){\circle{10}}
\put(0,0){\circle{10}}
\put(20,0){\circle{10}}

\put(0,5){\line(0,1){10}}
\put(20,5){\line(0,1){10}}
\put(-20,5){\line(0,1){10}}

\put(-5,0){\line(-1,0){10}}
\put(5,0){\line(1,0){10}}
}

\multiput(0,0)(-20,0){5}{
\put(0,0){\circle{10}}\put(-5,0){\line(-1,0){10}}
}

\multiput(0,0)(20,0){5}{
\put(0,0){\circle{10}}\put(5,0){\line(1,0){10}}
}

\put(-111,17){\color{red}$\cdots$}\put(97,17){\color{red}$\cdots$}
\put(-111,-3){$\cdots$}
\put(-40,5){\line(0,1){10}}\put(40,5){\line(0,1){10}}
\put(97,-3){$\cdots$}

\multiput(-80,-20)(20,0){9}{
\put(0,0){\color{red}\circle{10}}}

\multiput(-80,0)(20,0){9}{
\put(0,-5){\color{red}\line(0,-1){10}}}

\multiput(-80,-20)(20,0){10}{
\put(-5,0){\color{red}\line(-1,0){10}}}

\put(-111,-23){\color{red}$\cdots$}\put(97,-23){\color{red}$\cdots$}

\put(116,-20){\vector(1,0){17}}\put(131,-13){$s$}

\put(-30,-32){$\scriptstyle{0,-1}$}
\put(-6,-32){$\scriptstyle{0,0}$} 
\put(15,-32){$\scriptstyle{0,1}$}

\end{picture}
\end{equation}
The relevant T-system \cite{GKKV} is also formally of type A:
\begin{equation}\label{tgkv}
T_{a,s}(u-\textstyle\frac{i}{2})
T_{a,s}(u+\textstyle\frac{i}{2})
= T_{a,s-1}(u)T_{a,s-1}(u)
+T_{a-1,s}(u)T_{a+1,s}(u),
\end{equation}
where this time $(a,s)$ ranges over black 
as well as red nodes in (\ref{brf}).
The relation to the Y-system
$Y_{a,s}(u) = 
\frac{T_{a,s-1}(u)T_{a,s+1}(u)}
{T_{a-1,s}(u)T_{a+1,s}(u)}$ is as usual.
The diagram (\ref{brf}) is meant to 
capture the structure of the equations 
(\ref{ygkv}) and (\ref{tgkv})\footnote{Another, yet more 
intrinsic way of encoding the Y-system together with the T-system is by the 
quiver in the cluster algebra formulation in Section \ref{t:subsec:TY1}.}.

Recall that the Y-system for $U_q(sl(2|2))$ in Section \ref{ss:sup}
involves the variables $Y^{(a)}_m$ with $(a, m)$ ranging over 
$H_{2,2}$ (\ref{j:fh}) which is an L-shaped ``thin" hook.
This and its copy are embedded into (\ref{brf}) 
as $Y_{a,m}$ and $Y_{a,-m}$.
The extra variables $Y_{a,0}(u)$ on the middle  
vertical array $(a,0)_{a\ge 1}$ 
are the carriers of the ``momentum" (cf. (\ref{bel})). 
The two wings $s<0$ and $s>0$ correspond to 
$SU(2|2)_L$ and $SU(2|2)_R$ mentioned earlier. 
The range $m \in \Z$ for the 
``fusion degree" or ``string length" 
for $T_{a,m}$ and $Y_{a,m}$ 
is a natural convention in those systems equipped 
with doubled symmetry, e.g.
the $O(4)$ nonlinear sigma model ($SU(2)$ principal chiral field)
having the global $SU(2)_L\times SU(2)_R$ symmetry \cite{GKV1}.
   
\subsection{Formula for planar AdS/CFT spectrum}\label{ss:abe}

Now the planar AdS/CFT spectrum 
(with R-charge subtracted) is given in terms of 
the solutions to the Y-system in the previous subsection by the formula
\begin{equation}\label{bel}
\sum_{j=1}^{K_0}\epsilon_1(u_{0,j}) + 
\sum_{a\ge 1}\int_{-\infty}^\infty
\frac{du}{2\pi i}
\frac{\partial \epsilon^\ast_a(u)}{\partial u}
\ln(1+Y^\ast_{a,0}(u)).
\end{equation}  
Here $K_0$ is specified from the sector in question 
(see (\ref{rrdef})--(\ref{bbdef}))
and $\epsilon_a(u)$ is defined by 
$\epsilon_a(u) = a + \frac{2ig}{x(u+\frac{ia}{2})}
- \frac{2ig}{x(u-\frac{ia}{2})}$
in terms of $x(u)$ 
satisfying $\frac{u}{g} =x(u) + x(u)^{-1}$
and $|x(u\pm\frac{ia}{2})|>1$.
The parameter 
$g$ is related to the 't Hooft coupling $\lambda$
by $\lambda = (4\pi g)^2$.
The above choice of the branch is called physical kinematics.
On the other hand, $\epsilon^\ast_a(u)$ with $a\ge 1$ is defined by 
the same formula but with another branch called mirror kinematics 
(cf. \cite{AF3,GKKV,BJ}). 
The function $Y^\ast_{a,0}(u)$ is defined 
by the mirror kinematics.
Finally, the rapidities $u_{0,j}$ are determined by the Bethe equation
\begin{equation}\label{bey}
Y_{1,0}(u_{0,j}) = -1\qquad (j=1,\ldots, K_0).
\end{equation}
This description of the planar AdS/CFT spectrum 
has been claimed exact 
for any 't Hooft coupling (i.e. to all loop orders) 
and operators of any {\em finite} $L$ \cite{GKKV,GKV2}.

\subsection{Asymptotic Bethe ansatz}\label{ss:baba}

To be consistent with the ABA equation \cite{BSt2},
the Y-system (\ref{ygkv}) should split into 
the left and right wings in the limit $L\rightarrow \infty$. 
Compatibly with this, 
the middle series should behave as 
\begin{equation}\label{yxtt}
Y_{a\ge 1, 0}(u)\simeq
\left(\frac{x(u-\frac{ia}{2})}
{x(u+\frac{ia}{2})}\right)^L
\frac{\phi(u-\frac{ia}{2})}{\phi(u+\frac{ia}{2})}
T^L_{a,-1}(u)T^R_{a,1}(u),
\end{equation}
where $\phi$ is a function obeying the relation (\ref{por}).
The last two factors represent the T-functions for the 
decoupled $SU(2|2)_{L}$ and $SU(2|2)_{R}$.
They are constructed from the $a=1$ case \cite{T2,KSZ}
in a way analogous to (\ref{ldef1}),  (\ref{ldef2}) and (\ref{ldef3}).
Explicitly, the $a=1$ case is given as the dressed vacuum form
\begin{equation}\label{TLR}
\begin{split}
T^{L, R}_{1, \mp 1}(u)
= &\frac{R^{(+)}_0(u-\frac{i}{2})}{R^{(-)}_0(u-\frac{i}{2})}
\Biggl(
\frac{Q_{\pm 2}(u-i)Q_{\pm 3}(u+\frac{i}{2})}
{Q_{\pm 2}(u)Q_{\pm 3}(u-\frac{i}{2})}
+\frac{Q_{\pm 2}(u+i)Q_{\pm 1}(u-\frac{i}{2})}
{Q_{\pm 2}(u)Q_{\pm 1}(u+\frac{i}{2})}\\
&-\frac{R^{(-)}_0(u-\frac{i}{2})Q_{\pm 3}(u+\frac{i}{2})}
{R^{(+)}_0(u-\frac{i}{2})Q_{\pm 3}(u-\frac{i}{2})}
-\frac{B^{(+)}_0(u+\frac{i}{2})Q_{\pm 1}(u-\frac{i}{2})}
{B^{(-)}_0(u+\frac{i}{2})Q_{\pm 1}(u+\frac{i}{2})}
\Biggr),
\end{split}
\end{equation}
where $Q_l(u) = \prod_{j=1}^{K_l}(u-u_{l,j})$.
In addition we introduce\footnote{
The Bethe roots 
$u_{1,j}, u_{2,j}, u_{3,j}, u_{0,j}, u_{-3,j}, u_{-2,j}, u_{-1,j}$
and the T-functions $T^L_{1,-1}, T^R_{1,1}$ here 
denote 
$u_{1L,j}, u_{2L,j}, u_{3L,j}, u_{4,j}, u_{3R,j}, u_{2R,j}, u_{1R,j}$ 
and $T^L_{1,1}, T^R_{1,1}$ in \cite{GKKV}, respectively.
The notation for the Q-functions is also slightly modified accordingly. 
These Bethe roots further correspond to 
$u_{1,j}, u_{2,j}, u_{3,j}, u_{4,j}, u_{5,j}, u_{6,j}, u_{7,j}$
in \cite{BSt2}.} 
\begin{align}
R_l(u) &= \prod_{j=1}^{K_l}\frac{x(u)- x(u_{l,j})}{\sqrt{x(u_{l,j})}},
\quad 
R^{(\pm)}_l(u) 
= \prod_{j=1}^{K_l}
\frac{x(u)- x(u_{l,j}\mp \frac{i}{2})}
{\sqrt{x(u_{l,j}\mp \frac{i}{2})}}, \label{rrdef}\\
B_l(u) &= \prod_{j=1}^{K_l}
\frac{x(u)^{-1}- x(u_{l,j})}{\sqrt{x(u_{l,j})}},
\quad 
B^{(\pm)}_l(u) 
= \prod_{j=1}^{K_l}
\frac{x(u)^{-1}- x(u_{l,j}\mp \frac{i}{2})}
{\sqrt{x(u_{l,j}\mp \frac{i}{2})}} \label{bbdef}
\end{align}
for $-3\le l \le 3$.
They are factorized pieces of $Q_l(u)$ in that 
\begin{equation}
R_l(u)B_l(u) = (-g)^{-K_l}Q_l(u),\quad
R^{(\pm)}_l(u)B^{(\pm)}_l(u) 
= (-g)^{-K_l}Q_l(u\pm \textstyle{\frac{i}{2}}).
\displaystyle
\end{equation}
The numbers $K_l$ specify the relevant sectors.
As usual in the analytic Bethe ansatz (cf. Section \ref{s:aba}), 
analyticity of $T^{L,R}_{1,\pm 1}(u)$ leads to the equations
\begin{align}
1 &=\frac{Q_{\pm 2}(u_{\pm 1, k}+\frac{i}{2})
B^{(-)}_0(u_{\pm 1, k})}
{Q_{\pm 2}(u_{\pm 1, k}-\frac{i}{2})
B^{(+)}_0(u_{\pm 1, k})},
\quad
1=\frac{Q_{\pm 2}(u_{\pm 3, k}+\frac{i}{2})
R^{(-)}_0(u_{\pm 3, k})}
{Q_{\pm 2}(u_{\pm 3, k}-\frac{i}{2})
R^{(+)}_0(u_{\pm 3, k})},\\
-1 &= \frac{Q_{\pm 1}(u_{\pm 2, k}-\frac{i}{2})
Q_{\pm 2}(u_{\pm 2, k}+i)
Q_{\pm 3}(u_{\pm 2, k}-\frac{i}{2})}
{Q_{\pm 1}(u_{\pm 2, k}+\frac{i}{2})
Q_{\pm 2}(u_{\pm 2, k}-i)
Q_{\pm 3}(u_{\pm 2, k}+\frac{i}{2})}.
\end{align}
In addition, the cyclicity of the single trace operator in SYM 
is to be reflected as the ``zero momentum" condition
$\prod_{j=1}^{K_0}
\frac{x(u_{0,j}+\frac{i}{2})}{x(u_{0,j}-\frac{i}{2})}=1$.
Upon a convention adjustment, these relations 
coincide with the ABA equation in \cite[section 5.1]{BSt2}
except the most complicated one
\begin{equation}\label{mce}
-1=\left(\frac{x(u_{0,k}-\frac{i}{2})}{x(u_{0,k}+\frac{i}{2})}
\right)^L\!
\frac{\bigl(B^{(+)}_0B_1B_{-1}R_3R_{-3}/R^{(+)}_0\bigr)
(u_{0,k}+\frac{i}{2})}
{\bigl(B^{(-)}_0B_1B_{-1}R_3R_{-3}/R^{(-)}_0\bigr)
(u_{0,k}-\frac{i}{2})}
S(u_{0,k})^2,
\end{equation}
which involves the dressing factor $\sigma$ \cite{BES}
via $S(u) = \prod_{j=1}^{K_0}\sigma(x(u),x_{0,j})$.
The ABA equation (\ref{mce}) 
is to be reproduced in the present scheme 
as the large $L$ limit of the equation (\ref{bey}). 
In view of $T^{L,R}_{1,\mp 1}(u_{0,j}) = 
-\frac{Q_{\pm 3}(u_{0,j}+\frac{i}{2})}
{Q_{\pm 3}(u_{0,j}-\frac{i}{2})}$ and (\ref{yxtt}),
this amounts to postulating that $\phi$ therein
should satisfy the difference equation
\begin{equation}\label{por}
\frac{\phi(u-\frac{i}{2})}{\phi(u+\frac{i}{2})}
=\frac{B_0^{(+)}B_1B_{-1}}
{R^{(+)}_0B_3B_{-3}}(u+\textstyle{\frac{i}{2}})
\displaystyle
\frac{R^{(-)}_0B_3B_{-3}}
{B^{(-)}_0B_1B_{-1}}(u-\textstyle{\frac{i}{2}})
S(u)^2.
\end{equation}
The asymptotics (\ref{yxtt}) with (\ref{por})
specifies the large $L$ solution of the Y-system.

With regard to the finite $L$ effects,
the above formulation reproduces  
wrapping corrections at weak coupling for twist two 
operators obtained by other methods such as 
the L\"uscher formula.
For instance in the case of the Konishi 
operator $\mathrm{Tr}(D^2Z^2- DZDZ)$,
one gets the scaling dimension from ABA as
$E_{\rm ABA} = 4+12g^2-48g^4+336g^6
-(2820+288\zeta(3))g^8$. 
The above Y-system approach 
yields the result $E_{\rm ABA} + E_{\rm wrapping}$ 
with the correction 
$E_{\rm wrapping} = (324 + 864\zeta(3)-1440\zeta(5))g^8$
starting at four-loop in agreement with \cite{BJ}.

\subsection{\mathversion{bold}Area of minimal surface in AdS}
\label{ss:msa}

Now we turn to the second topic of this section.
The T and Y-systems play an essential role 
in calculating the action of classical open string solutions, 
i.e. the area of minimal surface, in AdS space.
Via the AdS/CFT correspondence, 
this yields the planar amplitudes of gluon scattering 
in ${\mathcal N}=4$ SYM at strong coupling.
The gluon momenta are incorporated in   
null polygonal configurations at the AdS boundary.
The first important step in this problem is 
to linearize the equation of motion 
of the AdS sigma model (Section \ref{ss:msa}).
Once this is achieved,  
the T and Y-systems come into the game naturally
through the Stokes phenomena of the auxiliary linear problem 
around the irregular singularity 
at the boundary of the worldsheet (Section \ref{ss:sty}).
This part is close in spirit to Section \ref{ss:ODE2}.
Extra complication can occur
when passing to the TBA-type nonlinear integral equations
most typically due to the complex nature of the driving terms 
(``complex mass" appearing in asymptotics of Y-functions).
They are determined by period integrals of the Riemann surface 
reflecting the null polygonal boundary 
and the cross ratios of gluon momenta.
The regularized area is formally expressed in the same 
form as the free energy in the conventional TBA analysis
(Section \ref{ss:af}).
Sections \ref{ss:msa}--\ref{ss:af}
are quick digest of these recent progress  
\cite{AM1,AM2, AMSV,HISS} along a simple version of $AdS_3$.

The $AdS_3$ is given in terms of the global coordinate 
$\vec{Y}=(Y_{-1},Y_0,Y_1,Y_2)\in \R^{2,2}$ as
\begin{equation}\label{ads3}
\vec{Y}\cdot \vec{Y} := -Y^2_{-1}-Y^2_{0}+Y^2_{1}+Y^2_{2} = -1.
\end{equation}
General product $\vec{A}\cdot \vec{B}$ in $\R^{2,2}$ is defined 
similarly with the signature $-1,-1,1,1$.
The equation of motion and the Virasoro constraint read
\begin{equation}\label{emvc}
\partial\bar{\partial}\vec{Y} 
- (\partial \vec{Y}\cdot \bar{\partial}\vec{Y})\vec{Y}=0,\qquad
\partial\vec{Y}\cdot \partial\vec{Y} = 
\bar{\partial}\vec{Y}\cdot \bar{\partial}\vec{Y} = 0,
\end{equation} 
where $\partial = \frac{\partial}{\partial z}, 
\bar{\partial} = \frac{\partial}{\partial \bar{z}}$
and $z$ is a complex coordinate parameterizing the worldsheet.
This classical motion of strings in $AdS_3$ is integrable.
In fact, it is transformed to 
a $\Z_2$-projected $SU(2)$ Hitchin system
through a Pohlmeyer type reduction \cite{Po, dVS}.
To see this, introduce the new variables $\alpha$ and $p$ by
\begin{align}
&e^{2\alpha(z,\bar{z})}
= \frac{1}{2}\partial \vec{Y}\cdot \bar{\partial}\vec{Y},\quad
N_a  = \frac{1}{2}\epsilon_{abcd}Y^b\partial Y^c 
\bar{\partial}Y^d,\\
&p = \frac{1}{2}\vec{N}\cdot \partial^2\vec{Y},
\quad
\bar{p} 
= -\frac{1}{2}\vec{N}\cdot \bar{\partial}^2\vec{Y}.
\label{ppdef}
\end{align}
Note that 
$\vec{N}\cdot \vec{Y} = 
\vec{N}\cdot \partial \vec{Y} = 
\vec{N}\cdot \bar{\partial}\vec{Y} = 0$ and 
$\vec{N}\cdot \vec{N} = 1$. 
The variable 
$\alpha=\alpha(z,\bar{z})$ is real and 
$\vec{N}$ is pure imaginary.
Moreover it can be shown 
from (\ref{ads3})-(\ref{ppdef})
that $p=p(z)$ is holomorphic.
The area is given by $4\int d^2ze^{2\alpha}$.
The $\alpha$ satisfies the
sinh-Gordon equation modified with $p$ as 
$\partial \bar{\partial}\alpha - e^{2\alpha}
+|p(z)|^2e^{-2\alpha} = 0$.
As this fact indicates, the equations (\ref{emvc})
are expressible as the flatness condition of the connections:
\begin{equation}\label{flat}
\partial B^L_{\bar z} - \bar{\partial} B^L_z
+[B^L_z, B^L_{\bar z}] = 0,\quad
\partial B^R_{\bar z} - \bar{\partial} B^R_z
+[B^R_z, B^R_{\bar z}] = 0,
\end{equation}
where the connections are given by
\begin{align}
&B^L_z = B_z(1),\quad
B^L_{\bar{z}} = B_{\bar{z}}(1),\quad
 B^R_z = UB_z(i)U^{-1},\quad
B^R_{\bar{z}}= UB_{\bar{z}}(i)U^{-1}, \label{BU}\\
&B_z(\zeta) = \begin{pmatrix}
\frac{1}{2}\partial\alpha & -\zeta^{-1}e^\alpha\\
-\zeta^{-1}e^{-\alpha}p(z) & 
-\frac{1}{2}\partial\alpha
\end{pmatrix},\quad
B_{\bar{z}}(\zeta) = \begin{pmatrix}
-\frac{1}{2}\bar{\partial}\alpha & 
-\zeta e^{-\alpha}\bar{p}(\bar{z})\\
-\zeta e^{\alpha}& 
\frac{1}{2}\bar{\partial}\alpha
\end{pmatrix},\label{BBz}
\end{align}
with $U = \begin{pmatrix}
0 & e^{\pi i/4}\\ e^{3\pi i/4} & 0 \end{pmatrix}$.
Here $\zeta$ is the spectral parameter.
Actually the relation
$\partial B_{\bar z}(\zeta) - \bar{\partial} B_z(\zeta)
+[B_z(\zeta), B_{\bar z}(\zeta)] = 0$  including $\zeta$ is satisfied.
Splitting the connection into $\zeta$ dependent part and the rest
as $B_z(\zeta) = {\mathcal A}_z + \zeta^{-1}\Phi_z$ and 
$B_{\bar{z}}(\zeta) = {\mathcal A}_{\bar{z}} + \zeta\Phi_{\bar{z}}$,
one finds that the flatness conditions form the 
Hitchin system with gauge field ${\mathcal A}$ 
and Higgs field $\Phi$.
The gauge group is $SU(2)$ but the system is 
$\Z_2$-projected in the sense that 
the above form (\ref{BBz}) belongs   
to the invariant subspace under the involution
${\mathcal A}_z \rightarrow \sigma^3{\mathcal A}_z\sigma^3, 
\Phi_z \rightarrow - \sigma^3\Phi_z\sigma^3$
and similarly for ${\mathcal A}_{\bar{z}}$ and $\Phi_{\bar{z}}$.
($\sigma^3$ is a Pauli matrix.)

With each zero curvature condition in (\ref{flat}), 
there is associated a pair of auxiliary linear problems 
whose compatibility yields it.
Thanks to the relations (\ref{BU}), 
one can combine and promote them into the $\zeta$-dependent versions
$(\partial + B_z(\zeta))\psi = 0$ and 
$(\bar{\partial} + B_{\bar{z}}(\zeta))\psi = 0$
or equivalently,
\begin{equation}\label{mal}
\left(d+\frac{\Phi_zdz}{\zeta} + {\mathcal A} + \zeta\Phi_{\bar{z}}d\bar{z}
\right)\psi = 0
\end{equation}
with ${\mathcal A} = {\mathcal A}_zdz+{\mathcal A}_{\bar{z}}d\bar{z}$ 
for $\psi=\psi(z,\bar{z};\zeta)$.
A useful property is that if $\psi(\zeta)$ is a flat section with 
spectral parameter $\zeta$, then 
so is $\sigma^3\psi(e^{\pi i}\zeta)$ 
by the $\Z_2$-symmetry.

Given two solutions $\psi, \psi'$ to (\ref{mal}),
define their $SL(2)$-invariant pairing as
$\langle\psi,\psi'\rangle 
= \epsilon^{\alpha \beta}\psi_\alpha \psi'_\beta$,
where $\psi = (\psi_1,\psi_2)^T$, etc.
This is a constant function on the worldsheet playing the 
role analogous to Wronskians in Section \ref{s:ODE}.
Let $\psi^L_a = (\psi^L_{1,a}, \psi^L_{2,a})^T\,(a=1,2)$ 
be the two solutions $\psi(z,\bar{z},\zeta\!=\!1)$ normalized as 
$\langle \psi^L_a, \psi^L_b\rangle = \epsilon_{ab}$.
Fix also the solutions $\psi^R_{\dot{a}}
= (\psi^R_{1,\dot{a}}, \psi^R_{2,\dot{a}})^T\, (\dot{a}=1,2)$ 
which are similarly normalized at $\zeta=i$. 
Then the original $AdS_3$ coordinate $\vec{Y}=(Y_{-1},Y_0,Y_1,Y_2)$ is 
reproduced from the auxiliary linear problem by
\begin{equation}
\begin{pmatrix}
Y_{-1}+Y_2 & Y_1-Y_0\\
Y_1+Y_0 & Y_{-1}-Y_2
\end{pmatrix}_{a,\dot{a}}
= \psi^L_{1,a}\psi^R_{1,\dot{a}} + \psi^L_{2,a}\psi^R_{2,\dot{a}}.
\end{equation}
This substantially achieves the linearization of the problem.

\subsection{Stokes phenomena, T and Y-system}\label{ss:sty}

Scattering amplitudes for $2n$ gluons 
correspond to open string solutions having polygonal shapes
with $2n$ cusps at the $AdS_3$ boundary.
This translates to the following boundary condition:
\begin{equation}\label{alp}
\alpha \rightarrow \frac{1}{4}\ln |p(z)|^2\;\;(z\rightarrow \infty),
\quad
p(z) = z^{n-2}+\cdots \,(\text{polynomial of degree $n-2$}).
\end{equation} 
We assume that $n$ is odd for simplicity.
{}From (\ref{BBz}), 
solutions of the auxiliary linear problem (\ref{mal}) 
as $|z|\rightarrow \infty$ behave as
\begin{equation}
\psi \sim 
\begin{pmatrix}
(\bar{p}/p)^{\frac{1}{8}}\\
\pm(p/\bar{p})^{\frac{1}{8}}
\end{pmatrix}
\exp\left(\pm\frac{1}{\zeta}\int\sqrt{p}dz \pm 
\zeta \int\sqrt{\bar{p}}d\bar{z}\right).
\end{equation}
Since $\exp(\frac{1}{\zeta}\int\sqrt{p}dz)
\sim \exp(\frac{z^{n/2}}{\zeta})$ holds asymptotically,
there are $n$ Stokes sectors which 
are separated by $n$ rays in the $z$ plane.
We label them consecutively anticlockwise.

Let $s_k(\zeta)$ be the small 
(subdominant in the terminology 
of Section \ref{s:ODE}) solution 
in the $k$th Stokes sector.
Then we have the properties like 
$\sigma^3s_k(e^{\pi i}\zeta) \propto s_{k+1}(\zeta)$,
$s_k(e^{2\pi i}\zeta) \propto s_{k+2}(\zeta)$ and
$\langle s_j,s_k\rangle(e^{\pi i}\zeta)
= \langle s_{j+1},s_{k+1}\rangle(\zeta)$.
Fixing the small solution $s_1(\zeta)$ in the first Stokes
sector, we define the others by 
$s_{k+1}(\zeta) = (\sigma^3)^ks_1(e^{k\pi i}\zeta)$.

Set $T_k(\zeta) = 
\langle s_0, s_{k+1}\rangle(e^{-\pi i(k+1)/2}\zeta)$
in the normalization
$\langle s_i, s_{i+1}\rangle(\zeta) = 1$.
Then from the simplest Pl\"ucker relation or Schouten identity 
$\langle s_i, s_j\rangle\langle s_k, s_l\rangle - 
\langle s_i, s_k\rangle \langle s_j, s_l\rangle + 
\langle s_i, s_l\rangle \langle s_j, s_k\rangle =0$,
one finds
\begin{equation}\label{tkr}
T_k(e^{\frac{\pi i}{2}}\zeta)
T_k(e^{-\frac{\pi i}{2}}\zeta) = 
T_{k-1}(\zeta)T_{k+1}(\zeta) + 1.
\end{equation}
This is a version of the level $n-2$ restricted T-system for $A_1$
where the conditions 
$T_0(\zeta)=1$ and $T_{n-1}(\zeta)=0$ 
are imposed\footnote{The latter is a slightly weaker condition than 
$T_{n-2}(\zeta)=1$ in the definition of Section \ref{ss:res1}.}.
Setting further $Y_k(\zeta) = T_{k-1}(\zeta)T_{k+1}(\zeta)$
as usual, one gets 
the level $n-2$ restricted Y-system 
(for $Y^{-1}$-variables in (\ref{yade}))
\begin{equation}\label{yz}
Y_k(e^{\frac{\pi i}{2}}\zeta)
Y_k(e^{-\frac{\pi i}{2}}\zeta) = 
(1+Y_{k-1}(\zeta))(1+Y_{k+1}(\zeta))
\end{equation}
with the boundary condition
$Y_0(\zeta)=Y_{n-2}(\zeta)=0$ in the $k$ direction.

\subsection{Asymptotics, WKB and TBA}\label{ss:awt}
As is well known, 
the relation (\ref{yz}) determines the Y-functions effectively
only with the information on their analyticity.  
By the definition, $Y_k(\zeta)$'s are analytic away from 
$\zeta^{\pm 1}=0$ where they possess essential singularities.
One can deduce the asymptotic behavior around them
using the WKB approximation regarding 
$\zeta^{\pm 1}$ as the Planck constant.
For example when $\zeta\rightarrow 0$,
the solutions of (\ref{mal}), 
after a simple similarity transformation 
making $\Phi_z$ into $\sqrt{p}\, \mathrm{diag}(1,-1)$,
behave as $\exp(\pm\frac{1}{\zeta}\int\!\sqrt{p}\, dz)$
times constant vectors.
Thus they are well approximated by performing 
the integral along the Stokes (steepest descent) lines defined by
$\Im{\rm m}\, (\sqrt{p(z)}dz/\zeta)=0$. 
At a generic point in the $z$ plane, 
there is one Stokes line passing through it.
Exceptions are zeros of $p(z)$ (turning points).
From a single zero, there emanate three Stokes lines.
They go toward infinity along certain directions 
corresponding to Stokes sectors or flow into another turning point.
The family of these infinitely many non-crossing lines 
constitute the WKB foliations.
See Figure 3.

\begin{figure}[h]
\includegraphics[width=0.45\textwidth]{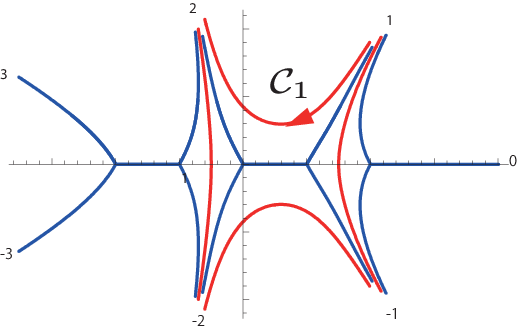}
\includegraphics[width=0.45\textwidth]{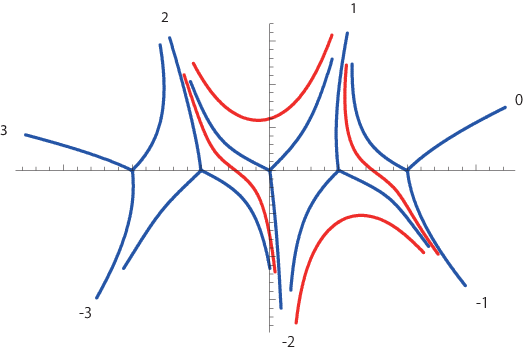}
\caption{Example of Stokes lines for $p(z)=z(z^2-1)(z^2-4)$.
The left and right figures correspond to 
$\mathrm{arg}(\zeta) = 0$ and $\frac{\pi}{3.1}$, respectively.
Blue lines are those emanating from turning points.
The number $k$ specifies the Stokes sector where $s_k$ is small.
For example, $\langle s_1, s_2\rangle \sim 
\exp(-\frac{1}{\zeta}\int_{{\mathcal C}_1}\sqrt{p}dz)$.
The integral $\int\sqrt{p}\,dz$ 
along the red lines anticlockwise yields asymptotics of 
$\ln Y_2(\zeta)$ as $\zeta \rightarrow 0$.
}

\end{figure}
First consider the case in which the zeros of $p(z)$ are 
aligned on the real axis.
Then one obtains the estimate like 
$\langle s_1,s_2\rangle \sim 
\exp(-\int_{C_1}\sqrt{p}\,dz/\zeta)$.
Therefore the Y-variables (without the normalization constraint on $s_i$)  
\begin{equation}\label{yss}
\begin{split}
Y_{2k}(\zeta) &=
\frac{\langle s_{-k},s_{k}\rangle\langle s_{-k-1},s_{k+1}\rangle}
{\langle s_{-k-1},s_{-k}\rangle\langle s_{k},s_{k+1}\rangle}(\zeta),\\
Y_{2k+1}(\zeta) &=
\frac{\langle s_{-k-1},s_{k}\rangle\langle s_{-k-2},s_{k+1}\rangle}
{\langle s_{-k-2},s_{-k-1}\rangle\langle s_{k},s_{k+1}\rangle}
(e^{\frac{\pi i}{2}}\zeta)
\end{split}
\end{equation}
have the asymptotics
\begin{equation}\label{lyz}
\ln Y_{2k}(\zeta) \sim \frac{Z_{2k}}{\zeta}+ \cdots,\quad
\ln Y_{2k+1}(\zeta) \sim 
\frac{Z_{2k+1}}{i\zeta}+ \cdots\;\;(\zeta \rightarrow 0),
\end{equation}
where $Z_k = -\oint_{\gamma_k}\sqrt{p}\,dz$ 
is the period integral along the 
cycle $\gamma_k$ going around the 
$k$th and $(k+1)$st largest zeros of $p(z)$ 
(cf. Fig. 5 in \cite{AMSV}).
The asymptotics 
as $\zeta \rightarrow \infty$ is similarly investigated. 
Together with the $\zeta \rightarrow 0$ case, 
the result is summarized as 
$\ln Y_k(e^\theta) = -m_k\cosh\theta + \cdots\,
(\theta\rightarrow \pm \infty)$,
where $m_{2k} = -2Z_{2k}$ and 
$m_{2k+1}= 2iZ_{2k+1}$ are both positive.
Now that the combination 
$\ln(Y_k(e^\theta)/e^{-m_k\cosh\theta})$
is analytic in the strip 
$|\Im{\rm m}\,\theta| \le \frac{\pi}{2}$ 
and decays as $|\theta| \rightarrow \infty$ within it,
the standard argument leads to the integral equation:
\begin{equation}\label{lymc}
\ln Y_k(e^{\theta}) = -m_k\cosh\theta +
\int_{-\infty}^\infty
\frac{\ln[(1+Y_{k-1}(e^{\theta'}))
(1+Y_{k+1}(e^{\theta'}))]d\theta'}
{2\pi\cosh(\theta-\theta')}
\end{equation}
for $1 \le k \le n-3\,(Y_0(\zeta)=Y_{n-2}(\zeta)=0)$. 
Up to the driving (mass) term, 
this has the same form with the integral equation  
in TBA or QTM analyses 
associated with the level $n-2$ 
restricted Y-system for $A_1$.
See for example (\ref{tba3}) and (\ref{QTMTBA}).

So far, we have considered the case where the zeros of 
$p(z)$ are on the real axis.
When they deviate from it, 
the T and Y-system remain unchanged.
On the other hand, the asymptotics is 
modified as 
$\ln Y_k(\zeta)\sim -\frac{m_k}{2\zeta}
\,(\zeta \rightarrow 0)$ and 
$\ln Y_k(\zeta)\sim -\frac{\bar{m}_k}{2}\zeta
\,(\zeta \rightarrow \infty)$, 
where $m_k=|m_k|e^{i\varphi_k}$ is complex in general.
Consequently, the integral equation (\ref{lymc}) 
is replaced with
\begin{equation}\label{lymc2}
\ln {\tilde Y}_k(e^{\theta}) = -|m_k|\cosh\theta +
\sum_{j=k\pm 1}\int_{-\infty}^\infty
\frac{\ln(1+{\tilde Y}_{j}(e^{\theta'}))d\theta'}
{2\pi\cosh(\theta-\theta'+i\varphi_k-i\varphi_j)},
\end{equation}
where ${\tilde Y}_k(e^{\theta}) = Y_k(e^{\theta+i\varphi_k})$.
This holds 
for $|\varphi_k-\varphi_{k\pm 1}|<\frac{\pi}{2}$.
If the phases go beyond this range (so-called wall crossing), 
the integral equation acquires extra terms 
corresponding to the contributions  
of the poles from the convolution kernel.
A simple illustration of such a situation has been given in 
\cite[appendix B]{AMSV}.

\subsection{Area and free energy}\label{ss:af}
The interesting part $A$ of the area is given by\footnote{
Our $\Phi_z$ here is ${\tilde \Phi}_z$ in \cite{AMSV}.}
\begin{equation}\label{area}
A = 2\int d^2z \mathrm{Tr}(\Phi_z\Phi_{\bar{z}}) 
= i\int \sqrt{p}\,dz \wedge \Phi^{11}_{\bar{z}}d\bar{z}
= -i\sum_{j,k=1}^{n-3}w_{jk}\oint_{\gamma_j} \sqrt{p}\,dz 
\oint_{\gamma_k}\Phi^{11}_{\bar{z}}d\bar{z},
\end{equation}
where the gauge $\Phi_z=\sqrt{p}\, \mathrm{diag}(1,-1)$
is taken and $\mathrm{Tr}\,\Phi_{\bar{z}}=0$ is used.
In the last equality we have dropped the contribution from infinity.
The matrix $(w_{j k})$ is 
the inverse of the intersection forms\footnote{The inverse exists 
under our assumption of $n$ being odd. 
The intersection form $\langle\;,\;\rangle$ here
should not be confused with the $SL(2)$-invariant pairing of spinors.}
$(\langle \gamma_j,\gamma_k\rangle)$ specified by 
$\langle \gamma_{2k},\gamma_{2k\pm 1}\rangle=1$.
Set $\hat{Y}_{2k}(\zeta) = Y_{2k}(\zeta)$ and 
$\hat{Y}_{2k+1}(\zeta) = Y_{2k+1}(e^{-\frac{\pi i}{2}}\zeta)$
somehow reconciling the shift in (\ref{yss}).
The factor $\oint_{\gamma_k}\Phi^{11}_{\bar{z}}d\bar{z}$
in (\ref{area}) also appears as the coefficient of $-\zeta$ 
in the small $\zeta$ expansion of 
$\ln \hat{Y}_k(\zeta)$ based on the
perturbative solution of (\ref{mal}). 
On the other hand, 
the small $\zeta=e^{\theta}$ expansion of (\ref{lymc2}) gives 
\begin{equation}
\ln \hat{Y}_k(\zeta) = \frac{Z_k}{\zeta}+
\zeta\Bigl[\bar{Z}_k + 
\sum_j\frac{\langle \gamma_k,\gamma_j\rangle}{\pi i}
\int\frac{d\zeta'}{\zeta'^2}\ln(1+\hat{Y}_j(\zeta'))\Bigr]
+ \cdots,
\end{equation}
where the appearance of 
$\langle \gamma_k,\gamma_j\rangle$ is the effect of 
using $\hat{Y}_k(\zeta)$ rather than $Y_k(\zeta)$.
Thus one can substitute $\oint_{\gamma_k}\Phi^{11}_{\bar{z}}d\bar{z}$
in (\ref{area}) by $[\ldots]$ here times $(-1)$.
As the result the area is expressed as 
$A= A_{\rm periods}+A'_{\rm free}$ with
\begin{align}
A_{\rm periods} = -i\sum_{j,k}w_{jk}Z_k\bar{Z}_j,\qquad
A'_{\rm free} = -\frac{1}{\pi}\sum_kZ_k
\int\frac{d\zeta}{\zeta^2}\ln(1+\hat{Y}_k(\zeta)).
\end{align}
Actually one should replace $A'_{\rm free}$ by 
the average $A_{\rm free}$ taking the contribution 
from large $\zeta$ into account.
Thus the final result reads $A= A_{\rm periods}+A_{\rm free}$ with
\begin{equation}
A_{\rm free} = \sum_k |m_k|\int_{-\infty}^\infty
\frac{d\theta}{2\pi}\cosh\theta \ln(1+{\tilde Y}_k(e^{\theta}))
\end{equation}
in terms of ${\tilde Y}_k(e^\theta)$ defined after (\ref{lymc2}).
This has the same form as the free energy in the conventional TBA.
See for example (\ref{fay}).

To summarize, the symmetry aspects of the problem 
(AdS, Virasoro constraints, null-cusp boundary) 
are incorporated into the restricted T and Y-systems.
Then, all the dynamical information 
(gluon momenta, Riemann surface, cycles) 
are remarkably integrated 
in the ``complex mass" parameters $m_1,\ldots, m_{n-3}$.

\subsection{Bibliographical notes}\label{ss:ads}

The subjects in this section are
currently in the course of rapid development.
For various aspects of the planar AdS/CFT spectrum,
see the literatures given in the end of Section \ref{ss:pacs}
and reference therein.
We have only dealt with the limited issues related to 
T and Y-systems.
The contents in Section \ref{ss:tyads}--\ref{ss:baba} are
mainly based on \cite{GKKV}.
For numerical studies, 
it is important to formulate the analyticity precisely and 
to derive the TBA (or other type of) integral equations 
including excited states.
We refer to \cite{BFT1, GKKV, AF3, GKV2, CFT} for this problem.
Similar analyses have been made in \cite{MZ2,BFT2,GM} 
for the AdS$_4$/CFT$_3$ duality proposed recently \cite{ABJM}.

Calculation of gluon scattering amplitudes at strong coupling 
using gauge/string duality was initiated in \cite{AM1} and 
developed in a series of works \cite{AM2, AGM, AMSV, HISS, Y, MaZ}.
For classical integrability of AdS sigma models and their connection 
to Hitchin system, see also \cite{BG}.
Auxiliary linear problem in Section \ref{ss:awt} 
is a special case of that for 
general $SU(2)$ Hitchin system \cite{GMN},
where a number of aspects in the Riemann-Hilbert problem have been 
discussed including WKB triangulations,  
the Fock-Goncharov coordinates,
the Kontsevich-Soibelman wall-crossing formula, TBA  and so forth. 
The contents of Section \ref{ss:msa}--\ref{ss:af} are mainly 
taken from \cite{AMSV}.
We have treated $n$ (number of gluons) odd case.
For the case $n$ even, see \cite{Y, MaZ}.
In \cite{MaZ}, further effect of operator insertion is studied, and 
the (slightly deformed) 
level 2 restricted Y-system for $D_n$ has been obtained.
For a similar appearance of the  $D$ type Y-system
in $A_1$ related lattice models, see Remark \ref{re:las}.
The generalized sinh-Gordon equation has also been studied 
in the context of generalized ODE/IM correspondence in \cite{LZ}. 

\section{Aspects as classical integrable system}\label{s:cis}

Besides the quantum integrable systems,
T and Y-systems also have interesting 
aspects as classical nonlinear difference equations. 
For instance, the T-system relation (\ref{ta}) is 
presented in the form 
\begin{equation}
\tau_1\tau_{23}-\tau_2\tau_{31} + \tau_3\tau_{12}=0
\end{equation}
with a suitable redefinition up to the boundary condition.  
Here the indices signify a shift of 
the independent vector variable in the respective directions 
($\tau_{ij}=\tau_{ji}$).
This is a version of Hirota-Miwa equation on tau functions 
in the theory of discrete KP equations \cite{Hi1,Hi2,Hi3,Mw}.
A simplest account for its integrability is the Lax representation,
namely, the compatibility of the linear system:
\begin{equation*}
\psi_i-\psi_j = \frac{\tau\tau_{ij}}{\tau_i\tau_j}\psi\quad (i<j).
\end{equation*} 
The Hirota-Miwa equation serves as a master equation generating 
a variety of soliton equations under
suitable specializations and boundary conditions.
See for instance \cite{Hi3,DJM1,DJM2}.
Apart from this, there are numerous aspects in type 
$A$ T-system, sometimes called octahedron recurrence, related to 
discrete geometry \cite{BSu, Do3, Sc},
Littlewood-Richardson rule \cite{KTW}, 
perfect matchings and partition functions on a network \cite{Sp,Di}
and so forth.
For types other than $A$ however, such results are relatively few.

Our presentation in this section is necessarily selective.
In Section \ref{ss:cl}, we explain that the T-system for ${\mathfrak g}$ 
is a discretized Toda field equation that has decent continuous limits
with a known Hamiltonian structure.
In Section \ref{ss:dg}, a connection of the 
Y-system for $A_\infty$ with discrete geometry is reviewed.

\subsection{Continuum limit}\label{ss:cl}
We present a simple continuous limit of the 
T-system for general ${\mathfrak g}$ known as 
the lattice Toda field equation \cite{IH}.
It is a difference-differential system containing 
continuous time and discrete space variables.
Further continuous limit on the latter yields 
the Toda field equation on 
$(1+1)$-dimensional continuous spacetime \cite{LS}.

We begin by making a slight change of variables in the T-system as
\begin{equation}\label{iran}
T^{(a)}_m(u) = \tau_a(u+\frac{m}{t_a}, s+\varepsilon\frac{m}{t_a})
\quad (1\le a \le r, u \in \Z/t, m \in \Z).
\end{equation}
Here $\varepsilon$ is a small parameter and 
$s$ is going to be the continuous time variable soon.
For the symbols $t, t_a$ and root system data, 
see around (\ref{eq:t1}).
We substitute (\ref{iran}) into the T-system (\ref{tga}) 
$T^{(a)}_m(u-\frac{1}{t_a})T^{(a)}_m(u+\frac{1}{t_a})
-T^{(a)}_{m-1}(u)T^{(a)}_{m+1}(u)=g^{(a)}_m(u)M^{(a)}_m(u)$ 
with $m \in t_a\Z$.
For each ${\mathfrak g}$ of rank $r$ there are $r$ such equations.
(The case $m \not\in t_a\Z$ leads to the same continuum limit 
as the one considered in the following.)  
For example, the $B_2$ case reads
\begin{align*}
&\textstyle
\tau_1(n-1,s)\tau_1(n+1,s)
-\tau_1(n-1,s-\varepsilon)\tau_1(n+1,s+\varepsilon) 
= g_1\tau_2(n,s),\\
&\textstyle
\tau_1(n-\frac{1}{2},s)\tau_1(n+\frac{1}{2},s)
-\tau_1(n-\frac{1}{2},s-\frac{\varepsilon}{2})
\tau_1(n+\frac{1}{2},s+\frac{\varepsilon}{2}) 
= g_2\tau_1(n-\frac{1}{2},s)\tau_1(n+\frac{1}{2},s),
\end{align*}
where we have chosen $g_a=g^{(a)}_{t_am}(u)$ to be a constant.
We take the continuum limit in the time variable $s$ 
keeping $n \in \Z/t$ 
as the coordinate of a one dimensional lattice without boundary.
Namely, we replace $g_a$ by $\varepsilon g_a/t_a$ and set 
$\varepsilon \rightarrow 0$.
The result reads
\begin{align*}
D_s\tau_1(n-1)\cdot \tau_1(n+1) &= g_1 \tau_2(n),\\
D_s\tau_2(n-\frac{1}{2})\cdot \tau_2(n-\frac{1}{2})
&= g_2 \tau_1(n-\frac{1}{2})\tau_1(n+\frac{1}{2}).
\end{align*}
Here we suppressed the time dependence as 
$\tau_a(n) = \tau_a(n,s)$, 
which we shall also do in the remainder of this subsection. 
$D_s$ denotes the Hirota derivative:
\begin{equation*}
D_s f \cdot g = \frac{\partial f}{\partial s}g - f \frac{\partial g}{\partial s}.
\end{equation*}
Similarly, the general ${\mathfrak g}$ case is given by
\begin{equation}\label{Dtt}
\begin{split}
&D_s\tau_a(n-\tai)\cdot \tau_a(n+\tai) 
= g_a {\mathcal M}_a(n),\\
&{\mathcal M}_a(n) := \prod_{b: C_{ab}=-1}\tau_b(n)
\prod_{b: C_{ab}=-2}
\tau_b(n-{\textstyle\frac{1}{2}})\cdot \tau_b(n+{\textstyle\frac{1}{2}})
\prod_{b: C_{ab}=-3}
\tau_b(n-{\textstyle\frac{2}{3}})\tau_b(n)
\tau_b(n+{\textstyle\frac{2}{3}}),
\end{split}
\end{equation}
where $n \in \Z/t$.
We call this the lattice Toda field equation for ${\mathfrak g}$.
In some case,  it actually splits into disjoint sectors.
For instance in types $ADE$, one has $t_a = t = 1$ for any $a\in I$, hence 
(\ref{Dtt}) closes among 
$\{\tau_a(n) | \,a \in I_{(-1)^n}\}$ or 
$\{\tau_a(n) | \,a \in I_{(-1)^{n+1}}\}$,
where $I_{\pm}$ is the bipartite decomposition of the Dynkin diagram nodes
$I= \{1,\ldots, r\} = I_+ \sqcup I_-$. 

One can rewrite (\ref{Dtt}) in a form that looks more like
Toda equation and explore its Hamiltonian structure.
As an illustration,  we first treat the $A_1$ case. 
Let us introduce the dynamical variables $x(n)$ and $\beta(n)$ by
\begin{equation}\label{xnb}
x(n) = \frac{\partial}{\partial s}
\ln \frac{\tau_1(n-1)}{\tau_1(n+1)},\quad
\beta(n) = \frac{x(n-1)}{x(n+1)}\quad (n \in \Z).
\end{equation}
Equation (\ref{Dtt}) for $A_1$ reads
\begin{equation}\label{dda}
\frac{\partial\tau_1(n-1)}{\partial s}\tau_1(n+1) - 
\tau_1(n-1)\frac{\partial \tau_1(n+1)}{\partial s} = g_1.
\end{equation}
This allows us to rewrite (\ref{xnb}) as
\begin{equation}\label{xt}
x(n) = \frac{g_1}{\tau_1(n-1)\tau_1(n+1)},\quad
\beta(n) = \frac{\tau_1(n+2)}{\tau_1(n-2)}.
\end{equation}
{}From the expression of $x(n)$ in  (\ref{xnb}) 
and $\beta(n)$ in (\ref{xt}), one gets
another form of the lattice Toda field equation for $A_1$: 
\begin{align}\label{lob}
\frac{\partial \ln\beta(n)}{\partial s}=-x(n-1)-x(n+1),
\end{align} 
which is a discrete analog of the Liouville equation.
It is derived as the equation of motion
\begin{equation}
\frac{\partial \beta(n)}{\partial s} = \{{\mathcal H}, \beta(n)\},
\end{equation}
with the following Hamiltonian and Poisson bracket:
\begin{align}
{\mathcal H } = \sum_{m \in \Z} x(m),
\qquad
\{x(m), x(n)\} = x(m)x(n)\,{\rm sgn}_2(n-m).
\end{align}
See (\ref{sgd}) for the definition of ${\rm sgn}_2(n)$.
We remark that (\ref{dda}), (\ref{lob}) and their relation 
explained in the above are 
difference-differential analog of 
the T-system, Y-system
and their transformation stated in 
Theorem \ref{th:ty1} for $A_1$, respectively.

All these features are generalized to ${\mathfrak g}$ straightforwardly.
The relevant dynamical variables are 
\begin{equation}\label{xxb}
x_a(n), \quad \beta_a(n) = \frac{x_a(n-\tai)}{x_a(n+\tai)}\qquad
(a \in I,\; n \in \Z/t),
\end{equation}
which are functions of the continuous time $s$.
We keep the notation $I, t, t_a, C, (\alpha_a|\alpha_b)$ around (\ref{eq:t1})
and set 
\begin{equation}\label{Bd}
B_{ab} = B_{ba} = \frac{t_b}{\max(t_a, t_b)}C_{ab} 
= \begin{cases}
2 & C_{ab}=2,\\
-1 & C_{ab}<0,\\
0 & C_{ab}=0.
\end{cases}
\end{equation}
$(B_{ab})$ is the Cartan matrix for simply laced Dynkin 
diagram obtained by forgetting 
the multiplicity of oriented edges in that for ${\mathfrak g}$.   
We specify the Poisson bracket of $x_a(n)$ as
\begin{equation}\label{xx}
\{x_a(m), x_b(n)\} = 
\frac{1}{2}B_{ab}\,x_a(m)x_b(n) 
\,{\rm sgn}_{B_{ab}}\Bigl(\max(t_a, t_b)(n-m)\Bigr),
\end{equation}
where ${\rm sgn}_k(v)$ with $k \in \{2,-1\}$ 
is the odd function of $v\in \R$ defined by\footnote{${\rm sgn}_0(v)$ 
is not necessary since the RHS of (\ref{xx}) contains the factor $B_{a b}$.}
\begin{equation}\label{sgd}
{\rm sgn}_k(v) = \begin{cases}
1 & \text{if $v>0$ and $v \in 2\Z+k$},\\
-1 &  \text{if $v<0$ and $v \in 2\Z+k$},\\
0 & \text{otherwise}.
\end{cases}
\end{equation}
Consequently, the Poisson bracket concerning $\beta_a(n)$ 
becomes local in that it is non vanishing only with 
finitely many opponents.
\begin{align}
\{x_a(m),\beta_b(n)\} &= \begin{cases}
-x_a(m)\beta_a(n)(\delta_{m,n+\tai}+\delta_{m,n-\tai}) & C_{ab}=2,\\
x_a(m)\beta_b(n)\sum_{j=C_{ab}+1}^{-C_{ab}-1}
\delta_{m+\frac{j}{t_a},n} & C_{ab}<0,\\
0 & C_{ab}=0,
\end{cases}\label{xb}\\
\{\beta_a(m), \beta_b(n)\} &=
\beta_a(m)\beta_b(n)(\delta_{m+(\alpha_a|\alpha_b), n}
-\delta_{m-(\alpha_a|\alpha_b), n}).\label{bb}
\end{align}
In (\ref{xb}), the $j$-sum is taken with the condition 
$j \equiv C_{ab}+1\mod 2$. 
The equation of motion with the Hamiltonian
\begin{equation}\label{bhb}
\frac{\partial \beta_a(n)}{\partial s} = \{{\mathcal H}, \beta_a(n)\},
\quad
{\mathcal H} = \sum_{a \in I, n \in \Z/t}x_a(n)
\end{equation}
leads to the differential-difference system:
\begin{equation}\label{bxx}
\begin{split}
\frac{\partial \ln\beta_a(n)}{\partial s} 
&= -x_a(n-\frac{1}{t_a})-x_a(n+\frac{1}{t_a})\\
&+\sum_{b: C_{ba}=-1}x_b(n)
+ \sum_{b: C_{ba}=-2}\Bigl(x_b(n-\frac{1}{2})+x_b(n+\frac{1}{2})\Bigr)\\
&+ \sum_{b: C_{ba}=-3}
\Bigl(x_b(n-\frac{2}{3})+x_b(n)+x_b(n+\frac{2}{3})\Bigr).
\end{split}
\end{equation}
For ${\mathfrak g}=A_1$ this reduces to (\ref{lob}).
The equation (\ref{bxx}) with $x_a(n)$ and $\beta_a(n)$ 
related as (\ref{xxb}) is another form of the lattice Toda field equation (\ref{Dtt}).
In fact, the transformation between (\ref{Dtt}) and (\ref{bxx}) 
is parallel with the $A_1$ case (\ref{xnb})--(\ref{lob}).
Generalizing (\ref{xnb}) we relate $x_a(n)$ and $\tau_a(n)$ by
\begin{equation}\label{xmtt}
x_a(n)= \frac{\partial}{\partial s}
\ln\frac{\tau_a(n-\frac{1}{t_a})}{\tau_a(n+\frac{1}{t_a})}\\
=\frac{g_a{\mathcal M}_a(n)}
{\tau_a(n-\frac{1}{t_a})\tau_a(n+\frac{1}{t_a})},
\end{equation}
where the latter equality is due to 
the lattice Toda field equation (\ref{Dtt}).
Substituting the latter form into (\ref{xxb}), we find
\begin{equation}\label{bot}
\beta_a(n) = \prod_{b \in I}
\frac{\tau_b(n+(\alpha_a|\alpha_b))}{\tau_b(n-(\alpha_a|\alpha_b))}.
\end{equation}
This can also been derived  from (\ref{sf}) by
noting the same structure in $A^{-1}_{a,z=q^{tn}}$(\ref{Ade}) 
and ${\mathcal M}_a(n)/
(\tau_a(n-\frac{1}{t_a})\tau_a(n+\frac{1}{t_a}))$ given by (\ref{Dtt}).
Anyway, $\frac{\partial\ln\beta_a(n)}{\partial s}$
is expressed as a linear combination of $x_a(n)$ by using the first formula in 
(\ref{xmtt}).
The result reproduces (\ref{bxx}). 

A further continuous limit on $n$ can be taken by letting  
\begin{equation}\label{cl}
x_a(n) \rightarrow 2\varepsilon \exp(\phi_a(z+\varepsilon n)),
\quad
\ln \beta_a(n) \rightarrow -\frac{2\varepsilon}{t_a}\phi'_a, 
\end{equation}
where $'=\frac{\partial}{\partial z}$.
Then the limit $\varepsilon\rightarrow 0$ of (\ref{bxx}) leads to
a version of the Toda field equation for $\phi_a=\phi_a(z,s)$:
\begin{equation}
\frac{\partial^2\phi_a}{\partial z\partial s} 
= \sum_{b \in I}t_at_b(\alpha_a|\alpha_b)e^{\phi_b}.
\end{equation}
The case ${\mathfrak g}=A_1$ is the Liouville equation.
Switching to $\psi_a$ by 
$\phi_a=\sum_{b\in I}C_{ab}\psi_b-\ln t_a$,  
one may rewrite it in the form
\begin{equation*}
\frac{\partial^2\psi_a}{\partial z\partial s}=
\exp\Bigl(\,\sum_{b\in I}C_{ab}\psi_b\Bigr)
\end{equation*}
studied in \cite{LS}.
An explicit construction of the general solution is known  
containing $2r$ arbitrary functions \cite{LS}.
We see that (\ref{bhb}) and (\ref{xb}) are lattice analog of the
Hamiltonian formulation of the Toda field equation:
\begin{equation*}
\frac{\partial \phi'_a}{\partial s} = \{{\mathcal H}, \phi'_a\},
\quad 
{\mathcal H}= \sum_{a\in I}\int dz e^{\phi_a(z)},
\quad
\{\phi_a(z), \phi'_b(z')\} = t_at_b(\alpha_a|\alpha_b)\delta(z-z').
\end{equation*}

The Poisson structures (\ref{xx})--(\ref{bb}) have an origin 
in the lattice analog of the $W$-algebras going back to \cite{TF}.
In particular, they may be deduced from the Poisson relations among 
appropriate constituent fields in the $q$-deformed $W$-algebra.
See for example \cite{BC, BCC, FRS, IH, I} and reference therein.
Here we only mention, as an example,  that (\ref{bb}) is a lattice analog of 
the Poisson relation
\begin{equation*}
\{A_a(z), A_b(w)\} = \left(\delta\Bigl(q^{(\alpha_a|\alpha_b)}\frac{w}{z}\Bigr)
-\delta\Bigl(q^{-(\alpha_a|\alpha_b)}\frac{z}{w}\Bigr)\right) 
A_a(z)A_b(w)
\end{equation*}
among the fields $A_a(z)$ corresponding 
to the exponential simple root $e^{\alpha_a}$ whose counterpart 
in the theory of $q$-character has appeared in (\ref{Ade}).
See equation (3.1) in \cite{I} and also 
equation (8.8) in \cite{FR2} for the logarithmic form.

\subsection{Discrete geometry}\label{ss:dg}
As we have seen in the previous subsection,
continuous limits of T-system lead to Toda type differential equations. 
On the other hand, geometric origins of 
many differential equations of such kind have been known 
from the days of Darboux.
Like the continuous case, 
it is natural to seek discrete geometry responsible for the 
integrability of discrete integrable equations.
In fact, if we let such geometric objects speak of themselves, 
they would say ``We exist, therefore it is 
integrable\footnote{V. V. Bazhanov, talk at Newton Institute, 
Cambridge, UK, March 2009.}".  
There are many results in this direction.
See for example \cite{BoP,BSu, Do3, Sc, Ve} and reference therein.
In a sense they provide a most natural framework to set up 
Lax formalisms of the integrable difference equations from
geometric points of view.
Here we only include a simple exposition 
of the basic example \cite{Do1,Do2} connecting 
Y-system for $A_\infty$ to a discrete analog of 
the Laplace sequence of conjugate nets.

We begin by recalling the appearance of the Toda field equation in 
projective differential geometry.
Consider a surface in the real projective space ${\mathbb P}^3$ 
which has the homogeneous coordinate vector 
${\bf z}={\bf z}(x,y)\in {\mathbb P}^3$.
A local coordinate $(x,y)$ of the surface 
is called a {\em conjugate net}  if 
\begin{equation}\label{zabc}
{\bf z}_{xy}+a(x,y){\bf z}_x + b(x,y){\bf z}_y + c(x,y){\bf z}=0
\end{equation}
is valid for some functions $a, b, c$, 
where the indices mean the derivatives.
Although ${\bf z}$ and ${\bf w}$ 
specify the same surface if they are related by 
${\bf z}=\lambda {\bf w}$, 
the above equation is not invariant but changed into
\begin{equation}
{\bf w}_{xy}+{\tilde a}(x,y){\bf w}_x + {\tilde b}(x,y){\bf w}_y 
+ {\tilde c}(x,y){\bf w}=0
\end{equation}
with ${\tilde a} = a+(\ln \lambda)_y$,
${\tilde b} = b+(\ln \lambda)_x$,
${\tilde c} = c+a(\ln \lambda)_x+b(\ln \lambda)_y+\lambda_{xy}/\lambda$.
A characteristic of a surface independent of the gauge $\lambda$ is
the Laplace invariant
\begin{equation}\label{hk}
h = a_x + ab-c, \quad k = b_y+ab-c,
\end{equation} 
satisfying ${\tilde h} = h$ and ${\tilde k}=k$.
In what follows we consider the generic situation that they are nonzero.

For the homogeneous coordinate vector ${\bf z}$ satisfying (\ref{zabc}), 
the Laplace transformation ${\mathcal L}_\pm$ is defined by 
\begin{equation}
{\mathcal L}_+({\bf z}) = {\bf z}_y + a{\bf z},\quad 
{\mathcal L}_-({\bf z}) = {\bf z}_x + b{\bf z}.
\end{equation}
This is compatible with the defining property (\ref{zabc}) of the 
conjugate net in that 
${\mathcal L}_+(\lambda{\bf w}) 
= \lambda({\bf w}_y+{\tilde a}{\bf w})$ and 
${\mathcal L}_-(\lambda{\bf w}) 
= \lambda({\bf w}_x+{\tilde b}{\bf w})$ hold with
${\tilde a}$ and ${\tilde b}$ given in the above equation.
Any component $z$ of ${\bf z}$
transforms as ${\mathcal L}_-\circ {\mathcal L}_+(z) = h z$
and 
${\mathcal L}_+\circ {\mathcal L}_-(z) = k z$,
meaning that ${\mathcal L}_+$ and ${\mathcal L}_-$ are 
inverse to each other as transformations in ${\mathbb P}^3$.
The family of surfaces in ${\mathbb P}^3$ 
generated from ${\bf z}^{(0)} = {\bf z}$ as  
${\bf z}^{(\pm n)}=({\mathcal L}_\pm)^n({\bf z})$ $(n\ge 1)$ is 
called a Laplace sequence.
Denote by $h_n, k_n$ the Laplace invariant associated with ${\bf z}^{(n)}$.
It is easy to see that ${\bf z}^{(\pm1)}$ satisfies (\ref{zabc})
with $a,b,c$ replaced by
$a^{(\pm 1)}, b^{(\pm 1)}, c^{(\pm 1)}$ given by
\begin{equation}\label{abc}
\begin{split}
&a^{(1)} = a-\frac{h_y}{h},\quad
b^{(1)} = b,\quad
c^{(1)} = ab - h + h \Bigl(\frac{b}{h}\Bigr)_y,\\
&a^{(-1)} = a,\quad
b^{(-1)} =  b-\frac{k_x}{k},\quad
c^{(-1)} = ab-k +  k\Bigl(\frac{a}{k}\Bigr)_x.
\end{split}
\end{equation}
Substituting this into (\ref{hk}), one can express $h_{\pm1}$ and $k_{\pm 1}$ 
in terms of $h_0=h$ and $k_0=k$. 
The result shows that the sequence of Laplace invariants 
satisfy a Toda field equation for $A_\infty$:
\begin{equation}
\frac{\partial^2\ln h_n}{\partial x\partial y}
=-h_{n-1} + 2h_n - h_{n+1},\quad h_n = k_{n+1}.
\end{equation} 

Now we move onto the discrete analog of these constructions.
The first step is to observe that (\ref{zabc}) implies the four infinitesimally 
neighboring points are coplanar.
This motivates us to introduce   
a map ${\bf x}: \Z^2 \rightarrow {\mathbb P}^3$
such that the 4 points 
${\bf x}(n,m),  {\bf x}(n+1,m), {\bf x}(n,m+1), {\bf x}(n+1,m+1)$ 
are coplanar for any $(n,m)\in \Z^2$.
Such a map is called  two dimensional {\em quadrilateral lattice},
which serves as a discrete analog of the conjugate net.
In the inhomogeneous coordinate of the projective space,
a two dimensional quadrilateral lattice is represented by a map
$x : \Z^2 \rightarrow \R^3$ satisfying 
the discrete analog of (\ref{zabc}) as follows:
\begin{equation}\label{ddz}
\Delta_1\Delta_2x = (T_1A) \Delta_1x + (T_2B)\Delta_2x.
\end{equation}
Here $\Delta_i = T_i-1$ and $T_i$ changes $n_i$ in any 
function $f(n_1,n_2)$ to $n_i+1$.
The functions $A, B$ on $\Z^2$ are ``gauge potentials" analogous to
$a, b$ in the continuum case.
The Laplace transformation, denoted by the same symbol as before, reads
\begin{equation}\label{llz}
{\mathcal L}_+(x)= x - \frac{\Delta_1 x}{B},\quad
{\mathcal L}_-(x)= x - \frac{\Delta_2 x}{A}.
\end{equation}
To see the geometric meaning of this, note that the four points  
$x, T_1x, T_2x, T_1T_2x$ form a quadrilateral on a plane due to (\ref{ddz}). 
The points $T_1{\mathcal L}_+(x)$ and $T_2{\mathcal L}_-(x)$ 
are intersections of the two lines extending the opposite sides of the 
quadrilateral. 

\unitlength 0.1in
\begin{picture}(30.0000,  18.0000)(0,-22)
\put(16.4400,-21.1500){\makebox(0,0){$x$}}%
\put(23.8800,-19.7900){\makebox(0,0){$T_1x$}}%
\put(17.0800,-13.6300){\makebox(0,0){$T_2x$}}%
\put(21.8800,-13.8700){\makebox(0,0)[lb]{$T_1T_2x$}}%
\put(19.7200,-6.4300){\makebox(0,0){$T_1{\mathcal L}_-(x)$}}%
\put(36.3600,-14.9100){\makebox(0,0)[lt]{$T_2{\mathcal L}_+(x)$}}%
%
\special{pn 8}%
\special{pa 1676 2036}%
\special{pa 1836 1396}%
\special{fp}%
%
\special{pn 8}%
\special{pa 1676 2036}%
\special{pa 2316 1876}%
\special{fp}%
%
\special{pn 8}%
\special{pa 1836 1396}%
\special{pa 1996 756}%
\special{dt 0.045}%
%
\special{pn 8}%
\special{pa 2316 1876}%
\special{pa 3596 1556}%
\special{dt 0.045}%
%
\special{pn 8}%
\special{pa 2156 1396}%
\special{pa 1996 756}%
\special{dt 0.045}%
%
\special{pn 8}%
\special{pa 3596 1556}%
\special{pa 2164 1420}%
\special{dt 0.045}%
%
\special{pn 8}%
\special{pa 1836 1396}%
\special{pa 2156 1420}%
\special{fp}%
%
\special{pn 8}%
\special{pa 2316 1876}%
\special{pa 2164 1420}%
\special{fp}%
\end{picture}%

As in (\ref{abc}), the postulate 
$\Delta_1\Delta_2{\mathcal L}_\pm(z) 
= T_1{\mathcal L}_\pm(A) \Delta_1{\mathcal L}_\pm(z) 
+ T_2{\mathcal L}_\pm(B)\Delta_2{\mathcal L}_\pm(z)$
fixes the Laplace transformation of the gauge potentials as
\begin{equation}\label{LLAB}
\begin{split}
{\mathcal L}_+(A) &= \frac{B}{T_2B}(1+T_1A)-1,\quad
{\mathcal L}_+(B) =T^{-1}_2\left(
\frac{T_1{\mathcal L}_+(A)}{{\mathcal L}_+(A)}(1+B)\right)-1,\\
{\mathcal L}_-(A) &=T^{-1}_1\left(
\frac{T_2{\mathcal L}_-(B)}{{\mathcal L}_-(B)}(1+A)\right)-1,
\quad
{\mathcal L}_-(B) = \frac{A}{T_1A}(1+T_2B)-1.
\end{split}
\end{equation}
It follows that the Laplace transformation is invertible, i.e.
${\mathcal L}_+\circ {\mathcal L}_- 
= {\mathcal L}_-\circ {\mathcal L}_+ = {\rm id}$. 
Introduce the Laplace sequence as the continuous case by 
$x^{(0)} = x$ and  
$x^{(\pm n)}=({\mathcal L}_\pm)^n(x)$ $(n\ge 1)$.

Now we are going to assign a cross ratio to each member of the 
Laplace sequence.
For the four colinear points $q_1, q_2, q_3, q_4$ in $\R^3$, 
we define the cross ratio as
\begin{equation*}
{\rm cr}(q_1,q_2,q_3,q_4) ={\rm cr}(q_2,q_1,q_4,q_3) 
= \frac{(q_3-q_1)(q_4-q_2)}{(q_3-q_2)(q_4-q_1)},
\end{equation*}
which is invariant under projective transformations.
Define the sequence of cross ratio by 
\begin{equation}
Y^{(n)} = -{\rm cr}
(x^{(n)},{\mathcal L}_+(x^{(n)}), 
T_1x^{(n)}, T_2{\mathcal L}_+(x^{(n)}))
\quad (n \in \Z),
\end{equation}
or equivalently, by setting 
$Y^{(0)}=Y$ and 
$Y^{(\pm n)} = ({\mathcal L}_\pm)^n(Y) (n \ge 1)$ with
$Y^{(0)}=
Y = -{\rm cr}(x,{\mathcal L}_+(x), T_1x, T_2{\mathcal L}_+(x))$.
The four points in ${\rm cr}$ are colinear.
By using (\ref{ddz})--(\ref{LLAB}) 
one can derive various formulas, e.g.
\begin{align*}
Y&= \frac{T_2B-(1+T_1A)B}{(1+B)(1+T_1A)}
= - \frac{{\mathcal L}_+(A)}{1+{\mathcal L}_+(A)}
\frac{B}{1+B},\\
Y^{(-1)} &=-{\rm cr}
(x,{\mathcal L}_-(x), T_2x, T_1{\mathcal L}_-(x)).
\end{align*}
The sequence $Y^{(n)}$ satisfies the functional relation \cite{Do1,Do2}
\begin{equation}
(T_1T_2Y^{(n)})Y^{(n)}=
T_1\left(\frac{1+Y^{(n-1)}}{1+(Y^{(n)})^{-1}}\right)
T_2\left(\frac{1+Y^{(n+1)}}{1+(Y^{(n)})^{-1}}\right).
\end{equation}
With a suitable identification, 
this coincides with the Y-system for $A_\infty$ 
(\ref{yade})
\begin{equation*}
Y^{(a)}_m(u-1)Y^{(a)}_m(u+1)
=\frac{
(1+Y^{(a-1)}_{m}(u))(1+Y^{(a+1)}_{m}(u))}
{(1+Y^{(a)}_{m-1}(u)^{-1})(1+Y^{(a)}_{m+1}(u)^{-1})}
\end{equation*}
with no boundary conditions on $a$ and $m$.

\subsection{Bibliographical notes}
The contents of Section \ref{ss:cl} and Section \ref{ss:dg} are 
mainly taken from \cite{IH, I} and \cite{Do1,Do2}, respectively.

\section{Q-system and Fermionic formula}\label{s:q}

\subsection{Introduction}\label{ss:qi}
Consider the T-system for ${\mathfrak g}$.
If one formally forgets the spectral parameter $u$ in 
$T^{(a)}_m(u)$, the resulting variable is conventionally denoted by $Q^{(a)}_m$
and the T-system reduces to the relation among them called {\em Q-system}.
In the context of $q$-characters,  $T^{(a)}_m(u)$ is the $q$-character 
$\chi_q(W^{(a)}_m(u))$ of the Kirillov-Reshetikhin module 
$W^{(a)}_m(u)$ (Theorem \ref{th:nh}).
Therefore, 
\begin{equation}\label{qrt}
Q^{(a)}_m = {\rm res}\, T^{(a)}_m(u) 
\end{equation}
is the usual character of ${\mathfrak g}$ obtained by the restriction defined in (\ref{res}).
Consider an arbitrary product of $Q^{(a)}_m$'s and the 
two kinds of decompositions 
(we assume $\nu^{(a)}_m\in \Z_{\ge 0}$ for the time being)
\begin{equation}\label{qdec}
\prod_{a,m}(Q^{(a)}_m)^{\nu^{(a)}_m} 
= \sum_\lambda b_\lambda \,\chi(V_\lambda)
=\sum_\lambda c_\lambda \,e^{\lambda}.
\end{equation}
Here $\chi(V_\lambda)$ denotes the (usual) character of the 
irreducible ${\mathfrak g}$-module $V_\lambda$ with highest weight $\lambda$.
The multiplicities $b_\lambda$ of the irreducible 
representation $V_\lambda$ (branching coefficients)
and the multiplicities $c_\lambda$ of weights $\lambda$ 
(dimensions of weight spaces)
are two basic quantities characterizing the decompositions.
It turns out that analyses of the Q-system provide them with 
{\em Fermionic formulas} 
$b_\lambda=\EuScript{M}_\lambda$ and 
$c_\lambda = \EuScript{N}_\lambda$.
They possess fascinating forms 
that symbolize the formal completeness 
of the {\em string hypothesis} 
in the Bethe ansatz at $q=1$ and $q=0$, respectively.

In Sections \ref{ss:mf} and \ref{ss:nf} we explain 
how $\EuScript{M}_\lambda$ and $\EuScript{N}_\lambda$ 
emerge from the Bethe ansatz 
along the simplest setting in ${\mathfrak g}=A_1$.
Precise statements for $A_1$ are formulated in Section \ref{ss:at} 
and the proof by a 
unified perspective of the multivariable Lagrange inversion method
is outlined in Section \ref{ss:mli}. 
All the essential ingredients are given by this point.
In Section \ref{ss:qs}, we introduce the Q-system for ${\mathfrak g}$
and write down the associated Fermionic formulas  
$\EuScript{M}_\lambda$ and $\EuScript{N}_\lambda$.
The main Theorem \ref{th:mnxbc} in the general case  is stated.
In Section \ref{ss:eqc}, 
the expansion of $Q^{(a)}_m$ into classical characters
is given for non exceptional algebras $A_r, B_r, C_r$ and $D_r$.
There are a lot of further aspects which are beyond the scope of this review.
They will be mentioned briefly in Section \ref{ss:xm}.
For simplicity we restrict ourselves to untwisted affine Lie algebras in this section.
Analogous results are also available in the twisted cases. 

\subsection{\mathversion{bold}Simplest example of $\EuScript{M}_\lambda$}
\label{ss:mf}

Recall the Bethe equation (\ref{be1}) for the 6 vertex model.
In the rational limit $q\rightarrow 1$,  it takes the form
\begin{equation}\label{rbe}
-\left(\frac{u_j + \sqrt{-1}}{u_j-\sqrt{-1}}\right)^L = 
\prod_{k=1}^n\frac{u_j-u_k + 2\sqrt{-1}}{u_j-u_k - 2\sqrt{-1}},
\end{equation}
where we have set all the inhomogeneity $w_j=0$ and replaced $u_j$
by $\sqrt{-1}u_j$.
The string hypothesis \cite{Be} is that 
the roots $u_1,\ldots, u_n$ are arranged as 
(called originally ``WellenKomlex" in \cite{Be}) 
\begin{equation}\label{eq:string}
\bigcup_{m \ge 1}\bigcup_{1\le \alpha\le N_m}\bigcup_{u_{m\alpha} \in \R}
\{ u_{m \alpha} + \sqrt{-1}(m+1-2i) + \epsilon_{m\alpha i}
\mid 1 \le i \le m \}
\end{equation}
for each partition $n = \sum_{m \ge1}mN_m\; (N_m \in \Z_{\ge 0})$.
Here $\epsilon_{m\alpha i}$ stands for a small deviation.
The $m$-tuple configuration (with negligible $\epsilon_{m\alpha i}$)
is called a length $m$ string with string center $u_{m\alpha}$.
The $N_m$ is the number of length $m$ strings.
The string hypothesis is not literally true 
as exemplified for instance when $n=2$ and $L>21$ (cf. \cite{EKS}).
Nevertheless, a formal count of 
the number of solutions to (\ref{rbe}) is done as follows \cite{Be,Kir1}.
First one rewrites the Bethe equation into the one for the string centers.
This is done by replacing $u_j$ by a member of a string 
$u_{m \alpha} + \sqrt{-1}(m+1-2i) + \epsilon_{m\alpha i}$
and taking the product over $1\le i \le m$.
The resulting equation in the logarithmic form  
$\ln ({\rm LHS}/{\rm RHS}) \in 2\pi\sqrt{-1}\Z$
is cast, if $\epsilon_{m\alpha i}$ is negligible,  into the form
$f_m(u_{m\alpha}) \in \Z$ or $\Z+\frac{1}{2}$
$(1\le \alpha \le N_m)$ which depends on $m$ and the partition $\{N_m\}$.
Explicitly,  $f_m(u)$ is given by
\begin{align}
f_{m}(u) &= L\theta_{m,1}(u) -\sum_{k\ge 1}
\sum_{\beta=1}^{N_k}(\theta_{m,k-1}+\theta_{m,k+1})
(u-u_{k\beta}),\\
\theta_{m,k}(u) & = \frac{1}{\pi}\sum_{\alpha=1}^{\min(m,k)}
\tan^{-1}\left(\frac{u}{|m-k|+2\alpha-1}\right).
\end{align}
Let us employ the principal branch 
$-\frac{\pi}{2}\le \tan^{-1}(u) \le \frac{\pi}{2}$.
Then from $\theta_{m,k}(\pm\infty) = \pm \min(m,k)/2$ and 
$(\theta_{m,k-1}+\theta_{m,k+1})(\pm \infty) 
= \pm (\min(m,k)-\delta_{m,k}/2)$,
we get $f_m(\pm \infty) = \pm (P_m+N_m)/2$.
Here $P_m$, called vacancy number, is given by
\begin{equation}\label{vn}
P_m = L - 2\sum_{k\ge 1}\min(m,k)N_k,
\end{equation}
and will play a significant role in the sequel.
The bold argument is then that if $P_m\ge 0$, the solutions
$\{u_{m\alpha}\}$ 
(up to permutations of $u_{m1},\ldots, u_{mN_m}$ for each $m$)  
are in one to one correspondence with 
the sequences $(I_1,\ldots, I_{N_m}) \in 
(\Z+\frac{P_m+N_m+1}{2})^{N_m}$ such that   
$-f_m(\infty)+\frac{1}{2} 
\le I_1<\cdots < I_{N_m} \le f_m(\infty)-\frac{1}{2}$.
There are $\binom{P_m+N_m}{N_m}$ such sequences for each $m$.
Accordingly if one admits the argument, the number of solutions is
\begin{equation}\label{mdef}
\EuScript{M}_n= \sum_{\{N_m\}} \prod_{m\ge 1}\binom{P_m+N_m}{N_m},
\end{equation}
where the sum extends over all the partitions of $n$, namely
those $N_m \ge 0$ satisfying $n = \sum_{m \ge1}mN_m$.
(We understand $\EuScript{M}_0=1$.)

What number should we expect for $\EuScript{M}_n$?
The quantum space for the rational 6 vertex model is
$(V_{\omega_1})^{\otimes L}$, where 
$V_{\omega_1}\simeq \C^2$ is the spin $\frac{1}{2}$
representation whose highest weight is 
the fundamental weight $\omega_1$.
As a result of the global $A_1=sl_2$ symmetry,
the Bethe vectors become by construction 
highest weight vectors in the quantum space \cite{FT}.
The sector labeled by $n$ carries the weight $(L-2n)\omega_1$.
Thus for the Bethe's string hypothesis to be complete, 
one should have $\EuScript{M}_n = b_n$ for $0 \le n \le L/2$, 
where $b_n$ is the branching coefficient in the irreducible decomposition 
$(V_{\omega_1})^{\otimes L} 
= \bigoplus_{0\le n \le L/2}b_n V_{(L-2n)\omega_1}$\footnote{This
argument lacks the consideration on the associated Bethe vectors.}.
Explicitly, $b_n = \binom{L}{n}-\binom{L}{n-1}$.
Note that the condition $0 \le n \le L/2$,
and (\ref{vn}) imply that 
$P_1\ge P_2 \ge \cdots \ge P_\infty = L-2n\ge 0$,
which automatically guarantees the condition $P_m \ge 0$.

\begin{example}\label{ex:ffm}
For $L=6$, one has 
$(V_{\omega_1})^{\otimes 6} = 
V_{6\omega_1}\oplus 5V_{4\omega_1}\oplus 9V_{2\omega_1}\oplus 5V_{0}$.
Accordingly one can check 
$(\EuScript{M}_0,\EuScript{M}_1, \EuScript{M}_2, \EuScript{M}_3)=
(1,5,9,5)$. 
In fact, the nontrivial cases are checked as
\begin{align*}
&\EuScript{M}_1 
= \underset{N_1=1}{\binom{4+1}{1}}= 5,\qquad
\EuScript{M}_2 
= \underset{N_2=1}{\binom{2+1}{1}} 
+ \underset{N_1=2}{\binom{2+2}{2}} = 9,\\
&\EuScript{M}_3 
= \underset{N_3=1}{\binom{0+1}{1}} 
+ \underset{N_1=N_2=1}{\binom{2+1}{1}\binom{0+1}{1}}
+\underset{N_1=3}{\binom{0+3}{3}} = 5.
\end{align*}
\end{example}

We postpone 
what can be proved mathematically in a more general setting 
to Section \ref{ss:at}.

\subsection{\mathversion{bold}Simplest example of 
$\EuScript{N}_\lambda$}
\label{ss:nf}
Here we return to the trigonometric Bethe equation (\ref{be1}).
After setting the inhomogeneity $w_j=0$,  
$q=e^{-2\pi\hbar}$ and replacing $u_j$ by $u_j/(\sqrt{-1}\hbar)$,
it reads
\begin{equation}\label{be2}
\left(
\frac{\sin\pi\!\left(u_j + \sqrt{-1}\hbar\right)}
{\sin\pi\!\left(u_j - \sqrt{-1}\hbar\right)}\right)^{L}
= - \prod_{k=1}^{n}
\frac{\sin\pi\!\left(u_j - u_k + 2\sqrt{-1}\hbar \right)}
{\sin\pi\!\left(u_j - u_k - 2\sqrt{-1}\hbar \right)}.
\end{equation}
In this convention,  
the analog of the string configuration (\ref{eq:string}) is
\begin{equation}\label{eq:string2}
\bigcup_{m \ge 1}\bigcup_{1\le \alpha\le N_m}
\bigcup_{u_{m\alpha} \in \R}
\{ u_{m \alpha} + \sqrt{-1}(m+1-2i)\hbar + \epsilon_{m\alpha i}
\mid 1 \le i \le m \},
\end{equation}
where $N_m$ is again the number of length $m$ strings.
Apart from $q=1$ treated in the previous subsection,
there is a point $q=0$,  i.e. the limit $\hbar \rightarrow \infty$
where one can make another formal 
but systematic counting of the string solutions \cite{KN2}.
Leaving the precise definitions and statements to \cite{KN2},
we just state here casually that at $q=0$
the Bethe equation (\ref{be2}) becomes
the following linear congruence equation on the string centers:
\begin{equation}\label{sce}
\sum_{k\ge 1}\sum_{\beta=1}^{N_k} 
A_{m\alpha,k\beta} u_{k\beta} \equiv
\frac{P_m + N_m + 1}{2} \quad \mathrm{mod}\ \Z.
\end{equation}
Here the coefficient $A_{m\alpha,k\beta}$ is given by
\begin{equation}\label{amat}
A_{m\alpha, k\beta} = \delta_{m k}\delta_{\alpha \beta}(P_m+N_m) +
2\min(m,k) - \delta_{m k}
\end{equation}
with the same $P_m$ as in (\ref{vn}).
Equation (\ref{sce}) is called the 
string center equation.
The concrete form of its RHS will not matter 
in the counting problem considered in what follows.
Given a string pattern $(N_m)$, 
one should actually regard the solutions to (\ref{sce}) as belonging to
\begin{equation*}
(u_{k 1}, u_{k2}, \ldots, u_{kN_k}) \in
\left(\R/\Z\right)^{N_k} /{\mathfrak S}_{N_k}
\end{equation*}
for each $k$, 
where ${\frak S}_{N}$ denotes the degree $N$ symmetric group.
This is because the Bethe vector is a symmetric function of 
$e^{2\pi\sqrt{-1}u_{k1}},\ldots, e^{2\pi\sqrt{-1}u_{kN_k}}$
for each $k$.
We say that a solution $(u_{k\beta})$ to (\ref{sce}) 
is {\em off-diagonal} if 
$u_{k 1}, u_{k2}, \ldots, u_{kN_k} \in \R/\Z$ are all distinct for each $k$.
This definition is motivated by the fact that the Bethe vectors vanish
unless the associated Bethe roots are all distinct \cite{TV}.

For $0 \le n \le L/2$ we define 
\begin{equation}\label{nd1}
\EuScript{N}_n = \sum_{\{N_m\}}
\sharp\{\text{off-diagonal solutions to 
the string center eq.(\ref{sce})}\},
\end{equation}
where the sum is taken over $N_m \in \Z_{\ge 0}$ 
satisfying $n= \sum_{m\ge 1}mN_m$ as in (\ref{mdef}).
(We understand $\EuScript{N}_0=1$.)
 
\begin{example}\label{ex:nff}
We derive $\EuScript{N}_n=\binom{L}{n}$ for $n=1,2$ as an illustration.
When $n=1$, the only possible string pattern 
$(N_m)$ is $N_m=\delta_{m1}$.
The equation (\ref{sce}) is just 
$Lu_{11}\equiv \text{const} \mod \Z$;
hence, there are $\EuScript{N}_1=L$ off-diagonal solutions.

For $n=2$ (hence $L\ge 4$), 
there are two possible string patterns (i) $N_m=\delta_{m2}$
and (ii) $N_m=2\delta_{m1}$.
In (i), equation (\ref{sce}) is $Lu_{21}\equiv \text{const} \mod \Z$,
which again yields $L$ off-diagonal solutions.
In (ii), equation (\ref{sce}) reads in the matrix notation as
\begin{equation*}
\begin{pmatrix}
L-1 & 1 \\
1 & L-1
\end{pmatrix}
\begin{pmatrix} u_{11} \\ u_{12} \end{pmatrix}
\equiv
\vec{c}\quad \mod \Z^2
\end{equation*}
for some $\vec{c}$.
The number of solutions equals the determinant $L(L-2)$ of the 
coefficient matrix, which is positive by the assumption $L\ge 4$.
However,  they contain the collision 
$(u_{11}=u_{12})$ $L$ times which should be excluded from 
the off-diagonal solutions.
Thus there are $(L(L-2)-L)/2$ off-diagonal solutions for (ii),
where the division by 2 is due to the identification by $\mathfrak{S}_2$.
Collecting the contributions from (i) and (ii), one gets
$\EuScript{N}_2 = L + (L(L-2)-L)/2 = L(L-1)/2$ as desired.
\end{example}

It is possible to generalize the calculations in Example \ref{ex:nff}
by a systematic application of the inclusion-exclusion principle.
The final result reads \cite{KN2}
\begin{equation}\label{ndef}
\begin{split}
\EuScript{N}_n &= \sum_{\{N_m\}}
\det_{m,k \in {\mathcal J}}(F_{m,k})
\prod_{m\in {\mathcal J}}\frac{1}{N_m}\binom{P_m+N_m-1}{N_m-1},
\\
F_{m,k} &= \delta_{mk}P_m + 2\min(m,k)N_k,
\end{split}
\end{equation}
where ${\mathcal J} = \{j \in \Z_{\ge 1}\mid N_j \ge 1\}$ and 
$P_m$ is defined by (\ref{vn}).
Again the sum in (\ref{ndef}) is taken in the same way as (\ref{nd1}).
As noted before Example \ref{ex:ffm},
the assumption $0 \le n \le L/2$ implies $P_m \ge 0$ $(m \ge 1)$.
By using this property 
it can be shown that $\det_{m,k \in {\mathcal J}}(F_{m,k}) >0$ and 
the RHS of the first equality in (\ref{ndef}) is a positive integer.

What number should we expect for $\EuScript{N}_n$?
Unlike the rational case in the previous subsection,
the 6 vertex model with $q\neq 1$ under the periodic boundary condition 
does not possess the global $sl_2$-symmetry.
Thus for the string solutions (\ref{eq:string2}) to be complete,
one should have $\EuScript{N}_n = c_n$,
where $c_n$ is the weight multiplicity of the quantum space
$(V_{\omega_1})^{\otimes L}$ with weight 
$(L-2n)\omega_1$\footnote{The same remark as the previous
footnote applies here.}.
Explicitly, $c_n = \binom{L}{n}$.
This has been confirmed for $n=1,2$ in Example \ref{ex:nff}.
The next case is checked as
\begin{equation*}
\EuScript{N}_3 = \underset{N_3=1}{L} + 
\underset{N_1=N_2=1}{
\begin{vmatrix}L-2 & 2\\ 2 & L-2\end{vmatrix}}
+\underset{N_1=3}{L\,\frac{1}{3}\binom{L-6+2}{2}}
=\frac{L(L-1)(L-2)}{6}.
\end{equation*}

One may wonder what happens for $n> L/2$ where $c_n$ still makes sense.
The answer will be given in the next subsection in a more general setting
together with the analogous result for $b_n$. 
The only preliminary we mention here is that 
such considerations necessarily involve the situation $P_m<0$ hence 
the binomial coefficients $\binom{X}{N}$ with $X<N$. 

\subsection{\mathversion{bold}Theorems for type $A_1$}\label{ss:at}

We have hitherto argued about three kinds of quantities 

(i) Number of string solutions in the Bethe ansatz,

(ii) Fermionic forms $\EuScript{M}_n$ and $\EuScript{N}_n$,

(iii) Representation theoretical data $b_n$ and $c_n$,

\noindent
especially without a much distinction between (i) and (ii). 
Here we redefine (ii) without recourse to (i) and 
formulate the theorems on the relations between (ii) and (iii).
We treat the general spin case 
$\bigotimes_{m\ge 1} (V_{m\omega_1})^{\otimes \nu_m}$ and 
present the Fermionic character formulas.
As power series formulas, they are actually valid for 
arbitrary $\nu_m \in \C$.
The proof of the theorem, which will be outlined 
in the next subsection, does not lean on the string hypotheses but 
is solely derived from the Q-system.
As such, it does not prove nor disprove 
the completeness of the string hypothesis.

Let $Q_m$ $({\mathcal Q}_m)$ 
be the character (normalized character) 
of the irreducible $m+1$ dimensional 
representation $V_{m\omega_1}$. Namely,
\begin{align}
Q_m &= \chi(V_{m\omega_1}) = y^m+y^{m-2}+\cdots + y^{-m} 
= \frac{y^{m+1}-y^{-m-1}}{y-y^{-1}}\quad (y=e^{\omega_1}),\\
{\mathcal Q}_m &= y^{-m}Q_m.\label{nch}
\end{align}
The $Q_m$ is a simplified notation for the variable 
$Q^{(1)}_m$ (\ref{qrt}) in the Q-system for $A_1$:
\begin{equation}\label{qsa1}
Q_m^2 = Q_{m-1}Q_{m+1}+1.
\end{equation}
See (\ref{qade}).
The $Q_m$ expressed as a function of $Q_1$ is the Chebyshev polynomial 
of the second kind.
In Section \ref{ss:mli},  we will utilize the one adapted to 
the normalized character (\ref{nch}).
\begin{equation}\label{qnc}
\frac{{\mathcal Q}_{m-1}{\mathcal Q}_{m+1}}{{\mathcal Q}_m^2}
+y^{-2m}{\mathcal Q}_m^{-2}=1.
\end{equation}

Let $\nu_m \in \C\,(m \in \Z_{\ge 1})$ be arbitrary except that 
$\nu_m = 0$ for all but finitely many $m$.
We define the branching coefficient $b_n$ 
and the weight multiplicity $c_n$ for all $n \in \Z_{\ge 0}$ by
\begin{equation}\label{bcdef}
\prod_{m\ge 1}({\mathcal Q}_m)^{\nu_m} 
=\frac{\sum_{n\ge 0}b_ny^{-2n}}{1-y^{-2}}
=\sum_{n \ge 0}c_ny^{-2n}. 
\end{equation}
By the definition, the normalized character 
${\mathcal Q}_m$ is a polynomial in $y^{-2}$ with unit constant term.
$({\mathcal Q}_m)^{\nu_m}$ denotes its $\nu_m$th power 
with unit constant term
$1+\nu_m({\mathcal Q}_m-1)+\frac{\nu_m(\nu_m-1)}{2}
({\mathcal Q}_m-1)^2+\cdots$, 
which is a polynomial or a power series in $y^{-2}$   
according as $\nu_m \in \Z_{\ge 0}$ or not. 
When $\nu_m \in \Z_{\ge 0}$ for any $m\ge 1$, 
this definition of $b_n$ agrees with the one for the branching coefficient
of $V_{(\sum_mm\nu_m-2n)\omega_1}$ 
in $\bigotimes_{m\ge 1}(V_{m\omega_1})^{\otimes \nu_m}$  
for $0 \le n \le \sum_mm\nu_m/2$.
The above $b_n$ is an extension of this by 
$b_n=-b_{-n+1+\sum_mm\nu_m}$, which is 
the skew symmetry under the Weyl group.

As for the Fermionic forms, we redefine 
$\EuScript{M}_n$ (\ref{mdef}) and $\EuScript{N}_n$ (\ref{ndef})
by replacing 
$P_m$ (\ref{vn}) and the binomial coefficient therein with the 
generalized ones\footnote{
In Sections \ref{ss:mf} and \ref{ss:nf}, the symbol 
$\binom{X}{N}$ was used only for $0 \le N \le X$.}: 
\begin{align}
P_m &= \sum_{k\ge 1}\min(m,k)(\nu_k-2N_k),\label{pdef2}\\
\binom{X}{N} & = \frac{\prod_{i=1}^N(X-i+1)}{N!} 
\qquad (X \in \C,\,N \in \Z_{\ge 0}).\label{abn} 
\end{align}
The sum over $\{N_m | \,m\in \Z_{\ge 1} \}$ is taken in the same way 
as (\ref{mdef}) and (\ref{ndef}).
Namely, it is the finite sum over those 
$N_m \in \Z_{\ge 0}$ satisfying $\sum_{m\ge 1}mN_m = n$.
There is {\em no} condition like $P_m \ge 0$ which does not make sense 
in the general setting $\nu_m \in \C$ under consideration. 
The generalized binomial (\ref{abn}) is nonzero except the $N$ points 
$X=0,1,\ldots, N-1$, and appears in the expansion
\begin{equation}\label{eq:exp}
(1-x)^{-\beta-1} = \sum_{N=0}^\infty \binom{\beta+N}{N}x^N,
\end{equation}
for any $\beta \in \C$.
With these definitions we have

\begin{theorem}[\cite{Kir1, KN2}]\label{th:mnbc}
The equalities (1) $\EuScript{M}_n = b_n$ and
(2) $\EuScript{N}_n = c_n$ 
hold for all $n \in \Z_{\ge 0}$.
Namely, the following power series formulas hold.
\begin{equation}\label{mnexp}
\prod_{m\ge 1}({\mathcal Q}_m)^{\nu_m} 
=\frac{\sum_{n\ge 0}\EuScript{M}_n y^{-2n}}{1-y^{-2}}
=\sum_{n \ge 0}\EuScript{N}_n y^{-2n}.
\end{equation}
\end{theorem}

The formulas (1) and (2) are due to \cite{Kir1} and \cite{KN2},
respectively.
The theorem reproduces the observations in Sections \ref{ss:mf} and \ref{ss:nf}
in the special case $\nu_m = L\delta_{m1}$ and $0 \le n \le L/2$,
where $P_m \ge 0$ for any $m\ge 1$ automatically holds.
However, even for this simple choice $\nu_m = L\delta_{m1}$, 
it further claims infinitely many nontrivial identities including  
$\EuScript{M}_n = 0$ for $n\ge L+2$ and 
$\EuScript{N}_n = 0$ and $n\ge L+1$.

\begin{example}\label{ex:mn}
Assume that $\nu_m = 0$ for $m\ge 4$.
Then LHS of (\ref{mnexp}) is 
$(1+y^{-2})^{\nu_1}(1+y^{-2}+y^{-4})^{\nu_2}
(1+y^{-2}+y^{-4}+y^{-6})^{\nu_3}$.
Setting $\gamma_m = \sum_{k=1}^3\min(m,k)\nu_k$,
we write down $\EuScript{M}_n$ (\ref{mdef}) 
and $\EuScript{N}_n$ (\ref{ndef}) for $n=1,2,3$.
\begin{align*}
&\EuScript{M}_1 = \gamma_1-1,\qquad 
\EuScript{M}_2 =
\gamma_2-3+\frac{1}{2}(\gamma_1-2)(\gamma_1-3),\\
&\EuScript{M}_3 = (\gamma_3-5) +(\gamma_1-3)(\gamma_2-5)
+\frac{1}{6}(\gamma_1-3)(\gamma_1-4)(\gamma_1-5),\\
&\EuScript{N}_1 = \gamma_1,\qquad \qquad
\EuScript{N}_2 =
\gamma_2+\frac{1}{2}\gamma_1(\gamma_1-3),\\
&\EuScript{N}_3 = 
\gamma_3+
\begin{vmatrix} \gamma_1-2 & 2 \\ 2 & \gamma_2-2\end{vmatrix}
+\frac{1}{6}\gamma_1(\gamma_1-4)(\gamma_1-5).
\end{align*} 
One can directly check these coefficients in 
the power series expansions (\ref{mnexp}).
For instance in the simplest case 
$\nu_m = 0$ hence $\gamma_m = 0$ for all $m\ge 1$,
all these coefficients vanish except $\EuScript{M}_1=-1$ 
as they should.
\end{example}

In the case $\nu_m \in \Z_{\ge 0}\,(m\ge 1)$,
$P_m$ in (\ref{pdef2}) can be a nonnegative integer 
for some $\{N_m\}$.
Then it makes sense to introduce the following variant 
of $\EuScript{M}_n$:
\begin{equation}\label{mbdef}
\overline{\EuScript{M}}_n = 
\sum_{\{N_m\}}\!\!\!{}^+ \prod_{m\ge 1}\binom{P_m+N_m}{N_m},
\end{equation}
where $P_m$ and $\binom{X}{N}$ are again specified by 
(\ref{pdef2}) and (\ref{abn}) as for $\EuScript{M}_n$.
The only difference from it is that 
the sum $\sum^+_{\{N_m\}}$ 
extends over those $N_m \in \Z_{\ge 0}$ 
satisfying $n = \sum_{m \ge1}mN_m$ with the extra condition
$P_m \ge 0$ if $N_m \ge 1$.

Given $\{\nu_m\}$, $n$ and $\{N_m\}$ 
satisfying $\sum_{m\ge 1}mN_m = n$,
let $m_0$ be the maximal $m$ such that $N_m \ge 1$.
Then we have $P_{m_0} = \sum_{k\ge 1}\min(m_0,k)\nu_k -2n 
\le \sum_{k\ge 1}k\nu_k - 2n$.
Thus we see  
$\overline{\EuScript{M}}_n=0$ if 
$n > \frac{1}{2}\sum_{k\ge 1}k\nu_k$.

\begin{theorem}[\cite{KR2,KiSS}]\label{th:mbb}
For any $\nu_m \in \Z_{\ge 0}$, the equality 
$\overline{\EuScript{M}}_n = b_n$ holds 
for $0 \le n \le \frac{1}{2}\sum_{m\ge 1}m\nu_m$.
\end{theorem}

As remarked after Theorem \ref{th:mnbc}, 
Theorem \ref{th:mbb} is equivalent to 
Theorem \ref{th:mnbc} (1) in the the special case 
$\nu_m = L\delta_{m1}$ and $0 \le n \le L/2$.
In general, they imply that 
the contributions to $\EuScript{N}_n$ 
involving $P_m <0$ cancel out.

\begin{example}
Take $\nu_m = 2\delta_{m 3}$ in Example \ref{ex:mn}. 
Then $(\gamma_1,\gamma_2,\gamma_3) = (2,4,6)$.
The three terms in $\EuScript{M}_3$ 
correspond to choosing nonzero $N_m$ as 
$N_3=1$, $N_1=N_2=1$ and $N_1=3$.
The relevant $P_m$'s are 
$P_3=0$, $P_1=P_2=-2$ and $P_1=-4$, respectively.
Thus $\overline{\EuScript{M}}_3$ is given by the first term 
only $\gamma_3-5=1$.
This coincides with $\EuScript{M}_3$ since the other two terms cancel.
\end{example}

\subsection{Multivariable Lagrange inversion}\label{ss:mli}

Here we outline the proof of Theorem \ref{th:mnbc}.
We describe an essential step of deriving (\ref{mnexp}) from 
(\ref{qnc}) in a generalized setting applicable to ${\mathfrak g}$ case \cite{KNT1}. 

Let $H$ denote a finite index set.
Let $w=(w_i)_{i\in H}$ and $v=(v_i)_{i\in H}$
be complex multivariables,
and let
$G=(G_{ij})_{i,j\in H}$
be a complex square matrix of size $|H|$.
We  consider a holomorphic map
${\mathcal D} \rightarrow \C^H$,
$v\mapsto w(v)$ with
\begin{align}
\label{eq:wv1}
w_i(v)&=v_i\prod_{j\in H} (1-v_j)^{-G_{ij}},
\end{align}
where ${\mathcal D}$ is some neighborhood of $v=0$
 in $\C^H$.
The Jacobian $(\partial w / \partial v)(v) $ is 1
at $v = 0$, so that
the map $w(v)$
is bijective around $v=w=0$.
Let $v(w)$ be the inverse map around $v=w=0$.
Inverting (\ref{eq:wv1}),
we obtain the following functional equation
for $v_i(w)$'s:
\begin{align}\label{eq:vw1}
v_i(w)
= w_i\prod_{j\in H}(1-v_j(w))^{G_{ij}}.
\end{align}
By introducing new functions
\begin{align}\label{eq:qv1}
{\mathcal Q}_i(w)=1-v_i(w),
\end{align}
the equation
(\ref{eq:vw1}) is written as
\begin{align}\label{eq:qsys1}
{\mathcal Q}_i(w)+w_i \prod_{j\in H}{\mathcal Q}_j(w)^{G_{ij}}=1.
\end{align}

{}From now on, we regard 
(\ref{eq:qsys1}) as equations for a family $({\mathcal Q}_i(w))_{i\in H}$ of
power series  of
$w=(w_i)_{i\in H}$
with the unit constant terms.
The procedure from (\ref{eq:wv1}) to (\ref{eq:qsys1}) 
can be reversed;
therefore,
the power series expansion of ${\mathcal Q}_i(w)$ in (\ref{eq:qv1})
gives the unique family 
$({\mathcal Q}_i(w))_{i\in H}$ of power series of
$w$ with the unit constant terms
which satisfies (\ref{eq:qsys1}).

We define (finite) Q-system to be the following equations for
a family $({\mathcal Q}_i(w))_{i\in H}$ of 
power series of $w$ with the unit constant terms:
\begin{align}\label{eq:qsys2}
\prod_{j\in H}
{\mathcal Q}_j(w)^{D_{ij}}+w_i 
\prod_{j\in H}{\mathcal Q}_j(w)^{G_{ij}}=1\quad (i \in H),
\end{align}
where $D=(D_{ij})_{i,j\in H}$ and $G=(G_{ij})_{i,j\in H}$ are arbitrary
complex matrices with $\det D\neq 0$.
Equation (\ref{eq:qsys1}), which is the special case
of (\ref{eq:qsys2}) with $D=I$ ($I$: the identity matrix),
is called a standard Q-system.
By setting ${\mathcal Q}'_i(w) = \prod_{j\in H}{\mathcal Q}_j(w)^{D_{ij}}$, 
(\ref{eq:qsys2}) is always 
transformed to the standard one (\ref{eq:qsys1})
with $G$ replaced by $G' = GD^{-1}$ and vice versa. 
Therefore, the Q-system (\ref{eq:qsys2}) also has the unique solution.

Given the Q-system (\ref{eq:qsys2})
and $\nu=(\nu_i)_{i\in H} \in \C^{H}$,
we define two power series of $w$
\begin{align}
\label{eq:cKdef1}
\EuScript{M}^\nu(w) = \sum_{N} \EuScript{M}(\nu,N) w^N,
\quad
\EuScript{N}^\nu(w) = \sum_{N} \EuScript{N}(\nu,N)w^N,
\end{align}
where $w^N=\prod_{i\in H}w_i^{N_i}$ 
and the sums run over $N =(N_i)_{i \in H} \in (\Z_{\ge 0})^H$. 
The coefficients are given by
\begin{align}
\label{eq:kdef1}
\EuScript{M}(\nu,N) &= \prod_{i\in H(N)}
\binom{P_i + N_i}{N_i},\\
\label{eq:rdef1}
\EuScript{N}(\nu,N) &=
\Bigl(\det_{H(N)} F_{ij}\Bigr)
\prod_{i\in H(N)} \frac{1}{N_i}
\binom{P_i+ N_i - 1}{N_i- 1},
\end{align}
where the binomial is defined by (\ref{abn}) and 
we have set $H(N)=\{\, i\in H\mid N_i\neq 0\, \}$,
\begin{align}\label{eq:padef1}
P_i&=P_i(\nu,N):=-
\sum_{j\in H}\nu_j(D^{-1})_{ji}
-\sum_{j\in H}N_j(GD^{-1})_{ji},\\
\label{eq:fdef1}
F_{ij}&=
F_{ij}(\nu,N):=
\delta_{ij}P_j+(GD^{-1})_{ij}N_j.
\end{align}
$\det_{H(N)}$ is a shorthand notation for $\det_{i,j\in H(N)}$.
In (\ref{eq:kdef1}) and (\ref{eq:rdef1}),
 $\det_\emptyset$ and $\prod_\emptyset$
mean 1; therefore, 
$\EuScript{M}^\nu(w)$ and 
$\EuScript{N}^\nu(w)$ are power series
with the unit constant terms.
See \cite[section 2]{KNT1} for the convergence radius.
Note a similarity to (\ref{mdef}) and (\ref{ndef}).

\begin{theorem}[\cite{KNT1}]\label{thm:fqmain1}
Let $({\mathcal Q}_i(w))_{i\in H}$  be the unique
solution of (\ref{eq:qsys2}).
For $\nu=(\nu_i)_{i\in H} \in \C^{H}$, the following formulas are valid:
\begin{align}\label{eq:qkr01}
\prod_{i\in H}
{\mathcal Q}_i(w)^{\nu_i}=
\frac{\EuScript{M}^\nu(w)}{\EuScript{M}^0(w)} 
=\EuScript{N}^\nu(w).
\end{align}
\end{theorem}

${\mathcal Q}_i(w)$ itself is obtained by setting $\nu_j=\delta_{ij}$.

\begin{example}\label{ex:h1}
Let $|H|=1$.
Then, (\ref{eq:qsys2}) is an equation
for a single power series $Q(w)$:
\begin{align*}
{\mathcal Q}(w)^D+w {\mathcal Q}(w)^G =1,
\end{align*}
where $D\neq 0$ and $G$ are complex numbers and 
Theorem \ref{thm:fqmain1} shows that
\begin{align*}
{\mathcal Q}(w)^\nu = 
\EuScript{N}^\nu(w) 
=\frac{\nu}{D}
\sum_{N=0}^\infty
\frac{\varGamma((\nu+NG)/D)(-w)^N}
{\varGamma((\nu+NG)/D-N+1)N!}.
\end{align*}
This power series formula is well known and
have a very long history
since Lambert  (e.g.\ \cite[pp.\ 306--307]{Ber}).
\end{example}

As noted before,  the Q-system (\ref{eq:qsys2}) is 
bijectively transformed to the standard one (\ref{eq:qsys1}).
Under the corresponding changes $D \rightarrow I$,
$\nu_i \rightarrow \sum_{j \in H}\nu_j(D^{-1})_{ji}$
and $G \rightarrow GD^{-1}$, 
quantities (\ref{eq:padef1}) and (\ref{eq:fdef1}) remain invariant, 
hence so are $\EuScript{M}(\nu,N)$ and $\EuScript{M}(\nu,N)$.
Thus we have only to prove 
Theorem \ref{thm:fqmain1}
for the standard case $D=I$, where 
${\mathcal Q}_i(w)$ is described by (\ref{eq:wv1})--(\ref{eq:qv1}). 
Therefore, Theorem \ref{thm:fqmain1} follows from
\begin{proposition}[\cite{KNT1} Proposition 2.8]\label{prop:pse2}
Let $v=v(w)$ be the inverse map of (\ref{eq:wv1}).
Let $\EuScript{M}^\nu(w)$ and $\EuScript{N}^\nu(w)$ be those 
for $D=I$ in (\ref{eq:padef1}) and (\ref{eq:fdef1}).
Then, the power series expansions
\begin{align}
\label{eq:pse1}
\det_H \Bigl(\frac{w_j}{v_i}
\frac{\partial v_i}{\partial w_j}(w)\Bigr)
\prod_{i\in H}
(1-v_i(w))^{\nu_i-1}
&=\EuScript{M}^\nu(w),\\
\label{eq:vr1}
\prod_{i\in H}(1-v_i(w))^{\nu_i}
&=
\EuScript{N}^\nu(w)
\end{align}
hold around $w=0$. 
\end{proposition}

This is a particularly nice example of the multivariable Lagrange
inversion formula (e.g.\ \cite{G}), where all the calculations can be 
carried through by a multivariable residue analysis.

\noindent
{\em Proof. The first formula (\ref{eq:pse1}).} 
\noindent
We evaluate the coefficient for 
$w^N$ on the LHS of (\ref{eq:pse1}) as follows:
\begin{align*}
&\,
{\rm Res}_{w=0} 
\frac{\partial v}{\partial w}(w)
\prod_{i\in H}
\Bigl\{ (1-v_i(w))^{\nu_i-1} (v_i(w))^{-1}
(w_i)^{1-N_i-1} \Bigr\}dw\\
=&\, {\rm Res}_{v=0} 
\prod_{i\in H}
\Bigl\{ (1-v_i)^{\nu_i-1}(v_i)^{-1}
\Bigl(v_i\prod_{j\in H}(1-v_j)^{-G_{ij}}\Bigr)^{-N_i}
 \Bigr\}dv\\
=&\, {\rm Res}_{v=0} 
\prod_{i\in H}
\Bigl\{ (1-v_i)^{-P_i(\nu,N)-1}
(v_i)^{-N_i-1}
 \Bigr\}dv =  \prod_{i\in H}
\binom{P_i(\nu,N) + N_i}{N_i}
=\EuScript{M}(\nu,N),
\end{align*}
where we used (\ref{eq:exp}) to get the last line.
Thus,  (\ref{eq:pse1}) is proved.

{\em The second formula (\ref{eq:vr1}).}
By a simple calculation, we have
\begin{align}\label{eq:k02}
\det_H \Bigl(\frac{v_j}{w_i}
\frac{\partial w_i}{\partial v_j}(v)\Bigr)
\prod_{i\in H}(1-v_i)
=\det_{ H}
\Bigl(\delta_{ij}+(-\delta_{ij}+G_{ij})v_i\Bigr)
= \sum_{J\subset H}
d_J \prod_{i\in J}v_i,
\end{align}
where $d_J:= \det_J (-\delta_{ij}+G_{ij})$,
and the sum is taken over
all the subsets $J$ of $H$.
Therefore,
the LHS of (\ref{eq:vr1})
is written as
\begin{align}
\label{eq:qig1}
\det_H \Bigl(\frac{w_j}{v_i}
\frac{\partial v_i}{\partial w_j}(w)\Bigr)
 \sum_{J\subset H}
 d_J 
\prod_{i\in H}
\Bigl\{
(1-v_i(w))^{\nu_i-1}
v_i(w)^{\theta(i\in J)}\Bigr\}.
\end{align}
By a similar residue calculation as above,
the  coefficient for $w^N$ of (\ref{eq:qig1})  is evaluated
as ($\theta(\text{\rm true}) = 1$ and
$\theta(\text{\rm false}) = 0$)
\begin{align*}
&\, 
\sum_{J\subset H}
d_J
{\rm Res}_{v=0} 
\prod_{i\in H}
\Bigl\{ (1-v_i)^{-P_i(\nu,N)-1}
(v_i)^{-N_i+\theta(i\in J)-1}
 \Bigr\}dv\\
= &\, 
\sum_{J\subset H(N)}d_J
\prod_{i\in H(N)}
\binom{P_i(\nu,N) + N_i-\theta(i\in J)}{N_i-\theta(i\in J)}\\
=&\,
\biggl(
\sum_{J \subset H(N)}
d_J
\prod_{i \in  J} N_i
\prod_{i \in H(N)\setminus J}(P_i+N_i)
\biggr)
\prod_{i\in H(N)} \frac{1}{N_i}
\binom{P_i+ N_i - 1}{N_i- 1}\\
=&\,
\det_{H(N)} \Bigl(\delta_{ij}(P_j+N_j)+(-\delta_{ij}+G_{ij})N_j
\Bigr)
\prod_{i\in H(N)} \frac{1}{N_i}
\binom{P_i+ N_i - 1}{N_i- 1}\\
=&\, \EuScript{N}(\nu,N).
\end{align*}

This completes the proof of Theorem \ref{thm:fqmain1}.
What is left to prove Theorem \ref{th:mnbc} from it?
Comparing the Q-systems (\ref{eq:qsys2}) and (\ref{qnc})
and also $P_m$ in (\ref{eq:padef1}) and  (\ref{pdef2}), 
we see that Theorem \ref{th:mnbc} formally corresponds to taking
\begin{equation}\label{hido}
\begin{split}
&H = \Z_{\ge 1},\quad w_i = y^{-2i},\\
&(D^{-1})_{ij} =-\min(i,j), \quad
D_{ij} = \delta_{i, j+1}+\delta_{i,j-1}-2\delta_{ij},
\quad
G_{ij} = -2\delta_{ij}
\end{split}
\end{equation}
in Theorem \ref{thm:fqmain1}, and claiming
$\EuScript{M}^0(w)=1-y^{-2}$ thereunder.
Since we started with the assumption that $H$ is a finite set,
it is nontrivial how to make sense of these choices and claims.
We refer to \cite{KNT1} for a proper treatment of 
such an infinite ($|H|=\infty$) Q-system 
as a projective limit of the finite Q-systems.
According a result therein,
Theorem \ref{th:mnbc} is shown, among other things, 
from the convergence property: the limit 
$\lim_{m\rightarrow \infty} {\mathcal Q}_m(w_i=y^{-2i})$ 
exists in $\C[[y^{-2}]]$.
 
\subsection{\mathversion{bold}Q-system 
and theorems for ${\mathfrak g}$}\label{ss:qs}
Here we present the Q-system and analog of 
Theorem \ref{th:mnbc} and Theorem \ref{th:mbb}
for general ${\mathfrak g}$.
We use the notations in Section \ref{ss:utw} 
such as $I$, $t$, $t_a$, 
$C=(C_{ab})$, $\alpha_a$ and $\omega_a$.
The unrestricted Q-system for ${\mathfrak g}$
is the following relations among the variables 
$\{Q^{(a)}_m\mid a\in I, m\ge 1 \}$,
where 
$Q^{(0)}_m=Q^{(a)}_0= 1$ if they occur
on the RHS.

For simply laced ${\mathfrak g}$,
\begin{align}\label{qade}
(Q^{(a)}_m)^2
=
Q^{(a)}_{m-1}Q^{(a)}_{m+1}
+
\prod_{b\in I: C_{ab}=-1}
Q^{(b)}_{m}.
\end{align}

For ${\mathfrak g}=B_r$,
\begin{equation}\label{brq}
\begin{split}
(Q^{(a)}_m)^2
&=
Q^{(a)}_{m-1}Q^{(a)}_{m+1}
+Q^{(a-1)}_{m}Q^{(a+1)}_{m}
\quad
 (1\leq a\leq r-2),\\
(Q^{(r-1)}_m)^2
&=
Q^{(r-1)}_{m-1}Q^{(r-1)}_{m+1}
+
Q^{(r-2)}_{m}Q^{(r)}_{2m},\\
(Q^{(r)}_{2m})^2
&=
Q^{(r)}_{2m-1}Q^{(r)}_{2m+1}
+
(Q^{(r-1)}_{m})^2,
\\
(Q^{(r)}_{2m+1})^2
&=
Q^{(r)}_{2m}Q^{(r)}_{2m+2}
+
Q^{(r-1)}_{m}Q^{(r-1)}_{m+1}.
\end{split}
\end{equation}

For ${\mathfrak g}=C_r$,
\begin{equation}\label{crq}
\begin{split}
(Q^{(a)}_m)^2
&=
Q^{(a)}_{m-1}Q^{(a)}_{m+1}
+Q^{(a-1)}_{m}Q^{(a+1)}_{m}
\quad
 (1\leq a\leq r-2),\\
(Q^{(r-1)}_{2m})^2
&=
Q^{(r-1)}_{2m-1}Q^{(r-1)}_{2m+1}
+
Q^{(r-2)}_{2m}
(Q^{(r)}_{m})^2,\\
(Q^{(r-1)}_{2m+1})^2
&=
Q^{(r-1)}_{2m}Q^{(r-1)}_{2m+2}
+
Q^{(r-2)}_{2m+1}
Q^{(r)}_{m}Q^{(r)}_{m+1},
\\
(Q^{(r)}_{m})^2
&=
Q^{(r)}_{m-1}Q^{(r)}_{m+1}
+
Q^{(r-1)}_{2m}.
\end{split}
\end{equation}

For ${\mathfrak g}=F_4$,
\begin{equation}\label{f4q}
\begin{split}
(Q^{(1)}_m)^2
&=
Q^{(1)}_{m-1}Q^{(1)}_{m+1}
+Q^{(2)}_{m},\\
(Q^{(2)}_m)^2
&=
Q^{(2)}_{m-1}Q^{(2)}_{m+1}
+
Q^{(1)}_{m}Q^{(3)}_{2m},\nonumber\\
(Q^{(3)}_{2m})^2
&=
Q^{(3)}_{2m-1}Q^{(3)}_{2m+1}
+
(Q^{(2)}_{m})^2
Q^{(4)}_{2m},\nonumber\\
(Q^{(3)}_{2m+1})^2
&=
Q^{(3)}_{2m}Q^{(3)}_{2m+2}
+
Q^{(2)}_{m}Q^{(2)}_{m+1}
Q^{(4)}_{2m+1},\nonumber\\
\nonumber
(Q^{(4)}_{m})^2
&=
Q^{(4)}_{m-1}Q^{(4)}_{m+1}
+Q^{(3)}_m.
\end{split}
\end{equation}

For ${\mathfrak g}=G_2$,
\begin{equation}\label{g2q}
\begin{split}
(Q^{(1)}_m)^2
&=
Q^{(1)}_{m-1}Q^{(1)}_{m+1}
+
Q^{(2)}_{3m},\\
(Q^{(2)}_{3m})^2
&=
Q^{(2)}_{3m-1}Q^{(2)}_{3m+1}
+
(Q^{(1)}_{m})^3,\\
(Q^{(2)}_{3m+1})^2
&=
Q^{(2)}_{3m}Q^{(2)}_{3m+2}
+
(Q^{(1)}_{m})^2
Q^{(1)}_{m+1},\\
(Q^{(2)}_{3m+2})^2
&=
Q^{(2)}_{3m+1}Q^{(2)}_{3m+3}
+
Q^{(1)}_{m}
(Q^{(1)}_{m+1})^2.
\end{split}
\end{equation}
These relations are uniformly written as
\begin{equation}\label{qsu}
(Q^{(a)}_m)^2 = 
Q^{(a)}_{m-1}Q^{(a)}_{m+1} 
+ (Q^{(a)}_m)^2
\prod_{(b,k) \in H}(Q^{(b)}_k)^{G_{am,bk}},
\end{equation}
by using the notations (\ref{hdef}) and (\ref{gak}).
We shall introduce the restricted Q-system in Section \ref{ss:rwq}.

As mentioned around (\ref{qrt}), 
these relations follow from the T-systems 
by forgetting the spectral parameter $u$.
Recall that ${\rm res } \,\chi_q(W^{(a)}_m(u))$ denotes
the classical character of the Kirillov-Reshetikhin module 
$W^{(a)}_m(u)$.
See (\ref{res}) for the definition of ${\rm res}$.
Since ${\rm res}$ removes the dependence on $u$,
we will simply write as
${\rm res } \,\chi_q(W^{(a)}_m)$ in what follows.
The following is a corollary of Theorem \ref{th:nh}.
\begin{proposition}\label{pr:rwq}
The substitution 
$Q^{(a)}_m={\rm res } \,\chi_q(W^{(a)}_m)$ 
satisfies the unrestricted Q-system.
\end{proposition}

{}From now on, we understand the symbol $Q^{(a)}_m$ as representing 
${\rm res } \,\chi_q(W^{(a)}_m)$.
By Theorem \ref{th:frm} (1),
the normalized character 
\begin{equation}
{\mathcal Q}^{(a)}_m = e^{-m\omega_a}Q^{(a)}_m
\end{equation}
is a polynomial in $e^{-\alpha_1},\ldots, e^{-\alpha_r}$
with unit constant term and 
coefficients from $\Z_{\ge 0}$.
In terms of ${\mathcal Q}^{(a)}_m$, the Q-system
is expressed as
\begin{align}\label{eq:qsys11}
\prod_{(b,k)\in H}
({\mathcal Q}^{(b)}_k)^{D_{am,bk}}
+ e^{-m\alpha_a}\prod_{(b,k)\in H}({\mathcal Q}^{(b)}_k)^{G_{am,bk}}=1
\end{align}
for $(a,m) \in H$. 
Here $H$, $D_{am,bk}$ and $G_{am,bk}$ are defined by
\begin{align}
H &= \{(a,m)| \,a \in I, m \in \Z_{\ge 1}\},\label{hdef}\\
D_{am,bk}&=-\delta_{ab}(2\delta_{mk}-\delta_{m,k+1}
-\delta_{m,k-1}),\\
\label{eq:krd2}
(D^{-1})_{am,bk}&= -\delta_{ab}\min(m,k).\\
G_{am,bk}=&
\begin{cases}
-C_{ba}
(\delta_{m,2k-1}+2\delta_{m,2k}+\delta_{m,2k+1})
&
t_a/t_b=2,\\
-C_{ba}
(\delta_{m,3k-2}+2\delta_{m,3k-1}+3\delta_{m,3k}&
t_a/t_b=3,\\
\qquad\qquad\qquad
+2\delta_{m,3k+1}+\delta_{m,3k+2})&\\
-C_{ab}\delta_{t_b m, t_a k}
& \text{otherwise}.\label{gak}
\end{cases}
\end{align}
For ${\mathfrak g}=A_1$, the data $H, D, G$ here reduce to (\ref{hido})
hence (\ref{eq:qsys11}) to (\ref{qnc}).
By an analysis parallel with $A_1$ case,
one can establish the power series formulas 
involving Fermionic forms.
They are read off (\ref{eq:cKdef1})--(\ref{eq:fdef1}) 
by formally replacing the single indices by double ones
as $i \rightarrow (a,m)$, $j \rightarrow (b,k)$, etc.
To be concrete, let $\nu=(\nu^{(a)}_m)_{(a,m)\in H} \in \C^H$, 
where $\nu^{(a)}_m =0$ for all but finitely many $(a,m)$.
For $N=(N^{(a)}_m)_{(a,m)\in H} \in (\Z_{\ge 0})^H$,
we define 
\begin{align}
\label{mdef1}
\EuScript{M}(\nu,N) &= \prod_{(a,m)\in H(N)}
\binom{P^{(a)}_m + N^{(a)}_m}{N^{(a)}_m},\\
\label{ndef1}
\EuScript{N}(\nu,N) &=
\Bigl(\det_{H(N)} F_{am, bk}\Bigr)
\prod_{(a,m)\in H(N)} \frac{1}{N^{(a)}_m}
\binom{P^{(a)}_m+ N^{(a)}_m - 1}{N^{(a)}_m- 1},
\end{align}
where the binomial is the generalized one (\ref{abn}).
We have set $H(N)=\{\, (a,m)\in H\mid N^{(a)}_m\neq 0\, \}$
and $\det_{H(N)}$ denotes $\det_{(a,m), (b,k)\in H(N)}$.
Define further 
\begin{align}\label{pdef1}
P^{(a)}_m&=
\sum_{k \ge 1}\min(m,k)\nu^{(a)}_k
-\sum_{(b,k)\in H}(\alpha_a |\alpha_b)\min(t_bm, t_ak)N^{(b)}_k,\\
\label{fde}
F_{am,bk}&=
\delta_{ab}\delta_{mk}P^{(a)}_m 
+(\alpha_a |\alpha_b)\min(t_bm, t_ak)N^{(b)}_k.
\end{align}
With these definitions we have

\begin{theorem}[\cite{KR3, HKOTY, KN3, KNT1, Her1}]\label{th:mnxbc}
The following power series formulas are valid:
\begin{equation}\label{qmn}
\begin{split}
\prod_{(a,m)\in H}({\mathcal Q}^{(a)}_m)^{\nu^{(a)}_m}
&= \frac{\sum_N\EuScript{M}(\nu,N)
e^{-\sum_{(a,m)\in H}mN^{(a)}_m\alpha_a}}
{\prod_{\alpha \in \Delta_+}(1-e^{-\alpha})}\\
&=\sum_N\EuScript{N}(\nu,N)
e^{-\sum_{(a,m)\in H}mN^{(a)}_m\alpha_a},
\end{split}
\end{equation}
where the sums run over 
$N =(N^{(a)}_m)_{(a,m) \in H} \in (\Z_{\ge 0})^H$
without any constraints.
The symbol $\Delta_+$ denotes the set of positive roots of ${\mathfrak g}$.
\end{theorem}
See Section \ref{ss:xm} how this theorem was established
by integrating many works.

Let us turn to the special case $\nu^{(a)}_m \in \Z_{\ge 0}$
for any $(a,m) \in H$.
Then the power series (\ref{qmn}) actually truncates to a polynomial,
and Theorem \ref{th:mnxbc} implies the Fermionic formulas for 
the branching coefficient $b_\lambda$ and 
the weight multiplicity $c_\lambda$ in (\ref{qdec}).
To write them down, we introduce
\begin{equation}
\EuScript{M}_\lambda =  \sum_N \EuScript{M}(\nu,N), \quad
\EuScript{N}_\lambda =  \sum_N \EuScript{N}(\nu,N)\qquad
(\lambda \in \sum_{a=1}^r\Z\omega_a),
\end{equation}
where the sums run over 
$N =(N^{(a)}_m)_{(a,m) \in H} \in (\Z_{\ge 0})^H$
satisfying the weight condition
\begin{equation}\label{wlc}
\lambda = \sum_{(a,m)\in H}
m\nu^{(a)}_m\omega_a - \sum_{(a,m)\in H}mN^{(a)}_m\alpha_a.
\end{equation}
Then the following is a corollary of Theorem \ref{th:mnxbc}:
\begin{alignat}{3}
&\prod_{a,m}(Q^{(a)}_m)^{\nu^{(a)}_m} 
= \sum_\lambda b_\lambda \,\chi(V_\lambda),&\quad
b_\lambda &= \EuScript{M}_\lambda
& \qquad &\text{for }
\lambda \in \sum_{a=1}^r \Z_{\ge 0}\, \omega_a,
\label{xbm}\\
&\prod_{a,m}(Q^{(a)}_m)^{\nu^{(a)}_m} 
=\sum_\lambda c_\lambda \,e^{\lambda},&\quad
c_\lambda &= \EuScript{N}_\lambda
&\qquad &\text{for } 
\lambda \in \sum_{a=1}^r \Z\, \omega_a.
\label{xcn}
\end{alignat}
As the generalization of (\ref{mbdef}), 
we further introduce 
\begin{equation}\label{mbx}
\overline{\EuScript{M}}_\lambda = \sum_N{}^{\!+}\;\EuScript{M}(\nu,N),
\end{equation}
where the sum $\sum_N^+$ extends over 
$N =(N^{(a)}_m)_{(a,m) \in H} \in (\Z_{\ge 0})^H$
satisfying (\ref{wlc}) and 
the extra condition that $P_m \ge 0$ whenever $N_m \ge 1$.
Then the following is the ${\mathfrak g}$ version of Theorem \ref{th:mbb}.

\begin{theorem}[\cite{KKR,KR2,KiSS,DK1}]\label{th:mbg}
For $\lambda \in \sum_{a=1}^r \Z_{\ge 0}\,\omega_a$,
the equality 
$b_\lambda = \overline{\EuScript{M}}_\lambda$ is valid.
\end{theorem}

\subsection{\mathversion{bold}$Q^{(a)}_m$ as a classical character}
\label{ss:eqc}

Here we present the expansion of $Q^{(a)}_m$ into classical characters.
Such an example has already been given in (\ref{g2e}) 
for the rank 2 algebras ${\mathfrak g}=A_2, B_2, C_2$ and $G_2$.  
Here are a few examples from $E_8$:
\begin{align*}
Q^{(1)}_1 &= \chi(V_{\omega_1}) + \chi(V_0),
\qquad 
Q^{(1)}_2 = \chi(V_{2\omega_1}) + \chi(V_{\omega_1}) 
+ \chi(V_0),\\
Q^{(2)}_1 &= \chi(V_{\omega_2})+2\chi(V_{\omega_1}) 
+ \chi(V_{\omega_7}) + \chi(V_0),\\
Q^{(3)}_1 &= \chi(V_{\omega_3})+2\chi(V_{\omega_8})
+4\chi(V_{\omega_7})+\chi(V_{\omega_1+\omega_7})\\
&+3\chi(V_{\omega_2})+\chi(V_{2\omega_1})+4\chi(V_{\omega_1})
+2\chi(V_{0}),
\end{align*} 
which satisfy a Q-system relation 
$(Q^{(1)}_1)^2 = Q^{(1)}_2 + Q^{(2)}_1$ for instance.
In general from (\ref{xbm}) and (\ref{wlc}),  the expansion takes the form   
\begin{equation}\label{chd}
Q^{(a)}_m = \chi(V_{m\omega_a}) + 
\overbrace{\sum_{\lambda <m\omega_a}
b_\lambda \chi(V_\lambda)}^{\text{called ``children"}},
\end{equation}
where $b_\lambda$ is obtained by 
specializing $\nu^{(a)}_m$ in 
(\ref{xbm}) or Theorem \ref{th:mbg}.
As we see in the above example, 
the description of the children is complicated in general 
for ${\mathfrak g}$ of exceptional types.
However, for non exceptional ${\mathfrak g}$, 
they can be described by simple combinatorial rules given below.
For simplicity we write $\chi(V_\lambda)$ as $\chi(\lambda)$.

For ${\mathfrak g}=A_r$, there is no children:
\begin{equation}\label{nca}
Q^{(a)}_m = \chi(m\omega_a).
\end{equation}
To check the relation $(Q^{(a)}_m)^2 = Q^{(a)}_{m-1}Q^{(a)}_{m+1}
+ Q^{(a-1)}_mQ^{(a+1)}_m$ is an easy exercise on Schur functions.
It is customary to depict the weights 
$m_1\omega_1+\cdots + m_r\omega_r$ $(m_i\in \Z_{\ge 0})$ 
as a Young diagram.
The rule is to regard each $\omega_a$ as a depth $a$ column.
Thus (\ref{nca}) is represented as the $a\times m$ rectangle Young diagram.
As we will see, in the other nonexceptional algebras,
the children for most $Q^{(a)}_m$ are 
described by removals of dominos from the $a\times m$ rectangle.  

For ${\mathfrak g}=C_r$, we have
\begin{equation}\label{qexc}
Q^{(a)}_m = \begin{cases}
\chi(k_1\omega_1 + \cdots + k_a\omega_a) \quad & 1 \le a \le r-1,\\
\chi (m\omega_r) \quad &a = r,
\end{cases}
\end{equation}
where the sum is taken over nonnegative integers
$k_1, \ldots, k_a$ that satisfy 
$k_1 + \cdots + k_a \le m, k_j \equiv m\delta_{j a}
\mod 2$ for all $ 1 \le j \le a$.
The summands correspond to the removals of horizontal dominos 
(shape $1\times 2$ Young diagram).

For ${\mathfrak g}=B_r$ and $D_r$, we have
\begin{equation}\label{qexbd}
\begin{split}
Q^{(a)}_m  &= 
\sum \chi(k_{a_0}\omega_{a_0} + \cdots  + k_{a -2}\omega_{a-2} 
+ \cdots + k_a\omega_a) \quad 1 \le a \le r^\prime,\\
r^\prime &= r \text{ for } B_r,  \; r' = r-2 \text{ for } D_r,\quad
a_0 \equiv a \;\;{\rm mod }\, 2, \;\; a_0 =0 \text{ or } 1,\\
Q^{(a)}_m &=\chi(m\omega_a) \quad a = r-1, r
\quad \text{ for } D_r.
\end{split}
\end{equation}
Here $\omega_0=0$. 
The sum extends over non-negative integers
$k_{a_0}, k_{a_0 + 2}, \ldots, k_a$ obeying the constraint
$t_a(k_{a_0} + k_{a_0+2} + \cdots + k_{a-2}) + k_a = m$.
The summands correspond to the removals of vertical dominos 
(shape $2\times 1$ Young diagram).

\subsection{Bibliographical notes and further aspects}\label{ss:xm}

The Q-system\footnote{They are named so in \cite{KNS2} after 
the notation $Q^{(a)}_m$ due to \cite{KR3,Kir3}, which was 
adopted to mean ``quantum character" \cite{Kirf}.} for ${\mathfrak g}$ 
first appeared in \cite{KR3,Kir3}.
In \cite{KR3}, it was claimed that (in a nowadays terminology) 
$Q^{(a)}_m={\rm res }\, \chi_q(W^{(a)}_m)$ satisfies the Q-system, 
and the generalization of Bethe's Fermionic formula 
$b_\lambda = \overline{\EuScript{M}}_\lambda$ 
(Theorem \ref{th:mbg}) holds.
These assertions became known as 
the Kirillov-Reshetikhin conjecture.
Together with the closely related formulas 
$b_\lambda = \EuScript{M}_\lambda$,
$c_\lambda = \EuScript{N}_\lambda$ and Theorem \ref{th:mnxbc},
they have now been established by the integration of numerous works since then.
Here we shall only mention the literatures that are most relevant to our 
presentation in this section.
More detailed accounts are available in \cite[section 5.7]{KNT1}
and \cite[section 1]{Her4}.

The method of multivariable residue analysis was 
initiated in \cite{Kir1,Kir2} for $A_1, A_r$ and 
extended to ${\mathfrak g}$ in \cite{HKOTY}.
The main conclusion from this approach is that the Fermionic formula
$b_\lambda = \EuScript{M}_\lambda$ follows from
the Q-system and a convergence property of ${\mathcal Q}^{(a)}_m$ 
as $m \rightarrow \infty$.
It was found in \cite{KN2, KN3} that these properties also lead to 
another version of the Fermionic formula 
$c_\lambda = \EuScript{N}_\lambda$.
The two stories $b_\lambda = \EuScript{M}_\lambda$ (``XXX type")
and $c_\lambda = \EuScript{N}_\lambda$ (``XXZ type")
were put in a unified perspective 
by a version of multivariable Lagrange inversion \cite{KNT1}
with a proper passage from the finite to infinite Q-systems.
Last but a crucial input that 
${\rm res}\, \chi_q(W^{(a)}_m)$ actually satisfies the 
Q-system for any ${\mathfrak g}$ was proved as a corollary of 
Theorem \ref{th:nh} \cite{N3,Her1} together with the convergence 
property \cite[Theorem 3.3(2)]{Her1}.
Thus, 
Theorem \ref{th:mnbc} (1) and (2) for $A_1$ 
are due to \cite{Kir1} and \cite{KN2}, respectively.
Its ${\mathfrak g}$ version, Theorem \ref{th:mnxbc}, is an outcome of 
\cite{KR3, HKOTY, KN3, KNT1, Her1}.

The identity $b_\lambda = \overline{\EuScript{M}}_\lambda$
(Theorem \ref{th:mbg}) 
has been proved by combinatorial methods in 
\cite{KKR,KR2,KiSS} for $A_r$.
Thanks to $b_\lambda = \EuScript{M}_\lambda$,
it suffice to show 
$\overline{\EuScript{M}}_\lambda = \EuScript{M}_\lambda$
for dominant $\lambda$.
A uniform proof of the latter for all ${\mathfrak g}$  
is given in \cite{DK1} by a generating function method.

The expansion of $Q^{(a)}_m$ into classical 
characters as in Section \ref{ss:eqc} 
also has a long history going back to \cite{OW}.
By many works e.g. \cite{CP1,CP2, C2,HKOTY},
such formulas have been established for 
all $Q^{(a)}_m$'s for $A_r, B_r, C_r, D_r$ 
and many ones from $E_{6,7,8}, F_4$ and $G_2$. 

We conclude with a few remarks on further aspects which have
not been discussed in this section.

(i) The series $\EuScript{M}(w)$ (\ref{eq:cKdef1}) 
has an interpretation
of the grand partition function of the ideal
gas with the Haldane exclusion statistics \cite{Wu}.
The finite $Q$-system  (\ref{eq:qsys2}) appeared
in \cite{Wu} as the thermal equilibrium condition
for the distribution functions 
of the same system\footnote{For the translation,
substitute $w_i={\mathcal Q}_i/(1-{\mathcal Q}_i)$
in equation (10) in \cite{Wu}.}.
See also \cite{IA} for another interpretation.
The one variable case (Example \ref{ex:h1})
also appeared in \cite{Su2} as the thermal equilibrium condition
for the distribution function of the Calogero-Sutherland model.
As an application of our second formula
in Theorem \ref{thm:fqmain1},
we can quickly reproduce the ``cluster expansion
formula'' in \cite[eq.\ (129)]{Ig}.
Setting $D=I$ in (\ref{eq:cKdef1})--(\ref{eq:fdef1}), we have
\begin{align}
\begin{split}
 &\ln {\mathcal Q}_i(w)
=
\Bigl[
\frac{\partial}{\partial \nu_i}
\EuScript{N}^\nu(w)\Bigr]_{\nu=0}\\
=&\
\sum_{N}
\det_{\genfrac{}{}{0pt}{1}{H(N)}{j,k\neq i}} F_{jk}(0,N)
\prod_{j\in H(N)}
\frac{1}{N_j}
\binom{P_j(0,N)+N_j-1}{N_j-1}
w^N,
\end{split}
\end{align}
where $\{Q_i(w)\}_{i\in H}$ is
the solution of (\ref{eq:qsys1}).
The Sutherland-Wu equation also
plays an important role
for the CFT spectra.
See \cite{BSc} and the references therein.

(ii) There are decent $q$-analogs of $b_\lambda$ and $c_\lambda$ 
by using the crystal base of $U_q(\hat{\mathfrak g})$ \cite{Kas}.
A typical one for $A_r$ is the Kostka-Foulkes polynomial \cite{Ma1}.
Correspondingly, there is a $q$-analog of the Fermionic formula 
$b_\lambda = \overline{\EuScript{M}}_\lambda$
known as ``$X=M$ conjecture" \cite{HKOTY, HKOTT},
which has been solved for $A_r$ \cite{KR2,KiSS} and some other cases.
There is also a conjectural $q$-analog of 
$\overline{\EuScript{M}}_\lambda = \EuScript{M}_\lambda$
\cite[eq. (4.21)]{HKOTY}.
These formulas have the level restricted versions and 
are related to RSOS models and CFT characters.
For a historical survey, see \cite[section1]{HKOTT} and \cite{Mc}.

(iii) The Q-system, Theorem \ref{th:mnxbc} and the 
expansion formula as in Section \ref{ss:eqc} 
have been generalized to 
twisted quantum affine Lie algebras $U_q(X^{(\kappa)}_N)$
\cite{HKOTT, KNT1, KNT2, Her4}.

\section{\mathversion{bold}Y-system and thermodynamic Bethe ansatz}
\label{s:y}

In this section we explain 
how the level $\ell$ restricted 
Y-system for ${\mathfrak g}$ (\ref{yade})--(\ref{yg}) 
emerges from 
the thermodynamic Bethe ansatz (TBA) equation 
associated with $U_q(\hat{\mathfrak g})$ at 
$q = \exp(\frac{\pi \sqrt{-1}}{t(\ell+h^\vee)})$.
(See (\ref{eq:t1}) and (\ref{hhd}) for $t$ and $h^\vee$.)
The TBA equation is relevant to level $\ell$ RSOS models and 
quoted from Section \ref{s:tbarsos}.
We also introduce the constant Y-system and explain its 
relation to the Q-system in the both unrestricted and 
level restricted versions.
Conjecturally, the level restricted Q-system 
allows a solution via a specialization 
of characters to the $q$-dimension with $q$ being the root of unity.
They play important roles in the 
dilogarithm identity related to conformal field theory and 
the TBA analysis of RSOS models.
We use the notation
\begin{align}
\ell_a &= t_a\ell,\quad L=\ell+h^\vee,\label{lad}\\
H_\ell &= \{(a,m)\mid a \in I, 1 \le m \le \ell_a-1, m \in \Z\}, \label{hlde}
\end{align}
where $t_a$ is defined in (\ref{eq:t1}) and 
$h^\vee$ is the dual Coxeter number of ${\mathfrak g}$ (\ref{hhd}).
The set $H_\ell$ is a level truncation of $H$ (\ref{hdef}).
We will further use 
\begin{align}
&t_{ab}=\max(t_a, t_b),\label{tabd}\\
&N_{ab} = 2\delta_{ab}-B_{ab},
\quad B_{ab} = B_{ba} = \frac{t_b}{t_{ab}}C_{ab} 
= \begin{cases}
2 & C_{ab}=2,\\
-1 & C_{ab}<0,\\
0 & C_{ab}=0.
\end{cases}\label{nabd}
\end{align}
This $B_{ab}$ is the same as (\ref{Bd}).

\subsection{\mathversion{bold}Y-system for ADE and 
deformed Cartan matrices}
\label{ss:ydc}

For simplicity we first 
deal with the simply laced algebras ${\mathfrak g}=A_r, D_r$ and $E_{6,7,8}$.
In Section \ref{ss:tbarsos}, 
we obtain the TBA equation for level $\ell\,(\ell \in \Z_{\ge 2})$ 
critical RSOS model in (\ref{tba3}).
It is the following nonlinear integral equation on the functions 
$\{\epsilon^{(a)}_m(u)\mid (a,m) \in H_\ell, u \in \R\}$: 
\begin{equation}\label{tba4}
\begin{split}
\frac{\epsilon \beta\gamma\delta_{p a}\delta_{s m}}
{4\cosh(\pi u/2)} =\beta\epsilon^{(a)}_m(u)
+\int_{-\infty}^\infty dv 
\frac{\ln\left[
\frac{\prod_{b\in I}
\bigl(1 + \exp(-\beta\epsilon^{(b)}_m(v))\bigr)^{N_{ab}}}
{\bigl(1 + \exp(\beta\epsilon^{(a)}_{m-1}(v))\bigr)
\bigl(1 + \exp(\beta\epsilon^{(a)}_{m+1}(v))\bigr)}
\right]}
{4\cosh(\pi(u-v)/2)}.
\end{split}
\end{equation}
Here $\beta, \gamma >0$, $\epsilon=\pm 1$ and $(p,s)\in H_\ell$ 
are model parameters  
specifying the temperature, 
normalization of energy, two critical regimes 
and representation $W^{(p)}_s$ (fusion type) 
with which the model is associated, respectively.
The physical meaning of 
$\epsilon^{(a)}_m(u)$ is the pseudo energy defined by 
$\exp(-\beta\epsilon^{(a)}_m(u)) = 
\rho^{(a)}_m(u) / \sigma^{(a)}_m(u)$ in terms of 
the color $a$ length $m$ 
string density $\rho^{(a)}_m(u)$ and hole density $\sigma^{(a)}_m(u)$.
More details can be found in Section \ref{ss:tbarsos}, but we do not need
those background here.

We assume that (\ref{tba4}) can be analytically continued 
off the real axis of $u$ until $|\im\, u|\le 1$.
Setting $u\rightarrow u \pm i \mp 0i$, take the sum of the resulting two
equations.  The LHS vanishes and the RHS is evaluated by means of 
\begin{equation}\label{rdd}
\frac{1}{4\cosh\frac{\pi}{2}(u-v+i-0i)}+
\frac{1}{4\cosh\frac{\pi}{2}(u-v-i+0i)} = \delta(u-v)
\end{equation}
as the convolution kernel.
By introducing the variable 
$Y^{(a)}_m(u) = \exp(-\beta\epsilon^{(a)}_m(u))$, 
the Boltzmann factor of the pseudo energy, 
the result is the logarithm of
\begin{equation}\label{hip}
Y^{(a)}_m(u-i)Y^{(a)}_m(u+i) 
= \frac{\prod_{b \in I}(1+Y^{(b)}_m(u))^{N_{ab}}}
{(1+Y^{(a)}_{m-1}(u)^{-1})(1+Y^{(a)}_{m+1}(u)^{-1})}.
\end{equation}
This is the Y-system for 
${\mathfrak g}=A_r, D_r$ and 
$E_{6,7,8}$ (\ref{yade}) in the convention that 
$Y^{(a)}_m(u+k)$ there becomes $Y^{(a)}_m(u+ik)$. 
It is level $\ell$ restricted since only $Y^{(a)}_m(u)$ 
with $(a,m) \in H_\ell$ are present.

Notice that the LHS of (\ref{tba4}) 
that had carried the model dependent information 
$\beta, \gamma, \epsilon$ and $(p,s)$ disappeared all together.
In this sense, the Y-system is a universal feature
of all the physical systems described by the TBA equation (\ref{tba4})
whose LHS is any $2i$-antiperiodic function of $u$.
Put it differently, the LHS 
encodes the specific properties in each model
that are coupled as a driving term to the universal structure
(Y-system).

Let us observe another aspect of the Y-system (\ref{hip}).
It is written as
\begin{equation}\label{pih}
\frac{(1+Y^{(a)}_{m}(u-i)^{-1})(1+Y^{(a)}_{m}(u+i)^{-1})}
{(1+Y^{(a)}_{m-1}(u)^{-1})(1+Y^{(a)}_{m+1}(u)^{-1})}
= 
\frac{(1+Y^{(a)}_m(u-i))(1+Y^{(a)}_m(u+i))}
{\prod_{b \in I}(1+Y^{(b)}_m(u))^{N_{ab}}}.
\end{equation}
The LHS and RHS of (\ref{hip}) possess parallel structures 
related to $A_{\ell-1}$ and ${\mathfrak g}$, respectively.
In the Fourier space they are encoded in the deformed Cartan matrices
with indices corresponding to the length $m$ and 
the color $a$, respectively.
To see it, define the Fourier transformation 
${\hat f}={\hat f}(x)$ of $f=f(u)$ by
\begin{equation}\label{ft}
f(u) = {1 \over 2 \pi}
\int_{-\infty}^\infty {\hat f}(x) e^{i u x} dx,\quad
{\hat f}(x) = \int_{-\infty}^\infty f(u) e^{-i u x} du.
\end{equation}
If we formally interpret the  
multiplication with $e^{cx}$ $(c \in \R)$ 
in the Fourier $(x)$ space as the difference operator 
$u\rightarrow u - ic$ in the ``real" $(u)$ space,
the logarithm of the RHS of (\ref{pih}) is assigned with 
the Fourier component 
$\sum_{b\in I}{\hat {\mathcal M}}_{ab}(x)
\widehat{\ln}\bigl(1+Y^{(b)}_m\bigr)$,
where
\begin{equation}\label{mka}
{\hat {\mathcal M}}_{ab}(x) = 2\delta_{ab}\cosh x -N_{ab} \quad
(\text{for ADE})
\end{equation}
is the deformed Cartan matrix of ${\mathfrak g}$.
Actually the Fourier transformation of the TBA equation (\ref{tba4})
contains $\sum_{b \in I}\frac{{\hat {\mathcal M}}_{ab}(x)}{2\cosh x}\,
\widehat{\ln}\bigl(1+Y^{(b)}_m\bigr)$ 
so that the identity (\ref{rdd}) works in the real space.
Parallel remarks apply to the LHS of (\ref{pih}).

We call the functions like ${\hat {\mathcal M}}_{ab}(x)$ 
TBA kernels as they emerge in the TBA calculation 
(Section \ref{ss:tbarsos})
and play important roles as 
building blocks of integral kernels  in the TBA equation.

\subsection{\mathversion{bold}TBA kernels}\label{ss:ker}

Here we summarize the definitions and useful properties 
of the TBA kernels for general ${\mathfrak g}$.
In place of (\ref{mka}),  
we redefine ${\hat {\mathcal M}}_{a b}(x)$ and 
introduce ${\hat {\mathcal K}}^{m n}_a(x)$ as
\begin{align}
{\hat {\mathcal M}}_{a b}(x) 
&=2\delta_{a b}\cosh\bigl({x\over t_a}\bigr) - N_{ab}
= B_{a b} + 2\delta_{a b}
\Bigl( \cosh\bigl(\frac{x}{t_a}\bigr) - 1 \Bigr),
\label{mab}\\
{\hat {\mathcal K}}^{m n}_a(x) &=
\delta_{m n} - 
\frac{\delta_{m, n-1} + \delta_{m,n+1}}
{2\cosh\bigl({x\over t_a}\bigr)}.
\label{kan}
\end{align}
For $(a,m), (b,k) \in H_\ell$, we further introduce
\begin{align}
{\hat {\mathcal A}}^{m k}_{a b}(x)&= 
{\sinh\bigl(\hbox{min}({m \over t_a}, {k \over t_b})x \bigr) 
 \sinh\bigl((\ell-\hbox{max}({m \over t_a}, {k \over t_b}))x \bigr) 
\over 
\sinh({x\over t_{a b}}) \sinh(\ell x) }, \label{aab}\\
{\hat {\mathcal K}}^{m k}_{a b}(x)&= 
{\hat {\mathcal A}}^{m k}_{a b}(x){\hat {\mathcal M}}_{a b}(x),
\label{kab}\\
{\hat{\mathcal J}}^{m k}_{a b}(x)
&= \sum_{n=1}^{\ell_a - 1}{\hat {\mathcal K}}^{m n}_a(x)
{\hat {\mathcal K}}^{n k}_{a b}(x)
= \frac{{\hat {\mathcal M}}_{a b}(x){\hat {\mathcal P}}^{mk}_{ab}(x)} 
{2\cosh\bigl({x\over t_a}\bigr)}, \label{jha}\\
{\hat {\mathcal P}}^{mk}_{ab}(x) &=
2\cosh\bigl({x\over t_a}\bigr)
\sum_{n=1}^{\ell_a-1}{\hat {\mathcal K}}^{m n}_a(x)
{\hat {\mathcal A}}^{n k}_{a b}(x) \label{jas}\\
&=
{\sinh(\frac{x}{t_a}) \over \sinh(\frac{x}{t_{a b}}) } 
\delta_{t_bm,  t_ak} 
+ \sum_{j=1}^{t_b - t_a}
{\sinh(\frac{jx}{t_b}) \over \sinh(\frac{x}{t_b}) }
\bigl(\delta_{\frac{t_b}{t_a}(m+1) - j, \, k}
+ \delta_{\frac{t_b}{t_a}(m-1) + j, \, k}
\bigr).\nonumber
\end{align}
The sum $\sum_{j=1}^{t_b - t_a}$ in (\ref{jas}) 
is to be understood as zero if $t_a \ge t_b$.
Since the latter expressions in (\ref{jha}) 
and (\ref{jas}) do not contain 
$\ell$, we can and do extend the definition of 
${\hat {\mathcal J}}^{m k}_{a b}(x)$
and $\hat{{\mathcal P}}^{m k}_{a b}(x)$
to all the nonnegative integers $m, k \ge 0$.
The inverse Fourier transform 
${\mathcal J}^{m k}_{a b}(u)$ 
is an even function of $u$ 
but ${\mathcal J}^{m k}_{a b}(u) 
\neq {\mathcal J}^{k m}_{b a}(u)$ 
in general as opposed to 
${\hat {\mathcal A}}^{m k}_{a b}(x) 
= {\hat {\mathcal A}}^{k m}_{b a}(x)$
and 
${\hat {\mathcal K}}^{m k}_{a b}(x) 
= {\hat {\mathcal K}}^{k m}_{b a}(x)$.
The ${\hat {\mathcal K}}^{m n}_a(x)$ in (\ref{kan}) 
should be distinguished from 
${\hat {\mathcal K}}^{m n}_{a a}(x)$ in (\ref{kab}).
The following relations are easily checked:
\begin{align}
&2\cosh\bigl({x\over t_a}\bigr)\sum_{n=1}^{\ell_a-1}
{\hat {\mathcal A}}_{a a}^{m n}(x)
{\hat {\mathcal K}}_a^{n k}(x) = \delta_{m k},\label{inv}\\
&2\cosh\bigl({x\over t_a}\bigr)\sum_{n=1}^{\ell_a-1}
{\hat {\mathcal A}}_{a a}^{m n}(x)
{\hat {\mathcal J}}_{a b}^{n k}(x)
= {\hat {\mathcal K}}_{a b}^{m k}(x)\label{mak},\\
&{\hat{\mathcal J}}^{m k}_{a b}(x)
= \delta_{ab}\delta_{mk} - 
\frac{N_{a b}{\hat {\mathcal P}}^{mk}_{ab}(x)} 
{2\cosh\bigl({x\over t_a}\bigr)}. \label{jky}
\end{align} 
All the TBA kernels
(\ref{mab})--(\ref{jas}) are deduced from
${\hat {\mathcal A}}_{a b}^{m n}(x)$ and
${\hat {\mathcal M}}_{a b}(x)$ by using these relations.
The basic ones ${\hat {\mathcal A}}_{a b}^{m n}(x)$ and
${\hat {\mathcal M}}_{a b}(x)$  
are obtained as 
\begin{align}
&\int_{-\infty}^\infty du e^{-iux}
\frac{\partial}{\partial u}
\Theta^m_a\bigl(u ,\frac{s}{t_a}\bigr)
={\hat {\mathcal A}}^{m s}_{a a}(x)\vert_{\ell\rightarrow L},
\label{fte1}\\
&\int_{-\infty}^\infty du e^{-iux}
\frac{\partial}{\partial u}\Theta^{m k}_{a b}(u,(\alpha_a| \alpha_b))
= -\delta_{ab}\delta_{mk}
+{\hat {\mathcal M}}_{a b}(x)
{\hat {\mathcal A}}^{m k}_{a b}(x)\vert_{\ell\rightarrow L},
\label{fte2}
\end{align}
where $\Theta^m_a\bigl(u ,\frac{s}{t_a}\bigr)$ (\ref{tet1}) and 
$\Theta^{m k}_{a b}(u,(\alpha_a| \alpha_b))$ (\ref{tet2}) are the logarithm 
of the LHS and the RHS of the Bethe equation under the 
string hypothesis, respectively. See (\ref{ber})--(\ref{tet2}).

When ${\mathfrak g}$ is simply laced, the TBA kernels simplify as
\begin{align}
{\hat {\mathcal A}}^{m k}_{a b}(x) 
&= {\sinh\bigl(\hbox{min}(m, k)x \bigr) 
 \sinh\bigl((\ell-\hbox{max}(m, k))x \bigr) 
\over 
\sinh x \sinh(\ell x) },\label{saab}\\
{\hat{\mathcal J}}^{m k}_{a b}(x)
&=
\frac{{\hat {\mathcal M}}_{a b}(x)\delta_{mk}} 
{2\cosh x}
=\left(\delta_{ab} - \frac{N_{a b}}{2\cosh x}\right)
\delta_{mk}, \label{sjha}\\
{\hat {\mathcal P}}^{mk}_{ab}(x) &= \delta_{mk}. \label{sjas}
\end{align}

\subsection{\mathversion{bold}Y-system for ${\mathfrak g}$ from TBA equation}
\label{ss:ytba}

Let us derive the level $\ell$ restricted Y-system 
for general ${\mathfrak g}$
from the TBA equation. 
We quote the latter obtained in (\ref{tba2})
with the notation $Y^{(a)}_m(u) = \exp(-\beta\epsilon^{(a)}_m(u))$:
\begin{equation}\label{tba5}
\begin{split}
\frac{\epsilon \beta\gamma\delta_{p a}\delta_{s m}}
{4t^{-1}_p\cosh(t_p\pi u/2)} &=-\ln Y^{(a)}_m(u)
-\int_{-\infty}^\infty dv 
\frac{\ln\left[\bigl(1 + Y^{(a)}_{m-1}(v)^{-1}\bigr)
\bigl(1 + Y^{(a)}_{m+1}(v)^{-1}\bigr)\right]}
{4t_a^{-1}\cosh(t_a\pi(u-v)/2)}\\
&+\sum_{(b,k) \in H_\ell}N_{ab}\int_{-\infty}^\infty dv 
\frac{\left[
{\mathcal P}^{mk}_{ab}\ast \ln\bigl(1 + 
Y^{(b)}_k\bigr)\right](v)}
{4t_a^{-1}\cosh(t_a\pi(u-v)/2)}.
\end{split}
\end{equation}
${\mathcal P}^{m k}_{a b}$ is defined via its Fourier component 
(\ref{jas}) and  $\ast$ denotes the convolution
\begin{equation}
(f_1 \ast f_2)(u) = \int_{-\infty}^\infty dv f_1(u-v)f_2(v).
\end{equation}

As the simply laced case, 
we assume that (\ref{tba5}) can be analytically continued 
off the real axis of $u$ until $|\im\, u| \le t^{-1}_a$.
Then the sum after the shifts $u \rightarrow u \pm t^{-1}_ai \mp 0i$ 
eliminates the LHS, giving
\begin{equation}\label{sry}
\begin{split}
\ln\left[Y^{(a)}_m(u-\textstyle{\frac{i}{t_a}})
Y^{(a)}_m(u+\textstyle{\frac{i}{t_a}})\right]
&= -\ln\left[\bigl(1 + Y^{(a)}_{m-1}(u)^{-1}\bigr)
\bigl(1 + Y^{(a)}_{m+1}(u)^{-1}\bigr)\right]\\
&+\sum_{(b,k) \in H_\ell}N_{ab}
\left[{\mathcal P}^{mk}_{ab}\ast 
\ln\bigl(1 + Y^{(b)}_k\bigr)\right](u).
\end{split}
\end{equation}
For simply laced algebras, 
$P^{mk}_{ab}(u) = \delta_{mk}\delta(u)$ by (\ref{sjas}),  
and we are done.
To illustrate the general case,
take ${\mathfrak g} = G_2$ with $(a,b)=(1,2)$ as an example.
Then $(t_a, t_b)=(1,3)$ and (\ref{jas}) reads
\begin{equation*}
\begin{split}
&{\hat {\mathcal P}}^{m k}_{a b}(x)  
= {\hat {\mathcal P}}^{m k}_{1 2}(x) 
= (e^{\frac{2x}{3}}+1+e^{-\frac{2x}{3}})\delta_{3m, k} 
+ \delta_{3m-2, k}+\delta_{3m+2, k}\\
&\qquad\qquad\qquad \qquad+(e^{\frac{x}{3}}+e^{-\frac{x}{3}})
(\delta_{3m-1, k}+\delta_{3m+1, k}),\\
&{\mathcal P}^{m k}_{1 2}(u) = 
(\delta(u-\textstyle{\frac{2i}{3}})+\delta(u) + 
\delta(u+\textstyle{\frac{2i}{3}}))\delta_{3m, k}
+\delta(u)(\delta_{3m-2, k}+\delta_{3m+2, k})\\
&\qquad \qquad
+(\delta(u-\textstyle{\frac{i}{3}})
+\delta(u+\textstyle{\frac{i}{3}}))
(\delta_{3m-1, k}+\delta_{3m+1, k}).
\end{split}
\end{equation*}
If $\ln(1+Y^{(2)}_k(v))$ is analytic 
in the strip $|\im\, v| \le \frac{2}{3}$\footnote{
Actually $|\im\, v| \le \frac{2}{3}$
for $\ln(1+Y^{(2)}_{3m}(v))$ and 
$|\im\, v| \le \frac{1}{3}$ for 
$\ln(1+Y^{(2)}_{3m\pm 1}(v))$ suffice.} and 
decays rapidly as $|\re\, v | \rightarrow \infty$,
one can shift the convolution integral 
$\int dv {\mathcal P}^{mk}_{12}(u-v)\ln(1+Y^{(2)}_k(v))$
off the real axis of $v$ to pick the support of delta functions.
In this way the last term in (\ref{sry}) 
gives the logarithm of 
\begin{equation*}
\begin{array}{l}
\textstyle 
(1+Y^{(2)}_{3m}\left(u-\frac{2i}{3}\right))
(1+Y^{(2)}_{3m}\left(u\right))
(1+Y^{(2)}_{3m}\left(u+\frac{2i}{3}\right))
(1+Y^{(2)}_{3m-2}(u))
(1+Y^{(2)}_{3m+2}(u))\\
\textstyle
\;\;\times(1+Y^{(2)}_{3m-1}\left(u-\frac{i}{3}\right))
(1+Y^{(2)}_{3m-1}\left(u+\frac{i}{3}\right))
(1+Y^{(2)}_{3m+1}\left(u-\frac{i}{3}\right))
(1+Y^{(2)}_{3m+1}\left(u+\frac{i}{3}\right)).
\end{array}
\end{equation*}
This is the numerator of the RHS in the first relation of 
the Y-system for $G_2$ (\ref{yg}) with the 
shift unit multiplied by $i$.

The general case is similar and (\ref{sry}) 
gives rise to the logarithmic form of the 
(restricted) Y-system for ${\mathfrak g}$.
On account of (\ref{jas}),  
in general it suffices to assume that 
$\ln(1+Y^{(a)}_m(u))$ is analytic in the strip
$|\im\, u| \le \frac{t_a-1}{t_a}$ and 
decays rapidly as $|\re\, u | \rightarrow \infty$.

If the analyticity argument can be left out, 
the Y-system is deduced more quickly 
from the TBA kernels in the Fourier space.
In fact, one can start with the TBA equation (\ref{tba}) 
without the LHS\footnote{
According to our previous argument,
it is actually more proper to suppress the LHS after multiplying 
$2\cosh(\frac{x}{t_a})$.}:
\begin{equation}\label{kyj}
\sum_{n=1}^{\ell_a - 1}
{\hat {\mathcal K}}^{m n}_a(x) \,{\widehat \ln} 
\bigl(1 + (Y^{(a)}_n)^{-1}\bigr) 
=\sum_{(b,k) \in H_\ell}
{\hat {\mathcal J}}^{m k}_{a b}(x)\,
{\widehat \ln}\bigl(1 + Y^{(b)}_k\bigr).
\end{equation}
Multiply with $2\cosh(\frac{x}{t_a})$ 
and use (\ref{kan}) and (\ref{jky}) to rearrange it slightly as
\begin{equation}\label{ynp}
\begin{split}
2\cosh\bigl(\frac{x}{t_a}\bigr)\,
\widehat{\ln}Y^{(a)}_m &= 
\sum_{(b,k)\in H_\ell}N_{ab}{\hat {\mathcal P}}^{mk}_{ab}(x)\,
\widehat{\ln}\bigl(1+Y^{(b)}_k\bigr)\\
&-\widehat{\ln}\left[\bigl(1+(Y^{(a)}_{m-1})^{-1}\bigr)
\bigl(1+(Y^{(a)}_{m+1})^{-1}\bigr)\right].
\end{split}
\end{equation}
This is the Y-system if 
$\cosh(\frac{x}{t_a})$ and 
${\hat {\mathcal P}}^{mk}_{ab}(x)$ (\ref{jas}) 
are regarded as 
the difference operators as mentioned after (\ref{ft}).

We have demonstrated that  
the Y-system is a difference equation
whose structure is governed by the TBA kernels.
On the other hand, recall that Theorem \ref{th:ty1} offers another route to 
obtain the Y-system by invoking its connection to the T-system.
It is yet to be understood why the two ``characterizations" 
of the Y-system coincide.

\subsection{Constant Y-system}\label{ss:cy}

In either unrestricted or level $\ell$ restricted Y-system, 
one can discard the dependence of 
$Y^{(a)}_m(u)$ on $u$. 
The resulting algebraic equation on 
$Y^{(a)}_m = Y^{(a)}_m(u)$ is called the 
unrestricted or level $\ell$ restricted constant Y-system\footnote{
The level $\ell$ restricted constant Y-system here
is the same with the one introduced in Section \ref{t:ss:di}.}.

The unrestricted constant Y-system for ${\mathfrak g}$ 
is the set of algebraic 
equations on $\{Y^{(a)}_m\mid (a,m) \in H\}$. 
($H$ is defined in (\ref{hdef}).)

\noindent
For simply laced ${\mathfrak g}$, it has the form 
\begin{equation}\label{cyade}
(Y^{(a)}_m)^2
=\frac{
\prod_{b\in I: C_{ab}=-1}
(1+Y^{(b)}_{m})}
{(1+(Y^{(a)}_{m-1})^{-1})(1+(Y^{(a)}_{m+1})^{-1})},
\end{equation}
where $(Y^{(a)}_0)^{-1}=0$. See (\ref{yade}).
The nonsimply laced case is similarly written down 
from (\ref{yb})-(\ref{yg}).
 
\noindent
For ${\mathfrak g}=B_r$,
\begin{equation}\label{cyb}
\begin{split}
(Y^{(a)}_m)^2
&=
\frac{(1+Y^{(a-1)}_{m})(1+Y^{(a+1)}_{m})}
{(1+(Y^{(a)}_{m-1})^{-1})(1+(Y^{(a)}_{m+1})^{-1})} \qquad 
 (1\leq a\leq r-2), \\
(Y^{(r-1)}_m)^2
&=
\frac{\textstyle
(1+Y^{(r-2)}_{m})
(1+Y^{(r)}_{2m-1})(1+Y^{(r)}_{2m})^2
(1+Y^{(r)}_{2m+1})}
{(1+(Y^{(r-1)}_{m-1})^{-1})(1+(Y^{(r-1)}_{m+1})^{-1})},\\
(Y^{(r)}_{2m})^2
&=
\frac{1+Y^{(r-1)}_{m}}
{(1+(Y^{(r)}_{2m-1})^{-1})(1+(Y^{(r)}_{2m+1})^{-1})},\\
(Y^{(r)}_{2m+1})^2
&=
\frac{1}{(1+(Y^{(r)}_{2m})^{-1})(1+(Y^{(r)}_{2m+2})^{-1})}.
\end{split}
\end{equation}
For ${\mathfrak g}=C_r$,
\begin{equation}\label{cyc}
\begin{split}
(Y^{(a)}_m)^2
&=
\frac{
(1+Y^{(a-1)}_{m})(1+Y^{(a+1)}_{m})}
{(1+(Y^{(a)}_{m-1})^{-1})(1+(Y^{(a)}_{m+1})^{-1})}
\qquad (1\leq a\leq r-2),\\
(Y^{(r-1)}_{2m})^2
&=
\frac{(1+Y^{(r-2)}_{2m})(1+Y^{(r)}_{m})}
{(1+(Y^{(r-1)}_{2m-1})^{-1})(1+(Y^{(r-1)}_{2m+1})^{-1})},\\
(Y^{(r-1)}_{2m+1})^2
&=
\frac{1+Y^{(r-2)}_{2m+1}}
{(1+(Y^{(r-1)}_{2m})^{-1})(1+(Y^{(r-1)}_{2m+2})^{-1})},\\
(Y^{(r)}_{m})^2
&=
\frac{
\textstyle
(1+Y^{(r-1)}_{2m-1})
(1+Y^{(r-1)}_{2m})^2
(1+Y^{(r-1)}_{2m+1})
\textstyle}
{(1+(Y^{(r)}_{m-1})^{-1})(1+(Y^{(r)}_{m+1})^{-1})}.
\end{split}
\end{equation}
For ${\mathfrak g}=F_4$,
\begin{equation}\label{cyf}
\begin{split}
(Y^{(1)}_m)^2
&=
\frac{1+Y^{(2)}_{m}}
{(1+(Y^{(1)}_{m-1})^{-1})(1+(Y^{(1)}_{m+1})^{-1})},\\
(Y^{(2)}_m)^2
&=
\frac{
\textstyle
(1+Y^{(1)}_{m})
(1+Y^{(3)}_{2m-1})(1+Y^{(3)}_{2m})^2(1+Y^{(3)}_{2m+1})
\textstyle}
{(1+(Y^{(2)}_{m-1})^{-1})(1+(Y^{(2)}_{m+1})^{-1})},\\
(Y^{(3)}_{2m})^2
&=
\frac{(1+Y^{(2)}_{m})(1+Y^{(4)}_{2m})}
{(1+(Y^{(3)}_{2m-1})^{-1})(1+(Y^{(3)}_{2m+1})^{-1})},\\
(Y^{(3)}_{2m+1})^2
&=
\frac{1+Y^{(4)}_{2m+1}}
{(1+(Y^{(3)}_{2m})^{-1})(1+(Y^{(3)}_{2m+2})^{-1})},\\
(Y^{(4)}_m)^2
&=
\frac{1+Y^{(3)}_{m}}
{(1+(Y^{(4)}_{m-1})^{-1})(1+(Y^{(4)}_{m+1})^{-1})}.
\end{split}
\end{equation}
For ${\mathfrak g}=G_2$,
\begin{equation}\label{cyg}
\begin{split}
(Y^{(1)}_m)^2
&=
\frac{(1+Y^{(2)}_{3m-2})(1+Y^{(2)}_{3m-1})^2(1+Y^{(2)}_{3m})^3
(1+Y^{(2)}_{3m+1})^2
(1+Y^{(2)}_{3m+2})}
{(1+(Y^{(1)}_{m-1})^{-1})(1+(Y^{(1)}_{m+1})^{-1})},\\
(Y^{(2)}_{3m})^2
&=
\frac{1+Y^{(1)}_m}
{(1+(Y^{(2)}_{3m-1})^{-1})(1+(Y^{(2)}_{3m+1})^{-1})},\\
(Y^{(2)}_{3m+1})^2
&=
\frac{1}
{(1+(Y^{(2)}_{3m})^{-1})(1+(Y^{(2)}_{3m+2})^{-1})},\\
(Y^{(2)}_{3m+2})^2
&=
\frac{1}
{(1+(Y^{(2)}_{3m+1})^{-1})(1+(Y^{(2)}_{3m+3})^{-1})}.
\end{split}
\end{equation}

The level $\ell$ restricted constant Y-system for $\mathfrak{g}$  
is obtained from (\ref{cyade})-(\ref{cyg})
by setting $(Y^{(a)}_{t_a\ell})^{-1}=0$ and 
naturally restricting the variables 
$\{Y^{(a)}_m\mid (a,m) \in H\}$ to 
$\{Y^{(a)}_m\mid (a,m) \in H_\ell\}$.
($H_\ell$ is defined in (\ref{hlde}).)

For the TBA analysis,  it is useful to 
recognize that the level $\ell$ restricted constant Y-system 
is expressed in terms of 
the $0$th Fourier component ($x=0$) of the TBA kernels.
We prepare the notations for them.
\begin{align}
{\bar C}^a_{m n}&=2{\hat {\mathcal K}}^{m n}_a(0),\quad
({\bar C}^a_{m n})_{1\le m, n \le \ell_a-1}
= \text{Cartan matrix of }\,A_{\ell_a-1}, \label{camn}\\
K^{m k}_{a b} &= {\hat {\mathcal K}}^{m k}_{a b}(0)
= \Bigl(\hbox{min}(t_bm, t_ak) - {m k\over \ell}\Bigr)
(\alpha_a \vert \alpha_b),\label{kkab}\\
P^{m k}_{a b} &= {\hat {\mathcal P}}^{m k}_{a b}(0)
=  \frac{t_{ab}}{t_a}\delta_{t_bm,  t_ak} 
+ \sum_{j=1}^{t_b - t_a}
j\bigl(\delta_{\frac{t_b}{t_a}(m+1) - j, \, k}
+ \delta_{\frac{t_b}{t_a}(m-1) + j, \, k}
\bigr),\label{pzd}\\
J^{k m}_{b a} &= {\hat {\mathcal J}}_{b a}^{k m}(0)
= \frac{1}{2}\sum_{n=1}^{\ell_b - 1} 
{\bar C}^{b}_{k n} K^{m n}_{a b}
= \delta_{ab}\delta_{mk}
-\frac{1}{2}N_{ab}P^{km}_{ba}
=-\frac{1}{2}G_{am, bk},\label{jba}
\end{align}
where (\ref{kab}) -- (\ref{jky}) are used.
$G_{am,bk}$ is defined in (\ref{gak}).
The sum $\sum_{j=1}^{t_b - t_a}$ in (\ref{pzd})
is to be understood as zero if $t_a \ge t_b$ as in (\ref{jas}).
Note that $K^{m k}_{a b}= K^{k m}_{b a}$ but 
$P^{m k}_{a b}\neq P^{k m}_{b a}$ 
and $J^{m k}_{a b}\neq J^{k m}_{b a}$ in general.
We have $P^{mk}_{ab} \in \Z$.
{}From (\ref{camn}), the specialization $x=0$ of (\ref{inv}) gives
\begin{equation}\label{akd}
\sum_{n=1}^{\ell_a-1}{\hat {\mathcal A}}^{m n}_{a a}(0)
{\bar C}^a_{n k}= \delta_{m k}.
\end{equation}

Using $N_{a b}$ and $P^{mk}_{ab}$ 
in the above, the level $\ell$ restricted constant Y-system
is expressed uniformly for all ${\mathfrak g}$ as 
\begin{equation}\label{yyn}
(Y^{(a)}_m)^2 = 
\frac{\prod_{(b,k) \in H_\ell}(1+Y^{(b)}_k)^{N_{ab}P^{mk}_{ab}}}
{(1+(Y^{(a)}_{m-1})^{-1})(1+(Y^{(a)}_{m+1})^{-1})}
\qquad ((a,m) \in H_\ell),
\end{equation}
where $(Y^{(a)}_0)^{-1}=0$.
This is easily seen from (\ref{ynp}).
The unrestricted version is similarly presented 
by replacing $H_\ell$ here with $H$.

The level $\ell$ restricted constant Y-system
is expressed in several guises:
\begin{align}
&\sum_{n=1}^{\ell_a - 1}
{\hat {\mathcal K}}^{m n}_a(0) 
\ln\bigl(1 + (Y^{(a)}_n)^{-1}\bigr) 
=\sum_{(b,k) \in H_\ell}
{\hat {\mathcal J}}^{m k}_{a b}(0)\ln \bigl(1 + Y^{(b)}_k\bigr),
\label{kyj0}\\
&f^{(a)}_m = 
\prod_{(b,k) \in H_\ell}(1-f^{(b)}_k)^{K^{mk}_{ab}},\qquad
\text{where}\;\;f^{(a)}_m = \frac{Y^{(a)}_m}{1+Y^{(a)}_m}.
\label{fsy}
\end{align}
The form (\ref{kyj0}) directly follows from 
(\ref{kyj}) and shows up naturally as the TBA equation 
in a certain asymptotic limit. See (\ref{bkab}).
On the other hand, 
(\ref{fsy}) is deduced from (\ref{camn}) and (\ref{jba}).
It is related to the conjectural $q$-series formula \cite{KNS1} for the 
string function $c^{\ell\Lambda_0}_\lambda(q)$ \cite{Ka} 
of the level $\ell$ vacuum module of 
$\hat{\mathfrak{g}}$ up to a power of $q$:
\begin{equation}\label{qkns}
\prod_{j=1}^\infty(1-q^j)^{-\mathrm{rank}\,\mathfrak{g}}
\sum_{\{N^{(a)}_m\}}\frac{q^{\frac{1}{2}\sum_{(a,m), (b,k) \in H_\ell}
K^{mk}_{ab}N^{(a)}_mN^{(b)}_k}}
{\prod_{(a,m) \in H_\ell}(1-q)(1-q^2)\cdots (1-q^{N^{(a)}_m})}.
\end{equation}
The outer sum is over 
$N^{(a)}_m \in \Z_{\ge 0}$
such that 
$\sum_{(a,m) \in H_\ell}mN^{(a)}_m\alpha_a \equiv \lambda
\mod \ell \sum_{a\in I}\Z\, t_a\alpha_a$. 
In fact, the crude approximation of the extremum condition 
on the summand is
\begin{equation*}
q^{\sum_{(b,k)}K^{mk}_{ab}N^{(b)}_k} = 1-q^{N^{(a)}_m},
\end{equation*}
which is cast into (\ref{fsy}) upon setting 
$q^{N^{(a)}_m} = 1-f^{(a)}_m$.

The level $\ell$ restricted constant Y-system is the set of 
$|H_\ell|$ algebraic equations on 
the same number of unknowns $\{Y^{(a)}_m\mid (a,m) \in H_\ell\}$.
With regard to its solution, the uniqueness of the positive real one
(Theorem \ref{t:thm:unique}) is fundamental.
The concrete construction of the solution is a subject of the 
subsequent sections \ref{ss:rwq} and \ref{ss:qru}.

\subsection{\mathversion{bold}Relation with Q-system.}
\label{ss:rwq}

Recall that the unrestricted Q-system for 
${\mathfrak g}$ (\ref{qsu}) is 
\begin{equation}\label{qtu}
(Q^{(a)}_m)^2 = 
Q^{(a)}_{m-1}Q^{(a)}_{m+1} 
+ (Q^{(a)}_m)^2
\prod_{(b,k) \in H}(Q^{(b)}_k)^{-2J^{ba}_{km}},
\end{equation}
where we have replaced 
the notation of the power $G_{am, bk}$ by (\ref{jba}).
Given $\ell \in \Z_{\ge 1}$, we define 
the level $\ell$ restricted Q-system for ${\mathfrak g}$ 
to be the relations 
obtained from (\ref{qtu}) by restricting the variables 
$Q^{(a)}_m$ to those with $(a, m) \in H_\ell$ 
by imposing $Q^{(a)}_{\ell_a}=1$.
Thus it reads
\begin{equation}\label{rqs}
(Q^{(a)}_m)^2 = 
Q^{(a)}_{m-1}Q^{(a)}_{m+1} 
+ (Q^{(a)}_m)^2
\prod_{(b,k) \in H_\ell}(Q^{(b)}_k)^{-2J^{ba}_{km}}
\quad \text{for }\, (a,m) \in H_\ell.
\end{equation}

\begin{proposition}\label{pr:qcy}
Suppose $Q^{(a)}_m$ satisfies the 
level $\ell$ restricted Q-system for ${\mathfrak g}$. Then
\begin{equation}\label{yqe}
Y^{(a)}_m = \frac{(Q^{(a)}_m)^2\prod_{(b,k) \in H_\ell}
(Q^{(b)}_k)^{-2J^{k m}_{b a}}}
{Q^{(a)}_{m-1}Q^{(a)}_{m+1}}
\end{equation}
is a solution of the level $\ell$ restricted 
constant Y-system for ${\mathfrak g}$.
The same holds between the unrestricted 
Q-system and the unrestricted constant Y-system if 
the product $\prod_{(b,k) \in H_\ell}$ in (\ref{yqe}) is 
replaced by $\prod_{b\in I, k\ge 1}$.
\end{proposition}

This is a corollary (constant version) of Theorem \ref{th:ty1}.
For instance in the restricted case, 
it can also be verified directly by noting 
\begin{equation}\label{yyq}
1+(Y^{(a)}_m)^{-1} = 
\prod_{(b,n) \in H_\ell}(Q^{(b)}_k)^{2J^{n m}_{b a}},\quad
1+Y^{(b)}_k = 
\prod_{n=1}^{\ell_b-1}(Q^{(b)}_n)^{{\bar C}^b_{kn}},
\end{equation}
where ${\bar C}^b_{kn}$ is defined by (\ref{camn}).
By virtue of (\ref{fsy}),  
the assertion is reduced to 
$2J^{n m}_{b a} = \sum_{n=1}^{\ell_b-1}
{\bar C}^b_{k n}K^{m k}_{a b}$, 
which indeed holds by (\ref{jba}).
For ${\mathfrak g}$ simply laced, (\ref{yqe}) reads
\begin{equation}\label{yqad}
Y^{(a)}_m = \frac{\prod_{b\in I: C_{ab}=-1}Q^{(b)}_m}
{Q^{(a)}_{m-1}Q^{(a)}_{m+1}}.
\end{equation}

\subsection{\mathversion{bold}$Q^{(a)}_m$ at root of unity}
\label{ss:qru}

We fix the level $\ell \in \Z_{\ge 1}$.
Let $\chi(V_\omega)$ be the 
character of the irreducible finite dimensional representation $V_\omega$ 
of ${\mathfrak g}$ with highest weight 
$\omega \in \sum_{a\in I}\Z_{\ge 0}\,\omega_a$.
We introduce the following specialization of $\chi(V_\omega)$:
\begin{equation}\label{qdi}
\dim_q\!V_\omega = \prod_{\alpha \in \Delta_+}
\frac{\sin\frac{\pi(\alpha |\omega+\rho)}{\ell+h^\vee}}
{\sin\frac{\pi(\alpha |\rho)}{\ell+h^\vee}},
\end{equation}
where $h^\vee$ is the dual Coxeter number (\ref{hhd}),
$\Delta_+$ is the set of positive roots of ${\mathfrak g}$ 
and $\rho = \frac{1}{2}\sum_{\alpha \in \Delta_+}\alpha 
= \sum_{a\in I}\omega_a$.
The quantity 
$\prod_{\alpha \in \Delta_+}
\frac{[(\alpha |\omega+\rho)]_{q^{t}}}
{[(\alpha |\rho)]_{q^{t}}}$ 
is a $q$-analog of the dimension of $V_\omega$.
Thus (\ref{qdi}) is the 
$q$-dimension at the root of unity 
$q=\exp(\frac{\pi \sqrt{-1}}{t(\ell+h^\vee)})$.

By Proposition \ref{pr:rwq}, 
we know that the classical character of the 
Kirillov-Reshetikhin module 
$Q^{(a)}_m={\rm res}\, \chi_q(W^{(a)}_m)$
satisfies the unrestricted Q-system.
As shown in (\ref{chd}) and (\ref{xbm}),  
${\rm res}\, \chi_q(W^{(a)}_m)$ 
is a linear combination of various $\chi(V_\omega)$'s.  
The specialization of ${\rm res}\, \chi_q(W^{(a)}_m)$
to the $q$-dimension will be denoted by 
$\dim_q \mathrm{res}\, W^{(a)}_m$.
By the definition,
$Q^{(a)}_m=\dim_q \mathrm{res}\, W^{(a)}_m$
still satisfies the unrestricted Q-system.
Furthermore, it seems to match the level truncation as follows.

\begin{conjecture}\label{conj:rq}
$Q^{(a)}_m = \dim_q \mathrm{res}\, W^{(a)}_m$ satisfies the 
level $\ell$ restricted Q-system.
More strongly, the following properties hold for any $a\in I$:
\begin{align}
&Q^{(a)}_m =Q^{(a)}_{\ell_a-m} 
\quad \text{for }\, 0 \le m \le \ell_a,\label{qori}\\
&Q^{(a)}_m < Q^{(a)}_{m+1}
\quad \text{for }\, 0 \le m < \left[\ell_a/2\right],\label{qin}\\
&Q^{(a)}_{\ell_a+j}=0 \qquad \text{for }\, 
1 \le j \le t_ah^\vee -1, \label{sao}
\end{align}
where $\left[\ell_a/2\right]$ is the largest integer not exceeding 
$\ell_a/2$ (not $q$-integer).
\end{conjecture}

\begin{remark}\label{re:qs}
Conjecture \ref{conj:rq} implies 
$Q^{(a)}_m >0$ for all $(a,m) \in H_\ell$.
Thus $Y^{(a)}_m$ constructed by (\ref{yqe}) 
with the substitution 
$Q^{(a)}_m = \dim_q \mathrm{res}\, W^{(a)}_m$
is real positive for all $(a,m) \in H_\ell$.
Therefore it must coincide with the unique solution 
characterized in Theorem \ref{t:thm:unique}.
\end{remark}

We note that (\ref{qori}) implies 
$Q^{(a)}_{\ell_a}=Q^{(a)}_{0} = 1$; therefore,
$j=1$ case of (\ref{sao}) as well because of   
the Q-system relation 
$(Q^{(a)}_{\ell_a})^2
=Q^{(a)}_{\ell_a-1}Q^{(a)}_{\ell_a+1} + 
\prod_{b(\neq a)}(Q^{(b)}_{\ell_b})^{-C_{ab}}$
and the fact that $Q^{(a)}_{\ell_a-1}\neq 0$ by (\ref{qin}).

\begin{example}\label{ex:rqy}
For $\mathfrak{g}=A_r$, one has 
$Q^{(a)}_m = \dim_q \mathrm{res}\, W^{(a)}_m 
= \dim_q V_{m\omega_a}$
from (\ref{nca}). Thus 
\begin{equation}
Q^{(a)}_m = \prod_{i=1}^a\prod_{j=1}^{r+1-a}
\frac{\sin\frac{\pi(m+i+j-1)}{\ell+r+1}}
{\sin\frac{\pi(i+j-1)}{\ell+r+1}}.
\end{equation}
The property (\ref{qori}) and $Q^{(a)}_m >0$ for 
$(a,m) \in H_\ell$ are easily checked.
Substitution of this into (\ref{yqad})
gives the real positive solution of the 
level $\ell$ restricted constant Y-system:
\begin{equation}
Y^{(a)}_m = \frac{
\sin\frac{\pi a}{\ell+r+1}
\sin\frac{\pi(r+1-a)}{\ell+r+1}}
{\sin\frac{\pi m}{\ell+r+1}
\sin\frac{\pi(\ell-m)}{\ell+r+1}},
\quad
1+Y^{(a)}_m = \frac{
\sin\frac{\pi (a+m)}{\ell+r+1}
\sin\frac{\pi(a+\ell-m)}{\ell+r+1}}
{\sin\frac{\pi m}{\ell+r+1}
\sin\frac{\pi(\ell-m)}{\ell+r+1}}.
\end{equation}
Obviously 
$(Y^{(a)}_0)^{-1} = (Y^{(a)}_\ell)^{-1}=0$ and 
$Y^{(a)}_m >0$ hold for $(a,m) \in H_\ell$.
When $r=1$, this reduces to $Y^{(1)}_m$ in Example \ref{t:ex:A1}.
\end{example}

One of the most remarkable features of the level $\ell$ 
restricted constant Y-system and Q-system 
is their connection with the dilogarithm identity (\ref{t:eq:DI})
in Theorem \ref{t:thm:DI}.
The LHS emerges from the TBA analysis (Section \ref{s:tbarsos}).
The $Y^{(a)}_m$  in the dilogarithm is characterized by 
the Y-system as in Theorem \ref{t:thm:unique} or 
constructed by the Q-system as in Remark \ref{re:qs}.

\subsection{Bibliographical notes}\label{ss:byq}

The idea of converting TBA equations 
into difference equations (Y-system) as described in this section 
was put into practice by \cite{Z1} for factorized scattering theories
describing integrable perturbations of conformal field theories.
The TBA equation treated there corresponds to 
the simply laced ${\mathfrak g}$ with level 
$\ell=2$ in the terminology here up to the driving term.
There are numerous Y-systems or related 
nonlinear integral equations in the similar TBA approaches 
to various integrable field theories, e.g.
\cite{RTV, BR3, FaZ, Ra, DR, DTT, DPT}.
The Y-systems considered here 
appear as typical building blocks
in these theories in many cases.

There are also exotic variants and applications of Y-systems 
related to Takahashi-Suzuki's continued fraction TBA \cite{TS}
in the context of polymers \cite{Z2}, 
the sine-Gordon model \cite{Tat} and 
the T-system for XXZ model \cite{KSS}.
Intricate examples of T and Y-systems are also worked out 
for the dilute $A_L$ models \cite{Suz5}.

With regard to the Q-system,  
there are conjectures concerning more general specialization 
than $\dim_q$ and related dilogarithm sum rules. 
See \cite[appendix A]{KNS2}, \cite[appendix D]{KNS3}, \cite{KN1} and 
\cite[section 1.4]{Kir4}.

\section{TBA analysis of RSOS models}\label{s:tbarsos}

We digest the TBA analysis of the 
$U_q(\hat{\mathfrak g})$ Bethe equation, 
which is a natural candidate for the level $\ell$ critical 
restricted solid-on-solid (RSOS) model 
associated with the representation $W^{(p)}_s$ of $U_q(\hat{\mathfrak g})$
($\ell \in \Z_{\ge 1}$, $(p,s) \in H_\ell$ (\ref{hlde})).
The basic features of the model have been sketched in Section \ref{ss:rsos}.
The derivation of high temperature entropy and 
central charges in two critical regimes is outlined.
The level $\ell$ restricted Q-system, the constant Y-system and the 
dilogarithm identity described in Sections \ref{t:ss:di}
and  \ref{ss:cy}--\ref{ss:qru}
play a fundamental role.

We make a uniform treatment for 
general ${\mathfrak g}$ elucidating 
the origin of the Y-system.
The results cover rational 
vertex models formally as the limit $\ell\rightarrow \infty$.
The TBA equation (\ref{tba2}) 
also applies to a number of situations
in other contexts, most notably, 
integrable perturbations of conformal field theories 
(cf. Section \ref{ss:byq}) with a suitable modification of the LHS. 

Apart from the relatively well known results in the ADE case,
a curious aspect in nonsimply laced ${\mathfrak g}$ 
is that the central charges in one of the regimes 
correspond to the Goddard-Kent-Olive construction
of Virasoro modules \cite{GKO} involving the embeddings 
\begin{equation*}
B^{(1)}_r \hookrightarrow D^{(1)}_{r+1},\;\;\;
C^{(1)}_r \hookrightarrow A^{(1)}_{2r-1},\;\;\;
F^{(1)}_4 \hookrightarrow E^{(1)}_{6},\;\;\;
G^{(1)}_2 \hookrightarrow B^{(1)}_{3}.
\end{equation*} 
See (\ref{cbep})--(\ref{cgep}).
These results have stimulated notable developments in 
crystal basis theory of quantum groups \cite{HKOTT}.
The content of this section is based on \cite{BR2} for ADE case
and \cite{Ku} for general ${\mathfrak g}$.

\subsection{TBA equation}\label{ss:tbarsos}

We keep the notations $t, t_a, \alpha_a, C$ 
in (\ref{eq:t1})--(\ref{aaw}) and 
$L, \ell_a, H_\ell$ in (\ref{lad})--(\ref{hlde}).
The Bethe equation is the following
for the unknowns 
$\{u^{(a)}_j \vert \, a \in I, 1 \le j \le n_a \}$:
\begin{equation}\label{ber}
\Biggl(\frac{\sinh{\pi\over 2L}\bigl(
u^{(a)}_j - \sqrt{-1}{s \over t_p}\delta_{a p}\bigr)}
{\sinh{\pi\over 2L}\bigl(
u^{(a)}_j + \sqrt{-1}{s \over t_p}\delta_{a p}\bigr)}
\Biggr)^N = \Omega_a \prod_{b=1}^r\prod_{k=1}^{n_b}
\frac{\sinh{\pi\over 2L}\bigl(
u^{(a)}_j - u^{(b)}_k - \sqrt{-1}(\alpha_a \vert \alpha_b)\bigr)}
{\sinh{\pi\over 2L}\bigl(
u^{(a)}_j - u^{(b)}_k + \sqrt{-1}(\alpha_a \vert \alpha_b)\bigr)}.
\end{equation}
Here $n_a=Ns(C^{-1})_{a p}$ as in (\ref{na}) 
with $(r_i,s_i)=(p,s)$ for all $i$, and 
$\Omega_a$ is a root of unity without which (\ref{ber}) is 
essentially the same as the Bethe equation for the vertex model (\ref{be})
at $q=\exp(\frac{\pi \sqrt{-1}}{tL})$\footnote{
$\Omega_a=e^{-\alpha_a({\mathcal H})}$ in the notation of (iii) in  
Section \ref{ss:dvfrs}.}.
The Bethe equation (\ref{ber}) is indeed valid \cite{BR2}
for $U_q(A^{(1)}_r)$ RSOS model \cite{JKMO}.

It is a well known mystery that the TBA analysis 
yields supposedly correct results in the end despite that 
it involves arguments 
that can hardly be justified mathematically\footnote{
A more reliable derivation based on T-system is given 
in Section \ref{ss:QTM}.}.
Our arguments in the sequel are no exception.

We employ a string hypothesis.
Suppose that $\{u^{(a)}_j \vert \, a \in I, 1 \le j \le n_a \}$ 
is approximately grouped as 
the union of 
$\{u^{(a)}_{m,i} + \sqrt{-1}t_a^{-1}(m+1-2n) \vert 1 \le n \le m, \,
1 \le i \le N^{(a)}_m, u^{(a)}_{m,i} \in \R \}$ and the rest. 
Here $u^{(a)}_{m,i}$ is the center of 
a color $a$ length $m$ string and 
$N^{(a)}_m$ is the number of such strings.
Then the hypothesis is that
$\lim_{N \rightarrow \infty}
\sum_{m=1}^{\ell_a}m N^{(a)}_m/n_a = 1$ for all 
$a \in I$.
It means that for color $a$, only those strings with length $\le \ell_a$
contribute to the thermodynamic quantities.
This is a peculiar feature in the RSOS model and  
one of the most significant effects of the phase factor $\Omega_a$.
Substituting the string forms into (\ref{ber}) and taking product over the 
internal coordinate of strings, one gets
\begin{equation}\label{bes}
N\delta_{ap}\Theta^m_a\bigl(u^{(a)}_{m,i},\frac{s}{t_a}\bigr) = 
I^{(a)}_{m,i} + 
\sum_{\begin{subarray}{c} b\in I\\  1\le k \le \ell_b \end{subarray}}
\sum_{j=1}^{N^{(b)}_k}
\Theta^{m k}_{a b}(u^{(a)}_{m,i}-u^{(b)}_{k,j},(\alpha_a| \alpha_b)).
\end{equation}
Here $I^{(a)}_{m,i} \in \Z + \text{constant}$, and 
$\Theta^m_a, \Theta^{m k}_{a b}$ are defined by
\begin{align}
&\Theta^m_a(u,\Delta) = \frac{1}{2\pi\sqrt{-1}}
\sum_{n=1}^m\ln\frac{\sinh\frac{\pi}{2L}
(u+\sqrt{-1}t_a^{-1}(m+1-2n)-\sqrt{-1}\Delta)}
{\sinh\frac{\pi}{2L}
(u+\sqrt{-1}t_a^{-1}(m+1-2n)+\sqrt{-1}\Delta)},\label{tet1}\\
&\Theta^{m k}_{a b}(u,\Delta) = \Theta^{k m}_{b a}(u,\Delta) 
= \sum_{j=1}^k\Theta^m_a(u+\sqrt{-1}t_b^{-1}(k+1-2j),\Delta).
\label{tet2}
\end{align}
One assumes that each solution satisfying 
$u^{(a)}_{m,1}<u^{(a)}_{m,2}<\cdots < u^{(a)}_{m,N^{(a)}_m}$
corresponds to an array such that 
$I^{(a)}_{m,1}<I^{(a)}_{m,2}<\cdots < I^{(a)}_{m,N^{(a)}_m}$,
and introduces the string density $\rho^{(a)}_m(u)$ 
and the hole density $\sigma^{(a)}_m(u)$ 
for $u \sim u^{(a)}_{m,i}$ with large enough $N$ by
\begin{equation}\label{dens}
\rho^{(a)}_m(u) = \frac{1}{N(u^{(a)}_{m,i}-u^{(a)}_{m,i-1})},
\quad
\sigma^{(a)}_m(u) = \frac{I^{(a)}_{m,i}-I^{(a)}_{m,i-1}-1}
{N(u^{(a)}_{m,i}-u^{(a)}_{m,i-1})}.
\end{equation}
Then (\ref{bes}) is converted into an integral equation.
A little inspection of it shows a characteristic property 
$\sigma^{(a)}_{\ell_a}(u)=0$, which enables one to eliminate 
the density of the ``longest strings" $\rho^{(a)}_{\ell_a}(u)$.
For such calculations, 
it is convenient to work in the Fourier components.
We attach $\,{\hat \;}\,$ to them. 
See (\ref{ft}).
We shall flexibly present formulas either 
in the Fourier or original variables.
By means of the basic formulas (\ref{fte1}) and (\ref{fte2}),
the resulting integral equation is expressed in the Fourier space 
as\footnote{The replacement 
$\ell \rightarrow L$ in (\ref{fte1}) and (\ref{fte2})
has become unnecessary here due to the elimination of 
$\rho^{(a)}_{\ell_a}(u)$.}
\begin{equation}\label{akr}
\delta_{p a} \hat{{\mathcal A}}^{s m}_{p a}(x) = 
{\hat \sigma}^{(a)}_m(x) + \sum_{(b,k) \in H_\ell} 
{\hat {\mathcal K}}^{m k}_{a b}(x)
{\hat \rho}^{(b)}_k(x) \quad\text{ for }\,\, 
(a,m) \in H_\ell.
\end{equation}
The ``TBA kernels" ${\mathcal A}^{mk}_{ab}(x)$, 
${\mathcal K}^{mk}_{ab}(x)$, etc and their useful properties  
are summarized in Section \ref{ss:ker}.
By (\ref{inv}) and (\ref{jha}), (\ref{akr}) is also written as
\begin{equation}\label{ksj}
\frac{\delta_{ap}\delta_{sm}}{2\cosh(\frac{x}{t_a})}
= \sum_{n=1}^{\ell_a-1}
{\hat {\mathcal K}}^{m n}_{a}(x){\hat \sigma}^{(a)}_n(x)
+\sum_{(b,k) \in H_\ell}
{\hat {\mathcal J}}^{m k}_{a b}(x){\hat \rho}^{(b)}_k(x)
\end{equation}
for $(a,m) \in H_\ell$.
The equation (\ref{akr}) or equivalently (\ref{ksj}) is 
the Bethe equation for the string and hole densities.

We will actually consider the thermodynamics of 
the ``quantum spin" chain associated with the 
row to row transfer matrix $T^{(p)}_s(u)$ of the RSOS model.
We chose its Hamiltonian density 
${\mathcal H}$ as
\begin{equation}\label{hbax}
{\mathcal H} = -\frac{\epsilon\gamma}{N}\frac{\partial}{\partial u}
\ln T^{(p)}_s(u)\vert_{u=u_0}
\quad (\epsilon= \pm 1),
\end{equation}
where $\gamma>0$ is a normalization constant and 
$\epsilon=\pm 1$ specifies the
two critical regimes in the RSOS model. 
The point $u_0$ is such that  
$T^{(p)}_s(u_0)$ becomes a cyclic shift
(generator of momentum) up to an overall multiple, i.e. 
(\ref{tame}) becomes 
$(\text{scalar})\prod_{i=1}^N
\delta_{\lambda_i, \mu_{i-1}}
\delta_{\alpha_i, \beta_{i-1}}$.
In view of Section \ref{ss:dvfrs},
it is natural to assume that the spectrum ${\mathcal E}$ 
of ${\mathcal H}$ is obtained from the derivative of the top term 
$Q_p(u-\frac{s}{t_p})/Q_p(u+\frac{s}{t_p})$ therein
up to an overall factor independent of the Bethe roots.
Thus up to an additive constant we get\footnote{
The sign $(-1)$ in (\ref{hbax}) is absent here since 
$T^{(p)}_s(u)$ is related to 
$\frac{\partial}{\partial v}\Theta^m_p(v,s/t_p)
\left\vert_{v=\sqrt{-1}u}\right.$.}
\begin{equation}\label{ener}
\begin{split}
{\mathcal E} &= \frac{\epsilon\gamma}{N}
\sum_{m=1}^{\ell_p}
\sum_{i=1}^{N^{(p)}_m}\frac{\partial}{\partial u}
\Theta^m_p\Bigl(u,\frac{s}{t_p}\Bigr)
\left|_{u=u^{(p)}_{m,i}}\right. \\
&\simeq
\epsilon\gamma\sum_{m=1}^{\ell_p}\int_{-\infty}^\infty du \,
\frac{\partial}{\partial u}\Theta^m_p\Bigl(u, \frac{s}{t_p}\Bigr) 
\rho^{(p)}_m(u)
= \frac{\epsilon\gamma}{2\pi}\sum_{m=1}^{\ell_p-1}
{\hat {\mathcal A}}^{sm}_{p p}{\hat \rho}^{(p)}_m
+\epsilon {\mathcal E}_0,
\end{split}
\end{equation}
where in the last step $\rho^{(p)}_{\ell_p}(u)$ is eliminated 
as was done for (\ref{akr}).
${\mathcal E}_0$ is a constant whose concrete form 
(\cite[(2.20)]{Ku}) is irrelevant in what follows. 
On the other hand, the eigenvalues of the momentum density 
${\mathcal P}$ is directly related to the top term itself, 
and is given as
\begin{equation}\label{mome}  
{\mathcal P} = \frac{2\pi}{N} 
\sum_{m=1}^{\ell_p}
\sum_{i=1}^{N^{(p)}_m}
\Theta^m_p\Bigl(u^{(p)}_{m,i},\frac{s}{t_p}\Bigr)
\simeq
2\pi\sum_{m=1}^{\ell_p}\int_{-\infty}^\infty du \,
\Theta^m_p\Bigl(u, \frac{s}{t_p}\Bigr) 
\rho^{(p)}_m(u).
\end{equation}

The Yang-Yang type entropy density 
${\mathcal S}$ \cite{YY} responsible 
for the arrangement of strings and holes is
\begin{equation}\label{yy}
\begin{split}
{\mathcal S} = \sum_{(a,m) \in H_\ell}
\int_{-\infty}^\infty du &
\bigl((\rho^{(a)}_m(u)+\sigma^{(a)}_m(u))
\ln (\rho^{(a)}_m(u)+\sigma^{(a)}_m(u))\\
&-\rho^{(a)}_m(u)\ln \rho^{(a)}_m(u)
-\sigma^{(a)}_m(u)\ln \sigma^{(a)}_m(u)\bigr).
\end{split}
\end{equation}
The thermal equilibrium condition at temperature $T=\beta^{-1}$
is obtained by demanding that 
the free energy density 
${\mathcal F} = {\mathcal E} - T{\mathcal S}$
be the extremum with respect to $\rho^{(a)}_m(u)$, 
namely $\delta{\mathcal F}/\delta \rho^{(a)}_m(u)=0$, 
under the constraint (\ref{akr}).
Setting $\sigma^{(a)}_m(u) / \rho^{(a)}_m(u) =
\exp(\beta\epsilon^{(a)}_m(u))$, 
the result reads $((a,m) \in H_\ell)$
\begin{equation}\label{tba}
\begin{split}
\frac{\epsilon \beta\gamma\delta_{p a}\delta_{s m}}
{4t^{-1}_p\cosh(t_p\pi u/2)} 
&=  \sum_{n=1}^{\ell_a - 1}\int_{-\infty}^\infty dv 
{\mathcal K}^{m n}_a(u-v) \ln 
\bigl(1 + \hbox{exp}(\beta\epsilon^{(a)}_n(v))\bigr) \\
- &\sum_{(b,k) \in H_\ell}\int_{-\infty}^\infty dv 
{\mathcal J}^{m k}_{a b}(u-v) 
\ln\bigl(1 + \hbox{exp}(-\beta\epsilon^{(b)}_k(v))\bigr).
\end{split}
\end{equation}
The nonlinear integral equation (\ref{tba})
is an example of the TBA equation,
which serves as the basis in studying thermodynamic quantities. 
By using (\ref{kan}) and (\ref{jky}) it can be slightly rearranged as
\begin{equation}\label{tba2}
\begin{split}
\frac{\epsilon \beta\gamma\delta_{p a}\delta_{s m}}
{4t^{-1}_p\cosh(t_p\pi u/2)} &=\beta\epsilon^{(a)}_m(u)
-\int_{-\infty}^\infty dv 
\frac{\ln\left[\bigl(1 + \hbox{exp}(\beta\epsilon^{(a)}_{m-1}(v))\bigr)
\bigl(1 + \hbox{exp}(\beta\epsilon^{(a)}_{m+1}(v))\bigr)\right]}
{4t_a^{-1}\cosh(t_a\pi(u-v)/2)}\\
&+\sum_{(b,k) \in H_\ell}N_{ab}\int_{-\infty}^\infty dv 
\frac{\left[
{\mathcal P}^{mk}_{ab}\ast \ln\bigl(1 + 
\exp(-\beta\epsilon^{(b)}_k)\bigr)\right](v)}
{4t_a^{-1}\cosh(t_a\pi(u-v)/2)}.
\end{split}
\end{equation}
When ${\mathfrak g}$ is simply laced,  one has 
${\mathcal P}^{mk}_{ab}(u) = \delta_{m k} \delta(u)$ 
from (\ref{sjas}) and (\ref{ft}).
Therefore (\ref{tba2}) simplifies considerably to 
\begin{equation}\label{tba3}
\begin{split}
\frac{\epsilon \beta\gamma\delta_{p a}\delta_{s m}}
{4\cosh(\pi u/2)} =\beta\epsilon^{(a)}_m(u)
-\int_{-\infty}^\infty dv 
\frac{\ln\left[
\frac{\bigl(1 + \exp(\beta\epsilon^{(a)}_{m-1}(v))\bigr)
\bigl(1 + \exp(\beta\epsilon^{(a)}_{m+1}(v))\bigr)}
{\prod_{b\in I}
\bigl(1 + \exp(-\beta\epsilon^{(b)}_m(v))\bigr)^{N_{ab}}}
\right]}
{4\cosh(\pi(u-v)/2)}.
\end{split}
\end{equation}

\subsection{High temperature entropy}\label{ss:hte}

The free energy density is expressed as
\begin{equation}\label{fay}
{\mathcal F} = \epsilon {\mathcal E}_0
-T\sum_{m=1}^{\ell_p-1}\int_{-\infty}^\infty du\,
{\mathcal A}^{sm}_{p p}(u)\ln
\bigl(1 + \hbox{exp}(-\beta\epsilon^{(p)}_m(u))\bigr)
\end{equation}
by means of (\ref{tba}), (\ref{inv}) and (\ref{mak}).
Let us evaluate the high temperature limit of the entropy density
\begin{equation}\label{sht}
{\mathcal S}_{\text{high}}= -\lim_{T\rightarrow \infty}
\frac{{\mathcal F}}{T}.
\end{equation}
When $T\rightarrow \infty$, the leading part of the asymptotic of 
$\epsilon^{(a)}_m(u)$ is expected to become independent of $u$.
Thus we set $Y^{(a)}_m = \exp(-\beta\epsilon^{(a)}_m(u))$ to be a constant  
and obtain from (\ref{fay}) that 
\begin{equation}\label{saly}
{\mathcal S}_{\text{high}}= 
\sum_{m=1}^{\ell_p-1}
{\hat {\mathcal A}}^{sm}_{p p}(0)\ln
\bigl(1 + Y^{(p)}_m\bigr).
\end{equation}
Here ${\hat {\mathcal A}}^{sm}_{p p}(0)$ is the $0$th 
Fourier component of ${\mathcal A}^{sm}_{p p}(u)$ given by (\ref{aab}).
Similarly the TBA equation (\ref{tba}) tends to
\begin{equation}\label{bkab}
\sum_{n=1}^{\ell_a-1}
{\hat{\mathcal K}}^{mn}_a(0)
\ln\bigl(1+Y^{(a)-1}_n\bigr)
=\sum_{(b,k) \in H_\ell}
{\hat{\mathcal J}}^{mk}_{ab}(0)
\ln\bigl(1+Y^{(b)}_k\bigr).
\end{equation}
This is the logarithmic form of the 
level $\ell$ restricted constant Y-system (\ref{kyj0}).
Thus we employ the solution $Q^{(a)}_m = \dim_q \mathrm{res}\, W^{(a)}_m$
explained in Remark \ref{re:qs} constructed
from the $q$-dimension at a root of unity (\ref{qdi}).
Substituting the latter formula in (\ref{yyq})
into (\ref{saly}) and applying (\ref{akd}),  we find
\begin{equation}\label{hteq}
{\mathcal S}_{\text{high}} = \ln Q^{(p)}_s.
\end{equation}
This is consistent with the dimension of the space of states
$\EuScript{H}(N)$ of the RSOS spin chain (\ref{sop2}).
Namely, (\ref{hteq}) implies 
\begin{equation}\label{hteq2}
\lim_{N\rightarrow \infty}
\left(\dim \EuScript{H}(N)\right)^{1/N} 
= \dim_q \mathrm{res}\, W^{(p)}_s,
\end{equation}
which agrees with (\ref{ehq}).

\subsection{Central charges}\label{ss:lcs}
The central charge $c$ of the underlying conformal field theory is
extracted from the low temperature asymptotics of the 
entropy as 
${\mathcal S}_{\text{low}} \simeq \frac{\pi c T}{3v_F}$ 
\cite{BCN, Af},
where $v_F$ is the Fermi velocity of the 
low lying massless excitations.
In each regime $\epsilon = \pm 1$, the result 
is expressed as 
\begin{equation}\label{cva}
c=\epsilon\frac{6}{\pi^2}
\sum_{(a,m)\in H_\ell}\left(L(f^{(a)}_m(\infty))
-L(f^{(a)}_m(-\infty))\right),
\end{equation}
where $L(x)$ is the Rogers dilogarithm (\ref{rodi}).
The number $f^{(a)}_m(\infty)$ is the positive real solution of 
$
\ln f^{(a)}_m(\infty) = 
\sum_{(b,k)\in H_\ell}K^{mk}_{ab}\ln(1-f^{(b)}_k(\infty))
$
in the both regimes $\epsilon=\pm 1$, where
$K^{mk}_{ab}$ is the 0th Fourier component of 
${\mathcal K}^{mk}_{ab}$ (\ref{kkab}). 
By Theorem \ref{t:thm:unique}, 
$f^{(a)}_m(\infty)$ equals $f^{(a)}_m$ in (\ref{fsy})  
constructed from the unique real positive solution of the 
level $\ell$ restricted constant Y-system for ${\mathfrak g}$.

One the other hand, the numbers $f^{(a)}_m(-\infty)$ are 
to satisfy formally the same equation
$
\ln f^{(a)}_m(-\infty) = 
\sum_{(b,k)\in H_\ell}K^{mk}_{ab}\ln(1-f^{(b)}_k(-\infty))
$
but with extra condition
$f^{(a)}_m(-\infty) = (1-\epsilon)/2$ 
for $(a,m) \in H^\epsilon_\ell$ in the regime $\epsilon=\pm 1$.
Here the subset $H^\pm_\ell$ of $H_\ell$ is specified as 
\begin{align}
H^+_\ell &= \{(p,m) \mid 1 \le m \le \ell_p-1\},\label{hps}\\
H^-_\ell &= 
\begin{cases}
\{\bigl(a,\frac{st_a}{t_p}\bigr)\mid a \in I\} & \frac{s}{t_p} \in \Z,\\
H(p,s) \cap H_\ell & \frac{s}{t_p} \not\in \Z,
\end{cases}\label{hms}\\
H(p,s) &= \{
\bigl(a,\frac{s-s_0}{t_p}\bigr),
\bigl(a,\frac{s-s_0}{t_p}+1 \bigr)\mid a \in I, t_a=1\}\nonumber\\
&\qquad\cup
\{(a, s-s_0), (a,s), (a, s-s_0+t_p)\mid a \in I, t_a=t_p\},\nonumber\\
s&\equiv s_0\mod t_p,\quad  1 \le s_0 \le t_p-1. \nonumber
\end{align}
Consequently, the equations governing the remaining 
$f^{(a)}_m(-\infty)$'s
are split into the subsets corresponding 
to the complement $H_\ell\setminus H^\epsilon_\ell$.
Their solutions are obtained by 
restricted constant Y-system 
associated with various subalgebras of ${\mathfrak g}$ and levels.
The detail can be found in \cite[section 3]{Ku}.
In any case, the dilogarithm identity (\ref{t:eq:DI}) suffices
to evaluate the sum (\ref{cva}).
Below we list the results using the RHS of (\ref{t:eq:DI})
\begin{equation}\label{pfc}
{\mathcal L}({\mathfrak g},\ell) 
= \frac{\ell\dim {\mathfrak g}}{\ell+h^\vee} 
- \mathrm{rank}\, {\mathfrak g}
\end{equation}
as the building block.

{\it Regime $\epsilon=+1$}.

\noindent
${\mathfrak g}=A_r$,
\begin{alignat*}{2}
c&= {\mathcal L}(A_r,\ell)
-{\mathcal L}(A_{p-1},\ell)
-{\mathcal L}(A_{r-p},\ell) & \quad&1\le p \le r.
\end{alignat*}
${\mathfrak g}=B_r$,
\begin{alignat*}{2}
c&= {\mathcal L}(B_r,\ell)
-{\mathcal L}(A_{p-1},\ell)
-{\mathcal L}(B_{r-p},\ell) & \quad&1\le p \le r-2,\\
&={\mathcal L}(B_r,\ell)
-{\mathcal L}(A_{p-1},\ell)
-{\mathcal L}(A_{r-p},2\ell) & &p=r-1, r.
\end{alignat*}
${\mathfrak g}=C_r$,
\begin{alignat*}{2}
c&= {\mathcal L}(C_r,\ell)
-{\mathcal L}(A_{p-1},2\ell)
-{\mathcal L}(C_{r-p},\ell) &  \quad &1\le p \le r.
\end{alignat*}
${\mathfrak g}=D_r$,
\begin{alignat*}{2}
c&= {\mathcal L}(D_r,\ell)
-{\mathcal L}(A_{p-1},\ell)
-{\mathcal L}(D_{r-p},\ell) & \quad&1\le p \le r-2,\\
&={\mathcal L}(D_r,\ell)
-{\mathcal L}(A_{r-1},\ell) & &p=r-1, r.
\end{alignat*}
${\mathfrak g}=E_6$,
\begin{alignat*}{2}
c&= {\mathcal L}(E_6,\ell)
-{\mathcal L}(D_{5},\ell)&  \quad&p=1,6,\\
&= {\mathcal L}(E_6,\ell)
-{\mathcal L}(A_{1},\ell)
-{\mathcal L}(A_{4},\ell) & &p=2,5,\\
&= {\mathcal L}(E_6,\ell)
-2{\mathcal L}(A_{2},\ell)
-{\mathcal L}(A_{1},\ell) & &p=3,\\
&= {\mathcal L}(E_6,\ell)
-{\mathcal L}(A_{5},\ell)& &p=4.
\end{alignat*}
${\mathfrak g}=E_7$,
\begin{alignat*}{2}
c&= {\mathcal L}(E_7,\ell)
-{\mathcal L}(D_{6},\ell)&  \quad&p=1,\\
&= {\mathcal L}(E_7,\ell)
-{\mathcal L}(A_{1},\ell)
-{\mathcal L}(A_{5},\ell) & &p=2,\\
&= {\mathcal L}(E_7,\ell)
-{\mathcal L}(A_{1},\ell)
-{\mathcal L}(A_{2},\ell)-{\mathcal L}(A_{3},\ell)& &p=3,\\
&= {\mathcal L}(E_7,\ell)
-{\mathcal L}(A_{4},\ell)
-{\mathcal L}(A_{2},\ell) & &p=4,\\
&= {\mathcal L}(E_7,\ell)
-{\mathcal L}(A_{1},\ell)
-{\mathcal L}(D_{5},\ell) & &p=5,\\
&= {\mathcal L}(E_7,\ell)
-{\mathcal L}(E_{6},\ell)& &p=6,\\
&= {\mathcal L}(E_7,\ell)
-{\mathcal L}(A_{6},\ell)& &p=7.
\end{alignat*}
${\mathfrak g}=E_8$,
\begin{alignat*}{2}
c&= {\mathcal L}(E_8,\ell)
-{\mathcal L}(E_{7},\ell)&  \quad&p=1,\\
&= {\mathcal L}(E_8,\ell)
-{\mathcal L}(A_{1},\ell)
-{\mathcal L}(E_{6},\ell) & &p=2,\\
&= {\mathcal L}(E_8,\ell)
-{\mathcal L}(A_{2},\ell)
-{\mathcal L}(D_{5},\ell) & &p=3,\\
&= {\mathcal L}(E_8,\ell)
-{\mathcal L}(A_{3},\ell)
-{\mathcal L}(A_{4},\ell) & &p=4,\\
&= {\mathcal L}(E_8,\ell)
-{\mathcal L}(A_{4},\ell)
-{\mathcal L}(A_{2},\ell)-{\mathcal L}(A_{1},\ell)& &p=5,\\
&= {\mathcal L}(E_8,\ell)
-{\mathcal L}(A_{6},\ell)
-{\mathcal L}(A_{1},\ell) & &p=6,\\
&= {\mathcal L}(E_8,\ell)
-{\mathcal L}(D_{7},\ell)& &p=7,\\
&= {\mathcal L}(E_8,\ell)
-{\mathcal L}(A_{7},\ell)& &p=8.
\end{alignat*}
${\mathfrak g}=F_4$,
\begin{alignat*}{2}
c&= {\mathcal L}(F_4,\ell)
-{\mathcal L}(C_{3},\ell)&  \quad&p=1,\\
&= {\mathcal L}(F_4,\ell)
-{\mathcal L}(A_{p-1},\ell)
-{\mathcal L}(A_{4-p},2\ell) & &p=2,3,\\
&= {\mathcal L}(F_4,\ell)
-{\mathcal L}(B_{3},\ell)& &p=4.
\end{alignat*}
${\mathfrak g}=G_2$,
\begin{alignat*}{2}
c&= {\mathcal L}(G_2,\ell)
-{\mathcal L}(A_{1},3\ell)&  \quad&p=1,\\
&= {\mathcal L}(G_2,\ell)
-{\mathcal L}(A_{1},\ell)& &p=2.
\end{alignat*}

{\it Regime $\epsilon=-1$}.
If $\frac{s}{t_p} \in \Z$,  the central charge is given by
\begin{equation}\label{cno}
c={\mathcal L}\left({\mathfrak g},\frac{s}{t_p}\right)
+{\mathcal L}\left({\mathfrak g},\ell-\frac{s}{t_p}\right)
-{\mathcal L}\left({\mathfrak g},\ell\right)
+ \mathrm{rank}\,{\mathfrak g}.
\end{equation}
This is the value
corresponding to the coset pair
\begin{alignat}{3}
&\;\;\hat{\mathfrak g} &\oplus \;\;&\hat{\mathfrak g} &\;\;\supset &\;\hat{\mathfrak g}
\label{cnop}\\ 
\text{level}\quad& \ell-\frac{s}{t_p} && \frac{s}{t_p} && \;\ell.
\nonumber
\end{alignat}
The situation $\frac{s}{t_p} \not\in \Z$ can take place in 
nonsimply laced algebras.
The central charges for such cases are given as follows.

\smallskip
${\mathfrak g} = B_r \; (p=r,\, 1\le s \le 2\ell-1,\, s \in 2\Z+1)$,
\begin{equation}\label{cbe}
c={\mathcal L}\left(B_r,\frac{s-1}{2}\right)
+{\mathcal L}\left(B_r,\ell-\frac{s+1}{2}\right)
-{\mathcal L}\left(B_r,\ell\right)+2r+1.
\end{equation}
This value corresponds to the following coset pair
via the embedding $B^{(1)}_r \hookrightarrow D^{(1)}_{r+1}$:
\begin{alignat}{4}
&\;\;\;\;B^{(1)}_r &\oplus \;\;\;&\;B^{(1)}_r
&\oplus \;\;&D^{(1)}_{r+1} &\;\;\supset &\;B^{(1)}_r
\label{cbep}\\ 
\text{level}\quad& \ell-\frac{s+1}{2} && \frac{s-1}{2}
&& \;\;1  && \;\ell.\nonumber
\end{alignat}

${\mathfrak g} = C_r \; (1\le p\le r-1,\, 1\le s \le 2\ell-1,\, s \in 2\Z+1)$,
\begin{equation}\label{cce}
c={\mathcal L}\left(C_r,\frac{s-1}{2}\right)
+{\mathcal L}\left(C_r,\ell-\frac{s+1}{2}\right)
-{\mathcal L}\left(C_r,\ell\right)+3r-1.
\end{equation}
This value corresponds to the following coset pair
via the embedding $C^{(1)}_r \hookrightarrow A^{(1)}_{2r-1}$:
\begin{alignat}{4}
&\;\;\;\;C^{(1)}_r &\oplus \;\;\;&\;C^{(1)}_r
&\oplus \;\;&A^{(1)}_{2r-1} &\;\;\supset &\;C^{(1)}_r
\label{ccep}\\ 
\text{level}\quad& \ell-\frac{s+1}{2} && \frac{s-1}{2}
&& \;\;1  && \;\ell.\nonumber
\end{alignat}

${\mathfrak g} = F_4 \; (p=3,4,\, 1\le s \le 2\ell-1,\, s \in 2\Z+1)$,
\begin{equation}\label{cfe}
c={\mathcal L}\left(F_4,\frac{s-1}{2}\right)
+{\mathcal L}\left(F_4,\ell-\frac{s+1}{2}\right)
-{\mathcal L}\left(F_4,\ell\right)+10.
\end{equation}
This value corresponds to the following coset pair
via the embedding $F^{(1)}_4 \hookrightarrow E^{(1)}_{6}$:
\begin{alignat}{4}
&\;\;\;\;\;F^{(1)}_4 &\oplus \;\;\;&\;F^{(1)}_4
&\oplus \;\;&E^{(1)}_{6} &\;\;\supset &\;F^{(1)}_4
\label{cfep}\\ 
\text{level}\quad& \ell-\frac{s+1}{2} && \frac{s-1}{2}
&& \;\;1  && \;\ell.\nonumber
\end{alignat}

${\mathfrak g} = G_2\; 
(p=2,\, 1\le s \le 3\ell-1,\, s\equiv s_0 \mod 3,\, s_0=1,2)$,
\begin{equation}\label{cge}
c={\mathcal L}\left(G_2,\frac{s-s_0}{3}\right)
+{\mathcal L}\left(G_2,\ell-\frac{s-s_0}{3}-1\right)
+{\mathcal L}\left(A_1,2\right)
-{\mathcal L}\left(G_2,\ell\right)+5.
\end{equation}
This value corresponds to the following coset pair
via the embedding $G^{(1)}_2 \hookrightarrow B^{(1)}_{3}$:
\begin{alignat}{4}
&\;\;\;\;\;\;\;G^{(1)}_2 &\oplus \;\;\;&\;\;G^{(1)}_2
&\oplus \;\;&B^{(1)}_{3} &\;\;\supset &\;G^{(1)}_2
\label{cgep}\\ 
\text{level}\quad& \ell-\frac{s-s_0}{3}-1 && \frac{s-s_0}{3}
&& \;\;1  && \;\ell.\nonumber
\end{alignat}
In (\ref{cbe}), (\ref{cce}), (\ref{cfe}), (\ref{cge}),
the contributions 
$2r+1,3r-1,10,5$ 
other than the dilogarithm ${\mathcal L}$
are equal to $|H(p,s)|$ in (\ref{hms}).

These values of the central charges and coset pairs 
are consistent with the analyses of 
RSOS models \cite{DJKMO1, JMO1, DJKMO2}
by Baxter's corner transfer matrix method \cite{Ba3}.
For $A_r$ level $\ell$, 
the central charges 
in regime $\epsilon = +1$ and $\epsilon=-1$ are transformed 
to each other via the interchange
$(r-1,\ell, p, s) \leftrightarrow  (\ell, r-1, s, p)$,
which is a manifestation of the level-rank duality 
\cite{JMO1, BR2, KN0}.

So far we have considered the $N$ site RSOS chain
with the homogeneous quantum space, namely
the one corresponding to 
$(W^{(p)}_s)^{\otimes N}$ in the dual picture of vertex models.  
One can extend the whole analysis to the inhomogeneous case
corresponding to $(W^{(p_1)}_{s_1}\otimes
\cdots \otimes W^{(p_k)}_{s_k})^{\otimes N}$.
Then the LHS of (\ref{tba}) becomes non vanishing 
for $(a,m)= (p_1,s_1),\ldots, (p_k,s_k)$, and 
$H^\epsilon_\ell$ in (\ref{hps}) and (\ref{hms}) gets replaced by 
$\cup_{i=1}^k (H^\epsilon_\ell\,\text{ for }\, (p_i,s_i))$. 
As the result, a broad list of central charges is realized, e.g. 
the coset pair 
$(\hat{\mathfrak g})^{\oplus k+1} \supset \hat{\mathfrak g}$ 
for ADE case in the regime $\epsilon = -1$.
For more details see \cite[section 4.2]{Ku}.
Such a generalization has also been consistently
incorporated into the crystal basis theory of 
one dimensional configuration sums \cite[section 3.2]{HKOTT}.

\section{T-system in use}\label{s:app}

Here we present various applications of the 
T and Y-systems to solvable lattice models. 

\subsection{Correlation lengths of vertex models}\label{correlationlength}

The correlation length $\xi$  is the simplest quantity to characterize 
ordered states.
It is evaluated from the energy gap, 
which needs a lengthy calculation in the Bethe ansatz approach.
As an application of the T-system for transfer matrices,
we will demonstrate a quick derivation of 
$\xi$ \cite{Kl1, KNS3} based on the ``periodicity at level $0$".

We consider the vertex models
associated with quantum affine algebra
$U_q(\hat{\mathfrak{g}})$. 
The row transfer matrix $T^{(a)}_m(u)$ is given by (\ref{tamg}).
We employ the parameterization $q={\rm e}^{-\lambda/t}$ 
with $\lambda>0$, where $t =1,2,3$ is defined in (\ref{eq:t1}).
To simplify the argument, we consider the 
homogeneous case $(r_i,s_i,w_i)=(p, s, 1)$ for all $i$, thus 
$T^{(a)}_m(u)$ acts on the quantum space 
$W^{(p)}_s(0)^{\otimes N}$.
We assume that $t_p=1$ and the system size $N$ is even.
Possible vertex configurations and the Boltzmann weights 
are explicitly given in (\ref{6v}) for $U_q(A^{(1)}_1)$ for instance.
The vertex weights associated to
$U_q(\hat{\mathfrak{g}})$ with $\mathfrak{g}$ other than $A_1$
have also been written down explicitly in 
some cases \cite{Baz, J2}.
Based on the concrete example from the $U_q(A^{(1)}_1)$ case,
we assume that there is a range of the spectral parameter $u$  
in which the model is in anti-ferroelectric order
in the sense that those features 
explained below are realized\footnote{In the parameterization 
(\ref{6v}) for $U_q(A^{(1)}_1)$ case,
the range is $-1<u<0$.  We assume the same range for 
general $U_q(\hat{\mathfrak{g}})$  
leaving the precise Boltzmann weights  
corresponding to it unspecified.}. 
For a more detailed account, see \cite[section 2.1]{KNS3}.

In the ordered regime, 
the ground state and the first excited state are almost degenerate.
The relevant energy gap is thus given by 
the energy difference between the ground state and the 2nd excited state(s).
Let $T_{\rm ground}$ and $T_{\rm 2nd}$ be the corresponding eigenvalues
of the transfer matrix. Consequently, 
$1/\xi= \ln (T_{\rm ground}/T_{\rm 2nd}) $.
We will show that $\xi$ is given as 
\begin{equation}
\xi= -\frac{1}{\ln k},
\label{j:xiV}
\end{equation}
where $k\,(0 < k <1)$ 
is determined by the data $U_q(\mathfrak{g})$ as
$$
\frac{K^\prime(k) }{K(k)} =\frac{\lambda h^{\vee} }{\pi},
$$
where $h^{\vee}$ is 
the dual Coxeter number of $\mathfrak{g}$ (\ref{hhd}) as before. 
$K(k) \, (K^\prime(k))$ stands for 
the complete elliptic integral of the first (second) kind with
modulus $k$.

Recall that the unrestricted T-system for $\mathfrak{g}$ (\ref{tga})
has the form
\begin{equation*}
\textstyle
T^{(a)}_m(u-\frac{1}{t_a})T^{(a)}_m(u+\frac{1}{t_a})
=T^{(a)}_{m-1}(u)T^{(a)}_{m+1}(u)+g^{(a)}_m(u)M^{(a)}_m(u),
\end{equation*}
where the scalar function $g^{(a)}_m(u)$ depends on 
the normalization of vertex weights.   
The factor $M^{(a)}_m(u)$ is a product of 
$T^{(b)}_{k}$'s.
We assume $m \in t_a \Z_{>0}$ and denote 
the eigenvalues of $T^{(a)}_m(u)$ also by the same symbol.
For the ground state in the anti-ferroelectric regime, 
the second term on the RHS is exponentially larger 
than the first.
So it is a good approximation to drop the first term on the RHS.
The same is true for the second excited state(s).
Let $L^{(a)}_m(u)$ be  the ratio of the eigenvalues
$$
L^{(a)}_m(u)= ( T^{(a)}_m(u) )_{\rm 2nd}
/(T^{(a)}_m(u))_{\rm ground}.
$$
Then the above argument implies that it satisfies
\begin{equation}
\textstyle
L^{(a)}_m(u-\frac{1}{t_a})L^{(a)}_m(u+\frac{1}{t_a})
=M^{(a)}_m(u)\vert_{\forall T^{(b)}_k(v) \rightarrow L^{(b)}_k(v)}.
\label{j:bulkT}
\end{equation}
This is regarded as the level zero restricted T-system.
{}From (\ref{tade})--(\ref{tg}), 
one can check that it closes among 
those $L^{(a)}_{m}(u)$'s with $m \in t_a \Z_{> 0}$.
Moreover it enforces the following periodicity. 
(See also (\ref{pe1}).)
\begin{proposition}[\cite{IIKNS}, Theorem 8.8]\label{pr:pe0}
Suppose that $L^{(a)}_m(u)$ satisfies (\ref{j:bulkT}). 
Then the relation
\begin{align*}
L^{(a)}_{ m}(u) L^{(\omega(a))}_{ m}(u +h^{\vee}) = 1
\end{align*}
is valid for $m \in t_a\Z_{> 0}$.
Here $\omega$ is the involution on the index set $I$ such that
$\omega(a)=a$ {\em except for} the following cases 
(see Fig \ref{fig:Dynkin})\footnote{
For $\mathfrak{g}=D_r$ ($r$: even), 
we set $\omega(a)=a$  for any $a\in I$.}:
\begin{alignat*}{2}
&\text{$\mathfrak{g}=A_r$,}&& \omega(a)=r+1-a,\\
&\text{$\mathfrak{g}=D_r$ ($r$: odd),}&\quad&  
\omega(r-1)=r,\ \omega(r)=r-1, \\
&\text{$\mathfrak{g}=E_6$,}&& \omega(1)=6,\  
\omega(2)=5,\  \omega(5)=2,\  \omega(6)=1.
\end{alignat*}
In particular, 
$L^{(a)}_{m}(u) 
= L^{(a)}_{m}(u + 2h^{\vee})$ holds.
\end{proposition}
See also \cite[appendix A]{KNS3} for some manipulation leading to the 
above result.
Below we only consider $a$ such that $\omega(a)=a$.
Obviously $L^{(a)}_m(u)$ has another periodicity in the imaginary direction 
$$
L^{(a)}_{ m}(u)=L^{(a)}_{ m}(u+\frac{2\pi i}{\lambda})
$$
because the vertex weights are rational functions of 
$z=q^{tu} = e^{-\lambda u}$.
We thus conclude that  $L^{(a)}_{ m}(u)$ is doubly periodic.
Introduce two further functions $h_1, h_2$ by
\begin{align*}
h_1(u,u_0)&=\sqrt{k} \hbox{ sn}\Bigl(
\frac{i\lambda K(k)}{ \pi}(u-u_0)\Bigr), \\
h_2(u,u_0)&=\sqrt{k} \hbox{ sn}\Bigl(
\frac{i\lambda K(k)}{ \pi}(u-u_0+h^{\vee})\Bigr).
\end{align*}
These are meromorphic, $2h^{\vee}$-periodic, 
$\frac{2\pi i }{\lambda}$-anti-periodic functions of $u$ and 
satisfy
$$
h_j(u,u_0)h_j(u+h^{\vee},u_0) = 1\quad (j = 1,2).
$$
We note also that $h_1(u,u_0) (h_2(u,u_0))$ 
has one simple zero (pole) and no poles (zeros)
in the rectangle 
$\Omega:=[0,h^{\vee})\times [0, 2\pi i/\lambda)$ 
for $u-u_0  \in \Omega$.
We denote  by $\{u_z\},\, \{u_p \}$ the set of zeros\footnote{In the
Bethe ansatz, these zeros show up as ``holes".} and poles of 
$L^{(a)}_m(u)$ in  $\Omega$, respectively.
The  ratio defined below 
is analytic and non-zero for $0 \le \re \,u < h^{\vee}$,
$$h(u) = 
\frac{L^{(a)}_m(u)} {
\prod_{u_z}h_1(u,u_z) 
\prod_{u_p}h_2(u,u_p)}.
$$
Furthermore  we have 
\begin{equation}
h(u)h(u+h^{\vee}) = 1.  
\label{j:hh1}
\end{equation}
The Liouville theorem and (\ref{j:hh1}) claim that $h(u)=\pm 1$.
We thus obtain the representation
$$
L^{(a)}_m(u) = \pm  
\prod_{u_z}\sqrt{k} \hbox{ sn}\Bigl(
\frac{i\lambda K(k)}{ \pi}(u-u_z)\Bigr) 
\prod_{u_p}\sqrt{k} \hbox{ sn}\Bigl(
\frac{i\lambda K(k)}{\pi}(u-u_p+h^{\vee} )\Bigr).
$$
The lower excited states are described by only two zeros.
The above expression is then simplified to
\begin{equation}
L^{(a)}_m(u)={\mathcal L}^{(a)}_m(u;u_1,u_2) := \pm  
k \hbox{ sn}\Bigl(
\frac{i\lambda K(k)}{ \pi}(u-u_1)\Bigr) \hbox{ sn}\Bigl(
\frac{i\lambda K(k)}{ \pi}(u-u_2)\Bigr) .
\label{j:Lamu}
\end{equation}
The locations of these zeros label the excitations.
The energy levels are  almost degenerate
with slight change in the locations of zeros.
Thus,  we observe the band structure of second excited states.
The correlation function $G(R)$ must sum up all
the contributions from the band \cite{JKM} as
$$
G(R) - G(\infty) \simeq \int du_1 \int du_2 \, \rho(u_1,u_2)
\Bigl({\mathcal L}^{(a)}_m(u; u_1,u_2)\Bigr)^R.
$$
By $\rho(u_1,u_2)$  we mean some weight function 
whose explicit form is not necessary for our argument. 
Substitution of (\ref{j:Lamu}) to the above leads to
$$G(R) - G(\infty) \simeq \hbox{const}\cdot k^R,$$
showing (\ref{j:xiV}).

\subsection{Finite size corrections}\label{finitesizecorrection}

Evaluation of finite size corrections 
to the energy spectra of the Hamiltonian or 
the free energy provides information on the critical behavior 
such as central charges 
and scaling dimensions \cite{Af, BCN,Car}.
Numerical approaches often suffer from the smallness of system size and other 
technical problems such as logarithmic corrections.
The evaluation of finite size corrections is 
a non trivial problem even for integrable models.  
The Bethe equation is highly transcendental and it
simplifies only in the thermodynamics limit  to an 
integral equation.
For an arbitrary given system size, 
it is not possible in general to 
find the exact locations of the Bethe roots.
Nevertheless, there are successful results 
in deriving finite size corrections
based on clever manipulations of Bethe equations 
\cite{dVW, Ha, ABBBQ, KBP}. 
Here we demonstrate yet another method utilizing 
the T-system in place of the Bethe equation following \cite{KP1, KP2}.

As a concrete example we treat 
a level $\ell$ critical RSOS model 
associated with $A^{(1)}_1$ in 
Section \ref{ss:rsos}--\ref{ss:tmrsos} 
($\ell \in \Z_{\ge 2}$).
Local states on lattice sites range over 
$\{1,2,\ldots, \ell+1\}$.
We consider the fusion model in which 
any neighboring pair of local states 
is $s$-admissible ($1 \le s \le \ell-1$).
See (\ref{adm1}) and (\ref{adm2}) for the definition of the
admissibility.
The transfer matrix $T_s(u)$ 
is defined by (\ref{tuba}) with 
$m,s_i$ and $v_i$ replaced by $s, s$ and $0$, respectively. 
We assume the system size $N$ is even
and treat the range 
$-2\le u\le 0$ (referred to as the regime III/IV
critical line \cite{ABF}) for simplicity.
We set 
$$
q={\rm e}^{i\lambda}, \qquad \lambda=\frac{\pi}{\ell+2}, 
$$
in the RSOS Boltzmann weights according to (\ref{bd}).  

Although we are concerned with such an isotropic model, 
the key in our approach is to embed it in 
a family of models in which  
the admissibility (fusion degree) conditions 
in the horizontal and vertical directions can be different.   
We consider the level $\ell$ fusion RSOS model \cite{DJKMO1}
in which neighboring states  
in the horizontal direction are $s$-admissible while
those in the vertical direction are $m$-admissible.
The corresponding transfer matrix is denoted by $T_m(u)$
and depicted in (\ref{tuba}) with  
$s_i=s$ and $v_i=0$.
The evaluation of  the
finite size correction to the largest eigenvalue of  
$T_s(u)$ utilizing the restricted T-system 
among $\{T_j(u)\}$
will be the main issue in the sequel.

First we need to fix the normalizations.
Let $W_{1,s}$ be the RSOS Boltzmann weights 
obtained by the $s$-fold fusion 
in the horizontal direction (cf.(\ref{wme})). 
Our normalization is such that 
$$
 W_{1,s}\!\left(\left. \begin{matrix}a+s-1& a-1\\ a+s & a
\end{matrix}\right|u\right) =
\frac{[u+s+1]_{q^{1/2}}}{[2]_{q^{1/2}}}.
$$
See (\ref{bd}) for the symbol $[u]_{q^{1/2}}$.
{}From now on 
we use $x=(u+1)i$ as the spectral parameter, and 
$T_m(u)$ will also be written as $T_m(x)$.
We furthermore define the normalized transfer matrices by 
$\tilde{T}_0(x)=1$ and
$$
\tilde{T}_m(x) =
\begin{cases}
T_m(x)& 1\le  m\le s,  \\
 \frac{T_{m}(x)}
{\prod_{j=1}^{m-s} \phi(x+(m-s+1-2j)i)}&
s+1\le m\le \ell,
\end{cases}
$$
where we have introduced
$$
\phi(x)= \Bigl(\frac{\sinh\frac{ \lambda x}{2}}
{ \sin \lambda}\Bigr )^N.
$$
Thanks to these normalizations 
$\tilde{T}_j(x)$ is of degree $N\min(j,s)$ 
in $[ix+\cdots]_{q^{1/2}}$ for $1 \le j  \le \ell$.
One then obtains  the level 
$\ell$ restricted T-system for $\mathfrak{g} =A_1$
\begin{equation}\label{iraq}
\tilde{T}_j(x-i) \tilde{T}_j(x+i) 
= f_j(x) \tilde{T}_{j-1}(x)\tilde{T}_{j+1}(x)
+g_j(x) \qquad (1\le j \le {\ell-1}).
\end{equation}
Here the scalar factors are given by 
$f_j(x)=\phi(x)^{\delta_{js}}$ and 
$$
g_j(x) =\prod_{k=0}^{{\rm min}(j,s)-1}
\phi(x+(s+j-2k) i)  \phi(x-(s+j-2k) i).
$$
Numerical calculations for small system sizes suggest the following 
analyticity of $\tilde{T}_j(x)$.
\begin{assumption}
$\tilde{T}_j(x)\,(1\le j \le  \ell)$  is analytic and nonzero  
in the strip $|\im\,x | \le 1$.
\end{assumption}
We then construct $Y_j(x)\,(1\le j \le \ell-1)$ 
by\footnote{We employ the inverse of (\ref{ty2}) to make
the resulting integral equation suitable for numerical investigations. }
\begin{equation}
Y_j(x)= \frac{f_j(x) \tilde{T}_{j-1}(x) \tilde{T}_{j+1}(x)}{g_j(x)}.
\label{j:Yjinv}
\end{equation}
This leads to the Y-system
\begin{equation}\label{yjx}
Y_j(x-i) Y_j (x+i) =(1+Y_{j-1}(x))(1+Y_{j+1}(x)) \qquad (1\le j \le \ell-1),
\end{equation}
where $Y_0(x)=Y_{\ell}(x) =0$.
The assumption on $T_j(x)$ is 
inherited to the analyticity of  $Y_j(x)$ except for $Y_s(x)$:
 $Y_s(x)$ has order $N$ zero at the origin due to $f_s(x)$.
We thus define the modified $Y$ by
\begin{equation}\label{ypm}
\tilde{Y}_j(x) =\frac{Y_j(x)}
{(\tanh \frac{\pi}{4} x)^{N \delta_{j s}}}.
\end{equation}
Then  the above assumption is rephrased as follows.
\begin{assumption}\label{anzc_gs}
$\tilde{Y}_j(x) \,(1\le j \le  \ell-1)$ 
is analytic and nonzero in the strip 
$|\im\, x | \le 1$.
Also, $1+Y_j(x)$ is analytic and nonzero 
in the strip $|\im\,x | \le \epsilon$ for small positive
$\epsilon$.
\end{assumption}
$Y$ and $\tilde{Y}$ satisfy
\begin{equation}
\tilde{Y}_j(x-i)\tilde{Y}_j (x+i) =(1+Y_{j-1}(x))(1+Y_{j+1}(x)),
\label{j:modifiedY}
\end{equation}
where a simple identity 
$\tanh\frac{\pi}{4}(x-i)
\tanh\frac{\pi}{4}(x+i) =1$ is used.
With the above analyticity assumption, 
one can apply the Fourier transformation
to the logarithmic derivative of 
the Y-system\footnote{The derivative here is not essential.  
It is done just in order to ensure the convergence.}.
After solving it with respect to
the logarithmic derivative of $\ln Y_j$, 
the inverse Fourier transformation followed by an 
integration converts the Y-system 
into the coupled integral equation $(1\le j \le \ell-1)$:
\begin{align}
&\ln Y_j(x)\!=\!\delta_{js}\ln \tanh^N \frac{\pi x}{4}\!+\!
\int_{-\infty}^{\infty}K(x-x')
\ln[(1+Y_{j-1}(x')) (1+Y_{j+1}(x'))]\frac{dx'}{2\pi},
\label{j:finiteTBA} \\
&K(x)\!=\!\frac{\pi}{2\cosh \frac{\pi x}{2} }.  \label{j:ker}
\end{align}
The integration constant turns out to be zero 
due to the asymptotic values
\begin{equation}
Y_j(\infty) = \frac{\sin (j \vartheta) \sin ((j+2)\vartheta)}
{\sin^2 \vartheta}=:\iota(j,\vartheta)
\label{j:Yasym}
\end{equation}
with $\vartheta=\frac{\pi}{\ell+2}$.
Up to the driving term, (\ref{j:finiteTBA}) coincides with
the thermodynamic Bethe ansatz (TBA) equation (\ref{tba3}) for 
$\mathfrak{g}=A_1$ 
although they originate from completely different contexts.
The asymptotic value (\ref{j:Yasym}) is an example of 
solutions to the constant Y-system. 
See Example \ref{t:ex:A1} and Example \ref{ex:rqy}.

Once $Y_j(x)$ is obtained from (\ref{j:finiteTBA}), 
the quantity $T_s(x)$ in question can be
evaluated by using the relation
\begin{equation}
T_s(x-i) T_s(x+i) =g_s(x)(1+Y_s(x)).
\label{j:Tpexpr}
\end{equation}
Note  $\tilde{T}_s(x)=T_s(x)$.
As numerical data tells $|Y_s(x)|\ll 1$,  
the bulk contribution 
$T_s^{\rm bulk}(x)$ is determined by 
$T^{\rm bulk}_s(x-i) T^{\rm bulk}_s(x+i) =g_s(x)$.
To separate the bulk part and finite size correction, 
let  $T_s(x)=T_s^{\rm bulk}(x)T_s^{\rm finite}(x)$.
Then (\ref{j:Tpexpr}) yields
\begin{align*}
\ln T_s^{\rm bulk}(x)&=  -N\int^{\infty}_{-\infty}
\frac{\sinh sk \cosh(\ell+1-s)k}{k\sinh 2k \sinh(\ell+2)k}
{\rm e}^{-ikx}dk,  \\
\ln T_s^{\rm finite}(x)&=
\int^{\infty}_{-\infty}K(x-x')\ln(1+Y_s(x'))\frac{dx'}{2\pi}.
\end{align*}

So far, all the relations are valid for arbitrary even $N$.  
We now proceed to the evaluation of $\ln T_s^{\rm finite}(x)$ 
in the large $N$ limit for $x \sim O(1)$.
The main contribution to the integrals in (\ref{j:finiteTBA}) 
comes from $x' \sim \pm  \frac{2}{\pi}\ln 2N $.
Thus it is convenient to introduce
\begin{equation*}
y_j^{\pm}(\theta) := 
\lim_{N \rightarrow \infty}
Y_j \Bigl(\pm \frac{2}{\pi}(\theta +  \ln 2N)\Bigr).
\end{equation*}
The evenness of the original $Y_j$ as a function of $x$ 
implies 
$y_j^{+}(\theta)=y_j^{-}(\theta)$.
We then arrive at simpler expressions for $N$ sufficiently large:
\begin{align*}
&\ln y^{\epsilon}_j(\theta)  =   - \delta_{js} {\rm e}^{-\theta}  +
\int_{-\infty}^{\infty}  K_{\theta}(\theta-\theta')
 \ln[ (1+y^{\epsilon}_{j-1}(\theta')) (1+y^{\epsilon}_{j+1}(\theta'))]
\frac{d\theta'}{2\pi} ,\\
& \ln T_s^{\rm finite}\Bigl(\frac{2\theta}{\pi}\Bigr)=
 \frac{2 \cosh \theta}{N} \int_{-\infty}^{\infty} 
 {\rm e}^{-\theta'} \ln(1+y_s^+(\theta')) \frac{d\theta'}{2\pi},
\end{align*}
where  
$K_{\theta}(\theta):=\frac{2}{\pi} K(\frac{2}{\pi}\theta)
=\frac{1}{\cosh \theta}$. 
The first equation exactly coincides with the TBA equation 
in the low temperature limit.
Thus the dilogarithm trick (cf. \cite[section 3.3]{KP2}, 
\cite[section 3.2]{KNS3}) 
is naturally applied to evaluate $\ln T_s^{\rm finite}(x)$.
The final result of the finite size correction 
to the largest eigenvalue of $T_s(x)$ is given by
\begin{align}
&\ln T_s^{\rm finite}\Bigl(\frac{2\theta}{\pi}\Bigr) \simeq 
\frac{\cosh \theta}{2 \pi N} 
  \sum_{j=1}^{\ell-1} \int_{y^+_j(-\infty)}^{y^{+}_j(\infty)}
      \Bigl( \frac{\ln(1+y)}{y} -\frac{\ln y}{1+y} \Bigr ) dy   \nonumber  \\
&= \frac{\cosh \theta}{\pi N}\sum_{j=1}^{\ell-1} 
\Bigl( L_+(y_j^+(\infty)) - L_+(y_j^+(-\infty)) \Bigr)  \nonumber\\
&=\frac{ \pi \cosh\theta  }{6N} 
\Bigl( 
\frac{3s}{s+2} -\frac{6s}{(\ell+2)(\ell+2-s)}
\Bigr)
=:\frac{ \pi \cosh\theta  }{6N}c.
\label{j:cres}
\end{align}
Here 
$L_+ (y) $ is related to the Rogers dilogarithm $L(y)$ in (\ref{rodi}) by
$$
L_+(y)=L(\frac{y}{1+y}) =L(1)-L(\frac{1}{1+y}).
$$
We have also used 
$y_j^{+}(\infty)= Y_j(\infty) = \iota(j, \frac{\pi}{\ell+2})$ 
as in (\ref{j:Yasym}) while  
$$
y_j^{+}(-\infty) =
\begin{cases}
  \iota(j, \frac{\pi}{s+2})&
1 \le j \le s-1, \\
\iota(j-s, \frac{\pi}{\ell+2-s})
&
s \le j \le  \ell-1. \\
\end{cases}
$$
Then the dilogarithm identity (\ref{t:eq:DIex}) is applied.
The quantity $c$ in the last expression in (\ref{j:cres}) 
is regarded as the central charge \cite{BCN}.
This value agrees with the TBA result (\ref{cno}) obtained from the 
low temperature specific heat 
with $\mathfrak{g}=A_1$ and 
$p=1, t_p=1$.

The above argument can be generalized to calculate 
the finite size correction
in excited states with suitable modifications.
The major difference from the ground state case is that 
Assumption \ref{anzc_gs} does not hold any longer.
Instead, we assume the following for low lying excited states.

\begin{assumption}
There are finitely many zeros 
$\{z^{(j)}_\alpha\}$ of $\tilde{T}_j(x)$ in the strip 
$|\im\, x| \le 1$.
\end{assumption}

Letting the zeros of $\tilde{T}_j(x)$ in the strip 
be $\{z^{(j)}_\alpha\}$, we modify (\ref{ypm}) as
\begin{equation*}
Y_j(x)=  
\tilde{Y}_j(x)(\tanh \frac{\pi}{4} x) ^{N \delta_{j s}} 
\prod_{\alpha}\tanh \frac{\pi}{4} (x-z^{(j-1)}_{\alpha} )
\prod_{\alpha'}\tanh \frac{\pi}{4} (x-z^{(j+1)}_{\alpha'} ),
\end{equation*}
which still satisfies (\ref{j:modifiedY}).
Then it is straightforward to derive 
the following equation valid for arbitrary $N$

\begin{align}
\ln Y_j(x)& = D_j+\delta_{j s}  \ln \tanh^N  \frac{\pi}{4} x   \nonumber \\
&+\sum_{\alpha}  \ln \tanh \frac{\pi}{4} (x-z^{(j-1)}_{\alpha} )
+
\sum_{\alpha'}  \ln \tanh \frac{\pi}{4} (x-z^{(j+1)}_{\alpha'} )  \nonumber \\
&+\int_{-\infty}^{\infty}K(x-x') \ln [(1+Y_{j-1}(x')) (1+Y_{j+1}(x')) ]  \frac{dx'}{2\pi}.
\label{finiteexcitedTBA}
\end{align}
The integration constant $D_j$ 
takes account of the branch of $\ln\tanh$ and it
must be fixed case by case.
For low lying excitations in the thermodynamic limit,  
it is reasonable to assume $|z^{(j)}_{\alpha}| \gg 1$.
Thus we employ the parameterization
$$ 
 z^{(j)}_{\alpha} =
 \begin{cases}
\frac{2}{\pi}(\theta^{(j)}_{\alpha, +} + \ln 2N) 
&  \text{for}\,   
z^{(j)}_{\alpha} \gg 1 \quad  (1\le \alpha \le n_+^{(j)}), \\
-\frac{2}{\pi}(\theta^{(j)}_{\alpha,-} +  \ln 2N) 
& \text{for}\, 
z^{(j)}_{\alpha} \ll -1 \quad  (1\le \alpha \le n_-^{(j)}),
\end{cases}
$$
where $n^{(j)}_{\pm}$ 
denotes the number of $z^{(j)}_{\alpha}$ near
$\pm \frac{2}{\pi} \ln 2N $.
Then (\ref{finiteexcitedTBA}) 
is reduced in the limit $N \rightarrow \infty$ to 
\begin{align}
\ln y^{\epsilon}_j(\theta)& 
=D^{\epsilon}_j - \delta_{js} {\rm e}^{-\theta} +
\sum_{\alpha}  \ln \tanh \frac{1}{2}
(\theta-\theta^{(j-1)}_{\alpha,\epsilon} )
+
\sum_{\alpha'}\ln \tanh \frac{1}{2}
(\theta-\theta^{(j+1)}_{\alpha',\epsilon} )   \nonumber\\
 +
&\int_{-\infty}^{\infty}
K_{\theta}(\theta-\theta') \ln [(1+y^{\epsilon}_{j-1}(\theta'))
(1+y^{\epsilon}_{j+1}(\theta'))]\frac{d\theta'}{2\pi}.
\label{excitedscalingTBA}
\end{align}
The constants $D^{\pm}_j$ can be in general different 
and depend on $n_{\pm} ^{(j)}$, etc.

The subsidiary conditions
$T_j(z^{(j)}_{\alpha})=0$ must also be satisfied.  
This is rephrased as  $Y_j(z^{(j)}_{\alpha} +i)=-1$ or 
equivalently 
$$
\ln  y^{\epsilon}_j(\theta^{(j)}_{\alpha,\epsilon}+\frac{\pi}{2}i) 
=( 2 I^{(j)}_{\alpha,\epsilon}+1 )\pi i
$$
in terms of the branch cut integers
$\{I^{(j)}_{\alpha,\pm} \}$.
Thanks to (\ref{excitedscalingTBA}),  
this is rewritten as
\begin{equation}\label{subsidTBA}
\begin{split}
&-\int_{-\infty}^{\infty} 
\frac{1}{\sinh(\theta^{(j)}_{\alpha,\epsilon}-\theta-i\epsilon')} 
\ln [(1+y_{j-1}^{\epsilon}(\theta))
(1+y_{j+1}^{\epsilon}(\theta))] \frac{d\theta}{2\pi} \\
&\phantom{abc}=( 2 I^{(j)}_{\alpha,\epsilon}+1 )\pi 
+ i D^{\epsilon}_j  -\delta_{js}
{\rm e}^{-\theta^{(j)}_{\alpha,\epsilon}} \\
&\phantom{abc}+i \sum_{\alpha'} 
\ln \tanh\Bigl(\frac{ \theta^{(j)}_{\alpha,\epsilon}
- \theta^{(j-1)}_{\alpha',\epsilon}}{2}+\frac{\pi}{4}i\Bigr)
+i\sum_{\alpha'} 
\ln \tanh\Bigl( \frac{ \theta^{(j)}_{\alpha,\epsilon}-
\theta^{(j+1)}_{\alpha',\epsilon}   }{2}+\frac{\pi}{4}i\Bigr),
\end{split}
\end{equation}
where $\epsilon'>0$ is infinitesimally small.
The finite part of the eigenvalue is now given by
$$
\ln T_s^{\rm finite}\Bigl(\frac{2\theta}{\pi}\Bigr)=
\sum_{\epsilon=\pm} \,
 \frac{\rm e^{ \epsilon \theta}}{N} 
\Bigl(  -  \sum_{\alpha}
{\rm e}^{ -\theta^{(s)}_{\alpha,\epsilon} } +
\int_{-\infty}^{\infty}
{\rm e}^{-\theta} \ln(1+y_s^{\epsilon}(\theta))
\frac{d\theta}{2\pi}\Bigr).
$$
Although the expressions are more involved than the ground state case, 
one can still apply the dilogarithm trick to evaluate the above.
In particular, (\ref{subsidTBA}) and the elementary relations 
$(\ln \tanh \frac{x}{2})'=1/\sinh x$ and 
$\ln \tanh(x+\frac{\pi i}{4}) 
+\ln \tanh(-x+\frac{\pi i}{4}) 
=\pi i$
are useful.
The final result reads
\begin{align}
 \ln T_s^{\rm finite}\Bigl(\frac{2\theta}{\pi}\Bigr)&  =
 \sum_{\epsilon=\pm} \frac{ {\rm e}^{\epsilon \theta}}{2\pi N} 
 \sum_{j=1}^{\ell-1} \Bigl( 
L_+(y_j^{\epsilon} (\infty)) - L_+(y_j^{\epsilon} (-\infty)) +
\frac{1}{2}D^\epsilon_j
\ln\frac{1+y^{\epsilon}_j(\infty)}
{1+y^{\epsilon}_j(-\infty)}  \nonumber \\
&\qquad\qquad\qquad-2\pi n^{(j)}_{\epsilon}i D^\epsilon_j  
- 2\pi^2\sum_{\alpha=1}^{n^{(j)}_\epsilon}
(2I^{(j)}_{\alpha,\epsilon} +1)\Bigr).
\label{j:resex}
\end{align}

The above derivation is based on the first principle.
However it lacks a general prescription to determine the integration
constants and to choose the branch cut integers.
With regard to this, an interesting observation has been made
in \cite{KP2, OPW}.
It is possible to  absorb the additional 
driving terms in (\ref{finiteexcitedTBA})
to integrals by adopting deformed contours ${\mathcal L}_j$ as
\begin{align*}
\ln Y_j(x)& = D_j+\delta_{js}\ln \tanh^N \frac{\pi}{4} x \\
+&\int_{{\mathcal L}_{j-1}} K(x-x') \ln (1+Y_{j-1}(x')) 
+ \int_{{\mathcal L}_{j+1}} K(x-x') \ln (1+Y_{j+1}(x'))
\frac{dx'}{2\pi}.
\end{align*}
Then the evaluation of the finite size correction goes parallel to the case of 
the largest eigenstate.  
The differences lie in the asymptotic values of $y_j^{\epsilon}(x)$
and the non trivial homotopy 
in the integration contours of ${\mathcal L}_j$.
The authors of  \cite{KP2, OPW} have found  
empirical rules for the choice of homotopy and integration constants
to reproduce known scaling dimensions from conformal field theories.

We have seen that the T-system 
provides an efficient tool in the analysis of finite size corrections.
It enables one to analytically calculate the central charge (\ref{j:cres}) 
in the ground state.
The scaling dimensions of relevant operators
can also be obtained by use of 
the result in excited states (\ref{j:resex}).
The above calculation of the 
finite size correction of the largest eigenvalue has been generalized 
to RSOS models associated with $\mathfrak{g}$ 
in \cite[section 3]{KNS3} up to  
analyticity argument on auxiliary functions.

\subsection{Quantum transfer matrix approach}\label{ss:QTM}

According to Matsubara, 
finite size corrections and low temperature asymptotics are 
dual pictures of the same physical characteristics of a two dimensional system
on an infinite cylinder of circumference $N=\beta$.
Here $N$ is the system size in the former picture and 
$\beta$ is the inverse temperature in the latter.
Our analyses of the $U_q(A^{(1)}_1)$ RSOS model 
in Section \ref{s:tbarsos} and Section \ref{finitesizecorrection}
have been done along these two points of view. 
What is remarkable there is that  
beyond the formal coincidence of the two pictures,  
the two entirely different approaches end up with essentially 
the same integral equation of TBA type.
One then expects a framework to treat the 
finite temperature problem in the same manner as the 
finite size corrections without recourse to string hypothesis.
As we will see in the sequel, the 
Quantum Transfer Matrix (QTM) approach \cite{Suzu2}
offers such a scheme.
For a further detail, see the recent reviews \cite{Kl4, GS}.

\begin{figure}[tb]
\psfrag{u}{$u$}
\psfrag{v}{$v$}
\psfrag{[u-v]}{\begin{small}$b(u,v)$\end{small}}
\psfrag{[u-v+1]} {\begin{small}$a(u,v)$\end{small}}  
\psfrag{quv}{\begin{small}$c(u,v)$\end{small}}
\psfrag{qvu}{\begin{small}$c^{-1}(u,v)$\end{small}}
\psfrag{+}{\begin{small}$+$\end{small}}
\psfrag{-}{\begin{small}$-$\end{small}}

\psfrag{a}{\begin{small}$\alpha$\end{small}}
\psfrag{b}{\begin{small}$\beta$\end{small}}
\psfrag{g}{\begin{small}$\gamma$\end{small}}
\psfrag{d}{\begin{small}$\delta$\end{small}}
\psfrag{Rabgd}{\begin{small}
$R_{\beta  \delta} ^{\alpha \gamma}(u,v)$\end{small}}

\begin{center}
     { \includegraphics[width=8cm]{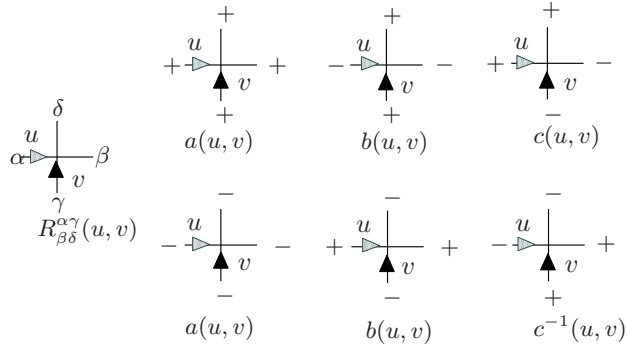}}
\end{center}
\caption{A graphic representation for 
$R_{\beta  \delta} ^{\alpha \gamma}(u,v)$. 
The spectral parameter $u$ ($v$) is associated
to horizontal (vertical) lines.}
\label{sixv_fig}
 \end{figure}

\begin{figure}[tb]
\psfrag{u}{$u$}
\psfrag{v}{$v$}
\psfrag{[v-u]}{\begin{small}$ b(v,u)$\end{small}}        
\psfrag{[v-u+1]} {\begin{small}$ a(v,u)$\end{small}}   
\psfrag{qvu} {\begin{small}$ c(v,u)$\end{small}}         
\psfrag{quv} {\begin{small}$ c^{-1}(v,u)$\end{small}}   	
\psfrag{+}{\begin{small}$+$\end{small}}				
\psfrag{-}{\begin{small}$-$\end{small}}				

\psfrag{a}{\begin{small}$\alpha$\end{small}}				
\psfrag{b}{\begin{small}$\beta$\end{small}}
\psfrag{g}{\begin{small}$\gamma$\end{small}}
\psfrag{d}{\begin{small}$\delta$\end{small}}
\psfrag{tRabgd}{\begin{small}
$\widetilde{R}^{\alpha \gamma}_{\beta \delta}(u,v)$\end{small}}

\begin{center}
     { \includegraphics[width=8cm]{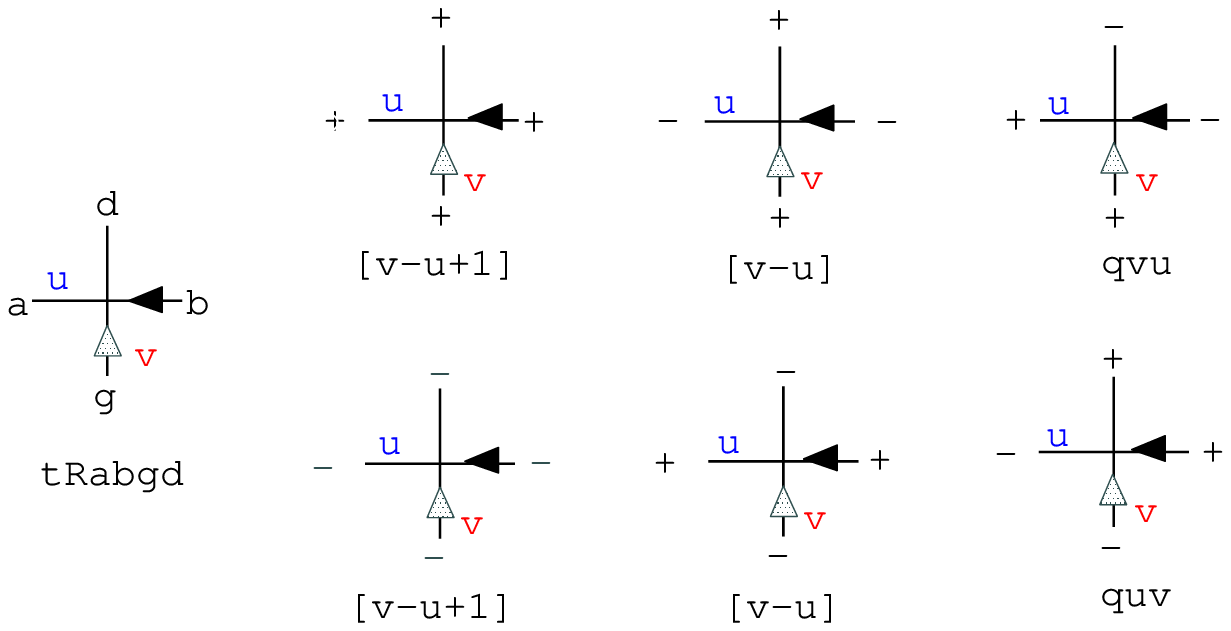}}
\end{center}
\caption{A graphic representation for
$\widetilde{R}^{\alpha \gamma}_{\beta \delta}(u,v)$. 
The spectral parameter $u$ ($v$) is associated
to horizontal (vertical) lines.}
\label{sixvr_fig}
\end{figure}

\begin{figure}
\psfrag{u}{$u$}
\psfrag{-u}{$-u$}
\psfrag{v}{$v$}
\psfrag{N}{$M$}
\psfrag{L}{$N$}
\psfrag{0}{$0$}
\begin{center}
     { \includegraphics[width=4cm]{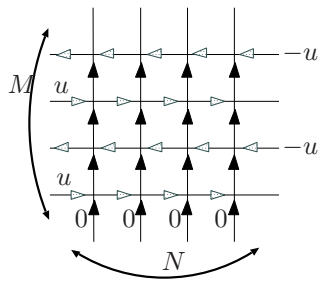}}
\end{center}
\caption{Fictitious two dimensional system}
\label{twoDz}
 \end{figure}

QTM utilizes the equivalence between
$d+1$ dimensional classical models 
and $d$ dimensional quantum system \cite{Suzu1}.  
To be concrete, we argue along 
the 1D spin $1/2$ XXZ model 
as a prototypical integrable lattice system.
\begin{equation}
\label{hamxxz}
{\mathcal H}=\frac{J}{4}\sum_{j=1}^N  
\Bigl(\sigma^x_j \sigma^x_{j+1} 
+ \sigma^y_j \sigma^y_{j+1} 
+ \Delta (\sigma^z_j \sigma^z_{j+1}+1) \Bigl) 
  =\sum_{j=1}^N  \hat{h}_{j, j+1},
\end{equation}
where $\sigma^a\, (a=x,y,z)$ are the Pauli matrices. 
The periodic boundary condition implies
$\sigma^a_{N+1}=\sigma^a_1$. 
The anisotropy is parameterized as $\Delta
= \cos \lambda$.  
The Hamiltonian acts on ``the physical space" $V_{\rm phys}:=
\bigotimes_{j=1}^N V_j$ where 
$V_j$ denotes the $j$th copy of 
$\C^2 = \C{\bf e}_+ \oplus \C{\bf e}_-$.
The main subject here is to calculate the partition function exactly
$$
 Z_{\rm 1d} (\beta,N) = {\rm Tr}_{V_{\rm phys}}
{\rm e}^{-\beta {\mathcal H}}.
$$
It would be nice if this task can be done for any finite $N$, 
although we do not have a satisfactory progress at present.
We thus concentrate on the evaluation of the free energy per site
in the thermodynamic limit
$$
f= - \lim_{N \rightarrow \infty} \frac{1}{\beta N} \ln Z_{\rm 1d}(\beta, N).
$$
We introduce the six vertex model on the 2D square lattice.
Let $R(u,v)$ be the $U_{q}(A^{(1)}_1)$ $R$ matrix 
(in a convention different from (\ref{6v})):
\begin{align*}
R(u,v) &= 
\begin{pmatrix}
 a(u,v) &               &                                           &                \\
            &    b(u,v) &            c(u,v)&             \\
&    c^{-1}(u,v)&      b(u,v)    &       \\
  &               &                 &                a(u,v)  
\end{pmatrix} \\
&a(u,v)=\frac{[2+u-v]_{q^{1/2}}}{  [2]_{q^{1/2}}   } ,
\qquad
b(u,v)=\frac{[u-v]_{q^{1/2}}}{  [2]_{q^{1/2}}  }, \\
&c(u,v)=q^{-\frac{u-v}{2}},  \qquad 
q={\rm e}^{i\lambda}.
\end{align*}
Define the matrix element 
$R_{\beta  \delta}^{\alpha \gamma}$ by 
\begin{equation*}
R(u,v) = \sum_{\alpha,\beta, \gamma, \delta=1,2}
R_{\beta  \delta} ^{\alpha \gamma}(u,v)  
E_{\alpha,\beta} \otimes E_{\gamma,\delta}.
\end{equation*}
The index $1 (2)$ refers to ${\bf e}_+  ({\bf e}_-)$ in
Fig.~\ref{sixv_fig}.
The arrows are assigned in order to 
distinguish this $R$ matrix from other $R$ matrices
that will appear below. 
By $R_{j,j+1}(u,v)$ we mean the $R$ matrix acting non trivially only
on the tensor product $V_{j}(u) \otimes V_{j+1}(v)$.
We introduce the row to row (RTR) transfer matrix $T_{\rm RTR}(u)
\in {\rm End}( V_{\rm phys})$ by
\begin{equation}
T_{\rm RTR}(u) = {\rm Tr}_{a} \,(R_{a,N}(u,0)
R_{a,N-1}(u,0)  \cdots   R_{a,1}(u,0)),
\label{defRTR}
\end{equation}
where the subscript ``a" stands for the auxiliary space.
With the lattice translation ${\rm e}^{i P}$ shifting the sites
by one, the Baxter-L{\"u}scher formula \cite{Ba1}
\begin{equation}
T_{\rm RTR}(u)  = {\rm e}^{i P} 
\bigl(1+ \frac{\lambda u}{J \sin \lambda} \, {\mathcal H} + O(u^2) \bigr)
\label{RTRandH}
\end{equation}
holds.
With a rotated $R$ matrix 
$\widetilde{R}^{\alpha \gamma}_{\beta \delta}(u,v) =
R^{\gamma \beta}_{\delta \alpha} (v,u)
$ (Fig.~\ref{sixvr_fig}), 
we introduce a rotated transfer matrix $\widetilde{T}_{\rm RTR}(u)
\in {\rm End}( V_{\rm phys})$ by
\begin{equation*}
\widetilde{T}_{\rm RTR}(u) = {\rm Tr}_{a}
\, \bigl(
\widetilde{R}_{a,N}(-u,0)  \widetilde{R}_{a,N-1}(-u,0)
\cdots  \widetilde{R}_{a,1}(-u,0)
 \bigr).
\end{equation*}
%
%
%
The expansion analogous to (\ref{RTRandH}) holds as
$
\widetilde{T}_{\rm RTR}(u) 
= {\rm e}^{-i P} 
\bigl(1+\frac{\lambda u}{J \sin \lambda} \, {\mathcal H} + O(u^2) \bigr).
$
We thus obtain an important identity
\begin{equation}
Z_{\rm 1d}(\beta, N) = {\rm Tr}_{V_{\rm phys}}
{\rm e}^{-\beta {\mathcal H}}
=\lim_{M \rightarrow \infty} 
{\rm Tr}_{V_{\rm phys}}  
\Bigl(T_{\rm double}(u=u_M)^{\frac{M}{2}}\Bigr),
\label{equivalence1}
\end{equation}
where $T_{\rm double}(u):= T_{\rm RTR}(u)
\widetilde{T}_{\rm RTR}(u)$ 
and 
\begin{equation}\label{umdef}
u_M = -\frac{\beta J\sin\lambda}{M\lambda}.
\end{equation}
The RHS of (\ref{equivalence1}) can be interpreted as a partition function
of a 2D classical system defined on $M \times N$ sites
(Fig.~\ref{twoDz}) 
$$
Z_{\rm 1d}(\beta, N) 
= \lim_{M \rightarrow \infty}
Z_{\rm 2d\, classical}(M, N, u_M).
$$
This embodies the equivalence between
$d+1$ dimensional  classical models and 
$d$ dimensional quantum system for $d=1$.  
Since the spectra of $T_{\rm double}(u)$ is gapless, 
we still need a trick to evaluate 
$Z_{\rm 2d\, classical}(M, N, u_M)$.

We follow the observation in \cite{Suzu2} and consider
the transfer matrix propagating in the horizontal direction, that is,
$T'_{\rm QTM}(u=u_M)$ 
which acts on a virtual space of size $M$.
It was shown that this transfer matrix possesses 
a gap between the largest ($\Lambda_0$) and the
other eigenvalues $\Lambda_j \, (j\ge 1)$. 
This is a crucial benefit, 
as one only has to consider the largest eigenvalue 
to evaluate the free energy in the thermodynamic limit
\begin{align*}
 &\lim_{N \rightarrow \infty}
Z^{\frac{1}{N}}_{\rm 2d\, classical} (M, N, u_M) = 
\lim_{N \rightarrow \infty}
\Bigl({\rm Tr}\, T'_{\rm QTM}
(u=u_M)^N \Bigr)^{\frac{1}{N}}\\
&=\lim_{N \rightarrow \infty}
(\Lambda_0^N + \Lambda_1^N +\cdots)^{\frac{1}{N}}
=\lim_{N \rightarrow \infty}
\Lambda_0 \Bigl(1+ \bigl( \frac{ \Lambda_1}{\Lambda_0}\bigr)^N 
+\cdots \Bigr)^{\frac{1}{N}}
\simeq
\lim_{N \rightarrow \infty} \Lambda_0.
\end{align*}
Although we have made use of the integrability for simplicity
in the above argument, the same conclusion can be  
proved in a more general setting.

\begin{figure}
\psfrag{u}{$u$}
\psfrag{v}{$v$}
\psfrag{[v-u]}{\begin{small}$b(v,u) $\end{small}}
\psfrag{[v-u+1]}{\begin{small}$a(v,u) $\end{small}}
\psfrag{qvu}{\begin{small}$c(v,u) $\end{small}}
\psfrag{quv}{\begin{small}$c^{-1}(v,u) $\end{small}}
\psfrag{+}{\begin{small}$+$\end{small}}
\psfrag{-}{\begin{small}$-$\end{small}}

\psfrag{a}{\begin{small}$\alpha$\end{small}}
\psfrag{b}{\begin{small}$\beta$\end{small}}
\psfrag{g}{\begin{small}$\gamma$\end{small}}
\psfrag{d}{\begin{small}$\delta$\end{small}}
\psfrag{tRabgd}{\small{$(R^{t})^{\alpha \gamma}_{\beta \delta}(u,v)$}}
\begin{center}
     { \includegraphics[width=7cm]{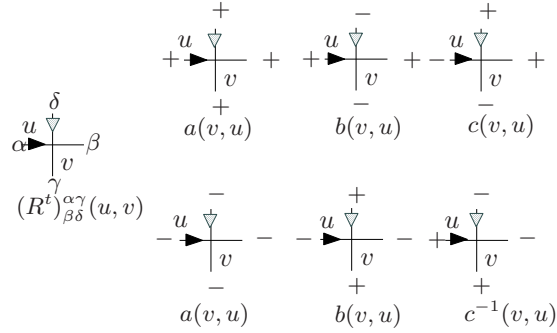}}
\end{center}
\caption{A graphic representation for 
$(R^{t})^{\alpha \gamma}_{\beta \delta}(u,v)$. 
The spectral parameter $u$ ($v$) is associated
to horizontal (vertical) lines. }
\label{sixvt_fig}
 \end{figure}

\begin{theorem}[\cite{Suzu2}]\label{Suzukitheorem}
Let $\Lambda_0$ be the largest eigenvalue of $T_{{\rm QTM}}$.
Then the free energy per site is  given by
\begin{equation}
f= -  \frac{1}{\beta} \lim_{M \rightarrow \infty}   \ln  \Lambda_0.
\label{formulaf}
\end{equation}
\end{theorem}
Two problems are still to be overcome.
First we must evaluate the largest eigenvalue of $T'_{\rm QTM}(u_M)$ 
in which interaction depends on the
fictitious system size $M$. 
Second we must take the ``Trotter limit" $M \rightarrow \infty$.   
Both of these are highly nontrivial.  
Nevertheless we stress the above formulation
makes it clear why the finite size correction and the finite temperature
problem can be treated in the same way.
To disentangle the difficulties, 
we introduce a slight generalization, 
a commuting QTM $T_{{\rm QTM}}(x,u)$, 
by assigning the parameter $ix$ in the ``horizontal" direction \cite{Kl2}.
We let the transposed $R$ matrix $R^{t}_{j,k}(u,v)$ \cite{Kl3} be
$
(R^{t})^{\alpha \gamma}_{\beta \delta}(u,v) 
= R^{\delta \alpha}_{\gamma \beta}(v,u).
$
See Fig.~\ref{sixvt_fig}.
%
Then $T_{{\rm QTM}}(x,u)$ is defined by
\begin{equation}
T_{{\rm QTM}} (x, u) 
={\rm Tr}_a \,( R_{a M}(ix, -u) R^{t}_{a,M-1} (ix,u) \cdots 
R_{a2}(ix, -u) 
R^t_{a1}(ix, u)).
\label{defQTM}
\end{equation}
The parameter $u$ will always be set to $u_M$ (\ref{umdef}), 
thus we drop its dependence hereafter.
It is the new parameter $x$ that will play the role of a spectral parameter
instead. 
By this we mean that two QTMs 
with different values of $x$ are intertwined by the same $R$ matrix
$$
R_{a,a'} (ix, iy) {\mathcal T}_a (x)  \otimes  {\mathcal T}_{a'} (x') 
=  {\mathcal T}_{a} (x')   \otimes  {\mathcal T}_{a'} (x)   R_{a,a'} (ix, iy).
$$
Here 
${\mathcal T}_a(x)$ denotes the monodromy matrix 
associated to $T_{{\rm QTM}}(x,u_M)$.
The proof is elementary.  
Now we are able to introduce the fusion hierarchy of commuting transfer matrices 
$T_j(x)$ which contains $T_{{\rm QTM}}(x, u_M)$ as the first member. 
(The $u_M$-dependence will be suppressed.)
By the construction, they satisfy the T-system
$$
T_j(x-i) T_j(x+i)  =T_{j-1}(x) T_{j+1}(x) + g_j(x),
$$
where $g_j(x)=T_0(x+(j+1)i)T_0(x-(j+1)i)$ with
\begin{equation}
T_0(x) =  \phi(x+(1+u_M)i) \phi(x-(1+u_M)i), 
\qquad \phi(x)= \Bigl(\frac{\sinh\frac{\lambda x}{2}}
{\sin \lambda} \Bigr)^{\frac{M}{2}}.
\label{defT0Phi}
\end{equation}
As in Section \ref{finitesizecorrection},   
we need assumptions on the analyticity of $T_j(x)$. 
For simplicity we consider the case $\lambda \rightarrow 0$ for a moment. 
Then the numerical analysis suggests
\begin{conjecture} \label{anzc_qtm}
The zeros of $T_j(x)$ are distributed almost on the line 
$|\im\,x |  = j+1 $ .
\end{conjecture}
We set
$Y_j(x)=T_{j-1}(x) T_{j+1}(x)/g_j(x)$ and 
introduce its modification 
\begin{equation}\label{yts1}
\tilde{Y}_j(x) = \frac{Y_j(x)}{(\tanh \frac{\pi}{4}(x-(1+u_M)i) 
\tanh \frac{\pi}{4}(x+(1+u_M)i))^{\frac{M}{2}} }.
\end{equation}
Note that $u_M$ is a small negative quantity.
Then the conjecture is translated to 
\begin{conjecture}
$\tilde{Y}_j(x)$ is analytic and nonzero in the strip $|\im\,x | \le 1$ and
$1+Y_{j+1}(x)$  is  analytic and nonzero in the strip $|\im\,x | \le \epsilon$ 
for small $\epsilon$.
\end{conjecture}
This immediately leads to the integral equation
\begin{align}
\ln Y_j(x) & = \delta_{j1}  
\frac{1}{2} 
 \ln \left[\tanh^M \frac{\pi}{4}(x-(1+u_M)i)
\tanh^M \frac{\pi}{4}(x+(1+u_M)i)\right]  \nonumber\\
&  +\int_{-\infty}^{\infty}  K(x-x') 
\ln [(1+Y_{j-1}(x'))(1+Y_{j+1}(x'))] \frac{dx'}{2\pi}, 
 \label{QTMTBA}
\end{align}
where $K(x)$ is defined in (\ref{j:ker}).
The $M$ enters only in the first line in (\ref{QTMTBA}).
Therefore the Trotter limit $M \rightarrow \infty$ 
can be taken {\it  analytically}, giving
\begin{equation}\label{yinte}
\begin{split}
\ln Y_j(x) & = \delta_{j1}D(x)\\
&+\int _{-\infty}^{\infty} 
K(x-x') \ln [(1+Y_{j-1}(x'))(1+Y_{j+1}(x')) ]
\frac{dx'}{2\pi}    \quad (j \ge 1).
\end{split}
\end{equation}
where $D(x)$ in the driving term is given by
\begin{equation}\label{Ddef}
D(x) = - \frac{\beta \pi J\sin\lambda}{2\lambda\cosh \frac{\pi}{2} x}.
\end{equation}
These are nothing but 
the Gaudin-Takahashi equations for the anti-ferromagnetic Heisenberg model.
Also, they coincide with (\ref{j:finiteTBA}) up to the driving term. 
The free energy per site is obtained from the solution to the above equations as
$$
f= -\frac{1}{\beta}  \int_{-\infty}^{\infty}
K(x')\ln(1+Y_1(x')) \frac{dx'}{2\pi}.
$$

Summarizing, we have seen that  T-system plays 
the central role for the quantitative studies 
on both finite size system and finite temperature system.
A wider range of the parameter 
$0<\lambda \le \frac{\pi}{2}$ is treated in \cite{KSS} under the 
restriction that the continued fractional expansion of 
$\pi/\lambda$ 
terminates at a finite stage.
A suitably chosen subset of the fusion QTMs are shown to 
satisfy a closed set of functional relations and it 
successfully recovers the well known
Takahashi-Suzuki continued fraction TBA equation \cite{TS}
without using string hypothesis.  
See \cite{KSS} for details.

\subsection{Simplified TBA equations}\label{ss:sTBA}
We continue our discussion on the 
XXZ spin chain at finite temperatures.
We retain the definitions of the symbols such as $\phi(x), T_j(x), u_M$, etc. 
in the previous subsection.
The TBA equation is a coupled 
set of integral equations with (finitely or infinitely) 
many unknown functions $Y_j(x)$.
It is known that  
equations change their forms drastically according to a 
small change in coupling constant $\lambda$ \cite{TS}.
On the other hand,  we expect only small changes in physical quantities.
Thus one may hope alternative formulations 
that are more stable against the change in $\lambda$.
Here we present one such approach 
which also originates from the T-system.
It is sometimes referred to as a simplified TBA equation \cite{Ta}.

The idea is complementary to the QTM method where 
one pays attention to the zeros of $T_j(x)$. 
In the simplified TBA, one is concerned with 
singularities of a renormalized $T_j(x)$. 
The latter is defined by 
\begin{equation}
\tilde{T}_j(x)= \frac{T_j(x)}{\phi(x+(j+1+u_M)i) \phi(x-(j+1+u_M)i) },
\end{equation}
where $\phi(x)$ is defined in (\ref{defT0Phi}).
Note $\tilde{T}_j(x)$ possesses poles of order 
$M/2$  at  $x \sim \pm (j+1)i$.
Accordingly, the first equation of the T-system reads
\begin{align}
\tilde{T}_1(x+i)\tilde{T}_1(x-i) &
=\tilde{T}_2(x) + b^{(M)}_1(x),   \label{T1T1} \\
b^{(M)}_1(x) &=
\frac{ \phi(x+(1-u_M)i)  \phi(x-(1-u_M)i) }
{\phi(x+(1+u_M)i)  \phi(x-(1+u_M)i)}.   \label{defTakahashib1} 
\end{align}
Let $\tau_j(x)$ be  $\tilde{T}_j(x)$ after the Trotter limit
$$\tau_j(x) = \lim_{M \rightarrow \infty} \tilde{T}_j(x). $$
Then  $\tau_1(x)$  develops singularity at  $x = \pm 2i$.  
By construction, it is periodic under 
$x \rightarrow x+ 2p_0i$, where $p_0=\pi/\lambda$.
We thus assume the expansion
\begin{equation}\label{tau1}
\tau_1(x)  =2 + \sum_{n \in \Z} \sum_{j=1}^{\infty}
\frac{c_j}{(x-2i-2p_0 n i)^j}\\
+\sum_{n \in \Z}
\sum_{j=1}^{\infty}   \frac{\bar{c}_j}{(x+2i-2p_0 n i)^j}.
\end{equation}
We utilize the T-system and information 
on the locations of singularities to fix  $c_j$ and $\bar{c}_j$.
Rewrite the Trotter limit of (\ref{T1T1}) as
\begin{align}
\tau_1(x+i)  &= \frac{b_1(x)}{\tau_1(x-i)} 
+ \frac{ \tau_2(x) }{\tau_1(x-i)},   \\
b_1(x)&= \lim_{M \rightarrow \infty}  b^{(M)}_1(x)
=\exp \Bigl(\frac{ \beta J \sin^2 \lambda }
{\cosh \lambda x-\cos \lambda}\Bigr).
\end{align}
The LHS possesses the singularities at $x = i, -3i$, 
while only the first term on the RHS possesses singularity at $x = i$. 
Consequently we have
$$
c_j = \oint_{y=i} \frac{b_1(y)}{\tau_1(y-i)}(y-i)^{j-1} \frac{dy}{2\pi i}
= \oint_{y=0} \frac{b_1(y+i)}{\tau_1(y)}y^{j-1} \frac{dy}{2\pi i}.
$$
The contour for the first integral is a small circle centered at $y=i$ 
and the same circle centered at $y=0$ for the second.
Similarly, by rewriting  (\ref{T1T1}) in the form
$\tau_1(x-i) 
= \frac{b_1(x)}{\tau_1(x+i)} + \frac{ \tau_2(x) }{\tau_1(x+i)} $, 
one finds
$$
{\bar c}_j = \oint_{y=0}\frac{b_1(y-i)}{\tau_1(y)}y^{j-1} \frac{dy}{2\pi i}.
$$
By substituting the expressions for $c_j, \bar{c}_j$ into (\ref{tau1}) 
and performing the summation over $j$ and $n$,
we arrive at the closed integral equation involving $\tau_1(x)$ only: 
\begin{align*}
\tau_1(x)= 2 &+ \frac{\lambda}{4\pi i}\Bigl( \oint_{y=0}
b_1(y+i) \coth \frac{\lambda}{2}(x-y-2i)
\frac{dy}{\tau_1(y)} \\
& + \oint_{y=0} b_1(y-i) \coth 
\frac{\lambda}{2}(x-y+2i) \frac{dy}{\tau_1(y)}\Bigr).
\end{align*}
Once the above equation is solved, 
the free energy is given by $f= -\frac{1}{\beta} \ln \tau_1(0)$.

It turned out the new equation
works efficiently to produce the high temperature expansion.
One assumes $\tau_1(x)$ in the form,
$$
\tau_1(x)=\exp \Bigl( \sum_{n=0}^{\infty} a_n(x)(\beta J)^n \Bigr).
$$
Then the coefficients $a_n(x)$ can be iteratively determined.

The simplified TBA equations are applied in many different contexts and
they successfully provide high temperature data of the models \cite{ShT,  TsSh}. 
The derivation of the simplified TBA equations requires less information on the 
analyticity.  
Therefore it is quite efficient 
when the analytic property is difficult to investigate.
The non-compact case is  such an example.
See \cite{Bel} for the applications to certain sectors of 
${\mathcal N}=4$ super Yang-Mills theory and 
\cite{BGOT} to thermodynamics of 
ladder compounds.

There is however a price to pay.
Any eigenvalue of  $T_j(x)$ satisfies the same equation after renormalization.
Therefore the equation itself can not select the right answer.
Rather, one has to know a priori the right goal to be achieved
and start from a sufficiently near point to the goal 
in numerical approaches.  
The convergence becomes also problematic 
in the low temperature regime and one
needs to apply, e.g. the Pad{\'e} approximation to improve the accuracy.

\subsection{Hybrid equations}

There is yet further approach to the finite size and the
finite temperature problems \cite{Kl3, JKS1, JKS2}.
It also makes use of a finite set of unknown functions and 
different types of integral equations 
from those derived in the previous sections.
Following \cite{DDV}, we refer to it as 
NLIE (NonLinear Integral Equation)\footnote{The equation first appeared
in the context of finite size problem in the XXZ model \cite{KBP}.
The simplest case is sometimes referred to as the DDV equation 
in the context of integrable field theories.}
just in order to distinguish it from the other nonlinear integral equations 
discussed hitherto.
It turns out that a hybridization of TBA and NLIE is possible \cite{Suz4}.
The hybrid approach is especially efficient in dealing with
thermodynamics of higher spin XXZ models as explained below.
  
We treat the integrable spin $s/2$ XXZ model whose Hamiltonian
${\mathcal H}$ is obtained from 
the fusion $R$ matrix in Section \ref{ss:vm} as
$$
{\mathcal H} = \sum_{i=1}^N h_{i,i+1},
\qquad
h_{i,i+1} \propto  \frac{d}{du} P R^{(k, k)}(q^u) |_{u=0},
$$
where $P$ is the transposition.
A simple generalization of the argument in Section \ref{ss:QTM} 
tells that the free energy per site is obtained 
from the largest value of QTM  $T_s(x=0)$
consisting of the $R$ matrix acting on $V_s \otimes V_s$.
As before we set 
$$q={\rm e}^{i \lambda}, \qquad \lambda = \frac{\pi}{p_0}$$ and
assume $s\le p_0 -1$. 
As in Section \ref{ss:QTM}, we introduce the auxiliary QTM $T_j(x)$.
This time, we prepare only a finitely many ones 
$\{T_j(x)\}_{j=1}^{\ell}$, where the integer $\ell$ is {\it arbitrary} 
as far as it is in the range
\begin{equation}\label{lcon}
s \le \ell \le 2p_0-s-2.
\end{equation}
With a suitable normalization, we have the T-system
\begin{align}
T_j(x+i) T_j(x-i) &= 
f_j(x)T_{j-1}(x) T_{j+1}(x) + g_j(x)
\quad(1\le j\le s-1),  \label{tptp} \\
g_j(x) &:= \prod_{m=0}^{\min(j,s)-1} 
\Phi (x-(s+j-2m)i)  \Phi (x+(s+j-2m)i),  \nonumber \\
\Phi (x)& :=\Bigl([x+(1+u)i]_{q^{\frac{1}{2}}}
[x-(1+u)i]_{q^{\frac{1}{2}}}\Bigr)^{M/2},  \nonumber
\end{align}
where $f_j(x)=\Phi(x)^{\delta_{js}}$.
This looks formally the same as (\ref{iraq}), although the meaning of $\ell$ 
is different here.
As usual we set $Y_j(x) = f_j(x)T_{j-1}(x) T_{j+1}(x)/g_j(x)$
and define its slight modification generalizing (\ref{yts1}) as 
$$
\tilde{Y}_j(x)=\frac{Y_j(x)}
{\bigl(\tanh\frac{\pi}{4}(x+(1+u)i) 
\tanh\frac{\pi}{4}(x-(1+u)i)  \bigr)^{\frac{M}{2} \delta_{js}}}.
$$
Then, the modified Y-system (\ref{j:modifiedY}) holds 
for $1\le j \le \ell-2$.

In addition we introduce the auxiliary functions
$\mathfrak{b}(x), \bar{\mathfrak{b}}(x)$.
They are defined by the combination of the terms
appearing in the dressed vacuum form of $T_{\ell}(x)$.
For general $n$, the dressed vacuum form reads
$T_n(x) = \sum_{m=1}^{n+1}\lambda^{(n)}_m(x)$, where
\begin{align*}
\lambda^{(n)}_m(x) &=\Phi^{(n)}_m(x)
\frac{Q(x+(n+1)i)Q(x-(n+1)i)}{Q(x+(2m-n-1)i) Q(x+(2m-n-3)i)},  \\
\Phi^{(n)}_m(x)&= 
\frac{\prod_{r=0}^{s-1} \Phi(x+(2m-n-s-1+2r)i)}
{\prod_{r=1}^{{\rm max}(s-n,0)} \Phi(x-(s+1-n-2r)i)}.
\end{align*}
Then the auxiliary functions are defined by
\begin{align*}
\mathfrak{b}(x) &=\frac{\lambda_1^{(\ell)}(x+i)
+ \cdots + \lambda_{\ell}^{(\ell)}(x+i) }
{\lambda^{(\ell)}_{\ell+1}(x+i)} \qquad  (-1 \le \im\, x  < 0 ),\\
\bar{\mathfrak{b}}(x) &
=\frac{\lambda_2^{(\ell)}(x-i)+ \cdots 
+ \lambda_{\ell+1}^{(\ell)}(x-i)}{\lambda^{(\ell)}_{1}(x-i)}
\qquad(0 < \im\, x \le 1),
\end{align*}
which are assumed to be analytic 
and nonzero in the strips indicated in the parentheses
for the largest eigenvalue of the QTM $T_s(x)$.
We also introduce
$$
\mathfrak{B}(x)=1+\mathfrak{b}(x), 
\qquad  \bar{\mathfrak{B}}(x)=1+\bar{\mathfrak{b}}(x)
$$
in each analytic strips. There are nice relations among them, e.g.
\begin{align*}
&Y_{\ell-1}(x-i)Y_{\ell-1}(x+i)=(1+Y_{\ell-2}(x))
\mathfrak{B}(x) \bar{\mathfrak{B}}(x), \\
&\mathfrak{b}(x) 
=\frac{\Phi(x)^{\delta_{\ell s}}}
{\prod_{r=1}^{s}\Phi(x+(\ell-s+2r) i)}
\frac{Q(x+(\ell+2)i)}{Q(x-\ell i) } T_{\ell-1}(x), \\
&\bar{\mathfrak{b}}(x)
=\frac{\Phi(x)^{\delta_{\ell s}}}
{\prod_{r=1}^{s}\Phi(x-(\ell-s+2r) i)}
\frac{Q(x-(\ell+2)i)}{Q(x+\ell i) } T_{\ell-1}(x), 
\end{align*}
which can be easily checked by using the definitions. 

By use of the analyticity assumptions, 
it is straightforward to derive the following equations
after the limit $M \rightarrow \infty$. 
\begin{align}
&\ln Y_j(x) =\delta_{js} D(x) +\int_{-\infty}^{\infty} K(x-x') 
\ln[(1+Y_{j+1}(x') ) (1+Y_{j-1}(x') )] \frac{dx'}{2\pi}, \nonumber \\
&\qquad \qquad  \qquad 1\le j \le \ell-2,  \label{hb1} \\
&\ln Y_{\ell-1} (x)=\delta_{\ell-1,s} D(x) +\int_{-\infty}^{\infty} K(x-x') 
\ln(1+Y_{\ell-2}(x'))\frac{dx'}{2\pi}  \nonumber  \\
&\phantom{cccc} +\int_{C_-}K(x-x')\ln\mathfrak{B}(x')\frac{dx'}{2\pi}  
+\int_{C^+}K(x-x')\ln\bar{\mathfrak{B}}(x')\frac{dx'}{2\pi},  
\label{hb2} \\
&\ln \mathfrak{b}(x)=\delta_{\ell s} D(x) +\int_{-\infty}^{\infty} K(x-x') 
\ln(1+Y_{\ell-1}(x'))\frac{dx'}{2\pi}    \nonumber \\
&\phantom{cccc}+\int_{C_-}F(x-x') \ln \mathfrak{B}(x') \frac{dx'}{2\pi}
- \int_{C_+}F(x-x'+2i)\ln \bar{\mathfrak{B}}(x') \frac{dx'}{2\pi} 
\qquad x \in C_- ,  \label{hb3}\\
&\ln\bar{\mathfrak{b}}(x) =\delta_{\ell s} D(x) 
+\int_{-\infty}^{\infty} K(x-x') 
\ln(1+Y_{\ell-1}(x'))\frac{dx'}{2\pi}   \nonumber  \\
&\phantom{cccc}+\int_{C_+} F(x-x')
\ln \bar{\mathfrak{B}}(x')\frac{dx'}{2\pi}
 - \int_{C_-}  F(x-x'-2i) \ln\mathfrak{B}(x')\frac{dx'}{2\pi}  \label{hb4}
\qquad x \in C_+ , 
\end{align}
where $C_{+} (C_{-})$ is a contour just  above (below) the real axis.
The kernel $K(x)$  is given in (\ref{j:ker})
and $F$ is related to the spinon $S$ matrix 
$$
F(x) = \int_{-\infty}^{\infty}
\frac{\sinh(p_0-\ell-1)k}{2\cosh k \sinh k(p_0-\ell)}{\rm e}^{-ikx} dk.
$$
The integration constants are found to be zero by comparing 
asymptotic values of the both sides and $D(x)$ is defined in (\ref{Ddef}).

Obviously (\ref{hb1}) is a reminiscence of 
the TBA type equation (\ref{yinte}),
while (\ref{hb3}) and  (\ref{hb4}) resemble 
NLIE were it not for the $\ln (1+Y_{\ell-1})$ term.  
In this sense we call the above equations hybrid.
They fix the values of $Y_s(x)$. 
The functional relations similar to (\ref{j:Tpexpr})
and the trick mentioned around (\ref{j:Tpexpr}) 
then yield the evaluation of the free energy per site.

\begin{remark}\label{re:las}
The number $\ell$ is arbitrary 
under the condition (\ref{lcon}).
This is quite different from ``genuine" TBA equations 
at special $\lambda$ \cite{TS,KSS},
where the number of equations is completely determined by $\lambda$.
When $\lambda\rightarrow 0$, we can formally put $\ell=\infty$,
which recovers the usual TBA equation in the rational limit
as argued in Section \ref{ss:QTM} for $s=1$.  
For $s=1$,  one can make $F(x)$ null by choosing $p_0=\ell+1$. 
The resulting system reproduces the known 
TBA equation corresponding to the 
level 2 restricted Y-system for $D_{\ell+1}$ for the XXZ chain.
See \cite[eq.(4.10)-eq.(4.12)]{KSS} for example.
For arbitrary $s \in \Z_{\ge 1}$, 
the choice $\ell=s$ recovers the result in \cite{Suz4}.
\end{remark}
  
The above equations are numerically stable 
and yield a quick convergence to the unique solution.
They are efficient in the analysis of the low temperature regime.
It is also known that with a suitable modification, 
one can derive the equations for excited states.
We again have to pay the price.
The systematic algorithm to construct the auxiliary functions is  
still lacking except for $\mathfrak{g}=A_1$ discussed here.  
This remains as an interesting future problem.

\section*{Acknowledgments}
The authors thank
Murray T. Batchelor,
Nikolay Gromov, 
Rei Inoue, 
Vladimir Kazakov, 
Shota Komatsu,
Robert Tateo,
Zengo Tsuboi and 
people from particle theory group at University of Tokyo Komaba 
for communications.
This work is supported by Grants-in-Aid for 
Scientific Research No.~21540209 and No.~20540370 
from JSPS.

\end{document}